\documentclass[prx,amsmath,amssymb,floatfix,twocolumn,amsmath,superscriptaddress,nofootinbib,tighten,
longbibliography]{revtex4-2}

\usepackage{comment}
\usepackage{graphicx}
\usepackage{xcolor}
 \usepackage[normalem]{ulem}

\usepackage{subcaption}
\usepackage{dcolumn}
\usepackage{bm}
\usepackage{cancel}

\usepackage[utf8]{inputenc}
\usepackage[T1]{fontenc}
\usepackage[english]{babel} 
\usepackage{mathrsfs}
\usepackage{graphicx}
\usepackage{amssymb}
\usepackage{amsbsy}
\usepackage{latexsym}
\usepackage{mathtools}
\usepackage{titlesec}
\usepackage{amsmath}
\usepackage{amsthm}
\usepackage[utf8]{inputenc}
\usepackage{lipsum}
\usepackage{booktabs}
\usepackage{tabularx}

\usepackage{ragged2e}
\usepackage{caption}
\DeclareCaptionJustification{justified}{\justifying}
\usepackage{chngcntr}
\usepackage{amsthm}

\usepackage{comment}
\makeatletter
\renewcommand*\env@matrix[1][*\c@MaxMatrixCols c]{%
 \hskip -\arraycolsep
 \let\@ifnextchar\new@ifnextchar
 \array{#1}}
\makeatother
\makeatletter
\renewcommand{\p@subsection}{}
\renewcommand{\p@subsubsection}{}
\makeatother

\def\b{\beta}

\renewcommand{\vec}[1]{\ensuremath{\mathbf{#1}}}
\def\dj{d\kern-0.4em\char"16\kern-0.1em}
\def \Dj {\mbox{\raise0.3ex\hbox{-}\kern-0.4em D}}

\usepackage{hyperref}

\renewcommand{\thesubsection}{\thesection.\Alph{subsection}}
\usepackage{cleveref}

\begin{document}


\title{Non-Hermitian Crystalline {Braid} Topology from Hermitian Projection:\\ A Zero-Mode Resonance Mechanism}

\author{Stefan {\DJ}or{\dj}evi{\'c}} 
\affiliation{Faculty of Physics, University of Belgrade, Studentski Trg 12-16, 11000 Belgrade, Serbia}
\author {Vladimir Juri\v{c}i\'c}
\affiliation{Departamento de F\'{i}sica, Universidad T\'{e}cnica Federico Santa Mar\'{i}a, Casilla 110, Valparaíso, Chile}

\selectlanguage{english}

\begin{abstract}
Non-Hermitian topological phases are typically engineered through gain and loss, nonreciprocity, or interaction with an environment. Here we show that they can instead emerge purely by
projecting a fully Hermitian, topologically trivial parent lattice onto
an embedded subsystem. The mechanism is general: when a zero mode of
the eliminated degrees of freedom couples to the retained subsystem,
the embedding self-energy develops a pole, the zero-frequency
description becomes singular, and topology is carried by the
finite-frequency projected Green's function. We realize the mechanism exactly in a trivial
nearest-neighbor square lattice with an embedded one-dimensional
zig-zag brane. In the periodic transverse geometry, the parity of the
eliminated complement selects the outcome: even sectors reduce to a
regular Schur complement and yield conventional SSH-type descendants,
whereas odd sectors host a sublattice-imbalance zero mode and follow
the resonant route. There, the complex bands braid through isolated
finite-frequency exceptional points (EPs), while a parity symmetry inherited
from the embedding, together with
\(\mathrm{TRS}^{\dagger}\), induces conjugated
pseudo-Hermiticity and quantizes the complex Berry phase. The stable bulk invariant of the nondegenerate phases is this quantized complex Berry phase; adjacent sectors are separated by parity-paired exceptional points whose half-integer vorticities encode the local exchange of complex-energy strands.The absence of the non-Hermitian skin effect ensures that the invariant
is defined directly on the ordinary Brillouin zone. A topolectrical implementation of the projected response predicts
momentum-resolved transmission minima at the exceptional-point
transition frequencies together with a characteristic low-frequency
resonant admittance, providing an experimentally testable signature of
the mechanism.
\end{abstract}

\maketitle
\tableofcontents
\selectlanguage{english}
\section{Introduction}

Non-Hermitian (NH) band theory has expanded the notion of topological
matter beyond closed quantum systems, revealing spectral structures,
symmetry classes, exceptional degeneracies, and boundary phenomena with
no direct Hermitian
counterpart~\cite{ElGanainy2018_NatPhys,Bergholtz2021_RMP}. In nearly
all realizations, non-Hermiticity is introduced \emph{microscopically}:
one engineers asymmetric couplings, gain and loss, radiative leakage,
measurement backaction, or explicit environmental
coupling~\cite{Bender1998_PRL,Rotter2009_JPhysA,ElGanainy2018_NatPhys}.
This viewpoint underlies the classification of point-gap and line-gap NH
phases, exceptional topology, and the NH bulk--boundary
correspondence~\cite{Gong2018_PRX,Shen2018_PRL,Kawabata2019_PRX,Yao2018_PRL,Kunst2018_PRL,non-Bloch_Band_theory,Class_cristaline_non_herm},
which itself builds on the classification of symmetry- and
crystalline-protected topological phases in the Hermitian
realm~\cite{Fu2011_PRL,Slager2013_NatPhys,ShiozakiSato2014_PRB,Shiozaki2016_PRB,Kruthoff2017_PRX,Bradlyn2017_Nature}.
A complementary, intrinsically NH perspective treats the
loops traced by complex eigenvalues---which we call complex-energy
strands---as braids or links, supplying braid- and knot-theoretic
spectral invariants with no Hermitian
analogue~\cite{Carlstrom2018_PRA,Yang2019_PRB,Wojcik2022_PRB,LiMong2019_arXiv}.
Here topology is encoded in the winding and exchange of
nondegenerate
strands as momentum traverses the Brillouin zone, and transitions occur
when strands meet at exceptional points (EPs).

In this paper we ask whether NH crystalline braid topology can arise from \emph{projection itself}, in an embedded subsystem of an otherwise fully Hermitian parent
lattice, without microscopic gain, loss, nonreciprocity, or explicitly
NH coupling. We answer in the affirmative. The mechanism, which is the
central result of the paper, is a zero mode of the eliminated complement
that, on coupling to the retained subsystem, renders the embedding
self-energy singular and forces topology onto the finite-frequency
projected Green's function.

\begin{figure}[t!]
    \centering
    \IfFileExists{brana.png}{%
        \includegraphics[width=0.4\textwidth]{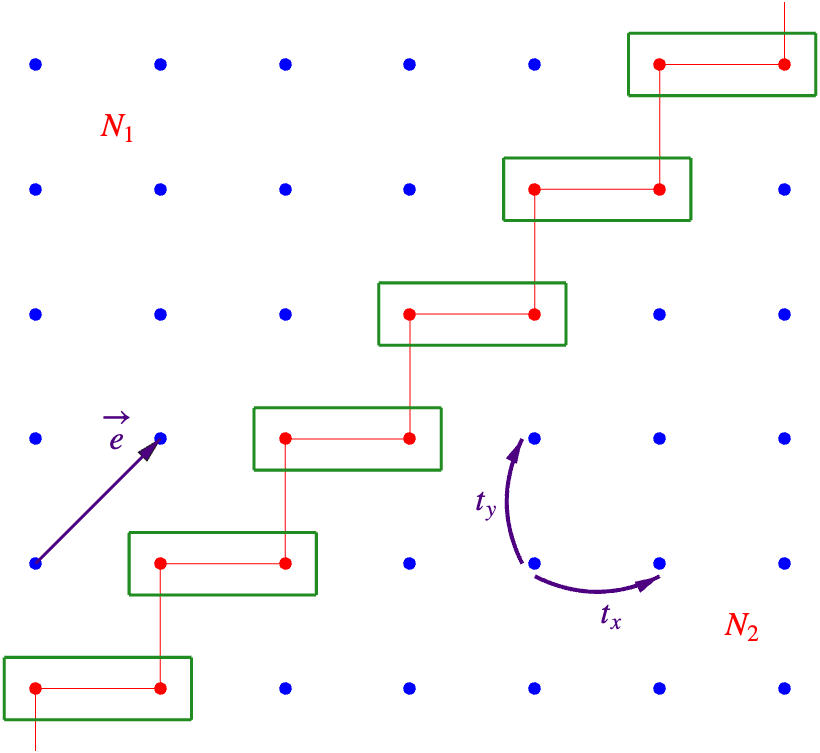}%
    }{%
        \fbox{\parbox[c][0.26\textheight][c]{0.36\textwidth}{%
        \centering Placeholder for \texttt{brana.png}}}%
    }
    \caption{Two-dimensional parent lattice with an embedded one-dimensional
``zig-zag'' brane (red); the remaining sites form the complement, and the
green rectangle marks the brane unit cell. Purple arrows show the
translation vector $\vec{e}$ [Eq.~(\ref{trans_op_m})] and the hopping
parameters $t_x$, $t_y$. The numbers $N_1$ and $N_2$ count the complement
chains to the left and right of the brane.}
    \label{brane}
\end{figure}

The setting is natural whenever a subset of degrees of freedom is
measured or controlled inside a larger structure. Surface states are
described by embedding self-energies~\cite{LopezSancho1985_JPhysF},
impurity problems by
\(T\)-matrices~\cite{Economou2006_GreenFunctions}, low-energy bands by
downfolding~\cite{Lowdin1951_JCP}, and subsystem dynamics by
projection~\cite{Feshbach1958_AnnPhys,Feshbach1962_AnnPhys,Rotter2009_JPhysA}.
This perspective connects directly to embedded topological matter, in
which lower-dimensional topological subsystems are placed inside
otherwise trivial higher-dimensional
environments~\cite{Tuegel2019_PRB,Velury2022_PRB}. Its most direct
foundation is the Schur-complement construction of projected brane
Hamiltonians~\cite{PTB}, later extended to crystalline, superconducting,
and higher-order topology in Hermitian
settings~\cite{Panigrahi2026_PRB,Tyner2024_SciRep,Tyner2025_arXiv}.

Our construction shares its starting point---Schur-complement
projection of a Hermitian parent onto an embedded brane---with the projected
topological branes of
Refs.~\cite{PTB,Panigrahi2026_PRB,Tyner2024_SciRep,Tyner2025_arXiv}, where the
projection is taken in the static $\omega \to 0$ limit, the self-energy stays
finite, and the brane inherits a Hermitian effective Hamiltonian; our even-$N$
sectors reproduce this regular route. The new ingredient is the resonant route: a
complement zero mode coupling to the brane makes $\Sigma(\omega)$ develop a
$1/\omega$ pole, so no static effective Hamiltonian exists and topology resides
irreducibly in the finite-frequency projected Green's function. Projection then yields finite-frequency EPs and a CPH-quantized complex Berry phase from a Hermitian, topologically trivial parent.

Our starting point is the observation that projection generically does
\emph{not} yield a static Hamiltonian. Integrating out degrees of
freedom from a Hermitian parent produces a \emph{frequency-dependent}
embedding self-energy, so that topology can reside in the full projected
Green's function rather than in any static effective
Hamiltonian~\cite{Gurarie2011_PRB,WangQiZhang2010_PRL}. This opens two
qualitatively distinct routes, and distinguishing them is the conceptual
core of our work. \emph{Regular} projection has a smooth low-frequency
limit: the self-energy is finite as \(\omega\!\to\!0\), the projected
problem reduces to a static Hermitian Schur complement, and one recovers
conventional SSH-type topological
descendants~\cite{Kunst2018_PRL,SSH,SSH_non_herm}. \emph{Resonant}
projection occurs when a zero mode of the eliminated complement couples
to the retained
subsystem~\cite{Sutherland-PRB1986,Inui-PRB1994}: the embedding
self-energy develops a pole, the zero-frequency limit becomes singular,
and no static effective Hamiltonian exists. The topology is then
necessarily a property of the finite-frequency projected Green's
function. We call this singular route \emph{zero-mode-resonant
projection}, and it is the route that produces NH crystalline braid topology
from a trivial Hermitian parent, realized here in the zig-zag geometry
of Fig.~\ref{brane}. Non-Hermiticity thus emerges from
the singular analytic structure of the projected Green's function,
generated by the zero-energy state in the eliminated complement.

Throughout, \(z\in\mathbb C\) denotes the spectral parameter. We use
\(z=i\omega\), with \(\omega>0\), for the imaginary-frequency slices
used in the topological analysis, and \(z=\Omega+i0^{+}\) for the
retarded response at real frequency \(\Omega\). At fixed \(i\omega\),
the projected inverse Green's function defines a NH response
problem whose CPH symmetry quantizes the Berry phase when the spectrum
is nondegenerate. After analytic continuation, its eigenvalues describe
the inverse retarded response; their zeros give resonance poles, while
exceptional crossings and spectral reconnections appear as real-frequency
response features.

In a topolectrical realization, eliminating internal nodes is Kron
reduction, the Schur-complement operation defining the projected
kernel~\cite{Dorfler2013_Kron}. The circuit implements the projected
theory when its reduced Laplacian obeys
\begin{equation}
J_{\rm eff}(k,\Omega)=\mathcal N(\Omega)
H_{\rm PTB}[k,z(\Omega)] ,
\end{equation}
with \(z(\Omega)\) and the nonzero scale \(\mathcal N(\Omega)\) fixed by
the unreduced network, as detailed in App.~\ref{app:circuit}.

Together with this response interpretation, two additional features make
the resulting phase well defined and observable. First, the projected model is free of the NH skin effect (NHSE): the generalized Brillouin zone coincides with the ordinary one, so the CPH complex Berry phase is a genuine Bloch-bulk invariant. Spatial
parity inherited from the Hermitian parent, together with
\(\mathrm{TRS}^{\dagger}\), induces conjugated pseudo-Hermiticity
(CPH), which quantizes the complex Berry phase, while parity
constrains the Abelian two-band spectral braid.

The resonant sector lies in class \(\mathrm{AI}^{\dagger}\) of the \(38\)-fold NH classification~\cite{Kawabata2019_PRX}, which is topologically trivial for one-dimensional line-gapped phases and therefore does not account for the finite-frequency crystalline Berry invariant realized here. The protection is instead inherited from the embedding geometry: parity relates the projected amplitudes at opposite momenta and organizes the exceptional-point pairs, while parity together with TRS\(^{\dagger}\) induces CPH and quantizes the biorthogonal complex Berry phase. An isolated chain does not generically enforce these constraints, rendering them accidental rather than robust. Crystalline constraints on NH bands beyond the internal classification have been explored only in restricted settings~\cite{Class_cristaline_non_herm, ShiozakiSato2014_PRB,Cristaline_top_ins_class_2}; here, they emerge automatically from projection, producing a quantized finite-frequency Berry invariant and braiding via parity-constrained exceptional spectral reconnections.

Having identified resonance and crystalline protection as the two
ingredients, we establish both routes---regular and resonant---in an
exactly solvable model: a topologically trivial nearest-neighbor square
lattice with an embedded one-dimensional ``zig-zag'' brane
(Fig.~\ref{brane}). The route is selected by the parity of the number of eliminated complement sites $N$. For periodic boundary conditions in the direction transverse to the brane, even-$N$ sectors are zero-mode free and realize regular SSH-type winding phases, whereas odd-$N$ sectors are sublattice imbalanced, host a complement zero mode, and follow the resonant route. The odd-\(N\) periodic sector is the central case: the zero mode produces
the singular self-energy, while embedding parity and TRS$^{\dagger}$
generate the crystalline constraints underlying both the quantized complex
Berry phase and the exceptional spectral braid. Isolated finite-frequency
EPs reconnect the complex bands and carry local braiding
charge. This central braiding picture---two complex
bands that reconnect through parity-paired EPs as the
frequency slice is lowered---is illustrated in
Fig.~\ref{fig:3D_spectrum}.
The same finite-frequency transitions carry direct response signatures:
in topolectrical realizations they appear as sharp features in
experimentally accessible response functions at the predicted critical
drive frequencies. In the resonant sector, the CPH complex Berry phase labels the nondegenerate bulk sectors, while parity-related EPs mediate the transitions between them and carry local braid charge. The boundary response remains termination-sensitive. This route is also experimentally timely:
embedded topology in a trivial bulk has recently been observed
acoustically via projective crystal symmetry~\cite{Teo2025PRL}, and our
construction generates such embedded topology dynamically, through the
projected finite-frequency Green's function.

\begin{table*}[t]
\centering
\begingroup
\small
\renewcommand{\arraystretch}{1.25}
\setlength{\tabcolsep}{2.5pt}
\newcommand{\tc}[2]{%
    \parbox[t]{#1}{\raggedright\strut #2\strut}%
}
\begin{tabular}{|c|c|c|c|c|}
\hline
\tc{0.13\textwidth}{Boundary/parity} &
\tc{0.10\textwidth}{Symmetry class} &
\tc{0.22\textwidth}{Parity/CPH constraint} &
\tc{0.30\textwidth}{Bulk invariant} &
\tc{0.17\textwidth}{Boundary response} \\
\hline\hline
\tc{0.13\textwidth}{OBC, \(N_1,N_2\) even} &
\tc{0.10\textwidth}{BDI\(^{\dagger}\)} &
\tc{0.22\textwidth}{termination dependent;
present for symmetric transverse terminations} &
\tc{0.30\textwidth}{SSH winding; real Berry
phase for \(N_1=N_2\)} &
\tc{0.17\textwidth}{NHSE-free; SSH type,
termination dependent} \\
\hline
\tc{0.13\textwidth}{OBC, at least one \(N_i\)
odd} &
\tc{0.10\textwidth}{BDI\(^{\dagger}\)} &
\tc{0.22\textwidth}{generically absent; present
only for symmetric transverse terminations} &
\tc{0.30\textwidth}{SSH/global Berry phase
(\(\mathcal{PT}_{+}\)-unbroken/broken); no
periodic braid count} &
\tc{0.17\textwidth}{NHSE-free; termination and
frequency dependent} \\
\hline
\tc{0.13\textwidth}{PBC, even \(N\)} &
\tc{0.10\textwidth}{BDI\(^{\dagger}\)} &
\tc{0.22\textwidth}{embedding parity, together with
TRS\(^{\dagger}\), induces CPH} &
\tc{0.30\textwidth}{SSH-type integer invariant} &
\tc{0.17\textwidth}{NHSE-free; SSH type} \\
\hline
\tc{0.13\textwidth}{PBC, odd \(N\)} &
\tc{0.10\textwidth}{AI\(^{\dagger}\)} &
\tc{0.22\textwidth}{embedding parity relates opposite momenta, constrains EP braid charges, and, with TRS\(^{\dagger}\), induces CPH.
} &
\tc{0.30\textwidth}{CPH-quantized complex Berry phase; local Abelian braid charge at exceptional-point transitions} &
\tc{0.17\textwidth}{NHSE-free;
termination-sensitive response} \\
\hline
\end{tabular}
\endgroup
\caption{Projection-induced symmetry and topology of the brane.
Depending on transverse boundary conditions and complement parity,
the projected brane realizes either regular SSH-type phases or a
singular zero-mode-resonant sector. The periodic odd-$N$ sector is
the central case: its internal class is AI$^{\dagger}$~\cite{Kawabata2019_PRX},
yet embedding parity together with TRS$^{\dagger}$ induces conjugated
pseudo-Hermiticity (CPH), which quantizes the complex Berry phase; parity
additionally relates opposite momenta, constraining the
exceptional-point (EP) reconnections and their local half-integer
braid charges. OBC~(PBC) refers to an open~(periodic) parent system
in the direction transverse to the brane. Under OBC, $N_1$ and $N_2$
are the numbers of complement chains on the two sides of the brane;
under PBC, $N$ is the total number of complement sites.}
\label{tab:projection_symmetry}
\label{tab:classification_summary}
\end{table*}

\subsection{Key results}

The results are summarized in Table~\ref{tab:classification_summary}. As defined above, the two routes are selected by the
parity of the eliminated complement: a zero-mode-free complement gives
the regular SSH-type descendant, while a complement zero mode coupling to
the brane gives the singular resonant sector in which the topology
resides on the finite-frequency projected Green's function. The crystalline Berry phase and the associated exceptional spectral
reconnections arise at finite frequency, where the projected response is
intrinsically NH.

Turning this mechanism into a quantized invariant requires four ingredients:
a complement zero mode coupled to the brane, a momentum-dependent pole
residue, an embedding symmetry inherited through projection, and a mismatch
between the internal NH classification and the resulting
finite-frequency topology. The periodic odd-\(N\) sector realizes all four
requirements. Although its internal class is \(\mathrm{AI}^{\dagger}\), embedding parity and TRS\(^{\dagger}\) generate the crystalline constraints that quantize the biorthogonal complex Berry phase. EPs mediate transitions between adjacent Berry sectors and carry local Abelian braid charge.

Two features establish the complex Berry invariant as a bulk observable.
First, the projected model is NH yet free of the skin effect;
the generalized Brillouin zone therefore collapses onto the ordinary
one, so the Berry phase is a true Bloch bulk quantity. Second, its
boundary manifestation is termination-sensitive rather than SSH-like.
The EPs mediate braid transitions between adjacent
crystalline Berry sectors. Finally, the finite-frequency transitions are experimentally accessible:
a minimal topolectrical realization, in which eliminating internal
circuit nodes is itself the Schur complement, predicts a discrete set of
transmission-zero and admittance features at the critical drive
frequencies, providing a concrete signature of the finite-frequency crystalline
topology and its exceptional-point-mediated braid transitions.

\subsection{Organization}

The paper is organized as follows. In
Sec.~\ref{sec:projection_framework}, we formulate the projected
Green's-function framework and discuss inherited topology. In
Sec.~\ref{sec:zero_mode_general}, we identify zero-mode-resonant
projection as the singular finite-frequency crystalline route, distinct
from the regular Schur-complement route. Section~\ref{sec:min-model}
introduces the minimal zig-zag embedding and shows how the parity of the
eliminated complement controls the projected structure.
Section~\ref{sec:spectral_structure} derives the projected Hamiltonians
for open and periodic boundary conditions and analyzes their spectra.
Section~\ref{sec:Main_top_inv} determines the topological invariants:
even periodic sectors realize SSH-type phases, whereas the odd periodic 
sector realizes a finite-frequency NH crystalline Berry phase
protected by embedding-induced CPH, with parity-constrained exceptional
spectral braiding at its transitions. Section~\ref{sec:BBC} discusses the
bulk invariant and boundary response, including the absence of the
NHSE and the termination-sensitive boundary
response of the odd-\(N\) crystalline sector.
Section~\ref{sec:discussion} summarizes the physical implications,
experimental routes, and broader scope of the projection mechanism.
Technical details and supporting figures are collected in the
Appendices.
\section{General projection framework and inherited topology}
\label{sec:projection_framework}

We start from a fully Hermitian parent lattice containing an embedded
lower-dimensional subsystem, which we call the brane. Projecting onto
the brane converts this closed Hermitian problem into a dynamical
effective theory: the eliminated degrees of freedom re-enter through a
frequency-dependent embedding self-energy. This self-energy is the
source of the effective non-Hermiticity. It arises entirely from virtual
propagation through the Hermitian complement, so the projected brane
inherits an open-subsystem structure from projection alone. The natural
object is therefore the finite-frequency Hamiltonian associated with
the projected Green's function.

We decompose the parent Hilbert space as
$\mathcal{H}=\mathcal{H}_{\rm brane}\oplus\mathcal{H}_{\rm comp}$, so
that the Hermitian parent Hamiltonian takes the block form
\begin{equation}
    H =
    \begin{pmatrix}
    H_{11} & H_{12}\\
    H_{21} & H_{22}
    \end{pmatrix},
    \label{eq:block_parent}
\end{equation}
where $H_{11}$ acts on the brane, $H_{22}$ on the complement, and
$H_{12},H_{21}$ couple the two sectors. Projecting the resolvent
$G(i\omega)=(i\omega-H)^{-1}$ onto the brane gives
\begin{equation}
    G_{\rm brane}(i\omega)
    =
    \bigl[i\omega-H_{11}-\Sigma(i\omega)\bigr]^{-1},
    \label{eq:Gb_general}
\end{equation}
with the exact embedding self-energy
\begin{equation}
    \Sigma(i\omega)
    =
    H_{12}(i\omega-H_{22})^{-1}H_{21}.
    \label{eq:self_energy_general}
\end{equation}
Thus the complement is retained exactly through $\Sigma(i\omega)$,
which resums all virtual processes from the brane into the surrounding
lattice and back.

We define the projected brane Hamiltonian as the inverse projected
Green's function,
\begin{align}
    H_{\rm PTB}(k,i\omega)
    &=
    -G_{\rm brane}^{-1}(k,i\omega) \nonumber\\
    &=
    H_{11}-i\omega
    -H_{12}(H_{22}-i\omega)^{-1}H_{21}.
    \label{eq:PTB_frequency}
\end{align}

Projection thus yields a finite-frequency inverse-response kernel on the
brane that is generally NH, while the full parent lattice
remains Hermitian. On the imaginary-frequency axis, \(z=i\omega\), this
kernel defines the response problem used for the topological analysis.
Its eigenvalues characterize the imaginary-frequency projected response.
After analytic continuation, \(z=\Omega+i0^{+}\),
physical resonances are determined by
\[
\det G_{\rm brane}^{-1}(k,\Omega+i0^{+})=0 .
\]
The imaginary-frequency spectrum therefore diagnoses the projected
topological response, while the retarded response encodes resonance
energies and lifetimes.

At each fixed nonzero frequency, the map
\(k\mapsto H_{\rm PTB}(k,i\omega)\) defines an NH band
problem with its own biorthogonal geometry, complex Berry phase, and
nondegenerate complex-energy strands. The topology studied here is
therefore a property of the finite-frequency projected Green's
function. Varying \(\omega\) moves between frequency slices; transitions
occur when the spectral gap closes at EPs, where the
complex bands coalesce and reconnect.

This framework naturally separates two projection regimes. Regular
projection has a smooth low-frequency limit and recovers the static
Schur-complement projected-brane construction of Hermitian branes 
~\cite{PTB}. Resonant projection instead occurs when
a zero mode of the eliminated complement couples to the brane, producing
a pole in the embedding self-energy. The zero-frequency limit then
becomes singular, and topology is carried by the finite-frequency
projected Green's function.

\subsection{Inherited topology}
\label{subsec:inherited_top}

The distinction between regular and resonant projection is already
visible in Eq.~\eqref{eq:PTB_frequency}. At large frequency,
\begin{equation}
    \bigl(H_{22}-i\omega\bigr)^{-1}
    \sim
    -\frac{1}{i\omega},
\end{equation}
so the embedding self-energy vanishes as \(1/\omega\). The projected
Hamiltonian then reduces, up to the trivial shift \(-i\omega\), to the
isolated brane Hamiltonian \(H_{11}\). This limit provides a common
reference point for both regular and resonant sectors.

The difference appears at low frequency. If the eliminated complement
has no zero mode, \(H_{22}^{-1}\) is well defined and the limit
\(\omega\to0\) yields the regular Hermitian Schur complement
\begin{equation}
    H_{\rm eff}(k)
    =
    H_{11}(k)
    -
    H_{12}(k)H_{22}^{-1}(k)H_{21}(k).
    \label{eq:regular_schur_complement}
\end{equation}
The projected problem then admits a static effective-Hamiltonian
description. Its topology may still be nontrivial, as in SSH-type descendants, but it is regular: classified by the zero-frequency Schur complement, with changes of invariant requiring a band-gap closing as in the standard SSH model.

By contrast, if the complement contains a zero mode coupled to the
brane, the low-frequency Schur complement becomes singular. The projected Hamiltonian then has no regular \(\omega\to0\)
representative. The system enters a singular projection regime in which
the finite-frequency projected Green's function becomes the natural
topological object: the zero-mode pole makes the zero-frequency limit
singular, so the topology cannot be reduced
to a regular static Schur complement.

\section{Zero-mode-resonant projection mechanism}
\label{sec:zero_mode_general}

{
We now formulate the resonant route in general terms, showing how it
produces a singular finite-frequency crystalline {braid} phase from a Hermitian
parent, beyond the regular Schur-complement route. The essential
ingredient is a zero mode of the eliminated complement.} When this zero mode couples to the retained
brane, it produces a resonant pole in the embedding self-energy. The
projected Hamiltonian then becomes singular in the zero-frequency limit,
while at any fixed nonzero frequency it defines an NH band problem. If the embedding also carries a spatial
symmetry, this finite-frequency problem can acquire a quantized
crystalline invariant. {This is the zero-mode-resonant projection mechanism: the singular
finite-frequency  route, distinct from regular SSH-type
projection.}

The mechanism follows directly from the exact self-energy inherited by
the brane (see App.~\ref{app:general_framework} for details),
\begin{equation}
    \Sigma(k,z)
    =
    H_{12}(k)\left[z-H_{22}(k)\right]^{-1}H_{21}(k),
    \label{eq:general_self_energy_main}
\end{equation}
where $z$ denotes the complex frequency. Suppose that the complement
Hamiltonian $H_{22}(k)$ contains a zero-mode subspace, associated with the projector
$P_0(k)$. Its resolvent then separates into a singular zero-mode
contribution and a regular part,
\begin{equation}
    \left[z-H_{22}(k)\right]^{-1}
    =
    \frac{P_0(k)}{z}
    +
    R_{\rm reg}(k,z),
\end{equation}
with $R_{\rm reg}(k,z)$ analytic at $z=0$. Consequently, the projected
self-energy acquires the pole structure
\begin{equation}
    \Sigma(k,z)
    =
    \frac{\Gamma(k)}{z}
    +
    \Sigma_{\rm reg}(k,z),\label{eq:general_resolvent_pole_main}
\end{equation}
where
\begin{equation}
    \Gamma(k) = H_{12}(k)\,P_0(k)\,H_{21}(k)
    \label{eq:general_resonance_matrix_main}
\end{equation}
is the resonance matrix. This matrix measures the amplitude for a brane
state to virtually enter the zero-mode sector of the complement and
return. 
With the convention
\(\Sigma(k,z)=H_{12}(k)[z-H_{22}(k)]^{-1}H_{21}(k)\), the projected
Hamiltonian is
\begin{equation}
    H_{\rm PTB}(k,z)=H_{11}(k)-z+\Sigma(k,z).
    \label{eq:HPTB_z_convention}
\end{equation}
Thus, when \(\Gamma(k)\neq0\),
\begin{equation}
    H_{\rm PTB}(k,z)
    =H_{11}(k)-z+
    \frac{\Gamma(k)}{z}+\Sigma_{\rm reg}(k,z),
\end{equation}
which has a singular \(z\to0\) limit. Its topology is therefore
naturally formulated at fixed nonzero frequency, as a property of the
projected Green's function rather than of a regular zero-frequency
Hamiltonian.

At imaginary frequency, \(z=i\omega\), the pole contributes
\begin{equation}
    \frac{\Gamma(k)}{i\omega}=-\frac{i}{\omega}\Gamma(k)
\end{equation}
to \(H_{\rm PTB}\). For a Hermitian parent,
\(H_{21}=H_{12}^{\dagger}\), and hence
\begin{equation}
    \Gamma(k)=H_{12}(k)P_0(k)H_{12}^{\dagger}(k)
\end{equation}
is positive semidefinite.
The resonant contribution is therefore a negative-semidefinite
anti-Hermitian term on the imaginary-frequency slice, whose magnitude
grows as \(1/\omega\) upon approaching zero frequency. This structure
characterizes the projected inverse response; its interpretation in
terms of damping or resonance widths requires analytic
continuation to the retarded problem.

{The zero-mode pole supplies the resonant finite-frequency
non-Hermiticity, while the spatial symmetry inherited from the embedding
organizes this resonant structure into a quantized crystalline invariant.} Let the parent lattice possess an
order-two spatial parity $\mathcal{P}$ that preserves the
brane--complement decomposition, inducing parity operators $P^{(b)}$
and $P^{(c)}$ on the two sectors, i.e., $\mathcal{P}=P^{(b)}\oplus P^{(c)}$. Since $\mathcal{P}$ maps
$k\to -k$, the zero-mode projector transforms as
$P^{(c)}P_0(k)=P_0(-k)P^{(c)}$. The resonance matrix then obeys
\begin{equation}
P^{(b)}\,\Gamma(k)\,\bigl[P^{(b)}\bigr]^{-1}
    =
    \Gamma(-k),
    \label{eq:general_parity_constraint_main}
\end{equation}
as derived in Appendix~\ref{subsec:symmetry-inheritance-embedding}. 
This constraint is inherited from the embedding geometry. It acts
on the finite-frequency projected Green's function and can quantize its
complex Berry phase, including cases in which the internal NH symmetry class has no corresponding one-dimensional invariant within the \(38\)-fold classification.

A sufficient mechanism for generating a crystalline invariant
beyond the internal 38-fold scheme is obtained
when four conditions are met. First, the projection is
resonant: the complement contains a zero mode that
couples to the brane, so that the self-energy develops
a pole. Second, the pole residue has nontrivial momentum
dependence. Third, the embedding supplies a spatial
symmetry, here an order-two parity, which constrains
the resonance matrix. Fourth, there is a mismatch between
the internal NH classification and the resulting
finite-frequency crystalline topology. The zero-mode pole
makes the zero-frequency limit singular, while embedding
parity together with TRS$^\dagger$ induces CPH and quantizes
the biorthogonal complex Berry phase. EPs mediate
transitions between the resulting Berry sectors and carry
local Abelian braid charge. The model-independent formulation
of these conditions is given in Appendix~\ref{app:general_framework}.

We now realize this mechanism exactly in the zig-zag model. There, the
parity of the eliminated complement selects between two qualitatively
different projections. Even complements produce regular SSH-type
projected Hamiltonians. Odd complements contain a sublattice-imbalance zero mode, generate the resonant pole described above, and, for periodic boundary conditions, yield a finite-frequency crystalline Berry phase protected by embedding-induced CPH, with parity-constrained exceptional spectral reconnections.

In the zig-zag model, the gap structure of the resonant sector
confirms this general mechanism: the zero-mode pole makes the
\(\omega=0\) limit singular, the complex strands remain nondegenerate between exceptional-point transition lines and touch only at the transitions, where the CPH Berry invariant can change and the degeneracy carries local braid charge.

\section{Minimal model}
\label{sec:min-model}

{
We now introduce the minimal Hermitian parent in which the projection
mechanism can be realized exactly. The starting point is a
topologically trivial two-dimensional anisotropic nearest-neighbor
lattice,
}
\begin{equation}
    \begin{aligned}
        H =&
        -t_x\sum_{n,m}
        \bigl[|n,m\rangle\langle n+1,m|+\mathrm{h.c.}\bigr]\\
        &-t_y\sum_{n,m}
        \bigl[|n,m\rangle\langle n,m+1|+\mathrm{h.c.}\bigr],
    \end{aligned}
\end{equation}
{
defined on the site basis
\(\{|n,m\rangle\,|\,n,m\in\mathbb{Z}\}\). Into this otherwise ordinary
lattice we embed the zig-zag brane shown in Fig.~\ref{brane}, whose
unit cell contains the two sites \((n,n)\) and \((n,n+1)\). This
geometry is minimal in the sense required by the mechanism: it retains
a one-dimensional translation symmetry along the brane, produces a
parity-sensitive complement, and derives the resonant finite-frequency
self-energy from the same two-dimensional Hermitian parent. In the
periodic transverse geometry, the embedding also preserves the spatial
parity that quantizes the complex Berry phase in the resonant sector.
For open transverse terminations, this parity need not be preserved,
which leads instead to the termination-sensitive boundary response
discussed below.

The brane is invariant under translations generated by
\(\vec e=\vec e_1+\vec e_2\). It is therefore natural to work in the
symmetry-adapted basis \(\{|n,n+m\rangle\}\) and introduce the
translation operators
}
\begin{equation}
    T_m=\sum_n |n,n+m\rangle\langle n+1,n+m+1|,
    \label{trans_op_m}
\end{equation}
{
which commute with \(H\). The problem then decomposes into independent
momentum sectors labeled by \(k\). In each sector, the Hamiltonian
retains the brane--complement block structure of
Eq.~\eqref{eq:block_parent}, and the projected brane Hamiltonian is
obtained exactly as the momentum-resolved Schur complement,
}
\begin{equation}
    \begin{aligned}
        H_{\rm PTB}(k,i\omega)
        =
        &H_{11}(k)-i\omega
        \\
        &-H_{12}(k)
        \bigl(H_{22}(k)-i\omega\bigr)^{-1}
        H_{21}(k).
    \end{aligned}\label{eq:PTB_k}
\end{equation}

{
In the \(|m\rangle\equiv |k,m\rangle\) basis, the brane corresponds to
\(m=0,1\), while the remaining sites form the complement. The
corresponding blocks are
}
\begin{align}
    H_{11}
    &=
    { q\,|1\rangle\langle 0|+q^*|0\rangle\langle 1|,}
    \label{H_11}\\
    H_{12}
    &=
    q\,|0\rangle\langle{-1}|
    +q^*|1\rangle\langle 2|,
    \label{H_12}\\
    H_{22}
    &=
    \sum_{m\notin\{0,\pm1\}}
    \bigl[q\,|m+1\rangle\langle m|+\mathrm{h.c.}\bigr],
    \label{H_22}
\end{align}
{
where
}
\begin{equation}
    q = -(t_y+t_x e^{ikd}),
    \label{q_deff}
\end{equation}
{
and \(d=\sqrt{a^2+b^2}\) is the period along the brane. Thus the
original two-dimensional hopping problem reduces, in each momentum
sector, to a one-dimensional chain in the transverse coordinate \(m\),
with an effective complex hopping amplitude \(q\) set by the
longitudinal momentum.

This representation exposes the structure responsible for the physics
below. The block \(\widetilde H_{11}\) describes the intrinsic brane
dynamics, \(\widetilde H_{22}\) describes propagation through the
complement, and \(\widetilde H_{12}\), together with
\(\widetilde H_{21}=\widetilde H_{12}^{\dagger}\), couples the two.
The finite-frequency projected Hamiltonian in Eq.~\eqref{eq:PTB_k}
therefore contains the effect of virtual propagation from the brane
into the surrounding Hermitian lattice and back. Its spectra,
topological invariants, and boundary response are all determined by
this exact Schur complement.
}

\subsection{Boundary conditions and complement parity}

We now specify the parent geometries used before projection. The
lattice is taken infinite along the brane and finite in the transverse
direction. Throughout, ``open'' (OBC) and ``periodic'' (PBC) refer to the boundary
conditions imposed on the Hermitian parent lattice in the direction
transverse to the brane, before projection. The lattice is infinite
along the brane and finite transversely: under OBC, \(N_1\) and \(N_2\)
complement chains lie on the two sides of the brane; under PBC, the
transverse direction closes into a ring with \(N\) complement sites. The projected brane
Hamiltonian is the Schur complement inherited from the chosen parent
geometry. Detailed derivations of the transverse spectra, determinant
formulas, gauge structure, and complement zero-mode condition are
collected in Appendix~\ref{app:parent_spectra_symmetry}.

For
OBC, the transverse coordinate ranges as \(-N_1\leq m\leq N_2+1\), the
brane occupies \(m=0,1\), and the total number of transverse chains is
\(K=N_1+N_2+2\). At fixed \(k\), the transverse parent Hamiltonian is a
finite nearest-neighbor chain with hopping amplitudes \(q\) and
\(q^*\). Its spectrum is
\begin{equation}
    E_n
    =
    2|q|\cos\left(\frac{n\pi}{K+1}\right),
    \qquad
    n=1,\ldots,K .
    \label{eq:OBC_parent_spectrum_main}
\end{equation}
as derived in Appendix~\ref{app:parent_spectra_symmetry}. For even
\(K\), the spectrum consists of paired SSH-like bands, gapped except at
the SSH critical point \(t_x=t_y\), where \(|q|=0\) at \(k=\pi\). For
odd \(K\), sublattice imbalance produces an exact zero-energy flat
band. This zero mode is the feature relevant for resonant projection
when it belongs to the eliminated complement, while the conventional
gapped BDI winding description must be applied with care in its
presence.

For even \(K\), the OBC parent retains the internal BDI symmetries
inherited from its bipartite nearest-neighbor structure and can be
written in chiral form,
\begin{equation}
    \widetilde{H}_{\rm OBC}(k)
    =
    \begin{pmatrix}
        0 & D_{\rm OBC}(k)\\
        D_{\rm OBC}^{\dagger}(k) & 0
    \end{pmatrix}.
    \label{eq:OBC_chiral_main}
\end{equation}
{
The corresponding one-dimensional chiral invariant is the winding of
\(\det D_{\rm OBC}(k)\)
\cite{Chiu2016_RMP,Asboth2016_ShortCourse}. In the present case,
}
\begin{equation}
    \det D_{\rm OBC}(k)=\left(q^{*}\right)^{K/2},
    \qquad
    \tilde{\nu}
    =
    \left\lfloor\frac{K}{2}\right\rfloor
    \frac{\arg q(k)\big|_{-\pi}^{\pi}}{2\pi}.
    \label{eq:OBC_parent_winding_main}
\end{equation}
Here the sign of the integer is fixed by the ordering of the chiral
basis and by the orientation chosen for the Brillouin-zone path; the
expression above is the convention used throughout the paper. 

The chiral block form and determinant entering this invariant are
derived in Appendix~\ref{app:parent_spectra_symmetry}. Thus, the OBC
parent can realize SSH-like topology associated with the transverse
truncation. This parent topology is distinct from the periodic
finite-frequency crystalline invariant of the projected resonant sector
discussed below.

For PBC, the transverse direction is closed
into a ring of \(K=N+2\) sites. We label the transverse coordinate as
\(m=0,1,\ldots,N+1\), choose the brane at \(m=0,1\), and denote by
\(N\) the number of eliminated complement sites. Since the brane
contains two sites per transverse slice, \(K=N+2\); hence \(K\) and
\(N\) have the same parity. In what follows we classify the projected
sectors by the complement parity \(N\), which is the quantity relevant
for the Schur complement. The transverse parent Hamiltonian is
circulant and is diagonalized by transverse momentum
\(p_n=2\pi n/K\). Its spectrum is
\begin{equation}
    E_n(k)
    =
    -2t_y\cos p_n
    -2t_x\cos(k+p_n).
    \label{eq:PBC_parent_spectrum_main}
\end{equation}
{
The circulant diagonalization leading to
Eq.~\eqref{eq:PBC_parent_spectrum_main} is given in
Appendix~\ref{app:parent_spectra_symmetry}. This cylindrical geometry
restores the closed transverse translation structure of the parent
lattice. For even \(K\), the parent belongs to BDI, but its allowed
winding vanishes: away from zeros of \(\det D(k)\), the trajectory of
\(\det D(k)\) is confined to a line in the complex plane and therefore
cannot wind around the origin. For odd \(K\), the periodic transverse
ring is frustrated and cannot be bipartitioned; chiral symmetry is
absent and the parent belongs to AI. In both cases, the PBC parent is
topologically trivial as a Hermitian two-dimensional cylinder.

The phase of \(q(k)\) can be gauged away locally in the bulk, but under
PBC it reappears as a total phase accumulated around the transverse
ring, equivalently as an Aharonov--Bohm flux through the cylinder. This
global obstruction distinguishes the periodic geometry from the open
one, but it does not generate a nonzero parent winding. The determinant
analysis and flux interpretation are summarized in
Appendix~\ref{app:parent_spectra_symmetry}.
}

\subsection{Symmetry inheritance}

{
We finally identify which symmetries of the parent survive projection.
Internal discrete symmetries act block-diagonally on the
brane--complement decomposition and are therefore inherited by the
projected Hamiltonian in the sense of Eq.~\eqref{inhereted_symmetry}. Spatial parity is more
delicate because it exchanges the two sides of the brane. Under OBC,
the brane inherits parity only for symmetric terminations,
\(N_1=N_2\). Under PBC, the transverse parent Hamiltonian is circulant,
and the brane can be chosen so that parity preserves the
brane--complement decomposition. The general symmetry-inheritance
criterion is given in Appendix~\ref{app:general_framework}.

This distinction is central for the resonant sector. The zero-mode pole
supplies the finite-frequency non-Hermiticity, but a crystalline
invariant requires the relevant spatial symmetry to survive projection.
The periodic odd-\(N\) geometry satisfies both conditions: the
eliminated complement is sublattice imbalanced and hosts a zero mode,
while the projected problem inherits the spatial parity that quantizes
the complex Berry phase. The complement determinant and zero-mode
condition are derived in Appendix~\ref{app:parent_spectra_symmetry}.
Open terminations generally do not realize the same periodic
crystalline invariant; instead, they lead to a termination-sensitive
boundary response.
}


\section{Projected brane Hamiltonians and spectral structure}
\label{sec:spectral_structure}

We now derive the projected brane Hamiltonians  together with the corresponding  
spectral structure that underlie the
parity-resolved topology discussed below; the topological characterization is developed in
Sec.~\ref{sec:Main_top_inv}.

The analysis is organized around the two boundary conditions. Open
boundaries give a projected inverse-response kernel whose
non-Hermiticity enters through diagonal imaginary-response terms and
whose spectra exhibit \(\mathcal{PT}_{+}\)-unbroken and
\(\mathcal{PT}_{+}\)-broken regimes.
Periodic boundaries introduce an additional parity-sensitive NH
contribution to the {off-diagonal part of the Hamiltonian}.
This even--odd distinction is the spectral origin of the crystalline
topology analyzed below. Detailed derivations, eigenvectors,
representative spectra, and limiting procedures are given in
Appendix~\ref{app:SecV_technical_spectra}.

According to Eq.~\eqref{eq:PTB_frequency}, the projected kernel
reduces at zero imaginary frequency to the Schur complement of
\(H_{22}\), provided this limit is well defined.
At finite imaginary frequency, it is NH because the
eliminated complement generates complex frequency-dependent matrix
elements. We denote its eigenvalues by
\[
    \bar{E}_{\pm}(k,i\omega)
    =E_{\pm}(k,i\omega)-i\frac{\hbar}{\tau_{\pm}(k,i\omega)},
\]
where \(E_{\pm}\) and \(\tau_{\pm}\) are the real energy and lifetime
of the inverse-response spectrum. Physical
resonance positions and widths are determined separately from the zeros
of the analytically continued retarded kernel.
We therefore employ right and left
eigenstates,
\begin{equation}
H|\psi_n\rangle=E_n|\psi_n\rangle,\qquad
\langle \tilde{\psi}_n|H=E_n\langle \tilde{\psi}_n|.
\end{equation}
Hereafter, \(n\in\mathbb Z\), translation symmetry is preserved along
the brane and \(m\in\{1,2,\ldots,N\}\) labels the degrees of freedom
integrated out in the transverse direction. For simplicity we set
\(d=1\).

\subsection{Open boundaries: \texorpdfstring{$\mathcal{PT}_{+}$}{PT}-unbroken and broken spectra}
\label{Sec:omega_to_0_OBC}

For open boundary conditions, \(H_{22}-i\omega\mathbb I\) separates
into two blocks \(A_{N_1}\) and \(A_{N_2}\) associated with the two
sides of the brane,
\begin{equation}
 H_{22}-i\omega\mathbb I=\begin{pmatrix}
 A_{N_1} & 0\\
 0 & A_{N_2}
 \end{pmatrix}.
 \label{H_22_open}
\end{equation}
Inverting these blocks gives
\begin{equation}
 \begin{aligned}
 \tilde{H}_{\rm PTB}^{\rm (OBC)}=&-i|q|\big[F_{N_{1}}(\phi)|0\rangle\langle0|+F_{N_{2}}(\phi)|1\rangle\langle1|\big]\\
 &+q|1\rangle\langle0|+q^{*}|0\rangle\langle1|,
 \label{H_PTB_open_BC}
 \end{aligned}
\end{equation}
where
\begin{equation}
 F_{N}(\phi)=\begin{cases}
 \dfrac{\sinh[(N+2)\phi]}{\cosh[(N+1)\phi]}, & N\;\text{even},\\[0.6em]
 \dfrac{\cosh[(N+2)\phi]}{\sinh[(N+1)\phi]}, & N\;\text{odd},
 \end{cases}
 \label{F_N_functions}
\end{equation}
and
\begin{align}
\phi&=\operatorname{arcsinh}\left(\frac{\omega}{2|q|}\right),
 \label{phi_deff}
\end{align}
while $q$ is defined in Eq.~\eqref{q_deff}. 
The corresponding spectrum is
\begin{equation}
 \bar{E}_{\pm}=|q|\left[-i\frac{F_{N_{1}}+F_{N_{2}}}{2}
 \pm\sqrt{1-\left(\frac{F_{N_{1}}-F_{N_{2}}}{2}\right)^{2}}\right].
 \label{Energy_open_BC}
\end{equation}
Equation~\eqref{Energy_open_BC} exhibits two regimes. When the square
root is real, the states are in a \(\mathcal{PT}_{+}\)-unbroken sector
and the real parts of the spectrum appear in \(\pm\) pairs. When the
square root becomes imaginary, the corresponding states enter a
\(\mathcal{PT}_{+}\)-broken sector and portions of the real spectrum
collapse into flat-band segments. The critical lines are determined by
\begin{equation}
 |F_{N_1}(\phi)-F_{N_2}(\phi)|=2,
\end{equation}
with the small-\(\phi\) estimates given in
Appendix~\ref{app:SecV_technical_spectra}.

For both $N_i$ ($i=1,2$) odd, the limit \(\omega\to0^+\) is singular at the level of the
full projected kernel because its diagonal imaginary-response terms
diverge. Resonance alone, however, does not produce the
periodic crystalline Berry phase. Odd open-boundary complements can
generate the same self-energy pole, but they generally do not inherit
the periodic embedding parity that 
together with TRS$^{\dagger}$, induces CPH. They therefore realize
zero-mode resonance without the crystalline symmetry needed to quantize
the complex Berry phase.

This distinction can be traced directly to the transverse blocks
\(A_N\). Each \(A_N\) is a bipartite chain of \(N\) sites with hopping
amplitude \(q\). For odd \(N\), the two sublattices are imbalanced, so a
chiral zero mode is enforced. Equivalently, Eq.~\eqref{eq:det_A_N}
implies \(\det A_N\to0\) as \(\omega\to0\), and the corresponding block
of the resolvent develops a pole. An odd open-boundary complement
therefore produces a divergent contribution to the embedding
self-energy,
\[
 \Sigma(i\omega)=H_{12}(i\omega-H_{22})^{-1}H_{21},
\]
and resonantly couples the brane to a zero mode of the surrounding
trivial lattice. The odd-\(N_i\) open-boundary sectors should therefore
be viewed as singular finite-frequency projected problems: their real
spectrum may display the SSH form
\begin{equation}
 H_{\rm SSH}=q|1\rangle\langle0|+q^{*}|0\rangle\langle1|,
 \label{SSH_hamiltonian}
\end{equation}
but this SSH structure is a low-frequency spectral diagnostic rather
than a regular zero-frequency Schur-complement Hamiltonian.

\begin{figure}[t]
 \centering
 \includegraphics[width=0.47\textwidth]{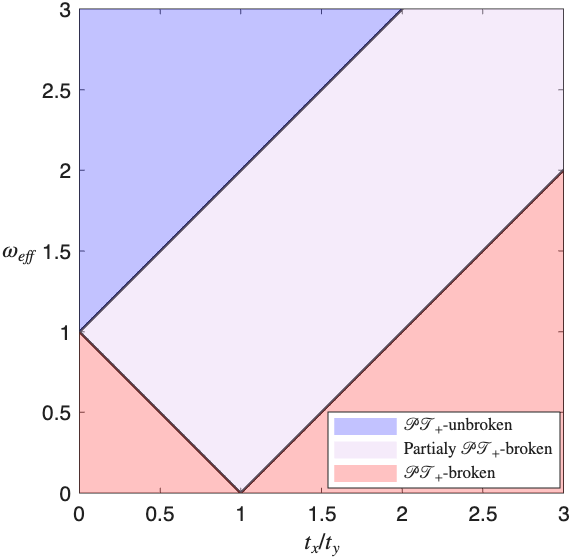}
 \caption{Phase diagram of the \(\mathcal{PT}_{+}\) symmetry for the
 open-boundary projected Hamiltonian. The value plotted on the
 \(y\)-axis is given by
 \(\omega_{eff} = \frac{\omega}{2 t_{y} \sinh \phi_{\mathrm{cr}}}\).
 If both \(N_{1}\) and \(N_{2}\) are even, only the
 \(\mathcal{PT}_{+}\)-unbroken phase is realized. When at least one of
 them is odd, \(\mathcal{PT}_{+}\)-broken regions appear. The critical
 lines are determined by the condition \(|F_{N_1}-F_{N_2}|=2\).}
 \label{fig:Phase_diagram_PT}
\end{figure}

Representative numerical support for this open-boundary structure is
provided in Fig.~\ref{fig:Phase_diagram_PT}. The figure shows that the
parity of the eliminated complement controls whether the
finite-frequency diagonal response terms preserve a fully
\(\mathcal{PT}_{+}\)-unbroken spectrum or generate partially and fully
broken regions. This is the spectral precursor of the open-boundary
comparison sectors used below. Additional open-boundary complex spectra are collected in the SM.

\subsection{Periodic boundaries: even--odd spectral distinction}

We next impose periodic boundary conditions on the complement. The
parent system is then placed on an infinite cylinder containing
\(N/2+1\) brane-like subsystems, including the original brane.
In contrast to the open-boundary case, all four relevant matrix
elements of the complement resolvent are nonzero:
\((\widetilde H_{22}-i\omega)^{-1}_{-1,-1}\),
\((\widetilde H_{22}-i\omega)^{-1}_{-1,2}\),
\((\widetilde H_{22}-i\omega)^{-1}_{2,-1}\), and
\((\widetilde H_{22}-i\omega)^{-1}_{2,2}\). The projected brane
Hamiltonian becomes
\begin{equation}
    \begin{aligned}
        \widetilde H_{\rm PTB}
        =&
        -i|q|F_N(\phi)
        \left(
        |0\rangle\langle 0|+|1\rangle\langle 1|
        \right) \\
        &+
        \left[
        q^{*}
        -
        \frac{(-q)^{N+1}}{|q|^N}G_N(\phi)
        \right]
        |0\rangle\langle 1| \\
        &+
        \left[
        q
        -
        \frac{(-q^{*})^{N+1}}{|q|^N}G_N(\phi)
        \right]
        |1\rangle\langle 0| .
    \end{aligned}
    \label{H_PTB_periodic_BC}
\end{equation}
Here \(F_N(\phi)\) is defined in Eq.~\eqref{F_N_functions}, while the
new auxiliary function \(G_N(\phi)\) is
\begin{equation}
    G_N(\phi)
    =
    \begin{cases}
        \displaystyle
        (-1)^{N/2}
        \frac{\cosh\phi}{\cosh[(N+1)\phi]},
        & N\ \mathrm{even},
        \\[1.1em]
        \displaystyle
        (-1)^{(N-1)/2}
        \frac{i\cosh\phi}{\sinh[(N+1)\phi]},
        & N\ \mathrm{odd}.
    \end{cases}
    \label{G_N_functions}
\end{equation}
The corresponding complex spectrum is
\begin{equation}
    \bar{E}_{\pm}
    =
    |q|
    \left[
    -iF_N
    \pm
    \sqrt{
    1+G_N^2+2(-1)^N G_N\cos\alpha
    }
    \right],
    \label{Energy_periodic_BC}
\end{equation}
with \(\alpha=(N+2)\arg q\). Equations~\eqref{G_N_functions}
and~\eqref{Energy_periodic_BC} make the even--odd distinction explicit.
For even \(N\), \(G_N\) is real and the square root is real, so the
periodic sector behaves as a regular SSH-type NH problem
with a real band splitting and a common imaginary shift. For odd \(N\),
\(G_N\) is purely imaginary, and the square root generically becomes
complex. {This is the spectral signature of the resonant
periodic sector: the complement zero mode produces a singular
finite-frequency self-energy, rather than merely a generic
NH correction.}

For even \(N\), the spectrum therefore reads
\begin{align}
E_{\pm}
&=\pm|q|\sqrt{1+G_{N}^{2}+2G_{N}\cos\alpha},
\label{Energy_periodic_BC_even_N}\\
\tau_{\pm}&=\frac{\hbar}{|q|F_N}.
\label{Lifetime_periodic_BC_even_N}
\end{align}

At strictly finite frequency, \(|G_N|<1\) for the even-\(N\) sector,
so the additional parity-induced finite-frequency gap does not close
except at singular points where the underlying hopping amplitude
\(|q|\) itself vanishes, i.e. the SSH critical point. The zero-frequency descendant, however, must
be treated through the effective off-diagonal element \(Q(k)\) below:
its gap can close whenever \(Q(k)=0\), including the SSH critical point
and possible interference zeros of the projection-induced term.
In the zero-frequency limit, \(F_N\to0\) and
\(G_N\to(-1)^{N/2}\), giving the Hermitian SSH-like descendant
\begin{equation}
 \tilde{H}_{\rm PTB}^{(\mathrm{even})}=Q|1\rangle\langle0|
 +Q^{*}|0\rangle\langle1|,
 \label{H_PTB_periodic_BC_even_N_omega_to_0}
\end{equation}
with
\begin{equation}
 Q=q\left(1+e^{iN\pi/2-i\alpha}\right).
 \label{Q_def}
\end{equation}
Thus
\begin{equation}
 E_{\pm}=\pm|Q|
 =\pm2|q|\left|\cos\left(\frac{\alpha}{2}-\frac{N\pi}{4}\right)\right|.
\end{equation}
{
This expression makes clear that the regular even-\(N\) topology is
controlled by the winding and zeros of the effective SSH-like hopping
\(Q(k)\), rather than by the resonant pole mechanism.}

For odd \(N\), \(G_N=i\tilde G_N\) is imaginary and both the diagonal and the off-diagonal
parts  of Eq.~\eqref{H_PTB_periodic_BC} are resonant and
singular in the low-frequency limit. The real and imaginary
components of the inverse-response eigenvalues are

\begin{align}
 E_{\pm}
 &=
 \pm|q|
 \sqrt{
 \frac{
 1-\tilde{G}_{N}^{2}
 +
 \sqrt{\left(1-\tilde{G}_{N}^{2}\right)^{2}
 +4\tilde{G}_{N}^{2}\cos^{2}\alpha}
 }{2}},
 \label{Energy_periodic_BC_odd_N}\\
 \frac{\hbar}{\tau_{\pm}|q|}
 &=
 F_{N}
 \pm
 \frac{\tilde{G}_{N}\sqrt{2}\cos\alpha}{
 \sqrt{
 1-\tilde{G}_{N}^{2}
 +
 \sqrt{\left(1-\tilde{G}_{N}^{2}\right)^{2}
 +4\tilde{G}_{N}^{2}\cos^{2}\alpha}
 }},
 \label{Lifetime_periodic_BC_odd_N}
\end{align}

with the finite-frequency gaps closing sequentially when
\(|\tilde G_N|=1\) and \(\alpha=(2\ell+1)\pi/2\), producing the phase
diagram analyzed in Sec.~\ref{sec:Main_top_inv}. In the singular
low-frequency limit,
\begin{equation}
 E_\pm \to \pm |q|\,|\cos\alpha|,
\end{equation}
while the two imaginary spectral components become strongly separated
as \(\omega\to0^{+}\). Their assignment exchanges across the
exceptional-point reconnection, consistently with the interchange of
the two complex-energy strands. This limiting behavior is used only as a
spectral diagnostic: the odd-\(N\) periodic topology is defined at finite
frequency, where the projected Hamiltonian is well defined.

\begin{figure}[t]
 \centering
 \includegraphics[width=0.45\textwidth]{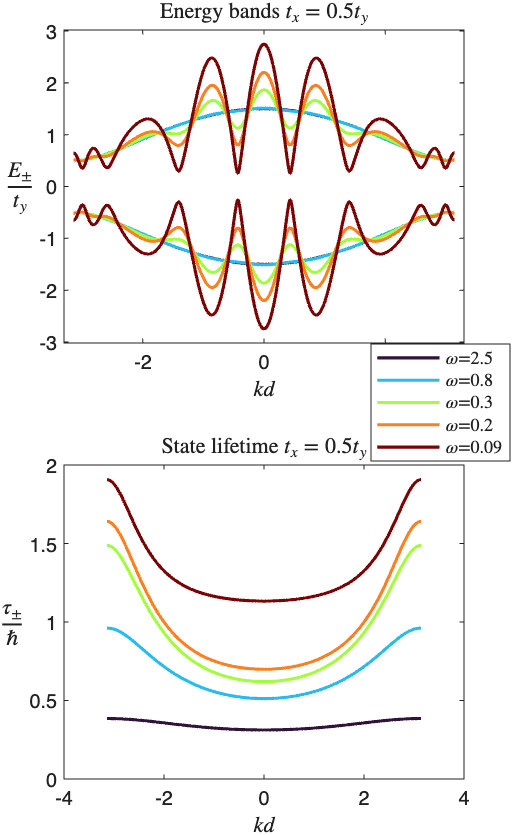}
 \caption{Real and imaginary components of the projected
 inverse-response spectrum for periodic boundary conditions with even
 \(N\), obtained from Eqs.~\eqref{Energy_periodic_BC_even_N} and
 \eqref{Lifetime_periodic_BC_even_N}. The bands retain a regular
 SSH-like structure; in the zero-frequency descendant, {ordinary gap closings are
 controlled by zeros of the effective hopping \(Q(k)\).}}
 \label{fig:E_tau_even_PBC_rad}
\end{figure}

\begin{figure}[t]
 \centering
 \includegraphics[width=0.45\textwidth]{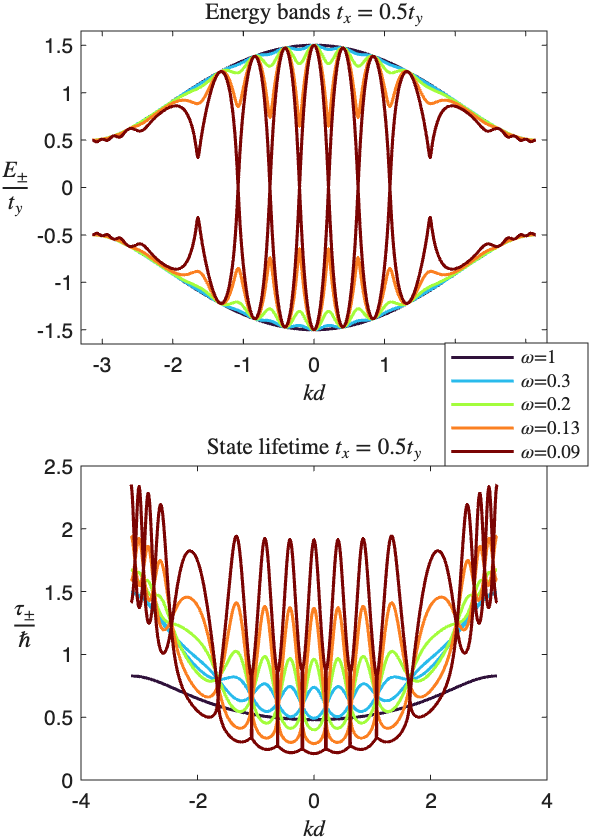}
 \caption{Real and imaginary components of the spectrum on the zig-zag brane for periodic boundary conditions with odd
 \(N\), obtained from Eqs.~\eqref{Energy_periodic_BC_odd_N} and
 \eqref{Lifetime_periodic_BC_odd_N}. As \(\omega\) is reduced,
 gaps close sequentially when \(|\tilde G_N|=1\) and
 \(\alpha=(2\ell+1)\pi/2\). {}}
 \label{fig:E_tau_odd_PBC_rad}
\end{figure}

The even--odd contrast in the periodic spectrum is illustrated in
Figs.~\ref{fig:E_tau_even_PBC_rad} and
\ref{fig:E_tau_odd_PBC_rad}. For even \(N\), the bands retain an
SSH-like structure and the projection-induced finite-frequency gaps
remain regular. For odd \(N\), by contrast, the zero-mode pole produces a resonant finite-frequency sector. Lowering \(\omega\) drives a sequence of exceptional-point transitions at critical frequencies \(\omega_\ell\); at each transition the two complex eigenvalues coalesce at \(k_\ell\), the Berry invariant may jump, and the touching carries local spectral-braiding charge. This finite-frequency topological structure is discussed in the next section.

We finally notice that in the large-system limit \(N\to\infty\),
\(G_N\to0\) and \(F_N\to e^{\phi}\), while the periodic Hamiltonian
then reduces to the open-boundary form, as expected when the compact
transverse direction becomes infinitely large.
The corresponding lifetime inverse-response component is
\begin{equation}
 \tau(E)=\frac{2\hbar}{\omega+\sqrt{\omega^2+4E^2}},
 \label{tau_large_system}
\end{equation}
whereas the real part of the Hamiltonian reduces to the
simple SSH model. This is the continuum large-complement limit of the
projected brane spectrum.


\section{Crystalline Berry topology and exceptional spectral braiding}
\label{sec:Main_top_inv}

We now turn from the projected spectra to their topological content.
The even-\(N\) sectors follow the regular projection route and realize
SSH-type descendants. The periodic odd-\(N\) sector is intrinsically
finite-frequency because of the complement zero-mode pole. Its stable
bulk index is a crystalline, CPH-quantized complex Berry phase, while
its critical frequencies host parity-related EPs with
local spectral-braiding charge.

The symmetry hierarchy is essential. In the projected two-band space,
embedding parity is represented by \(\mathcal{P}=\sigma_x\) and satisfies
\begin{equation}
    \sigma_x H(k,i\omega)\sigma_x=H(-k,i\omega).
    \label{eq:main_parity_constraint}
\end{equation}
For
\begin{align}
    H(k,i\omega)&=H_0(k,i\omega)\mathbb I
    +Q_1(k,i\omega)|0\rangle\langle1|\nonumber\\
    &\quad+Q_2(k,i\omega)|1\rangle\langle0|,
    \label{eq:H_Q1_Q2_form}
\end{align}
parity directly gives
\begin{equation}
    H_0(k)=H_0(-k),\qquad
    Q_1(k)=Q_2(-k).
    \label{eq:main_Q_parity}
\end{equation}
Together with TRS\(^{\dagger}\), parity induces CPH,
\begin{equation}
    \sigma_x H(k,i\omega)^*\sigma_x=H(k,i\omega)^\dagger,
    \label{eq:main_CPH}
\end{equation}
which constrains the biorthogonal eigenstates and quantizes their
complex Berry phase. Equation~\eqref{eq:main_Q_parity}, rather than CPH
alone, relates the two off-diagonal amplitudes and pairs exceptional
points at opposite momenta.

For the periodic projected Hamiltonian,
\begin{align}
    H_0(k,i\omega)&=-i|q|F_N(\phi),\\
    Q_1(k,i\omega)&=q^*-
    \frac{(-q)^{N+1}}{|q|^N}G_N(\phi),\\
    Q_2(k,i\omega)&=q-
    \frac{(-q^*)^{N+1}}{|q|^N}G_N(\phi)=Q_1(-k,i\omega).
\end{align}
The eigenvalues and spectral discriminant are
\begin{align}
    E_\pm&=H_0\pm\sqrt{Q_1Q_2},\\
    \Delta(k,i\omega)&=[E_+(k,i\omega)-E_-(k,i\omega)]^2
    =4Q_1Q_2.
    \label{eq:spectral_discriminant_braid}
\end{align}
An EP occurs when one off-diagonal factor vanishes while
the other remains nonzero.

To distinguish eigenvector topology from spectral braiding, we introduce
the off-diagonal windings on a closed contour \(\mathcal C\) that avoids
spectral degeneracies,
\begin{equation}
    w_a[\mathcal C]
    =
    \frac{1}{2\pi i}
    \oint_{\mathcal C}d\ln Q_a,
    \qquad
    a=1,2.
    \label{eq:main_offdiag_windings}
\end{equation}
The spectral-braid invariant and the CPH complex Berry invariant are,
respectively,
\begin{align}
    \nu_{\rm br}[\mathcal C]
    &=
    \frac{1}{4\pi i}
    \oint_{\mathcal C}d\ln\Delta
    =
    \frac{1}{2}
    \left(
        w_1[\mathcal C]+w_2[\mathcal C]
    \right),
    \label{eq:braid_invariant_normalized}
    \\
    \widetilde{\nu}[\mathcal C]
    &=
    \frac{1}{2}
    \left(
        w_1[\mathcal C]-w_2[\mathcal C]
    \right).
    \label{eq:positive_Z_invariant}
\end{align}
They therefore probe complementary combinations of the same
off-diagonal winding data.

This formulation applies to both regular and exceptional
sectors. When the spectrum is nondegenerate, the contour
\(\mathcal C\) is the physical Brillouin zone. Embedding parity then
implies
\begin{equation}
    w_2[\mathrm{BZ}]
    =
    -w_1[\mathrm{BZ}],
\end{equation}
and consequently
\begin{equation}
    \nu_{\rm br}[\mathrm{BZ}]
    =
    0,
    \qquad
    \widetilde{\nu}[\mathrm{BZ}]
    =
    w_1[\mathrm{BZ}].
    \label{eq:main_parity_Berry_braid}
\end{equation}
The quantized labels of the nondegenerate finite-frequency regions are
therefore given by the CPH complex Berry invariant.

At \(\omega=\omega_\ell\), a parity-related pair of EPs
on the Brillouin-zone contour reconnects the two complex-energy strands. The contour is then displaced infinitesimally away
from the degeneracy, or chosen locally around it, within the same
winding construction. A simple EP carries the local
braiding charge
\begin{equation}
    \nu_{\rm EP}
    =
    \frac{1}{4\pi i}
    \oint_{\gamma_{\rm EP}}d\ln\Delta
    =
    \pm\frac{1}{2},
    \label{eq:main_local_EP_charge}
\end{equation}
where \(\gamma_{\rm EP}\) encloses the degeneracy.

The half-integer charge has a direct spectral meaning. At the critical frequency \(\omega=\omega_\ell\), continuing the momentum once around \(\gamma_{\rm EP}\) takes the eigenvalue \(E_+\) into \(E_-\), while a second circuit returns it to its initial branch. The EP therefore reconnects the two complex-energy bands and carries the local charge associated with their strand exchange. 

For any fixed \(\omega\neq\omega_\ell\), the two spectral strands form nondegenerate closed trajectories over the compact Brillouin zone. Embedding parity forces their total discriminant winding on the physical Brillouin zone to vanish, while the CPH Berry invariant labels the corresponding bulk sector. As \(\omega\) passes through \(\omega_\ell\), the parity-related pair of EPs reconnects the two strands and the Berry invariant can change. Infinitesimal contours passing on opposite sides of the degeneracy define the local winding and fix the orientation of the associated local band exchange. The full derivation is given in Appendix~\ref{sec:CPH_sym}.

\begin{figure*}[htbp]
\centering
\includegraphics[width=\textwidth]{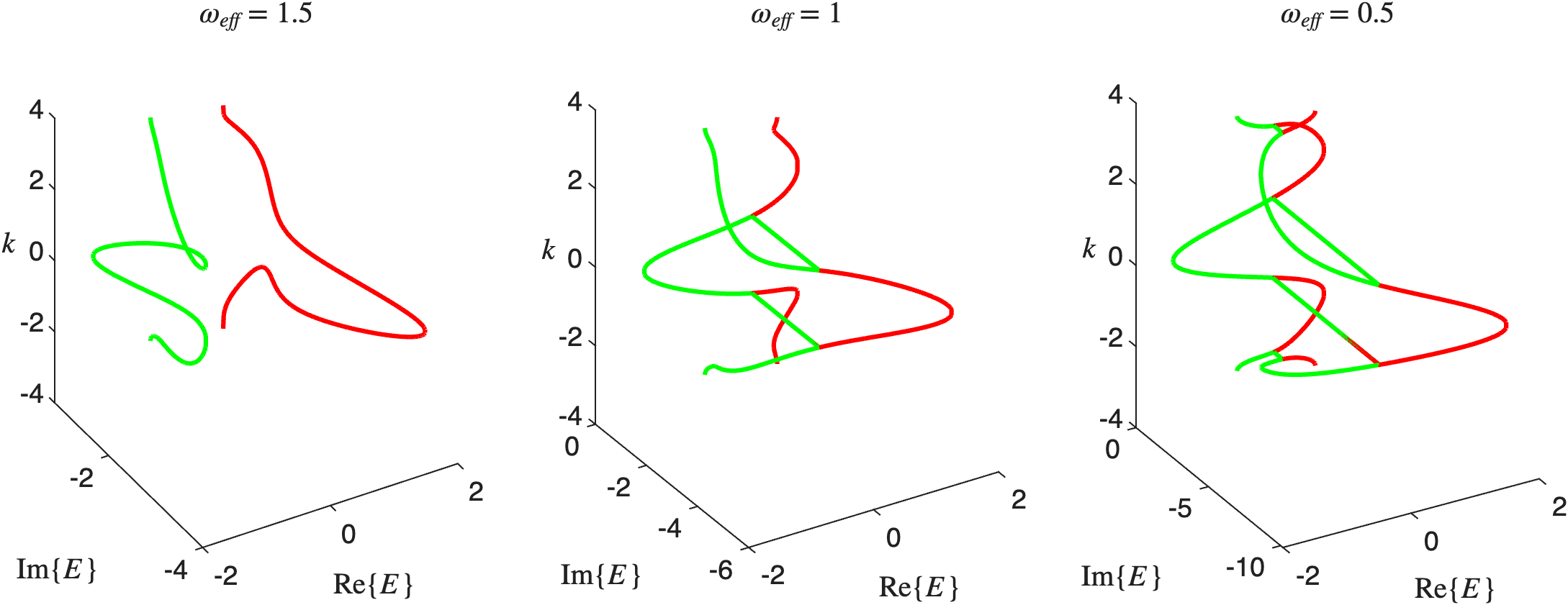}
\captionsetup{justification=justified,singlelinecheck=false}
\caption{Two-band braiding of the periodic odd-\(N\) spectrum.
The horizontal plane represents complex energy, while the vertical axis is the crystal momentum. The three panels follow the red trajectory in Fig.~\ref{fig:Phase_diagram_N=3}. Before the first transition, the two complex bands form separate closed trajectories. At the first critical frequency, a parity-related pair of EPs enters the Brillouin zone where the two
strands coalesce and reconnect. The middle panel shows the resulting fixed-frequency band connectivity between the two transitions. At the second critical frequency, the same pair of EPs exits the Brillouin zone and restores the original connectivity, after which the bands again form separate closed trajectories. The nondegenerate panels are labeled
by the CPH complex Berry invariant, while the exceptional-point
crossings carry local Abelian braid charge. The trajectory orientation
is set by increasing momentum.}
\label{fig:3D_spectrum}
\end{figure*}

For the odd-\(N\) projected Hamiltonian, the CPH Berry invariant takes
the explicit form
\begin{equation}
    \widetilde\nu_\pm
    =-\frac{1}{2\pi}
    \bigl(\arg\{q\}+\theta\bigr)\Big|_{-\pi}^{\pi},
    \label{top_inv_odd}
\end{equation}
where
\begin{equation}
    e^{2i\theta}=
    \frac{1+\widetilde G_N^2e^{-2i\alpha}}
    {\sqrt{1+\widetilde G_N^4
    +2\widetilde G_N^2\cos(2\alpha)}}.
    \label{Theta_N_ODD}
\end{equation}
The first term in Eq.~\eqref{top_inv_odd} is the SSH winding of
\(q=-t_y-t_xe^{ik}\), while the second is the finite-frequency
crystalline contribution generated by projection. As \(\omega\) is
lowered, exceptional degeneracies occur at
\(\alpha_\ell=(2\ell-1)\pi/2\). Across such a transition,
\begin{equation}
    2\theta\Big|_{-\pi}^{\pi}
    =-2\pi(N_+-N_-),
    \label{2theta}
\end{equation}
so the CPH Berry invariant changes by an integer. The quantities
\(N_+\) and \(N_-\) count the oriented EP-mediated changes in regions
where \(\alpha(k)\) increases or decreases. The local exceptional-point vorticity and
the jump of the Berry invariant are related manifestations of the same
critical degeneracy.

The integer quantization in Eq.~\eqref{2theta} follows from a simple
counting argument. The phase \(2\theta(k)\) defined by Eq.~\eqref{Theta_N_ODD} is a smooth,
single-valued function of \(k\) away from the degeneracies, and the
Brillouin zone is a closed loop, so \(2\theta\) must return to itself up
to an integer multiple of \(2\pi\); the net change
\(2\theta|_{-\pi}^{\pi}\) is therefore quantized in units of \(2\pi\),
which is what fixes the right-hand side to \(-2\pi(N_+-N_-)\) with
\(N_\pm\in\mathbb{Z}\). The individual EPs each carry a
half-integer vorticity \(\nu_{\rm EP}=\pm\tfrac12\) [Eq.~\eqref{eq:main_local_EP_charge}],
but embedding parity pairs them at \(\pm k\): a transition brings in a
parity-conjugate pair whose two half-integer contributions add to a
single integer step of the global invariant. The non-monotonicity of
\(\alpha(k)\)---which is why a naive count of sign changes would
overcount---is precisely accounted for by the oriented difference
\(N_+-N_-\), so that only the net oriented winding survives. Thus the
half-integer charges of EPs are the local content, while their
parity-related, oriented sum is the integer jump of the Berry invariant.

The periodic odd-\(N\) sector therefore lies beyond the internal
one-dimensional classification of class AI\(^{\dagger}\): embedding
parity supplies the opposite-momentum constraint, and parity together
with TRS\(^{\dagger}\) induces CPH and quantizes the biorthogonal Berry
phase. The zero-mode pole drives the exceptional transitions separating these crystalline Berry sectors, while each critical degeneracy carries local spectral-braiding charge.

A complementary perspective on the topological phase transitions is presented in Appendix \ref{APP:band_braiding}. There, it is demonstrated that the complex Berry phase invariant can be interpreted as a linking number between two knots that represent the two energy bands. In Fig.~\ref{fig:3D_spectrum}, these knots are shown in different colors (red and green). Hence, it is evident that the two continuous curves, i.e., the two spectral strands, are distinct entities from the two energy bands. In the first panel, the two bands are completely unlinked, with each band associated with a single spectral strand. When the topological transition driven by EPs takes place, each of the two energy bands traverses both complex spectral strands; in other words, they become linked. This situation is depicted in the second panel. 

The braiding of the energy bands is captured by the phase of $\sqrt{Q_{2}/Q_{1}}$, which quantizes the complex Berry phase. In contrast, the braiding of the spectral strands is encoded in the phase of $Q_{1}Q_{2}$. Since the two strands never become linked, this invariant remains zero whenever it is well defined.

\subsection{Zero-mode resonance and finite-frequency exceptional-point transitions}
\label{subsec:positive_classification}

The odd-\(N\) periodic sector realizes the resonant mechanism in its
minimal form. After projection onto a fixed momentum \(k\) along the
brane, the eliminated complement contains an odd-length bipartite
chain. Its two sublattices are imbalanced, with \((N+1)/2\) and
\((N-1)/2\) sites, and therefore the complement hosts one chiral zero
mode~\cite{Sutherland-PRB1986,Inui-PRB1994}. Equivalently,
Eq.~\eqref{eq:det_A_N} implies that the determinant of the complement
block vanishes as \(\omega\to0\). Since this zero mode couples to the
brane, the complement resolvent develops the pole described in
Eq.~\eqref{eq:gen_Sigma_decomp}, and the embedding self-energy becomes
singular. The projected Hamiltonian therefore has no regular
zero-frequency Schur-complement representative.

This singularity fixes the topological setting and
connects directly to the exceptional-point transition mechanism. The singular term
\(-i\Gamma(k)/\omega\) in the embedding self-energy reduces the real
part of the discriminant \(\Delta = 4Q_1Q_2\) as \(\omega\) decreases.
Consequently, discriminant zeros, which correspond to exceptional
points in the complex \(k\)-plane, are driven toward the real
\(k\)-axis. At each critical frequency \(\omega_\ell\), the imaginary
part of an EP vanishes, it enters the physical Brillouin
zone at momentum \(k_\ell\), and the two complex energy strands touch.
The complex-energy strands reconnect at this momentum and the CPH Berry
invariant can jump by an integer. The touching itself carries the local
vorticity \(\nu_{\rm EP}=\pm1/2\), which encodes  the associated
spectral braid. Thus, the zero-mode pole brings EPs into the physical spectrum and establishes finite frequency as the natural topological setting of the resonant sector. The off-diagonal windings encode two complementary manifestations of this topology: the quantized complex Berry phase and the exceptional-point braiding charge.

\subsection{Phase diagram, crystalline Berry invariant, and exceptional spectral braiding}

Equation~\eqref{2theta} gives the finite-frequency complex Berry phase invariant
\begin{equation}
    \tilde{\nu}_{\pm}
    =
    \frac{1}{2}\left[N_{+}(\omega,t_x)-N_{-}(\omega,t_x)\right]
    -\Theta(t_x-t_y),
    \label{nu_sol_odd_N_periodic_BC}
\end{equation}
where \(\Theta(t_x-t_y)\) is the Heaviside step
function, equal to unity for \(t_x>t_y\) and zero otherwise, and
constitutes the SSH winding contribution. 

The phase boundaries in the \((t_x/t_y,\omega_{eff})\) plane are
exceptional-point transition lines. Along each line the two eigenvalues
coalesce at an isolated momentum, the local exceptional-point vorticity is nonzero, and
the CPH Berry invariant changes. Between adjacent transition lines the
spectrum is nondegenerate and the crystalline complex Berry phase does not change. The number of distinct finite-frequency Berry sectors at
fixed \(t_x\) grows with \(N\), reflecting the increasing number of
exceptional-point transitions encountered as \(\omega\) is lowered.

For \(t_x<t_y\), the loop traced by \(q(k)\) does not enclose the
origin, so the SSH contribution vanishes. The angle \(\arg\{q\}\)
oscillates around \(\pi\), and the maximum number of exceptional-point transitions
accumulated in the increasing-\(\alpha\) region is
\begin{equation}
    N_{+}^{(\mathrm{max})}
    =
    \left\lceil
    \frac{1}{2}
    \left\lfloor
    \frac{2}{\pi}(N+2)
    \arctan\frac{t_{x}}{\sqrt{t_{y}^{2}-t_{x}^{2}}}
    \right\rfloor
    \right\rceil,
    \label{N_plus_max}
\end{equation}
for a fixed value of \(t_x<t_y\). Here, \(\lfloor x\rfloor\) denotes the floor function, i.e., the
greatest integer not exceeding \(x\), while \(\lceil x\rceil\) denotes
the ceiling function, i.e., the smallest integer not smaller than
\(x\). As \(t_x\to t_y^-\), the argument of the arctangent
diverges, \(\arctan[t_x/\sqrt{t_y^2-t_x^2}]\to\pi/2\), so
\(N_{+}^{(\mathrm{max})}\to\lceil (N+2)/2\rceil=(N+3)/2\) for odd \(N\):
the count of accessible transitions saturates as the SSH critical point
is approached. This is the maximal number of exceptional-point
crossings, and it matches---transition by transition---the
\(t_x>t_y\) asymptotic invariant
\[\tilde\nu^{\,\omega\to0^+}_{\pm}=(N+1)\Theta(t_x-t_y)\] of
Eq.~\eqref{nu_lim_low_freq} once all crossings generated by
\(|\tilde G_N|=1\) have occurred, so the \(t_x<t_y\) and \(t_x>t_y\)
counts join continuously at \(t_x=t_y\).  For \(t_x>t_y\), the loop traced by
\(q(k)\) winds once around the origin, so \(\arg\{q\}\in[0,2\pi]\) and
therefore \(\alpha=(N+2)\arg\{q\}\in[0,2\pi(N+2)]\). This is the
regime in which the SSH winding and the finite-frequency crystalline
braid contributions combine.

\begin{figure}[t]
    \centering
    \includegraphics[width=0.47\textwidth]{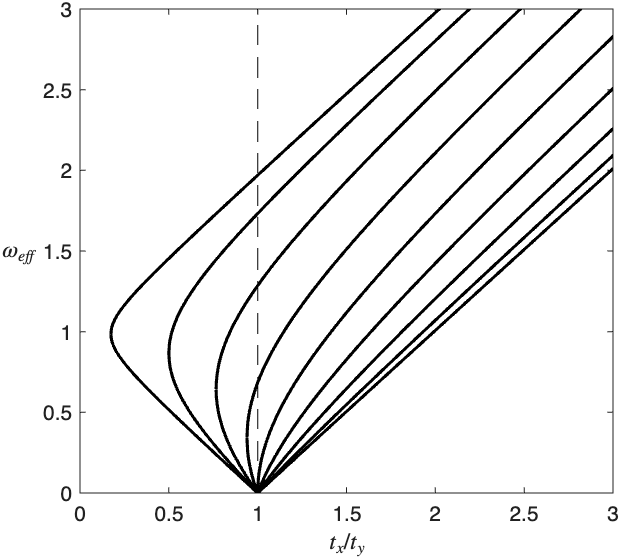}
    \caption{Phase diagram for periodic boundary conditions with odd
    \(N\). Each nondegenerate region is characterized by its
    integer value of \(\widetilde\nu\); the separating exceptional-point
    lines carry local charges \(\nu_{\rm EP}=\pm\tfrac12\).
    The phase boundaries are given by Eq.~\eqref{crit_lines}.
    The value plotted on the \(y\)-axis is given by
    \(\omega_{\mathrm{eff}} = \omega/[2 t_{y} \sinh
    \phi_{\mathrm{cr}}]\). For
    fixed \(t_x\), the number of distinct phases is controlled by
    \(N\): it is given by Eq.~\eqref{N_plus_max} for \(t_x<t_y\),
    while for \(t_x>t_y\) the SSH winding adds the contribution shown
    in Eq.~\eqref{nu_sol_odd_N_periodic_BC}.}
    \label{fig:Phase_diagram_odd_N}
\end{figure}

\begin{figure}[t]
    \centering
    \captionsetup{justification=justified,singlelinecheck=false}
    \includegraphics[width=0.47\textwidth]{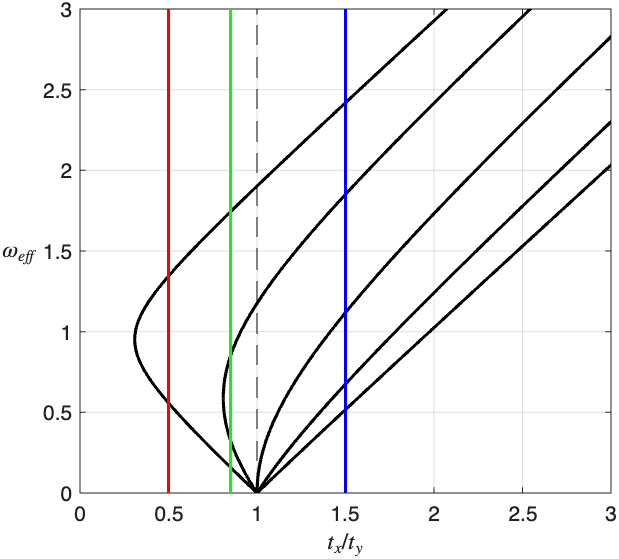}
    \caption{Phase diagram for periodic boundary conditions with
    \(N=3\). The nondegenerate regions are labeled by the integer
    CPH Berry invariant \(\widetilde\nu\), while the critical lines carry
    local exceptional-point vorticity \(\nu_{\rm EP}=\pm\tfrac12\).
    The critical lines separating different phases are given
    by Eq.~\eqref{crit_lines}. The value plotted on the \(y\)-axis is
    \(\omega_{\mathrm{eff}} = \omega/[2 t_{y} \sinh
    \phi_{\mathrm{cr}}]\). Distinct phases are distinguished by
    quantized values of the crystalline complex Berry invariant. As the
    imaginary-frequency slice is lowered, the invariant changes through
    a sequence of finite-frequency exceptional-point transitions. Crossing the
    dashed line by increasing \(t_x\) corresponds to the SSH critical
    point and reduces the invariant by one unit. The red, green, and
    blue lines indicate representative cuts through the phase diagram,
    which are analyzed in detail in
    Appendix~\ref{app:SecV_technical_spectra}.}
    \label{fig:Phase_diagram_N=3}
\end{figure}

Introducing
\begin{equation}
    \vartheta_\ell=\frac{2\ell-1}{2(N+2)}\pi,
    \label{eq:Braiding-tran-lines}
\end{equation}
with \(\vartheta_\ell\) measured from the negative real axis of the
circle traced by \(q(k)\), and using \(\omega=2|q|\sinh\phi_{\rm cr}\),
the exceptional-point transition lines are
\begin{equation}
    \frac{\omega_{\ell}^{\pm}}{2t_{y}\sinh\phi_{\rm cr}}
    =
    \cos\vartheta_\ell
    \pm
    \sqrt{\frac{t_x^2}{t_y^2}-\sin^2\vartheta_\ell}.
    \label{crit_lines}
\end{equation}
Physical transitions correspond only to positive real
\(\omega_\ell^\pm\). The two branches account for the two possible
intersections associated with each allowed gap-closing angle. The
resulting finite-frequency phase diagram for \(N=7\) is shown in
Fig.~\ref{fig:Phase_diagram_odd_N}, while the \(N=3\) case in
Fig.~\ref{fig:Phase_diagram_N=3} shows the same mechanism. Taken
together, these two phase diagrams make clear that the integer
invariant is controlled by the number of exceptional-point transitions ---
finite-frequency discriminant zeros of the projected Green's function
--- crossed as \(\omega\) is lowered.

\begin{figure*}[t]
    \centering
    \begin{subfigure}[b]{0.45\textwidth}
        \centering
        \includegraphics[width=\textwidth]{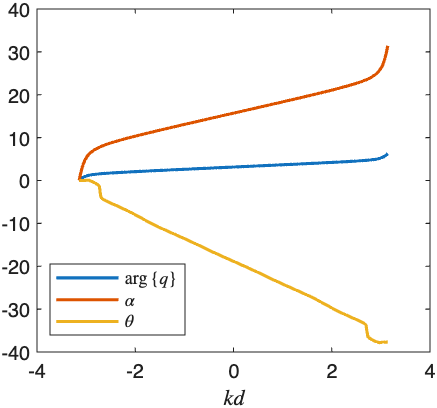}
        \label{fig:Phases_tx_gt_ty}
    \end{subfigure}
    \hspace{1cm}
    \begin{subfigure}[b]{0.45\textwidth}
        \centering
        \includegraphics[width=\textwidth]{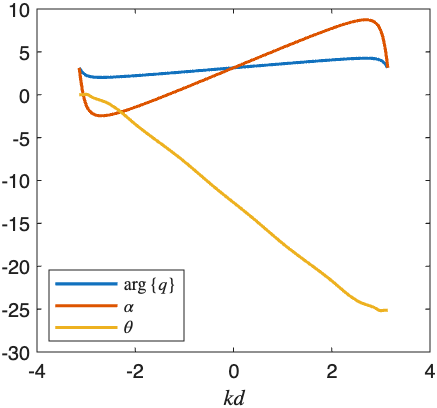}
        \label{fig:Phases_tx_lt_ty}
    \end{subfigure}
    \caption{Winding structure of the phases entering the odd-\(N=3\)
    periodic crystalline Berry invariant in Eq.~\eqref{top_inv_odd}. Left: the case \(t_x>t_y\).
    Right: the case \(t_x<t_y\). See  Eq.~\eqref{Theta_N_ODD} for the definition of $\alpha$ and $\theta$.}
    \label{fig:Phases_oddN_angles}
\end{figure*}

The winding mechanism behind the invariant is displayed in
Fig.~\ref{fig:Phases_oddN_angles}. The figure tracks \(\arg\{q\}\),
\(\alpha\), and \(\theta\) across the Brillouin zone and shows
explicitly how the finite-frequency phase \(\theta\) supplies the
crystalline contribution to Eq.~\eqref{top_inv_odd}. Representative
complex-spectrum evolutions across the same transitions are collected
in Appendix~\ref{app:SecV_technical_spectra}.

The odd-\(N\) periodic phase for \(t_x>t_y\) is resonant: the
projected Hamiltonian has no regular \(\omega\to0^{+}\) operator limit
in this sector. The finite-frequency invariant nevertheless has a
well-defined asymptotic form. Once all exceptional-point transitions generated by
\(|\tilde{G}_N|=1\) have occurred, the limiting value is
\begin{equation}
    \tilde{\nu}_{\pm}^{\omega\to0^{+}}
    =
    (N+1)\Theta(t_x-t_y).
    \label{nu_lim_low_freq}
\end{equation}
This expression is the low-frequency limit of the finite-frequency crystalline Berry invariant. At the SSH critical point
\(t_x=t_y\), the hopping amplitude \(|q|\) vanishes at \(k=\pi\), the
discriminant \(\Delta\) is zero at that momentum, and the
complex energy strand separation collapses. The corresponding half-integer
values of the Berry phase,
\begin{equation}
    \nu_{\pm}^{\mathrm{cr}}=z-\frac{1}{2},\qquad
    z\in\left\{0,1,2,\ldots,\frac{N+3}{2}\right\},
\end{equation}
are transition diagnostics rather than invariants of a
gapped phase. The half-integer values occur only at the transition line
\(t_x=t_y\), where the spectrum is degenerate and the Berry phase is not
defined. They therefore mark the critical points separating the
integer-valued regions in Fig.~\ref{fig:Phase_diagram_N=3}. The quantized phases are the nondegenerate sectors on either side of this
line, where the invariant takes integer values. The corresponding transition frequencies are obtained by
setting \(t_x=t_y\) in Eq.~\eqref{crit_lines}, which gives
\(\omega_\ell^{\pm}/(2t_y\sinh\phi_{\rm cr}) = \cos\vartheta_\ell \pm
|\cos\vartheta_\ell|\), so the physical branch yields
\(\omega_\ell = 4t_y\sinh\phi_{\rm cr}\cos\vartheta_\ell\) for
\(\vartheta_\ell < \pi/2\). Thus even a nearest-neighbor square
lattice, which is topologically trivial as a two-dimensional bulk, can
project into a brane with highly nontrivial finite-frequency topology
for periodic boundary conditions with odd \(N\). See also
Figs.~\ref{fig:Phase_diagram_odd_N},
\ref{fig:Phase_diagram_N=3}, and
Appendix~\ref{app:SecV_technical_spectra}.

\subsection{Comparison sectors: SSH topology and open-boundary phases}

The remaining parity and boundary-condition sectors provide the
SSH-type comparison cases against which the odd-\(N\) periodic
crystalline phase should be understood. The regular comparison sectors
are the even-\(N\) open and periodic projected Hamiltonians. They
belong to BDI\(^{\dagger}\) class and remain spectrally gapped away
from the SSH transition. Their invariant can be written as
\begin{equation}
    \nu_{\pm}
    =
    -\frac{1}{2\pi}
    \bigl(\arg\{q\}\pm\theta\bigr)\Big|_{-\pi}^{\pi},
    \label{top_inv_even}
\end{equation}
where the plus sign applies to open boundary conditions and the minus
sign to periodic boundary conditions. At finite Matsubara frequency the
phase \(\theta\) carries no winding, so both cases reduce to the SSH
invariant
\begin{equation}
    \nu_{\pm}=-\Theta(t_x-t_y).
    \label{nu_sol_even_finate_omega}
\end{equation}
For even \(N\) with periodic boundary conditions, the zero-frequency
descendant has \(G_N=(-1)^{N/2}\), so the SSH hopping $q$ is replaced by
the effective hopping \(Q\). This gives the enhanced SSH-type invariant
\begin{equation}
    \nu_{\pm}=\frac{N}{2}\Theta(t_x-t_y).
    \label{nu_sol_even_zero_omega_periodic_BC}
\end{equation}
This enhancement remains within the SSH mechanism and is distinct from
the crystalline Berry topology and exceptional spectral reconnections of the odd-\(N\) periodic sector.

For open boundary conditions with odd \(N\), the projected Hamiltonian
also remains in BDI\(^{\dagger}\), but the resonant zero mode can drive
\(\mathcal{PT}_{+}\)-broken regimes. In the
\(\mathcal{PT}_{+}\)-unbroken regime the invariant again reduces to
Eq.~\eqref{nu_sol_even_finate_omega}. In the partially
\(\mathcal{PT}_{+}\)-broken regime, the appropriate global Berry phase
is
\begin{align}
    \frac{1}{2}(\nu_{+}+\nu_{-})
    &=
    -\frac{1}{2\pi}
    \bigl(\arg\{q\}+\theta\bigr)\Big|_{-\pi}^{-k_{\rm cr}}
    -\frac{1}{2\pi}
    \arg\{q\}\Big|_{-k_{\rm cr}}^{k_{\rm cr}}
    \nonumber\\
    &\quad
    -\frac{1}{2\pi}
    \bigl(\arg\{q\}+\theta\bigr)\Big|^{\pi}_{k_{\rm cr}},
    \label{top_inv_odd_1}
\end{align}
but \(\theta(k)=\theta(-k)\) and has zero net winding, so the result
again reproduces the SSH invariant.When the \(\mathcal{PT}_{+}\)-broken region occupies part or all of the
Brillouin zone, the real parts of the eigenvalues collapse to zero and
the bands lie on the imaginary axis; the usual gapped-band invariant
then loses its meaning, and the Berry phase must be read as a global
biorthogonal quantity. This global Berry phase remains quantized and
SSH-type, distinct from the periodic odd-\(N\) crystalline invariant
protected at finite frequency by embedding-induced CPH. A symmetric
transverse truncation, \(N_1=N_2\), is the exception: parity is then
inherited from the Hermitian parent, so the OBC Hamiltonian also
possesses CPH, the states do not enter the \(\mathcal{PT}_{+}\)-broken
phase, and the complex Berry phase stays quantized. Thus OBC sectors
and even-\(N\) periodic sectors realize SSH-type projected topology,
whereas crystalline Berry topology beyond the \(38\)-fold
classification, accompanied by parity-related exceptional spectral
reconnections, appears only in the periodic odd-\(N\) sector.

The two routes from Hermitian projection are
therefore topologically distinct in a precise sense: regular sectors
produce band-gapped Hermitian Schur-complement descendants with
SSH-type winding and a bulk-edge correspondence that is conventional in
symmetric terminations and termination-sensitive otherwise (see
Sec.~\ref{sec:BBC}), while the resonant sector produces the
finite-frequency crystalline Berry phase with exceptional spectral braiding established in
Sec.~\ref{subsec:positive_classification}, with no regular
zero-frequency description available.

\section{\texorpdfstring{{ Bulk invariant and boundary response}}{Bulk invariant and boundary response}}
\label{sec:BBC}

We now examine the boundary response of the projected phases, separating
the SSH-type sectors from the resonant crystalline Berry sector. The
essential point is that the resonant sector is NH but
skin-effect free: its CPH complex Berry phase is a Bloch bulk quantity,
while its boundary manifestation is termination-sensitive. Exceptional
spectral braiding occurs at the transition frequencies and is encoded by
local exceptional-point vorticities.
The boundary tested here is a termination of the
projected one-dimensional brane along its translation direction,
distinct from the transverse parent geometry used before projection to
define the brane Hamiltonian.

In the Hermitian SSH model, the bulk invariant and the protected edge modes are locked together by bulk–boundary correspondence. In an NH projected problem one must separate
this relation from a third issue: whether the bulk bands are
reconstructed by the NHSE~\cite{MartinezAlvarezBarriosFoaTorres2018,YaoWang2018,YokomizoMurakami2019}.
The projected brane model separates all three. It is NH but
NHSE-free; its complex Berry phase is a Bloch bulk invariant; and
a conventional SSH-type edge correspondence is recovered only in the
SSH-type sectors. 
In the resonant odd-\(N\) periodic sector the finite-frequency CPH
complex Berry invariant remains a robust bulk quantity, while its
boundary manifestation is termination-sensitive. The exceptional-point
vorticities characterize the critical spectral reconnections but are
not an additive component of the Berry invariant. The
termination-dependent edge modes originate from the SSH-related part of
the Berry winding rather than from a separate bulk braid count.

We begin with the NHSE diagnostic. Introducing the non-Bloch variable
\(\beta=e^{ik}\) and
\begin{equation}
 \mu=\frac{1}{2}\left(\beta+\beta^{-1}\right),
 \qquad
 |q|^{2}=t_x^{2}+t_y^{2}+2t_x t_y\mu,
 \label{eq:main_mu_beta_BBC}
\end{equation}
the projected eigenvalue equations become polynomial equations in
\(\mu\). For open boundary conditions,
\begin{equation}
 \left(E+i|q|F_{N_{1}}\right)
 \left(E+i|q|F_{N_{2}}\right)=|q|^{2},
 \label{eq:main_OBC_bulk_eq_BBC}
\end{equation}
whereas for periodic boundary conditions,
\begin{align}
& \left(q-\frac{(-q^{*})^{N+1}}{|q|^{N}}G_{N}\right)
 \left(q^{*}-\frac{(-q)^{N+1}}{|q|^{N}}G_{N}\right)\nonumber\\
 &=\left(E+i|q|F_N\right)^{2}.
 \label{eq:main_PBC_bulk_eq_BBC}
\end{align}
The detailed reduction is given in Appendix~\ref{app:GBZ_NHSE}, in
particular Eqs.~\eqref{a1}--\eqref{a2} for open boundaries and
Eqs.~\eqref{b1}--\eqref{b2} for periodic boundaries. Using
\(|q|=\omega/(2\sinh\phi)\), the non-Bloch characteristic equation can
be written as
\begin{equation}
 \prod_{s=1}^{M}
 \left[
 \beta^{2}
 -
 \frac{
 \dfrac{\omega^{2}}{4\sinh^{2}\phi_s(E)}-t_x^{2}-t_y^{2}}
 {t_x t_y}\,\beta
 +1
 \right]=0,
 \label{eq:main_BBC_beta_equation}
\end{equation}
which is Eq.~\eqref{beta_eigen_value_problem} of
Appendix~\ref{app:GBZ_NHSE}. Each quadratic factor is palindromic in
\(\beta\), so its two roots obey
\begin{equation}
 \beta_s^{+}\beta_s^{-}=1.
 \label{eq:main_reciprocal_roots_BBC}
\end{equation}
The generalized-Brillouin-zone condition
\cite{non-Bloch_Band_theory} requires the two middle roots to have
equal modulus, \(|\beta_M^-|=|\beta_M^+|\). Together with
Eq.~\eqref{eq:main_reciprocal_roots_BBC}, this gives
\(|\beta|=1\).

The generalized Brillouin zone therefore coincides with the ordinary
one, and the projected model is
NHSE-free~\cite{Yao2018_PRL,Kunst2018_PRL}: its non-Bloch
characteristic polynomial is palindromic, so the complex Berry phase is
a genuine Bloch bulk invariant rather than a skin-effect artifact.
Since the generalized Brillouin zone is the unit circle, the CPH complex
Berry invariant \(\widetilde\nu\) is likewise a Bloch bulk invariant,
stable between exceptional-point transitions and changing when the
spectrum becomes defective. The total parity-symmetric discriminant
winding vanishes on the nondegenerate physical BZ; local EP vorticities
characterize the spectral braid at the transitions.

The real-space origin of this result lies in the structure of the
projection-induced hoppings. The inverse-Fourier derivation of the
coordinate-space Hamiltonians and of the induced hopping amplitudes is
given in Appendix~\ref{App:H_coord_rep}, while the corresponding
non-Bloch and boundary-mode analysis is presented in
Appendix~\ref{app:BBC_details}. For open boundary conditions, the
coordinate-space Hamiltonian is
\begin{align}
 &H_{\rm PTB}^{\rm OBC}
 =\sum_{n,m}\left[
 \varepsilon_{n-m}^{(N_1)}|n,0\rangle\langle m,0|
 +\varepsilon_{n-m}^{(N_2)}|n,1\rangle\langle m,1|
 \right]
 \nonumber\\
 &\quad
 -\sum_n\left[
 t_y|n,0\rangle\langle n,1|
 +t_x|n+1,0\rangle\langle n,1|+{\rm h.c.}
 \right],
 \label{eq:main_H_OBC_coord_summary}
\end{align}
which is Eq.~\eqref{H_OBC_coord_rep}. The coefficients
\(\varepsilon_{n-m}^{(N)}\), given in Eq.~\eqref{eps_sol_1}, decay
exponentially with distance. For periodic boundary conditions one
obtains, in addition, nonlocal inter-sublattice hoppings,
\begin{align}
 H_{\rm PTB}^{\rm PBC}
 &=\sum_{n,m}\varepsilon_{n-m}^{(N)}
 \left(|n,0\rangle\langle m,0|+|n,1\rangle\langle m,1|\right)
 \nonumber\\
 &\quad
 -\sum_{n,m}t_{n-m}^{(N)}
 \left(|m,0\rangle\langle n,1|+|n,1\rangle\langle m,0|\right)
 \nonumber\\
 &\quad
 -\sum_n\left[
 t_y|n,0\rangle\langle n,1|
 +t_x|n+1,0\rangle\langle n,1|+{\rm h.c.}
 \right],
 \label{eq:main_H_PBC_coord_summary}
\end{align}
which is Eq.~\eqref{H_PBC_coord_rep}, with the hoppings given explicitly
in Eq.~\eqref{tau_n-n'_odd}. In all sectors the non-Bloch
characteristic polynomial in Eq.~\eqref{eq:main_BBC_beta_equation} is
palindromic, so the NHSE is absent, but the underlying reason differs by
sector. Under open boundaries, and under periodic boundaries with even
\(N\), the projection-induced hoppings are reciprocal, which makes the
polynomial palindromic. Under periodic boundaries with odd \(N\) the
hoppings are nonreciprocal and no reciprocity argument applies; there
the absence of the NHSE is the  consequence of the
embedding-induced CPH~\cite{Li2024-NHSEBlockade}, a combined-parity symmetry that is present under periodic boundaries but not under open ones.

{
We next summarize the boundary-mode calculation, keeping the detailed
determinant analysis in App.~\ref{app:BBC_details}. For open boundaries along the
brane, the edge-state ansatz}
\begin{equation}
 a_n=\sum_i A_i\lambda_i^n,
 \qquad
 b_n=\sum_i B_i\lambda_i^n
 \label{eq:main_edge_ansatz_BBC}
\end{equation}
{
is inserted into the coordinate-space Schr\"odinger equations. The
bulk part gives the dispersion relation in Eq.~\eqref{dispersion_relation}; the finite-frequency
long-range hopping generates the consistency constraints in 
Eq.~\eqref{conditions_z_s}; and the physical boundary imposes
Eq.~\eqref{boundary_conditions}. The boundary energies are therefore
determined by}
\begin{equation}
 \det \mathcal M(E)=0,
 \label{eq:main_edge_det_BBC}
\end{equation}
{
with \(\mathcal M\) given explicitly in Eq.~\eqref{mode_eq_OBC}. Near
the SSH transition, one decay root remains continuously connected to
the SSH decay factor. The perturbative reduction of
Eq.~\eqref{eq:main_edge_det_BBC} gives an edge-mode criterion in terms
of a parameter \(\eta\), with the localized solution present when
\(|\eta|<1\). The derivation of \(\eta\), its determinant
representation, and representative brane configurations are given in
Appendix~\ref{app:BBC_details}.}

This analysis shows that SSH-type sectors obey the expected edge-mode
correspondence. For even complements, the projected edge modes remain
continuously connected to SSH boundary modes over the finite-frequency
range considered. When an odd complement is present, the same zero-mode
resonance responsible for the singular self-energy can modify the
boundary spectrum at low frequency. Depending on the brane position and
termination, the edge mode can disappear, merge into the bulk, or move
into a parameter region that is topologically trivial from the SSH
viewpoint. Representative complex spectra for open terminations are
shown in SM, Fig.~S15.

{
Periodic boundary conditions in the transverse direction lead to a
different edge problem once the brane is opened along its length. The
coordinate-space equations are given in Eq.~\eqref{difference_eq_PBC},
and the corresponding dispersion relation is Eq.~\eqref{disp_rel_PBC}.
The palindromic characteristic polynomial 
structure persists, so the absence of the NHSE
remains valid. The boundary response, however, is parity dependent. For
even \(N\), the projected Hamiltonian is continuously connected to an
SSH-type problem, and the usual edge-mode correspondence is recovered.
For odd \(N\), the bulk invariant is instead the CPH-quantized
finite-frequency complex Berry phase of Sec.~\ref{sec:Main_top_inv}.
This invariant is a well-defined Bloch bulk invariant because the
generalized Brillouin zone is the unit circle. It does not, however,
imply a universal SSH-like edge-counting rule. In the
\(\mathcal{PT}_{+}\)-broken sector, boundary-localized modes can appear
at large Matsubara frequency and be absorbed into the bulk as
\(\omega\) decreases, as shown in Appendix~\ref{app:PBC_edge_problem}. The mode remains isolated and gapped in the complex spectrum but is no longer edge-localized. Representative complex periodic-boundary spectra are shown in SM, 
Fig.~S16.}

The boundary analysis therefore separates three notions that coincide in
the Hermitian SSH problem: a Bloch bulk invariant, absence of skin
accumulation, and protected edge-mode counting. In the SSH-type sectors,
these notions remain tied together. In the resonant odd-\(N\) periodic
sector, the robust bulk index is the finite-frequency CPH-quantized
complex Berry invariant. Exceptional spectral braiding occurs at the
critical frequencies through local EP vorticities. The boundary
manifestation of the Berry phase depends on the termination and on the
symmetry sector of the boundary states. The resonant sector therefore displays a
termination-sensitive boundary response rather than a universal
SSH-like edge-counting rule.

The odd-\(N\) phase is most directly diagnosed through its bulk
finite-frequency structure: the discriminant zeros in
Eq.~\eqref{crit_lines} locate the exceptional-point transitions, the
complex Berry phase labels the crystalline phases, and the local EP
vorticities characterize the spectral reconnections. The absence of the
NHSE ensures a well-defined Bloch bulk description.
Boundary spectra then reveal how
this bulk structure appears for a chosen termination. In topolectrical
and photonic platforms, these finite-frequency transitions should appear
as response features at the predicted frequency and wave vector. They
occur in both \(\mathcal{PT}_{+}\)-unbroken and
\(\mathcal{PT}_{+}\)-broken regimes and persist in finite systems, as
shown SM, Figs.~S15 and S16.

For completeness we summarize the content of those supplementary
spectra, so that the boundary-response conclusions can be read without
reference to the Supplemental Material. The open-boundary spectra
(SM, Fig.~S15) confirm that, for even complements, the projected edge
mode tracks the SSH boundary mode continuously across the
finite-frequency range, whereas for odd complements the zero-mode
resonance can detach the boundary mode, merge it into the bulk, or shift
it into an SSH-trivial region as \(\omega\) is lowered---exactly the
termination sensitivity described above. The periodic-boundary spectra
(SM, Fig.~S16) confirm that gapless boundary modes on the imaginary axis
are guaranteed only for even \(N\) (preserving \(\mathcal{PT}_+\)),
while for odd \(N\) any boundary-localized mode present at large
\(\omega\) is reabsorbed into the bulk as \(\omega\) decreases, leaving
the bulk crystalline Berry invariant as the robust diagnostic. In all
cases the numerically computed finite-lattice spectra agree with the
analytic bulk results and show no skin accumulation, consistent with the
palindromic non-Bloch polynomial.

\section{Discussion and outlook}
\label{sec:discussion}

The zig-zag construction provides an exact minimal realization of
projection-induced NH topology from a fully Hermitian
parent. The starting point is a two-dimensional nearest-neighbor lattice that is topologically trivial as a bulk system: it has zero Chern number and no protected two-dimensional boundary modes. The bare brane
Hamiltonian contains only the conventional SSH-type structure, whereas
the resonant finite-frequency crystalline complex Berry invariant and
the accompanying exceptional spectral reconnections appear only after
the complement is integrated out. Thus, the projected topology is  a property of the
brane--complement decomposition encoded in the projected Green's
function.

Projection generates topology through two routes. In regular sectors the
projected Green's function has a smooth zero-frequency limit and reduces
to a Hermitian Schur complement, yielding SSH-type descendants; this
route dominates the large-complement limit, including \(N\to\infty\),
where the real part of the projected Hamiltonian recovers the
conventional SSH form. Resonance produces a distinct finite-frequency
route: when a complement zero mode couples to the brane, the self-energy
develops a pole and the projected Hamiltonian has no regular
zero-frequency representative.
In the odd-\(N\) periodic sector, nondegenerate finite-frequency
slices carry a stable CPH complex Berry invariant. At isolated critical
frequencies, EPs reconnect the complex-energy strands and
carry local half-integer braid vorticity. The zero-mode pole thus makes
finite-frequency crystalline Berry topology with exceptional spectral
braiding the natural description of the resonant sector.

The central result is the periodic sector with odd-parity complement.
There the complement carries a sublattice-imbalance zero mode whose
resonant coupling to the brane produces the singular finite-frequency
self-energy, while the embedding supplies a spatial parity that,
together with TRS\(^{\dagger}\), induces CPH of the projected two-band
Hamiltonian.
Embedding parity directly relates  the
EPs, while parity together with TRS$^{\dagger}$ induces
CPH and quantizes the complex Berry phase. The Berry invariant lives on
the finite-frequency projected Green's function and is well defined on
nondegenerate slices. It is distinct from the total discriminant
winding; at the critical frequencies, local half-integer EP vorticities
encode the spectral braid.

The other sectors clarify why the periodic odd-\(N\) case is special.
Periodic even-\(N\) sectors are regular: in the absence of a complement
zero mode, the projection reduces to an SSH-type Schur-complement
descendant. Open-boundary sectors remain in BDI\(^{\dagger}\) class and
display parity-controlled \(\mathcal{PT}_{+}\)-unbroken and
\(\mathcal{PT}_{+}\)-broken regimes, including flat-band segments
generated solely by coupling to the surrounding trivial bulk. Although
odd open-boundary complements can also host a resonant pole, they
generally do not inherit the periodic embedding parity that protects
the crystalline Berry phase and pairs the EPs. Thus, the crystalline Berry phase and the paired EPs are selected by the
conjunction of zero-mode resonance and the spatial constraint imposed by
the periodic embedding.

The bulk--boundary analysis further separates three notions that
coincide in the Hermitian SSH problem: a Bloch bulk invariant, absence
of skin accumulation, and protected edge-mode counting. The projected
model is NH but without skin effect: its coordinate-space
hoppings are such that the non-Bloch characteristic polynomial is
palindromic, and the generalized Brillouin zone coincides with the
ordinary Brillouin zone. Hence the complex Berry phase is a genuine
Bloch bulk invariant rather than a skin-effect artifact. In the
SSH-type sectors, the usual edge-mode correspondence is recovered. In
the resonant odd-\(N\) periodic sector, the invariant is instead a
finite-frequency crystalline complex Berry bulk invariant. It is robust as a bulk
quantity, but its boundary manifestation depends on the termination and
on the symmetry sector of the boundary states, leading to a
termination-sensitive boundary response rather than a universal
SSH-like edge-counting rule.

The spectral variable has a platform-dependent meaning. In the
electronic Green's-function formulation, \(z=i\omega\) labels an
imaginary-frequency slice of the projected response, not a temperature.
Although finite-temperature response functions sample fermionic
Matsubara frequencies, \(\omega_n=(2n+1)\pi k_BT/\hbar\), the invariant
defined here is frequency resolved. Thus,
Figs.~\ref{fig:Phase_diagram_odd_N} and~\ref{fig:Phase_diagram_N=3}
are finite-imaginary-frequency topological maps. Real drive frequencies
enter only after analytic continuation, \(z=\Omega+i0^+\), or through a
platform-specific circuit mapping \(z(\Omega)\).

This finite-frequency topology has a direct response interpretation.
After continuation to the retarded axis, the projected inverse Green's
function gives the brane response. Its complex response eigenvalues
determine the resonance poles through
\begin{equation}
    \det G_{\rm brane}^{-1}(k,\Omega+i0^+)=0 ,
\end{equation}
with the real and imaginary parts encoding resonance frequencies and
inverse lifetimes. Exceptional-point transitions found on the
imaginary-frequency axis continue to real-frequency response features,
where brane resonances approach, coalesce, reconnect, and modify
transmission or impedance. The phase boundaries in
Eq.~\eqref{crit_lines} therefore identify the spectral values at which
these EPs occur; the corresponding drive frequencies
\(\Omega_\ell\) are fixed by \(z(\Omega_\ell)=i\omega_\ell\).

Topolectrical
circuits~\cite{Glazman-2015,Lee2018_Topolectrical,Imhof2018_NatPhys}
provide a natural implementation. Kirchhoff equations are governed by a
circuit Laplacian, and eliminating internal nodes is Kron reduction, the
same Schur-complement operation that defines the projected
kernel~\cite{Chua1987,Dorfler2013_Kron,Caliskan2014_Kron}. A circuit
satisfying Eq.~\eqref{eq:circuit_mapping} encodes the projected response
in its reduced admittance matrix. Since topological circuit bands,
boundary resonances, NH spectral braiding, and skin modes can
be reconstructed from admittance
measurements~\cite{Chen2026_NonAbelianHN}, the predicted
exceptional-point reconnections should appear as changes in
response-eigenvalue trajectories and as pronounced impedance or
transmission features. Appendix~\ref{app:circuit} specifies the \(N=3\) reduced admittance, the
associated transition values, and the constraints for a five-node
realization. Momentum-resolved minima in the transfer response provide a
testable signature of the exceptional-point reconnections, while exact
transmission zeros require the port-conversion conditions derived there.

Classical-wave platforms provide a natural setting because the external
drive directly probes finite-frequency slices of the projected response.
In photonic or acoustic resonator arrays, a zig-zag chain of defect
modes can realize the brane, while the surrounding lattice supplies the
eliminated complement. Recent acoustic experiments demonstrate that lower-dimensional
topological structures can be embedded in an otherwise trivial
bulk~\cite{Teo2025PRL}. Within our scenario, finite-frequency projection generates the embedded topology: a
resonant zero mode of the complement produces complex bands,
exceptional-point transitions, and an embedding-inherited crystalline
Berry phase.

After analytic continuation, the projected complex bands determine the
resonance poles of the brane Green's function. Varying momentum,
boundary phase, or synthetic flux traces their trajectories in the
complex-frequency plane. Across an exceptional-point transition, these
trajectories coalesce and reconnect, accompanied by a jump of the CPH
Berry invariant and local half-integer spectral vorticity. Frequency
sweeps should therefore reveal changes in the defect-channel
transmission near the predicted transition lines. Phase-resolved
transmission, reflection, or surface-impedance measurements, analogous
to probes of photonic Zak phases and boundary response in one-dimensional
systems~\cite{Xiao2014_PRX,Wang2016_PRB}, provide direct experimental
access to the projected finite-frequency topology.

The stable phase invariant is the CPH-quantized complex Berry winding.
Each simple EP carries a local vorticity \(\pm 1/2\),
with the singular braid oriented by a one-sided deformation of the
Brillouin-zone contour. In the present two-band problem the braid
structure is Abelian; embedding parity relates the exceptional-point
pairs and constrains their charges, while non-Abelian braiding requires
multiband spectra.

This structure suggests a broader classification of projected
finite-frequency topology. The relevant data are the resonant zero modes
of the complement, the spatial symmetries that survive projection, and
the Green's-function invariants that control response and boundary
signatures. Varying the parent lattice, brane orientation, codimension,
or crystalline symmetry changes the complement Green's function and
therefore the projected invariant. Disorder, interactions,
superconducting pairing, and driving can be incorporated at the parent
level before projection.

The complex spectrum provides an additional geometric diagnostic. As
discussed in App.~\ref{app:BBC_details}, the projected bands trace
oriented loops in the complex-energy plane as momentum winds through the
Brillouin zone. Discriminant zeros mark exceptional-point transitions,
where spectral sheets touch and exchange locally; across these
reconnections the Berry index can jump, and each simple touching carries
half-integer braid vorticity. This connects the projected-brane
construction to the knot-theoretic description of NH band
topology, where complex spectra, exceptional links, Fermi--Seifert
surfaces, and eigenvalue braids provide natural topological
structures~\cite{Carlstrom2018_PRA,Yang2019_PRB,Wojcik2022_PRB}.

Projection of a Hermitian parent thus yields an effective
NH topological response. In the resonant sector, the
zero-mode pole makes the zero-frequency limit singular, while the
nondegenerate finite-frequency slices carry a crystalline Berry phase
separated by exceptional spectral reconnections. Embedding parity
organizes the exceptional-point pairs, and together with
TRS\(^{\dagger}\) induces CPH, thereby quantizing the complex Berry
phase beyond the regular SSH-type descendants.

\begin{figure*}[htbp]
 \centering
 \includegraphics[width=1\textwidth]{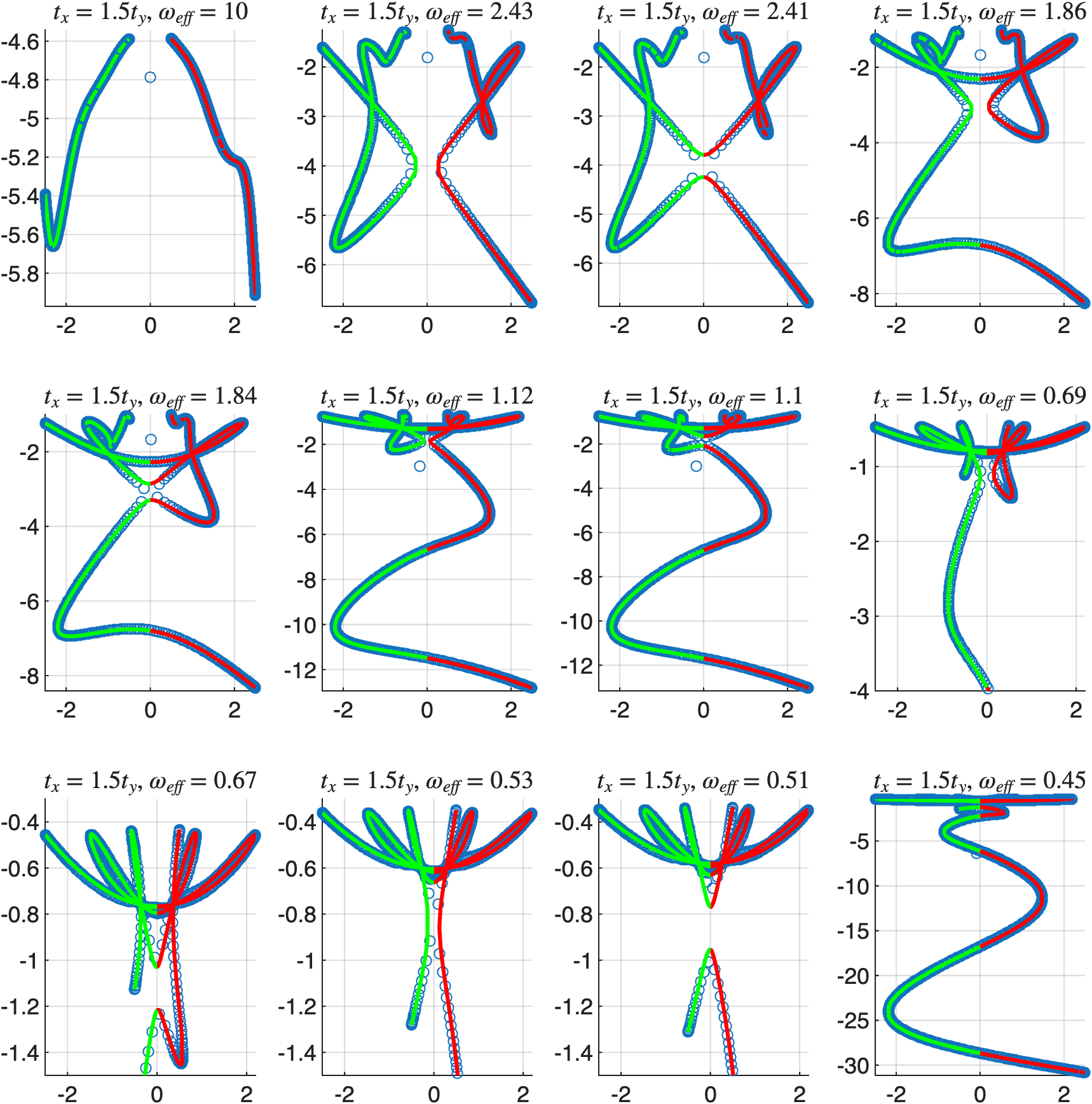}
 \captionsetup{justification=justified,singlelinecheck=false}
 \caption{This figure shows the change in the complex energy spectrum of the Hamiltonian with periodic boundary conditions with $N=3$ when the system moves through different phases of the \(N=3\) finite-frequency phase diagram. The system is tuned to the topologically nontrivial phase by setting $t_{x}=1.5t_{y}$ which corresponds to the blue cut through the phase diagram. From left to right, the panels correspond to decreasing values of the Matsubara frequency $\omega_{eff}=\frac{\omega}{2t_{y}\sinh\phi_{cr}}$. For large $\omega$, the spectrum remains relatively simple. As $\omega$ is lowered, the spectrum exhibits a much more intricate structure, which is analyzed in detail in Appendix \ref{app:top_inv}. The exact eigenvalues derived analytically from (\ref{Energy_periodic_BC}) are shown in red for the upper band and in green for the lower band, whereas the numerically obtained eigenvalues are represented by blue circles. The numerical calculations were carried out on a lattice with $K=500$ unit cells. A degenerate gapped edge mode appears as an isolated blue circle in about half of the panels (it in fact exists in all of them, but the lower part of the spectrum is sometimes omitted to better highlight the evolution of the spectrum as $\omega$ decreases), showing a finite-size spectral feature associated with the parity-quantized bulk invariant rather than a conventional protected edge-mode correspondence. Notice that as $\omega$ is reduced the gap becomes larger.}
 \label{fig:Spektar_N=3_t_x>t_y}
\end{figure*}
\begin{figure*}[htbp]
 \centering
 \includegraphics[width=1\textwidth]{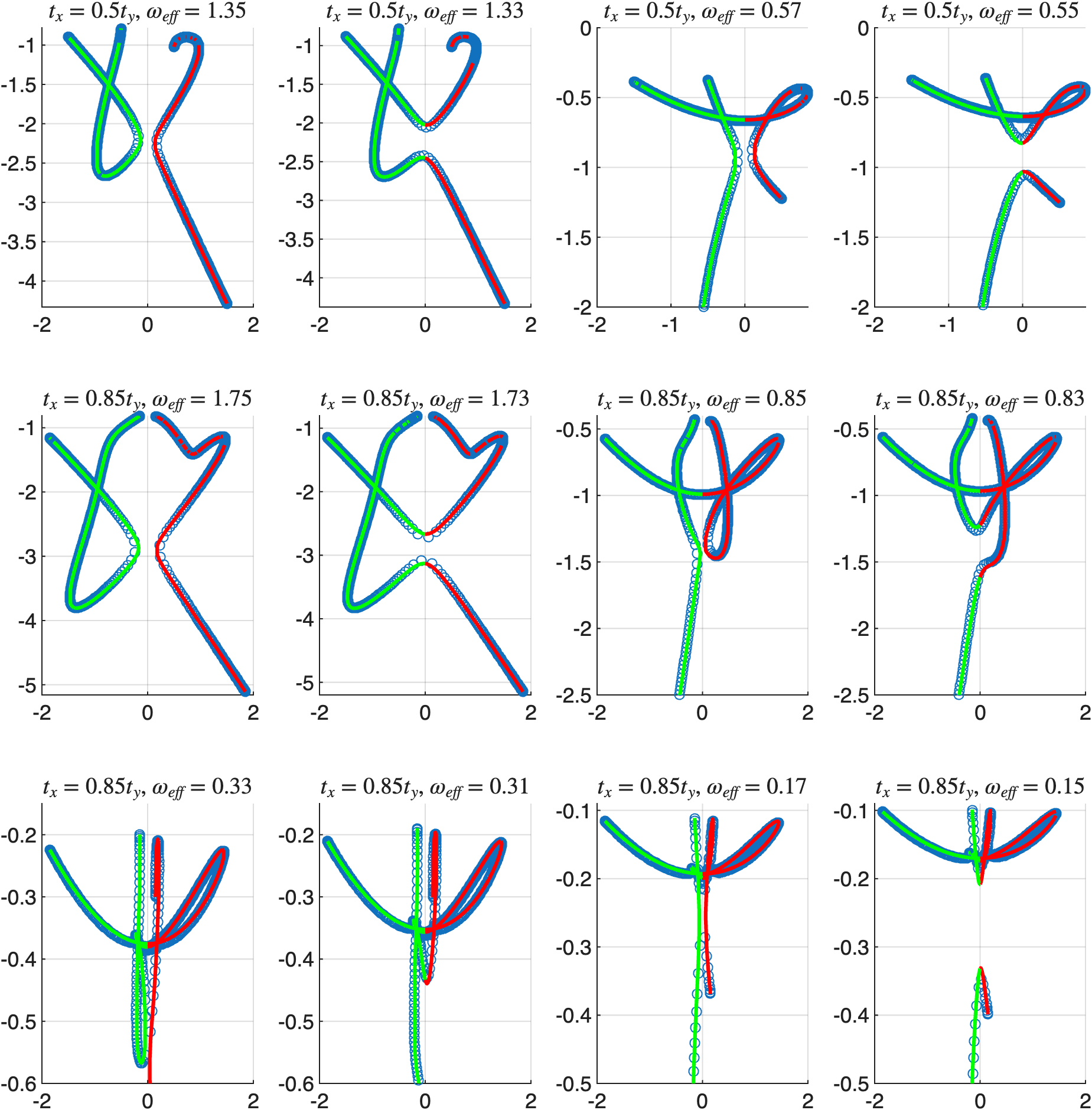}
 \captionsetup{justification=justified,singlelinecheck=false}
 \caption{This figure illustrates how the complex energy spectrum of the Hamiltonian with periodic boundary conditions and $N=3$ evolves as the system moves through different phases of the \(N=3\) finite-frequency phase diagram. The parameters are chosen such that the system lies in the topologically trivial phase. The first row corresponds to the red cut through the phase diagram, specified by $t_{x}=0.5t_{y}$, while the second and third rows correspond to the green cut, defined by $t_{x}=0.85t_{y}$. From left to right, the panels display results for progressively smaller Matsubara frequencies $\omega_{eff}=\frac{\omega}{2t_{y}\sinh\phi_{cr}}$. For large $\omega$, the spectrum remains relatively simple. As $\omega$ decreases, the spectrum exhibits a much more intricate structure, which is analyzed in detail in Appendix \ref{app:top_inv}. The exact eigenvalues obtained analytically from (\ref{Energy_periodic_BC}) are shown in red for the upper band and in green for the lower band, whereas the numerically computed eigenvalues are indicated by blue circles. The numerical simulations were carried out on a lattice with $K=500$ unit cells. No corresponding degenerate gapped mode is present here, consistent with the interpretation of the odd-$N$ periodic response as refined and symmetry-sensitive rather than a conventional edge-mode bulk--boundary correspondence.}
 \label{fig:Spektar_N=3_t_x<t_y}
\end{figure*}

\section*{Acknowledgments}
We thank Bitan Roy for the critical reading of the manuscript. The authors also thank Ana \Dj or\dj evi\'c for useful insights throughout the writing of this paper. The authors acknowledge the funding provided by the Faculty of Physics of the University of Belgrade, through a grant number 451-03-47/2023-01/200162 from the Ministry of Education, Science, and Technological Development of the Republic of Serbia and by the Science Fund of the Republic
of Serbia, Serbian Scientific Cooperation Program with the Diaspora:
Support for Visits of Diaspora Scientists, under Project TopArtGravity
No.~220. V. J. acknowledges the support by Fondecyt (Chile) Grant No. 1230933. 


\appendix

\renewcommand{\thesubsection}{\Alph{section}.\arabic{subsection}}
\renewcommand{\thesubsubsection}{\Alph{section}.\arabic{subsection}.\arabic{subsubsection}}

\setcounter{figure}{0}
\renewcommand{\thefigure}{A\arabic{figure}}
\section*{APPENDIX}

Appendices~\ref{Calc_det}--\ref{app:topo-invariant} give the construction of the projected brane Hamiltonian, the zero-mode-resonant projection framework, the spectral ingredients entering the Berry phase, and the classification of the corresponding invariants. Appendix~\ref{app:BBC_details} gives {the spectral-braid interpretation together with} the technical details of bulk--boundary correspondence. {  Predicted topolectrical signatures are derived in
App.~\ref{app:circuit}.} Further spectral data,  finite-frequency phase diagrams, density-of-states calculations, and additional edge-mode spectra are collected in the Supplemental Material. 

\section{Construction of the brane Hamiltonian}\label{Calc_det}
Here we describe the procedure for computing the brane Hamiltonian (\ref{eq:PTB_frequency}) for both open and periodic boundary conditions. In the case of the open boundary conditions, both matrices $A_{N_{1}}$ and $A_{N_{2}}$, appearing in (\ref{H_22_open}) have the following block-diagonal form, with the $2\times2$ blocks on the diagonal:
\begin{equation}
 A_{N}=\begin{pmatrix}
 -i\omega & q^{*} & 0 & 0 & \cdots & 0 &0\\
 q & -i\omega & q^{*} & 0 & \cdots & 0& 0\\
 0 & q & -i\omega & q^{*} & \cdots & 0 & 0\\
 \vdots & \vdots & \vdots & \vdots & \ddots & \vdots & \vdots\\
 0 & 0 &0 & 0 & \cdots & -i\omega & q^{*}\\
 0 & 0 &0 & 0 & \cdots & q &-i\omega
 \end{pmatrix}_{N\times N}\label{matric_A_N}
\end{equation}
where $q\equiv -\left(t_{y}+t_{x}e^{ikd}\right)$ is defined in (\ref{q_deff}). Since the elements of the inverse matrix that are needed:
\begin{equation}
 \langle{-1}|\left(\Tilde{H}_{22}-i\omega\right)^{-1}|-1\rangle,\langle{2}|\left(\Tilde{H}_{22}-i\omega\right)^{-1}|2\rangle\neq0,
\end{equation}
correspond to the first and last diagonal matrix elements of the inverse matrix of (\ref{matric_A_N}), and are mutually equal. These elements are expressed in terms of the determinants of matrices $A_{N}$ in the following manner:
\begin{equation}
 \left(A_{N}\right)_{N,N}=\frac{\det{A_{N-1}}}{\det{A_{N}}},\label{A_N_N}
\end{equation}
which means that the problem is reduced to the calculation of the determinant of matrix $A_{N}$. 

In the case of the periodic boundary conditions, the $\tilde{H}_{22}-i\omega$ matrix does not decompose into two diagonal blocks. Rather, it has the following form:
\begin{equation}
 \Tilde{H}_{22}-i\omega\mathbb{I}=\begin{pmatrix}[cccc|cccc]
 -i\omega & q^{*} & & & & & & q\\
 & \ddots & & \\
 & q & -i\omega& q^{*}\\
 & & q & -i\omega & 0\\\hline
 & & & 0 & -i\omega & q^{*}\\
 & & & & q & -i\omega & q^{*}\\
 & & & & & & \ddots\\
 q^{*} & & & & & & q & -i\omega
 \end{pmatrix}.\label{H_22_periodic}
\end{equation}
This is an $N\times N$ matrix. To find the needed matrix elements $(\Tilde{H}_{22}-i\omega)^{-1}_{-1,-1}$, $(\Tilde{H}_{22}-i\omega)^{-1}_{-1,2}$, $(\Tilde{H}_{22}-i\omega)^{-1}_{2,-1}$, $(\Tilde{H}_{22}-i\omega)^{-1}_{2,2}$ we first transform (\ref{H_22_periodic}) with the help of an auxiliary matrix $\Sigma$:
\begin{equation}
 \Sigma=\begin{pmatrix}[c|c]
 0 & \mathbb{I}\\\hline
 \mathbb{I} & 0
 \end{pmatrix}\label{Sigma_matrix}
\end{equation}
We also introduce the $\Sigma$-transformation generated by the matrix $\Sigma$, in the following manner:
\begin{equation}
 \begin{pmatrix}[c|c]
 A & B\\\hline
 C & D
 \end{pmatrix}\mapsto\Sigma\begin{pmatrix}[c|c]
 A & B\\\hline
 C & D
 \end{pmatrix}\Sigma=\begin{pmatrix}[c|c]
 D & C\\\hline
 B & A
 \end{pmatrix}
\end{equation}
The $\Sigma$-transformation of the Hamiltonian $\Tilde{H}_{22}-i\omega\mathbb{I}$ is:
\begin{equation}
 \Sigma\left(\Tilde{H}_{22}-i\omega\mathbb{I}\right)\Sigma=\begin{pmatrix}
 -i\omega & q^{*} & 0 & \cdots & 0 & 0\\
 q & -i\omega & q^{*} & \cdots & 0 & 0\\
 \vdots & \vdots & \vdots &\ddots & \vdots & \vdots\\
 0 & 0 & 0 & \cdots &-i\omega & q^{*}\\
 0 & 0 & 0 & \cdots & q & -i\omega
 \end{pmatrix},
\end{equation}
which is once again the same matrix $A_{N}$ that we encounter in the case of open boundary conditions (\ref{matric_A_N}). Since $\Sigma^{2}=\mathbb{I}$, the transformation of the inverse matrix is the inverse of the transformed matrix. 

The  matrix elements of the $2\times2$ blocks of the complement resolvent can be written
in terms of the determinants of the tridiagonal matrices \(A_N\) as
\begin{align}
 \left(\widetilde{H}_{22}-i\omega\right)^{-1}_{-1,-1}
 &=
 \left(\widetilde{H}_{22}-i\omega\right)^{-1}_{2,2}
 =
 \frac{\det A_{N-1}}{\det A_N},
 \label{eq:diag_element}
 \\
 \left(\widetilde{H}_{22}-i\omega\right)^{-1}_{2,-1}
 &=
 \frac{\left(-q^{*}\right)^{N-1}}{\det A_N},
 \label{eq:offdiag_21}
 \\
 \left(\widetilde{H}_{22}-i\omega\right)^{-1}_{-1,2}
 &=
 \frac{\left(-q\right)^{N-1}}{\det A_N}.
 \label{eq:offdiag_12}
\end{align}
The determinant \(\det A_N\) obeys the recurrence relation
\begin{equation}
    \det A_{N+2}
    +
    i\omega\,\det A_{N+1}
    +
    |q|^2 \det A_N
    =
    0,
\end{equation}
with initial conditions
\begin{equation}
    \det A_1=-i\omega,
    \qquad
    \det A_2=-\omega^2-|q|^2 .
\end{equation}
Solving this recurrence gives
\begin{equation}
    \det A_N
    =
    |q|^N
    \frac{
    \sinh\!\left[(N+1)\left(\phi-i\frac{\pi}{2}\right)\right]
    }{
    \sinh\!\left(\phi-i\frac{\pi}{2}\right)
    },
    \label{eq:det_A_N}
\end{equation}
where
\begin{equation}
    \phi=\operatorname{arcsinh}\left(\frac{\omega}{2|q|}\right).
\end{equation}

For real \(\omega\) and \(|q|>0\), \(\phi\) is real. Setting
\begin{equation}
    z=\phi-i\frac{\pi}{2},
\end{equation}
and using
\begin{equation}
    \frac{\sinh[(N+1)z]}{\sinh z}
    =
    U_N(\cosh z),
\end{equation}
where \(U_N\) is the Chebyshev polynomial of the second kind, we find
\begin{equation}
    \cosh z
    =
    -i\frac{\omega}{2|q|}.
\end{equation}
Hence Eq.~\eqref{eq:det_A_N} becomes
\begin{equation}
    \det A_N
    =
    |q|^N
    U_N\!\left(-i\frac{\omega}{2|q|}\right).
\end{equation}
Equivalently, using the polynomial representation of \(U_N\),
\begin{equation}
    \det A_N
    =
    (-i)^N
    \sum_{k=0}^{\lfloor N/2\rfloor}
    \binom{N-k}{k}
    |q|^{2k}\,
    \omega^{N-2k}.
    \label{eq:det_poly}
\end{equation}

Equation~\eqref{eq:det_poly} makes the parity dependence explicit. For
\(N=2r\),
\begin{equation}
    \det A_{2r}
    =
    (-1)^r
    \sum_{k=0}^{r}
    \binom{2r-k}{k}
    |q|^{2k}\,
    \omega^{2r-2k},
\end{equation}
which is purely real. In particular,
\begin{equation}
    \det A_{2r}(0)=(-1)^r |q|^{2r},
\end{equation}
so the determinant remains finite at zero frequency. By contrast, for
\(N=2r+1\),
\begin{equation}
    \det A_{2r+1}
    =
    (-1)^r(-i)
    \sum_{k=0}^{r}
    \binom{2r+1-k}{k}
    |q|^{2k}\,
    \omega^{2r+1-2k},
\end{equation}
which is purely imaginary and contains only odd powers of \(\omega\).
Thus, near \(\omega=0\),
\begin{equation}
    \det A_{2r+1}
    =
    (-1)^r(-i)(r+1)|q|^{2r}\omega
    +
    O(\omega^3).
    \label{eq:det_odd_small}
\end{equation}
Therefore, for odd \(N\), \(\det A_N\) vanishes linearly at zero
frequency, producing a pole in the complement resolvent.

We now evaluate this pole explicitly. Let
\begin{equation}
    N=2r+1,
    \qquad r=0,1,2,\ldots .
\end{equation}
Then
\begin{equation}
    \det A_{N}
    =
    \det A_{2r+1}
    =
    (-1)^r(-i)(r+1)|q|^{2r}\omega
    +
    O(\omega^3),
\end{equation}
while
\begin{equation}
    \det A_{N-1}
    =
    \det A_{2r}
    =
    (-1)^r |q|^{2r}
    +
    O(\omega^2).
\end{equation}
Substitution into Eqs.~\eqref{eq:diag_element}--\eqref{eq:offdiag_12} then gives
the $2\times2$ block of the
singular part of the complement resolvent is
\begin{widetext}
\begin{equation}
\left(\widetilde{H}_{22}-i\omega\right)^{-1}
=
\frac{2i}{(N+1)\omega}
\begin{pmatrix}
1 &
(-1)^{(N-1)/2}
\left(q/|q|\right)^{N-1}
\\[4pt]
(-1)^{(N-1)/2}
\left(q^{*}/|q|\right)^{N-1}
&
1
\end{pmatrix}
+
O(\omega).
\label{eq:endpoint_resolvent}
\end{equation}
\end{widetext}
Thus, for odd \(N\), all matrix elements of  $2\times2$ blocks display the same
\(1/\omega\) singularity. This pole is the resolvent manifestation of
the chiral zero mode of the eliminated complement in agreement with the general structure of the principal part of the complement resolvent, see Eq.~\eqref{eq:gen_Heff}.


\section{General framework for zero-mode-resonant projection topology}
\label{app:general_framework}

This appendix summarizes the model-independent structure behind
zero-mode-resonant projection. We derive the projected Green's
function, the zero-mode pole in the embedding self-energy, the
positivity of the resonance matrix, and the spatial-symmetry constraint
that allows the finite-frequency projected problem to carry a
crystalline invariant beyond the internal \(38\)-fold classification.

\subsection{Projection, zero-mode pole, and resonance matrix}

Let
\(\mathcal{H}=\mathcal{H}_{b}\oplus\mathcal{H}_{c}\) be a
brane--complement decomposition of a Hermitian parent Hamiltonian, with
block form given in Eq.~\eqref{eq:block_parent}. The exact projected
brane Green's function is
\begin{equation}
    G_b(k,z)
    =
    \bigl[z-H_{11}(k)-\Sigma(k,z)\bigr]^{-1},
    \label{eq:gen_Gb}
\end{equation}
where
\begin{equation}
    \Sigma(k,z)
    =
    H_{12}(k)\bigl[z-H_{22}(k)\bigr]^{-1}H_{12}^{\dagger}(k)
    \label{eq:gen_Sigma}
\end{equation}
is the embedding self-energy. With this convention,
\[
    H_{\mathrm{eff}}(k,z)
    =
    -G_b^{-1}(k,z)
    =
    H_{11}(k)-z+\Sigma(k,z),
\]
which reduces to \(H_{\mathrm{PTB}}(k,i\omega)\) of
Eq.~\eqref{eq:PTB_frequency} for \(z=i\omega\).

If the complement has a zero-mode subspace
\(\mathcal{H}_{0}(k)=\ker H_{22}(k)\), with projector \(P_0(k)\), its
resolvent separates into a pole and a regular part. Consequently,
\begin{equation}
    \Sigma(k,z)
    =
    \frac{\Gamma(k)}{z}
    +
    \Sigma_{\mathrm{reg}}(k,z),
    \label{eq:gen_Sigma_decomp}
\end{equation}
where \(\Sigma_{\mathrm{reg}}\) is analytic at \(z=0\), and
\begin{equation}
    \Gamma(k)
    =
    H_{12}(k)P_0(k)H_{12}^{\dagger}(k)
    \label{eq:resonance_matrix}
\end{equation}
is the resonance matrix.

The matrix \(\Gamma(k)\) is positive semidefinite, since
\[
    \Gamma(k)=A(k)A^{\dagger}(k),
    \qquad
    A(k)=H_{12}(k)P_0(k),
\]
and
\[
    {\rm rank}\,\Gamma(k)\leq \dim\mathcal{H}_0(k).
\]
Moreover, \(\Gamma(k)=0\) if and only if
\(P_0(k)H_{12}^{\dagger}(k)=0\), i.e., when the complement zero modes
do not couple to the brane.

At imaginary frequency \(z=i\omega\), with
\(\omega\in\mathbb{R}\setminus\{0\}\), the projected Hamiltonian takes
the singular form
\begin{equation}
    H_{\mathrm{eff}}(k,i\omega)
    =
    H_{11}(k)-i\omega
    -\frac{i}{\omega}\Gamma(k)
    +
    \Sigma_{\mathrm{reg}}(k,i\omega).
    \label{eq:gen_Heff}
\end{equation}
For \(\omega>0\), the pole term has eigenvalues
\(-i\lambda_j/\omega\), with \(\lambda_j\geq0\), and therefore gives a
definite anti-Hermitian contribution, consistent with
\(\bar E=E-i\hbar/\tau\). If \(\Gamma(k_0)\neq0\), this singular term
cannot be removed by an \(\omega\)-independent unitary change of brane
basis. A resonant projection therefore has no regular zero-frequency
Schur-complement representative, and its topology is naturally defined
for the projected Green's function at fixed finite frequency.

\subsection{Symmetry constraints inherited from the embedding}
\label{subsec:symmetry-inheritance-embedding}

We now record how parent symmetries constrain the resonance matrix. Let
\(U_g\) be a unitary internal symmetry that preserves the
brane--complement decomposition,
\begin{equation}
    U_g=U_g^{(b)}\oplus U_g^{(c)} .
    \label{inhereted_symmetry}
\end{equation}
If \(U_g\) commutes with the parent Hamiltonian, then
\[
    [H_{11},U_g^{(b)}]=0,\,
    [H_{22},U_g^{(c)}]=0,\,
U_g^{(b)}H_{12}=H_{12}U_g^{(c)} .
\]
Since \([H_{22}(k),U_g^{(c)}]=0\), one also has
\([P_0(k),U_g^{(c)}]=0\), and hence
\[
    U_g^{(b)}\Gamma(k)(U_g^{(b)})^{-1}
    =
    \Gamma(k).
\]
Thus internal symmetries inherited blockwise from the parent constrain
\(\Gamma(k)\) covariantly within the brane sector.

Spatial symmetries impose a stronger constraint because they relate
different momenta. Let \(\mathcal{P}\) be an order-two parity symmetry,
\[
    \mathcal{P}H(k)\mathcal{P}^{-1}=H(-k),
\]
which preserves the brane--complement decomposition,
\[
    \mathcal{P}=P^{(b)}\oplus P^{(c)} .
\]
Then \(P^{(c)}\) maps zero modes of \(H_{22}(k)\) to zero modes of
\(H_{22}(-k)\), or
\begin{equation}
    P^{(c)}P_0(k)=P_0(-k)P^{(c)} .
    \label{eq:P0_parity}
\end{equation}
Together with
\(P^{(b)}H_{12}(k)=H_{12}(-k)P^{(c)}\), this yields
\begin{equation}
    P^{(b)}\Gamma(k)\bigl(P^{(b)}\bigr)^{-1}
    =
    \Gamma(-k).
    \label{eq:parity_Gamma}
\end{equation}
Equation~\eqref{eq:parity_Gamma} is the central crystalline constraint:
it is inherited from the full embedding, not from the isolated brane
Hamiltonian alone.

\subsection{Sufficient conditions and relation to the \(38\)-fold scheme}

The finite-frequency topology of \(H_{\mathrm{eff}}(k,i\omega)\) is
controlled by the map \(k\mapsto H_{\mathrm{eff}}(k,i\omega)\), subject
to its internal NH symmetry class and to the crystalline
constraint in Eq.~\eqref{eq:parity_Gamma}. A sufficient route beyond
the internal \(38\)-fold scheme requires four ingredients: a resonant pole,
\(\Gamma(k_0)\neq0\); nontrivial momentum dependence of the resonance;
an embedding symmetry constraining \(\Gamma(k)\); and an internal class
which does not itself provide the observed one-dimensional invariant.

In the minimal model, even-\(N\) complements give regular
\(\mathrm{BDI}^{\dagger}\) SSH-type sectors. Odd-\(N\) periodic
complements are resonant and yield a projected
\(\mathrm{AI}^{\dagger}\) Hamiltonian, for which the internal
classification supplies no corresponding invariant in the present
finite-frequency setting. The missing structure is embedding parity
\(P^{(b)}=\sigma_x\), which directly enforces
$Q_1(k)=Q_2(-k)$ and, together with TRS$^{\dagger}$, induces CPH and
quantizes the complex Berry phase. The parity relation independently
constrains the pairing and local vorticities of the EPs.

In a CPH-adapted representation, the invariant is
\begin{equation}
    \tilde\nu
    =
    -\frac{1}{2\pi}
    \oint_{\mathrm{BZ}}
    \partial_k\,\arg\{Q(k,i\omega)\}\,\mathrm{d}k
    \in\mathbb{Z},
    \label{eq:gen_inv}
\end{equation}
where \(Q(k,i\omega)\) denotes the phase entering
\(\sqrt{Q_2/Q_1}\) in the biorthogonal eigenvectors. In the zig-zag
model this becomes \(\widetilde\nu_\pm\) of
Eq.~\eqref{top_inv_odd}. This is the CPH Berry invariant. It must be
distinguished from the discriminant braid winding, which is the sum,
rather than the difference, of the off-diagonal windings. The Berry
invariant is defined at fixed noncritical \(\omega\neq0\), changes at
finite-frequency exceptional degeneracies, and relies on embedding
parity rather than on the internal \(38\)-fold class alone.


\section{Parent spectra, determinant formulas, and complement parity}
\label{app:parent_spectra_symmetry}

In this Appendix we collect the technical details behind the compact
discussion of Sec.~\ref{sec:min-model}. We derive the transverse parent
spectra for open and periodic boundary conditions, give the determinant
formulas entering the chiral winding analysis, and identify the
zero-mode condition of the eliminated complement. General symmetry
inheritance and the parity quantization of the complex Berry phase are
discussed separately in Appendix~\ref{app:general_framework}.

\subsection{Open transverse geometry}

At fixed momentum \(k\) along the brane, the two-dimensional parent
Hamiltonian reduces to a one-dimensional transverse chain with hopping
amplitudes \(q\) and \(q^{*}\). For open boundary conditions, the
transverse parent Hamiltonian with \(K\) sites can be written as
\begin{equation}
    H_{\rm OBC}(k)
    =
    -\sum_{m=0}^{K-1}
    \left[
        q\,|m+1\rangle\langle m|
        +
        q^{*}|m\rangle\langle m+1|
    \right],
    \label{eq:app_H_OBC_parent}
\end{equation}
where
\begin{equation}
    q=|q|e^{i\varphi_q}.
\end{equation}
The phase of \(q\) can be removed by the local gauge transformation
\begin{equation}
    |m\rangle \mapsto e^{-im\varphi_q}|m\rangle .
    \label{eq:app_OBC_gauge}
\end{equation}
The Hamiltonian is then equivalent to a real nearest-neighbor chain
with hopping magnitude \(|q|\). Its eigenvalues are therefore
\begin{equation}
    E_n
    =
    2|q|\cos\left(\frac{n\pi}{K+1}\right),
    \qquad n=1,\ldots,K ,
    \label{eq:app_OBC_spectrum}
\end{equation}
up to the conventional ordering of the levels. For odd \(K\), the state
with \(n=(K+1)/2\) is pinned to zero energy. In the real gauge, this
zero mode has support on only one sublattice,
\begin{equation}
    |\psi_0\rangle
    =
    \sqrt{\frac{2}{K+1}}
    \sum_{\ell=1}^{(K+1)/2}
    (-1)^{\ell-1}
    |2\ell-1\rangle ,
    \label{eq:app_OBC_zero_mode}
\end{equation}
with the sublattice choice depending on the endpoint convention. This
sublattice-imbalance zero mode is the origin of the resonant pole when
the corresponding zero mode belongs to the eliminated complement and
couples to the brane.

For even \(K=2M\), the OBC parent is bipartite. Grouping odd and even
transverse sites into two sublattices brings the Hamiltonian to chiral
form,
\begin{equation}
    H_{\rm OBC}(k)
    =
    \begin{pmatrix}
        0 & D_{\rm OBC}(k)\\
        D_{\rm OBC}^{\dagger}(k) & 0
    \end{pmatrix}.
    \label{eq:app_OBC_chiral_form}
\end{equation}
With the ordering convention used in the main text, the block
\(D_{\rm OBC}(k)\) is triangular up to nearest-neighbor off-diagonal
entries, and its determinant is
\begin{equation}
    \det D_{\rm OBC}(k)
    =
    \left(q^{*}\right)^M,
    \qquad M=\frac{K}{2},
    \label{eq:app_det_D_OBC}
\end{equation}
up to an overall sign that does not affect the winding. The
one-dimensional chiral invariant can be written as the winding of
\(\det D_{\rm OBC}(k)\)~\cite{Chiu2016_RMP,Asboth2016_ShortCourse},
\begin{align}
    \tilde{\nu}
    &=
    \frac{1}{2\pi i}
    \int_{-\pi}^{\pi} dk\,
    \partial_k
    \ln \det D_{\rm OBC}(k)
    \nonumber\\
    &=
    \left\lfloor\frac{K}{2}\right\rfloor
    \frac{\arg q(k)\big|_{-\pi}^{\pi}}{2\pi},
    \label{eq:app_OBC_winding}
\end{align}
where the overall sign depends on the ordering convention for the two
sublattices. Thus the open transverse geometry realizes an SSH-like
parent topology associated with the boundary truncation.

\subsection{Periodic transverse geometry}

For periodic boundary conditions, the transverse direction is closed
into a ring. The parent Hamiltonian is
\begin{equation}
    H_{\rm PBC}(k)
   =
    -\sum_{m=1}^{K}
    \left[
        q\,|m+1\rangle\langle m|
        +
        q^{*}|m\rangle\langle m+1|
    \right],
    \label{eq:app_H_PBC_parent}
\end{equation}
where $|K+1\rangle\equiv |1\rangle $. It  is circulant and  diagonalized by transverse momenta
\begin{equation}
    p_n=\frac{2\pi n}{K},
    \qquad n=0,\ldots,K-1.
\end{equation}
With the convention of the main text, the spectrum is
\begin{equation}
    E_n(k)
    =
    -2t_y\cos p_n
    -2t_x\cos(k+p_n).
    \label{eq:app_PBC_spectrum}
\end{equation}
The overall sign depends on the convention used for \(q\) and does not
affect the symmetry or winding conclusions.

For even \(K=2M\), the periodic ring is bipartite and the Hamiltonian
can be written in chiral form. The off-diagonal block is a circulant
matrix. Up to an overall sign convention,
\begin{equation}
    \det D_{\rm PBC}(k)
    =
    \left(q^{*}\right)^M
    -
    \left(-q\right)^M .
    \label{eq:app_det_D_PBC}
\end{equation}
Writing \(q=|q|e^{i\varphi_q}\), this becomes
\begin{equation}
    \det D_{\rm PBC}(k)
    =
    |q|^M
    \left[
        e^{-iM\varphi_q}
        -
        (-1)^M e^{iM\varphi_q}
    \right].
    \label{eq:app_det_D_PBC_phase}
\end{equation}
For even \(M\), this expression is purely imaginary up to an overall
convention-dependent phase; for odd \(M\), it is purely real up to the
same type of convention-dependent phase. Therefore, away from zeros of
\(\det D_{\rm PBC}(k)\), its trajectory is confined to a line in the
complex plane and cannot wind around the origin. The even-\(K\) PBC
parent is thus topologically trivial despite belonging to a chiral
symmetry class.

For odd \(K\), the periodic transverse ring contains an odd cycle and
cannot be bipartitioned. Chiral symmetry is absent, and the parent
belongs to the corresponding nonchiral class. This is distinct from the
eliminated complement after projection, whose parity controls whether a
zero mode appears in the self-energy.

\subsection{Gauge structure and flux interpretation}

The distinction between OBC and PBC can also be understood from the
phase of \(q\). Under OBC, the local transformation in
Eq.~\eqref{eq:app_OBC_gauge} removes \(\varphi_q\) from every hopping
matrix element. Under PBC, the same transformation removes the phase
locally but leaves a residual phase on the boundary link,
\begin{equation}
    |K\rangle\langle 1|
    \mapsto
    e^{iK\varphi_q}|K\rangle\langle 1| .
    \label{eq:app_PBC_boundary_phase}
\end{equation}
Thus the phase of \(q(k)\) becomes a total flux
\begin{equation}
    \Phi(k)=K\varphi_q(k)\quad {\rm mod}\;2\pi
    \label{eq:app_flux}
\end{equation}
threading the transverse ring. This Aharonov--Bohm flux distinguishes
the periodic geometry from the open one. It does not, however, produce
a nonzero parent winding in the even-\(K\) PBC case, as shown by
Eq.~\eqref{eq:app_det_D_PBC_phase}.

\subsection{Complement determinant and zero-mode condition}

The resonant projection mechanism depends on zero modes of the
eliminated complement, not simply on zero modes of the full parent
strip. After the two brane sites are separated from the transverse
problem, the complement block is a nearest-neighbor chain with \(N\)
sites and hopping amplitudes \(q\) and \(q^{*}\). At zero frequency its
matrix has the tridiagonal form
\begin{equation}
    A_N =
    -\sum_{m=1}^{N-1}
    \left[
        q\,|m+1\rangle\langle m|
        +
        q^{*}|m\rangle\langle m+1|
    \right].
    \label{eq:app_complement_matrix}
\end{equation}
Let
\begin{equation}
    \mathcal{D}_N=\det A_N.
\end{equation}
 We notice that the vanishing of
\(\mathcal{D}_N\) for odd \(N\)  signals the complement zero mode and
hence the resonant pole.
The determinant obeys the recursion relation
\begin{equation}
    \mathcal{D}_N
    =
    -|q|^2\mathcal{D}_{N-2},
    \qquad
    \mathcal{D}_0=1,\qquad \mathcal{D}_1=0 .
    \label{eq:app_complement_recursion}
\end{equation}
Therefore
\begin{equation}
    \mathcal{D}_{2M}
    =
    (-|q|^2)^M,
    \qquad
    \mathcal{D}_{2M+1}=0 .
    \label{eq:app_complement_determinant}
\end{equation}
Even-\(N\) complements are zero-mode-free for \(|q|\neq0\), whereas
odd-\(N\) complements host an exact zero mode protected by sublattice
imbalance. The corresponding normalized zero mode can be chosen, in
the gauge where \(q\) is real and positive, as
\begin{equation}
    |\chi_0\rangle
    =
    \sqrt{\frac{2}{N+1}}
    \sum_{\ell=1}^{(N+1)/2}
    (-1)^{\ell-1}|2\ell-1\rangle ,
    \qquad N\ {\rm odd}.
    \label{eq:app_complement_zero_mode}
\end{equation}
Restoring the phase of \(q\) amounts to applying the inverse gauge
transformation of Eq.~\eqref{eq:app_OBC_gauge}. When this zero mode
has nonzero matrix element with the brane through \(H_{12}\) and
\(H_{21}\), it generates the pole
\begin{equation}
    \Sigma(k,z)
    =
    \frac{H_{12}(k)P_0(k)H_{21}(k)}{z}
    +
    \Sigma_{\rm reg}(k,z),
    \label{eq:app_resonant_self_energy}
\end{equation}
where \(P_0(k)=|\chi_0(k)\rangle\langle\chi_0(k)|\) is the complement
zero-mode projector. This is the zero-mode-resonant contribution used
in the main text.


\section{Spectral analysis of the brane Hamiltonian}\label{app:SecV_technical_spectra}

This appendix collects the algebraic details supporting Sec.~\ref{sec:spectral_structure}. It contains the eigenvectors used in the Berry-phase calculation and the small-frequency estimates of the $\mathcal{PT}_{+}$ transition lines. Additional spectral figures are collected in Section S.4. of the Supplemental Material.

\subsection{Open-boundary eigenvectors and \texorpdfstring{$\mathcal{PT}_{+}$}{PT} regimes}

In the $\mathcal{PT}_{+}$-unbroken regime of Eq.~(\ref{Energy_open_BC}), the right and left eigenstates of Eq.~(\ref{H_PTB_open_BC}) can be written as
\begin{align}
 &|\pm\rangle=\frac{1}{\sqrt{2}}\begin{pmatrix}
 1\\
 \pm\,e^{i\left(\arg\{q\}\pm\theta\right)}
 \end{pmatrix},\\
 &|\tilde{\pm}\rangle=\frac{1}{\sqrt{2}}\begin{pmatrix}
 1\\
 \pm\,e^{i\left(\arg\{q\}\mp\theta\right)}
 \end{pmatrix},
 \label{States_open_BC_PT_unbroken}
\end{align}
where $\theta=\arcsin\left(\frac{F_{N_1}-F_{N_2}}{2}\right)$.
The transition to the $\mathcal{PT}_{+}$-broken regime occurs when $|F_{N_1}-F_{N_2}|=2$. For small $\phi$, the auxiliary function behaves as $F_N(\phi)\sim[(N+1)\phi]^{-1}$ for odd $N$ and as $F_N(\phi)\sim (N+2)\phi$ for even $N$. This gives the estimates
\begin{equation}
 \phi_{\rm cr}\approx\frac{|N_1-N_2|}{2(N_1+1)(N_2+1)}
\end{equation}
when both $N_1$ and $N_2$ are odd, and
\begin{equation}
 \phi_{\rm cr}\approx\frac{1}{2(N_1+1)}
\end{equation}
when $N_1$ is odd and $N_2$ is even, with the analogous expression obtained by exchanging $N_1$ and $N_2$. The states in the $\mathcal{PT}_{+}$-broken regime are as follows:
\begin{align}
    |\pm\rangle&=\frac{1}{\sqrt{1+e^{\pm2\theta}}}\begin{pmatrix}
        1\\i\mathrm{sgn}\{F_{N_{1}}-F_{N_{2}}\}e^{i\arg\{q\}\pm\theta}
    \end{pmatrix},\label{States_open_BC_PT_broken}\\
    |\tilde{\pm}\rangle&=\frac{1}{\sqrt{1+e^{\pm2\theta}}}\begin{pmatrix}
        1\\-i\mathrm{sgn}\{F_{N_{1}}-F_{N_{2}}\}e^{i\arg\{q\}\pm\theta}
    \end{pmatrix},
\end{align}
where $\theta=\mathrm{sgn}\{F_{N_{1}}-F_{N_{2}}\}\mathrm{arccosh}\left(\frac{F_{N_{1}}-F_{N_{2}}}{2}\right)$.

\subsection{Periodic-boundary eigenvectors}

For even $N$, the right and left eigenstates of Eq.~(\ref{H_PTB_periodic_BC}) are
\begin{align}
 |\pm\rangle&=\frac{1}{\sqrt{2}}\begin{pmatrix}
 1\\ \pm e^{i\left(\arg\{q\}-\theta\right)}
 \end{pmatrix},\nonumber\\
 |\tilde{\pm}\rangle&=\frac{1}{\sqrt{2}}\begin{pmatrix}
 1\\ \pm e^{i\left(\arg\{q\}+\theta\right)}
 \end{pmatrix},
 \label{States_periodic_BC_PT_unbroken_N_even}
\end{align}
where
\begin{equation}
 e^{i\theta}=\frac{1+G_{N}e^{i\alpha}}{\sqrt{1+G_{N}^{2}+2G_{N}\cos\alpha}}.
 \label{Phase_theta_periodic_BC_N_even}
\end{equation}
For odd $N$, the right and left eigenstates are
\begin{equation}
\begin{aligned}
 |\pm\rangle&=\frac{1}{\sqrt{1+\rho^{2}}}\begin{pmatrix}
 1\\
 \pm\rho e^{i(\arg\{q\}+\theta)}
 \end{pmatrix},\\
 |\tilde{\pm}\rangle&=\frac{1}{\sqrt{1+\rho^{2}}}\begin{pmatrix}
 \rho\\
 \pm e^{i(\arg\{q\}+\theta)}
 \end{pmatrix},
\end{aligned}
\label{States_periodic_BC_N_odd}
\end{equation}
with
\begin{align}
 \rho&=\sqrt[4]{\frac{1+\tilde{G}^{2}_{N}-2\tilde{G}_{N}\sin\alpha}{1+\tilde{G}^{2}_{N}+2\tilde{G}_{N}\sin\alpha}},\\
 e^{2i\theta}&=\frac{1+\tilde{G}^{2}_{N}e^{-2i\alpha}}{\sqrt{1+\tilde{G}^{4}_{N}+2\tilde{G}^{2}_{N}\cos(2\alpha)}}.
 \label{Phase_theta_periodic_BC_N_odd}
\end{align}
For even values of $N$, these states possess $\mathcal{PT}_{+}$ symmetry, whereas for odd $N$ they do not. These eigenvectors are the input for the Berry-phase calculation in Sec.~\ref{sec:Main_top_inv}.

\subsection{Complex energy spectrum and the topological invariant}\label{app:top_inv}

In this section we present additional numerical results on the
various topological phases obtained in Sec.~\ref{sec:Main_top_inv} and
shown in Figs.~\ref{fig:Spektar_N=3_t_x>t_y}
and~\ref{fig:Spektar_N=3_t_x<t_y}, with emphasis on how moving across
the finite-frequency phase diagram affects the complex energy spectrum.
These phases are directly connected to the quantized complex Berry
phase in Eq.~(\ref{complex_berry_phase}), which in turn is linked to the
electrical polarization.

To illustrate how the topological phases are reflected in the
complex energy spectrum of the Hamiltonian under PBC, we take \(N=3\) as a representative example; the
corresponding phase diagram in Fig.~\ref{fig:Phase_diagram_N=3} serves
as the benchmark. These figures show that the quantization of the
complex Berry phase is not strictly tied to the SSH transition at the
critical point \(t_{x}=t_{y}\): nontrivial phases also appear for
\(t_{x}<t_{y}\). The full topological invariant~(\ref{top_inv_odd}) is
\begin{equation}
 \tilde{\nu}_{\pm}=-\frac{1}{2\pi}\bigg{(}\arg\{q\}+\theta\bigg{)}\bigg{|}_{-\pi}^{\pi}.
\end{equation}
We decompose it into two contributions: the standard SSH part,
\(-\frac{1}{2\pi}\arg\{q\}\big|_{-\pi}^{\pi}\), which controls the
appearance of gapped edge modes, and the contribution from the
\(\theta\) angle, \(-\frac{1}{2\pi}\theta\big|_{-\pi}^{\pi}\), which
generates the finite-frequency crystalline Berry phases and changes at the exceptional-point transitions. Gapped edge modes are
present only for \(t_{x}>t_{y}\) (see
Fig.~\ref{fig:Spektar_N=3_t_x>t_y}), corresponding to
\(-\frac{1}{2\pi}\arg\{q\}\big|_{-\pi}^{\pi}=-1\).

More interesting is the \(\theta\)-dependent part of the invariant.
For \(t_{x}<t_{y}\), at fixed \(t_{x}\) and varying \(\omega\), three
scenarios arise. For sufficiently small \(t_{x}\) the \(\theta\) term
makes no contribution. For intermediate \(t_{x}\) the system first
passes from a trivial phase to a topological phase with index \(1\) and
then reenters a trivial phase as \(\omega\) is lowered; this is the red
cut through the \(N=3\) phase diagram. For \(t_{x}\) close to the SSH
critical point there are four transitions as \(\omega\) is lowered,
\(-\frac{1}{2\pi}\theta|_{-\pi}^{\pi}=0\to1\to2\to1\to0\), the green cut
through the \(N=3\) phase diagram in Fig.~\ref{fig:Phase_diagram_N=3}.

The phases along the cuts of
Fig.~\ref{fig:Phase_diagram_N=3} are characterized by the CPH complex
Berry invariant \(\widetilde\nu\). Embedding parity enforces
\(Q_1(k)=Q_2(-k)\), so for a nondegenerate physical BZ the two
off-diagonal windings satisfy \(w_2=-w_1\). The oriented discriminant
winding, and hence the total two-strand braid exponent, therefore
vanishes, while the Berry invariant remains
\(\widetilde\nu=w_1\).

The spectral plots nevertheless contain genuine exceptional-point
braiding information at the transition frequencies. As \(\omega\) is
varied, parity-related EPs approach the physical BZ. At
a critical value the two eigenvalues and eigenvectors coalesce, the
Berry invariant is undefined, and a simple EP carries the local charge
\begin{equation}
    \nu_{\rm EP}=\frac{1}{4\pi i}
    \oint_{\gamma_{\rm EP}}d\ln\Delta=\pm\frac12.
\end{equation}
Immediately on the two sides of the transition the spectrum is again
nondegenerate, but the CPH Berry winding differs by an integer. Because the Brillouin zone forms a closed loop, encircling the EP exchanges the two spectral sheets, producing a nontrivial permutation of the strands, or monodromy. A sign of the  braid is obtained by displacing the BZ
infinitesimally above or below the EP, as described in
Sec.~\ref{sec:CPH_sym}.

The visually intertwined trajectories in
Figs.~\ref{fig:Spektar_N=3_t_x<t_y} and
\ref{fig:Spektar_N=3_t_x>t_y} should therefore be interpreted as the
evolution of the complex spectrum through parity-related EPs. The phase labels in Fig.~\ref{fig:Phase_diagram_N=3} are the quantized
values of \(\widetilde\nu\), obtained from the winding of
\(\arg{q}+\theta\). The EPs mark the transitions where
this Berry invariant changes and carry the associated local braid
vorticities.

\subsection{A complementary view: complex-energy band braiding versus exceptional-point transitions}\label{APP:band_braiding}

In this subsection, we characterize the topological phases in terms of the linking number \(L\) of
the two complex-energy bands, which matches the topological index
\(\tilde{\nu}\). Throughout our discussion, EPs
always enter and exit the Brillouin zone in \(\pm k\) pairs, a consequence of
the embedding-induced CPH (\(k \to -k\)) symmetry constraint. We make an explicit connection between these two perspectives: one based on EPs and the other on the braiding (linking) of complex-energy bands.

The first four panels of Fig.~\ref{fig:Spektar_N=3_t_x<t_y}
correspond to the red cut. We begin with
\(-\frac{1}{2\pi}\theta|_{-\pi}^{\pi}=0\): the complex energy spectrum
displays a line gap along the imaginary axis, with the states arranged
into two distinct lines in the complex plane, corresponding to the two
complex-energy strands of Fig.~\ref{fig:3D_spectrum}. We now invoke knot theory
to explain the origin of the invariant. At each Matsubara frequency
there are two energy bands, which we regard as two knots that may become
linked. The linking number counts how many times one knot winds around
the other, with orientation set by the direction of increasing crystal
momentum.

In the first panel of Fig.~\ref{fig:Spektar_N=3_t_x<t_y} the two
energy bands are fully separated, so for large \(\omega\) the linking
number is \(L=0\) and each band coincides with one strand. As
\(\omega\) decreases, the bands approach and the EPs
travel across the complex momentum plane. When the strands touch, a
\(\pm k\) pair of EPs enters the Brillouin zone and
becomes visible in the physical spectrum; at this critical value the
invariant is undefined, while on either side of it the invariant is
quantized with values differing by one. Beyond the transition the two
EPs lie inside the Brillouin-zone loop and act as
singularities generating two branch cuts. These cuts intersect the
Brillouin-zone circle at two points and appear as straight lines
parallel to the imaginary axis in Fig.~\ref{fig:3D_spectrum}. Tracking
the bands around the Brillouin zone, one is forced to follow the cuts,
which connect the two strands, so the bands swap strands. This is the
braid transition: the bands become mutually linked, the linking number
increases to \(L=1\), and the knots form a Hopf link. The transition is
shown in the first two panels of Fig.~\ref{fig:Spektar_N=3_t_x<t_y}.

Lowering \(\omega\) further gives the configuration in the third
panel. The detailed shapes of the strands differ, but the topology is
unchanged, and the system approaches a second braid transition. Beyond
it the two EPs leave the Brillouin-zone loop along a
different path, so the branch cuts intersect the circle at two further
points and a second family of straight lines emerges in
Fig.~\ref{fig:3D_spectrum}. Crossing these cuts disentangles the bands,
the linking number drops by one, and the system reaches the final-panel
configuration with \(L=0\), reentering the trivial phase.

The green cut across the \(N=3\) phase diagram, shown in the
remaining panels of Fig.~\ref{fig:Spektar_N=3_t_x<t_y}, leads to the
same conclusion. We again begin with two separated bands and \(L=0\).
In the second and third panels a Hopf link appears, \(L=1\). The third
panel shows the system nearing the second transition, after which it
develops into a more complex configuration, in the final panel of the
second row: one knot winds twice around the other, forming a
Solomon-link structure with \(L=2\), and four EPs lie
within the Brillouin-zone circle. The system approaches the third
transition in the first panel of the third row, and the resulting
change is shown in the next panel: a winding of the opposite
orientation appears as a \(\pm k\) pair of EPs leaves the
circle, reducing \(L\) to \(1\). A further pair exits in the final two
panels, lowering \(L\) to \(0\) and fully disentangling the bands.

One counts the windings in the same way for \(t_{x}>t_{y}\) (see
Fig.~\ref{fig:Spektar_N=3_t_x>t_y}). The difference is that all windings
proceed in the positive direction, so the number of EPs
on the Brillouin-zone circle grows monotonically and the linking number
increases accordingly. The quantization of the complex Berry phase is
therefore fixed by the linking number of the two knots, controlled by
the evolution of \(\arg\{q\}\), \(\alpha\), and \(\theta\) analyzed in
Sec.~\ref{sec:Main_top_inv}; this is the angle-winding mechanism encoded
by the finite-frequency invariant in Eq.~(\ref{top_inv_odd}).
Physically, the topological invariant equals half the number of EPs
accumulated within the Brillouin-zone circle in the complex momentum
plane.

\section{Berry phase and topological invariants}\label{app:topo-invariant}

Here we present a brief overview of Berry phases in a two-dimensional Hilbert space, considering non-unitary time evolution. For more details, see \cite{Beri_faza_non_herm,Beri_faza_non_herm_1}. When the Hamiltonian is NH two sets of eigenstates are defined:
\begin{align}
 H|\psi_{n}\rangle&=E_{n}|\psi_{n}\rangle,\\
 \langle{\tilde{\psi}_{n}}|H&=E_{n}\langle{\tilde{\psi}_{n}}|.
\end{align}
Here, $\psi_{n}$ denotes the right eigenstates, meaning the eigenstates of $H$, which belong to the Hilbert space $\mathcal{H}$. In contrast, $\tilde{\psi}_{n}$ refers to the left eigenstates, that is, the eigenstates of the Hermitian adjoint $H^{\dag}$, which reside in the dual Hilbert space $\mathcal{H}^{*}$. Assuming that the complex energies are non-degenerate, the left and right states corresponding to the different energies are mutually orthogonal $\langle{\tilde{\psi}_{n}}|\psi_{m}\rangle=0$, $n\neq m$. We proceed with the normalization of the left and right states. First, the right states are normalized in the usual way, requiring $\langle\psi_{n}|\psi_{n}\rangle = 1$. For the left states, there are two possible conventions. One can either enforce the bi-orthogonality relation $\langle\tilde{\psi}_{n}|\psi_{m}\rangle = \delta_{nm}$, or normalize them analogously to the right states, imposing $\langle\tilde{\psi}_{n}|\tilde{\psi}_{n}\rangle = 1$. In the latter case, the scalar product between left and right states is not fixed a apriori. This is the option we adopt.

Next, we move on to the theoretical background on the NH Berry phase. Let us assume that the Hamiltonian depends on a set of parameters $\vec{k}=(k_{1},k_{2},...,k_{n})$. Then the states and the energies depend on those parameters as well $|\psi_{n}(\vec{k})\rangle$ $E_{n}=E_{n}(\vec{k})$. Next, the parameters are varied adiabatically, $\vec{k} = \vec{k}(t)$. For Hermitian systems, the adiabatic theorem ensures that if the parameters change sufficiently slowly, the state evolves only by acquiring a phase factor. Extending the adiabatic theorem to NH systems is not straightforward; nevertheless, we assume here that it remains applicable. The evolution of the state over time is given by:
\begin{equation}
 |\psi(t)\rangle=U(t)|\psi_{n}(\vec{k}(t))\rangle.
\end{equation}
Solving the time-dependent Schrödinger equation for the right eigenstates, that is:
\begin{equation}
 i\hbar\frac{\mathrm{d}|\psi(t)\rangle}{\mathrm{d}t}=E_{n}(\vec{k}(t))|\psi(t)\rangle.\label{Sredinger}
\end{equation}
The non-Hermiticity of the system is reflected in the fact that there two scenarios. We may choose to act on the left either with $\langle\psi_{n}|$ or with $\langle\tilde{\psi}_{n}|$. In the first scenario, we arrive at the following equation for the phase $U(t)$:
\begin{equation}
 \frac{\mathrm{d}U(t)}{\mathrm{d}t}=U(t)\left[-\frac{i}{\hbar}E_{n}(\vec{k})-\langle\psi_{n}(\vec{k})|\partial_{i}|\psi_{n}(\vec{k})\rangle\frac{\mathrm{d}k^{i}}{\mathrm{d}t}\right],\label{right_U(t)_eq}
\end{equation}
where $\partial_{i}\equiv\frac{\partial}{\partial k^{i}}$. The first term in (\ref{right_U(t)_eq}) is the always present dynamical phase, while the second therm gives rise to the Berry phase when the evolution is preformed along the closed loop in the parameter space. In the usual manner we define the Berry connection:
\begin{equation}
 \mathcal{A}^{(n)}_{i}(\vec{k})=i\langle\psi_{n}(\vec{k})|\partial_{i}|\psi_{n}(\vec{k})\rangle,\label{Real_Berry_conection}
\end{equation}
and the Berry phase is given by the following closed loop integral:
\begin{equation}
 U_{\mathrm{B}}^{(n)}[\Gamma]=\exp{\left(i \oint_{\Gamma}\mathcal{A}^{(n)}_{i}\mathrm{d}k^{i}\right)},\label{real_berry_phase}
\end{equation}
which only depends on the path $\Gamma$ in the parameter space. We refer to these two as the real Berry connection and the real Berry phase respectively. Choosing to act with $\langle\tilde{\psi}_{n}|$ from the left on the equation (\ref{Sredinger}), we get:
\begin{equation}
 \frac{\mathrm{d}U(t)}{\mathrm{d}t}=U(t)\left[-\frac{i}{\hbar}E_{n}(\vec{k})-\frac{\langle\tilde{\psi}_{n}(\vec{k})|\partial_{i}|\psi_{n}(\vec{k})\rangle}{\langle\tilde{\psi}_{n}(\vec{k})|\psi_{n}(\vec{k})\rangle}\frac{\mathrm{d}k^{i}}{\mathrm{d}t}\right],
\end{equation}
Again, we get the dynamical phase and another term which we associate with a new complex Berry connection:
\begin{equation}
 \tilde{\mathcal{A}}^{(n)}_{i}(\vec{k})=i\frac{\langle\tilde{\psi}_{n}(\vec{k})|\partial_{i}|\psi_{n}(\vec{k})\rangle}{\langle\tilde{\psi}_{n}(\vec{k})|\psi_{n}(\vec{k})\rangle},\label{complex_Berry_conection}
\end{equation}
which gives rise to the complex Berry phase:
\begin{equation}
 \tilde{U}^{(n)}_{\mathrm{B}}[\Gamma]=\exp{\left(i \oint_{\Gamma}\tilde{\mathcal{A}}^{(n)}_{i}\mathrm{d}k^{i}\right)}.\label{complex_berry_phase}
\end{equation}
In the dual space $\mathcal{H}^{*}$ the time evolution is governed by the Hermitian conjugate $H^{\dag}$. In a similar way, we define two more Berry connections associated to the time evolution in this space:
\begin{align}
 \mathcal{A}^{(n)*}_{i}(\vec{k})&=i\langle\tilde{\psi}_{n}(\vec{k})|\partial_{i}|\psi_{n}(\vec{k})\rangle,\\
 \tilde{\mathcal{A}}^{(n)*}_{i}(\vec{k})&=i\frac{\langle\psi_{n}(\vec{k})|\partial_{i}|\tilde{\psi}_{n}(\vec{k})\rangle}{\langle\psi_{n}(\vec{k})|\tilde{\psi}_{n}(\vec{k})\rangle}.
\end{align}
These two Berry connections are not of interest in further discussion.
\subsection{Classification of NH topological insulators}\label{sec:Classification}
Here we give a brief overview of the {\(38\)-fold NH classification of topological phases}~\cite{Kawabata2019_PRX}. The key distinction in the NH case is that $H^{*}$ and $H^{T}$ are no longer identical which results into ramification of symmetries. Apart form time-reversal symmetry (TRS) $\mathcal{T}_{+}$, particle-hole symmetry (PHS) $\mathcal{C}_{-}$ and chiral symmetry (CS) $\Gamma$ defined by:
\begin{align}
 \text{TRS:}&\hspace{2mm}\mathcal{T}_{+}\tilde{H}^{*}(\vec{k})\mathcal{T}^{-1}_{+}=\tilde{H}(-\vec{k}),\hspace{2mm}\mathcal{T}_{+}\mathcal{T}_{+}^{*}=\pm1,\label{TRS}\\
 \text{PHS:}&\hspace{2mm}\mathcal{C}_{-}\tilde{H}^{T}(\vec{k})\mathcal{C}^{-1}_{-}=-\tilde{H}(-\vec{k}),\hspace{2mm}\mathcal{C}_{-}\mathcal{C}_{-}^{*}=\pm1,\label{PHS}\\
 \text{CS:}&\hspace{2mm}\Gamma\tilde{H}^{\dag}(\vec{k})\Gamma^{-1}=-\tilde{H}(\vec{k}),\hspace{2mm}\Gamma^{2}=1,\label{CS}
\end{align}
since $H^{*}\neq H^{T}$ we can define their counterparts, another time-reversal symmetry (TRS$^{\dag}$) $\mathcal{C}_{+}$, another particle-hole symmetry (PHS$^{\dag}$) $\mathcal{T}_{-}$ and sublattice symmetry (SLS) $\mathcal{S}$:
\begin{align}
 \text{TRS$^{\dag}$:}&\hspace{2mm}\mathcal{C}_{+}\tilde{H}^{T}(\vec{k})\mathcal{C}^{-1}_{+}=\tilde{H}(-\vec{k}),\hspace{2mm}\mathcal{C}_{+}\mathcal{C}_{+}^{*}=\pm1,\label{TRS_dag}\\
 \text{PHS$^{\dag}$:}&\hspace{2mm}\mathcal{T}_{-}\tilde{H}^{*}(\vec{k})\mathcal{T}^{-1}_{-}=-\tilde{H}(-\vec{k}),\hspace{2mm}\mathcal{T}_{-}\mathcal{T}_{-}^{*}=\pm1,\label{PHS_dag}\\
 \text{SLS:}&\hspace{2mm}\mathcal{S}\tilde{H}(\vec{k})\mathcal{S}^{-1}=-\tilde{H}(\vec{k}),\hspace{2mm}\mathcal{S}^{2}=1,\label{SLS}
\end{align}
where all matrices $\mathcal{T}_{\pm}$, $\mathcal{C}_{\pm}$, $\Gamma$ and $\mathcal{S}$ are unitary. Non-Hermiticity also unifies symmetries in a following way. If a Hamiltonian $\tilde{H}(\vec{k})$ obeys the TRS condition (\ref{TRS}), then the Hamiltonian $i\tilde{H}(\vec{k})$ instead fulfills the PHS$^{\dag}$ condition. This motivates the notation: we denote TRS by $\mathcal{T}_{+}$ and PHS$^{\dag}$ by $\mathcal{T}_{-}$. The analog argument holds for PHS and TRS$^{\dag}$, hence the notation $\mathcal{C}_{\pm}$. In the Hermitian case, sublattice symmetry is equivalent to the chiral symmetry. That is no longer the case when non-Hermiticity is permitted. Finally, there is one more {symmetry constraint} relevant for NH physics, pseudo-Hermiticity, defined by:
\begin{equation}
\text{pH:}\hspace{2mm}\eta\tilde{H}^{\dag}(\vec{k})\eta^{-1}=\tilde{H}(\vec{k}),\hspace{2mm}\eta^{2}=1.\label{pH}
\end{equation}
Positivity of $\eta$ implies the real spectrum of NH Hamiltonian, similarly to parity-time symmetry, discussed in the subsequent section.

With all the newly defined symmetries of NH systems, the original 10 AZ classes become generalized to 38 classes in total - 10 AZ classes, 6 AZ$^{\dag}$ classes, together with 22 classes exhibiting additional SLS symmetry apart form AZ symmetry. The AZ$^{\dag}$ classes with SLS symmetry are equivalent to the AZ classes. In addition, all of the AZ classes with pH, are equivalent to the corresponding AZ classes with SLS. The final classification is expressed by assigning the system to one of these 38 classes and specifying the type of complex energy gap it displays. In NH systems, two types of gaps can appear in the energy spectrum. A point gap $P$ is present when all complex energy eigenvalues are non-zero. A line gap $L$ arises when either the real parts of all eigenvalues are non-vanishing, or their imaginary parts are all non-zero. Accordingly, one distinguishes two variants of line gaps, denoted as $L_{r/i}$.

Note that this classification does not include the spatial symmetries, like the parity (inversion) symmetry:
\begin{equation}
 \mathcal{P}\tilde{H}(\vec{k})\mathcal{P}=\tilde{H}(-\vec{k})\label{parity_symetry}.
\end{equation}
A full classification of NH topological phases that incorporates spatial symmetries, analogous to the Hermitian case \cite{Cristaline_top_ins_class_1,Cristaline_top_ins_class_2}, has not yet been established. However, a classification of order-two spatial symmetries for point-gapped AZ and AZ$^{\dag}$ classes has been recently discussed~\cite{Class_cristaline_non_herm}.

In NH physics, the nontrivial point gap (when the Hamiltonian spectrum forms closed loops in complex plane) often implies the NHSE. This phenomenon occurs when an extensive number of bulk eigenstates, proportional to the system size, collapse and localize at the boundaries of the system. Then the point gap protects the skin states. On the other hand, the line gap may or may not imply the gapped edge modes. The traditional bulk-boundary correspondence may break in the presence of NH line gap. It can be fixed using generalized generalized Brillouin zone theory, firstly discussed in terms of the NH SSH model \cite{Yao2018_PRL} and later generalized to two-dimensional Chern invariants and non-Bloch band theory \cite{Yao2018_PRL,non-Bloch_Band_theory}.
\subsection{Topological invariants in one dimensional crystals}\label{sec:TOP_INV}
We now recall the symmetry-based classification of one-dimensional
crystals. In a crystal, the relevant parameter space is the Brillouin
zone, parametrized by the crystal momentum. In one dimension this
reduces to a single momentum \(k\), and the topological invariant is
obtained from the Berry connection integrated over the Brillouin zone. These invariants are generally winding numbers associated with different mappings. For example, in a Hermitian case which exhibit two bands and possess chiral symmetry, the inverse space Hamiltonian is given by:
\begin{equation}
 \tilde{H}(k)=E_{x}(k)\sigma_{x}+E_{y}(k)\sigma_{y},
\end{equation}
where $\sigma_{x,y}$ denote the Pauli matrices. The topological invariant reads:
\begin{equation}
 \nu_{\psi}=\frac{1}{\pi}\int_{-\pi}^{\pi}\mathrm{d}k\langle{\psi}|\sigma_{z}i\partial_{k}|\psi\rangle,\label{top_inv}
\end{equation}
where $|\psi\rangle$ is the energy eigenstate. The expression (\ref{top_inv}) represents the winding number of the embedding:
\begin{equation}
 \mathbb{S}^{1}\to\mathbb{R}^{2}\setminus\{0\}:k\mapsto(E_{x}(k),E_{y}(k)),
\end{equation} 
that is, how many times it winds around the origin in the punctured plane.
When a system exhibits the non-unitary evolution we have at least four different ways to define topological invariant. In the framework of NH topological band theory \cite{Top_band_theory}, these four Berry phases coincide whenever they take quantized values. Of interest are the two defined using real Berry connection (\ref{Real_Berry_conection}) and complex Berry connection (\ref{complex_Berry_conection}):
\begin{align}
 \nu_{n}&=\frac{1}{\pi}\int_{-\pi}^{\pi}\mathcal{A}^{(n)}(k)\mathrm{d}k,\label{nu_real}\\
 \tilde{\nu}_{n}&=\frac{1}{\pi}\int_{-\pi}^{\pi}\tilde{\mathcal{A}}^{(n)}(k)\mathrm{d}k.\label{nu_complex}
\end{align}
The natural question to ask is which of these two quantities is quantized and as such represents the topological invariant of the system. The other question is what physical quantity is that topological invariant related to. In a case of two-dimensional systems it is always proportional to the Hall conductivity, resulting into the quantized Hall conductivity. While, in one-dimensional Hermitian systems the relevant quantity is electric polarization \cite{Polarizacija_1,Polarizacija_2,Polarizacija_3}. According to \cite{Electric_polarization} polarization is the quantized in the NH systems as well. The correct result is given in terms of the global complex Berry connection:
\begin{equation}
 p=\frac{1}{2\pi}\sum_{n}\int_{-\pi}^{\pi}\tilde{\mathcal{A}}^{(n)}(k)\mathrm{d}k=\frac{1}{2}\sum_{n}\tilde{\nu}_{n}.\label{polarization}
\end{equation}
By global we refer to the total obtained by adding the complex Berry phases associated with each individual band. Next, we investigate how different symmetries can affect the quantization of Berry phases (\ref{nu_real}) and (\ref{nu_complex}). More details on the {\(38\)-fold NH classification of topological phases} can be found in Ref.~\cite{Kawabata2019_PRX}.
\subsubsection{Parity-time symmetry}\label{sec:PT_sym}
In a system which exhibits a two-dimensional Hilbert space, the $\mathcal{P}\mathcal{T}_{+}$ symmetry is represented by the $\sigma_{x}$ matrix, and defined via \cite{SSH_non_herm,PT1,PT2}:
\begin{equation}
 \sigma_{x}\tilde{H}(k)\sigma_{x}=\tilde{H}(k)^{*}.\label{PT_symetry}
\end{equation}
In general, the NH 2D Hamiltonian exhibiting this symmetry has the following form:
\begin{equation}
 \tilde{H}=\begin{pmatrix}
 H_{0} & Q^{*}\\
 Q &H_{0}^{*}
 \end{pmatrix},
\end{equation}
where $Q$ and $H_{0}=R-i\Gamma$ are two arbitrary complex numbers. The eigenvalues are:
\begin{equation}
 E_{\pm}=R\pm\sqrt{|Q|^{2}-\Gamma^{2}}.
\end{equation}
From the energy spectrum we observe that this symmetry exhibits two distinct phases. When the entire spectrum is real, corresponding to the condition $|Q|>|\Gamma|$, the system is in the so‑called $\mathcal{P}\mathcal{T}_{+}$ "unbroken" phase. In contrast, for $|Q|<|\Gamma|$, the system enters the $\mathcal{P}\mathcal{T}_{+}$ "broken" phase. The right and left eigenstates in the unbroken phase are as follows:
\begin{equation}
 |\pm\rangle=\frac{1}{\sqrt{2}}\begin{pmatrix}
 1\\\pm e^{i\left(\arg{\{Q\}}+\theta\right)}
 \end{pmatrix},|\tilde{\pm}\rangle=\frac{1}{\sqrt{2}}\begin{pmatrix}
 1\\\pm e^{i\left(\arg{\{Q\}}-\theta\right)}
 \end{pmatrix},\label{PT_symetric_states}
\end{equation}
where $\theta=\arcsin{\left(\frac{\Gamma}{|Q|}\right)}$. Both right and left states have the $\mathcal{P}\mathcal{T}_{+}$ symmetry, that is:
\begin{equation}
 |\pm\rangle=\sigma_{x}|\pm\rangle^{*}\hspace{2mm}\land\hspace{2mm}|\tilde{\pm}\rangle=\sigma_{x}|\tilde{\pm}\rangle^{*},\label{PT_symetric_states_unbroken}
\end{equation}
after the suitable choice of the corresponding phase factors. Substituting the $\mathcal{P}\mathcal{T}_{+}$-symetric states (\ref{PT_symetric_states}) into the expression for the real Berry phase (\ref{nu_real}) results in:
\begin{equation}
 \nu_{\pm}=-\frac{1}{2\pi}\bigg{(}\arg\{Q\}+\theta\bigg{)}\bigg{|}_{-\pi}^{\pi},
\end{equation}
implying the quantization of the real Berry phase. On the other hand, substituting this result into the expression (\ref{nu_complex}) gives:
\begin{equation}
 \tilde{\nu}_{\pm}=-\frac{1}{2\pi}\int_{-\pi}^{\pi}\mathrm{d}k\frac{e^{2i\theta}}{1+e^{2i\theta}}\partial_{k}\bigg{(}\arg\{Q\}+\theta\bigg{)},
\end{equation}
suggesting that this quantity is not a topological invariant. However, adding the contributions from all bands results into the following expression for the electric polarization (\ref{polarization}):
\begin{equation}
 p=\frac{1}{2}\left(\tilde{\nu}_{+}+\tilde{\nu}_{-}\right)=\frac{1}{2}\nu_{\pm},
\end{equation}
implying that the polarization is still quantized but with a topological invariant associated to the real Berry phase. Next, we move on to the broken phase. The eigenstates in the broken phase are as follows:
\begin{align}
 |\pm\rangle&=\frac{1}{\sqrt{1+e^{\pm2\theta}}}\begin{pmatrix}
 1\\i\mathrm{sgn}\{\Gamma\}e^{i\arg\{Q\}\pm\theta}
 \end{pmatrix},\\
 |\tilde{\pm}\rangle&=\frac{1}{\sqrt{1+e^{\pm2\theta}}}\begin{pmatrix}
 1\\-i\mathrm{sgn}\{\Gamma\}e^{i\arg\{Q\}\pm\theta}
 \end{pmatrix},
\end{align}
where $\theta=\mathrm{sgn}\{\Gamma\}\mathrm{arccosh}\left(\frac{\Gamma}{|Q|}\right)$. In this phase, the states satisfy the following relations:
\begin{equation}
 |\mp\rangle=\sigma_{x}|\pm\rangle^{*}\hspace{2mm}\land\hspace{2mm}|\tilde{\mp}\rangle=\sigma_{x}|\tilde{\pm}\rangle^{*}.\label{PT_symetric_states_broken}
\end{equation}
Calculating the real and complex Berry phases yields:
\begin{align}
 \nu_{\pm}=&-\frac{1}{\pi}\int_{-\pi}^{\pi}\mathrm{d}k\frac{e^{\pm\theta}}{e^{\pm\theta}+e^{\mp\theta}}\partial_{k}\bigg{(}\arg\{Q\}\bigg{)},\\
 \tilde{\nu}_{\pm}=&-\frac{1}{\pi}\int_{-\pi}^{\pi}\mathrm{d}k\frac{e^{\pm\theta}}{e^{\pm\theta}-e^{\mp\theta}}\partial_{k}\bigg{(}\arg\{Q\}\bigg{)}\nonumber\\
 &-\frac{1}{2\pi i}\ln{\bigg{|}\tanh{\theta}}\bigg{|}_{-\pi}^{\pi}.
\end{align}
The second term in the $\tilde{\nu}_{\pm}$ vanishes since $\theta$ is periodic function of $k$. We therefore find that neither of the Berry phases by itself produces an integer-valued invariant. However, when we compute the global Berry phase, whether using real or complex phases, we obtain a well-defined topological index, which once again corresponds to the electric polarization:
\begin{equation}
 p=\frac{1}{2}(\tilde{\nu}_{+}+\tilde{\nu}_{-})=\frac{1}{2}(\nu_{+}+\nu_{-})=-\frac{1}{2\pi}\bigg{(}\arg\{Q\}\bigg{)}\bigg{|}_{-\pi}^{\pi}.
\end{equation}
To conclude, the topological invariant is given by the global Berry phase when the $\mathcal{P}\mathcal{T}_{+}$ symmetry is present, which is stated in \cite{Beri_faza_non_herm} and in \cite{SSH_non_herm} for the case of the $\mathcal{P}\mathcal{T}_{+}$ symmetric SSH model. This means that in the $\mathcal{P}\mathcal{T}_{+}$ symmetric models the invariant is not a property of the band. Rather, it is a property of the whole system. In the unbroken phase it can be calculated using the real Berry phase, but it is a consequence of the real eigenvalues. Alternatively, we give another definition for this same winding number, inspired by the presence of the $\mathcal{P}\mathcal{T}_{+}$ symmetry:
\begin{equation}
 \nu_{\mathcal{P}\mathcal{T}_{+}}=\frac{1}{\pi}\int_{-\pi}^{\pi}\mathrm{d}k\frac{\langle\pm|i\partial_{k}\sigma_{x}K|\pm\rangle}{\langle\pm|\sigma_{x}K|\pm\rangle},\label{nu_PT}
\end{equation}
where $K$ stands for the complex conjugation. The topological index (\ref{nu_PT}) gives the same result as the global topological index associated with the complex Berry phase. Notice that in the unbroken phase $\sigma_{x}K|\pm\rangle=|\pm\rangle$ and expression (\ref{nu_PT}) reduces to the topological index defined with the help of the real Berry connection (\ref{nu_real}).
\subsubsection{Chiral symmetry}\label{sec:Chiral_sym}
For the two-band NH Hamiltonians considered here, chiral symmetry is naturally represented by\cite{Kawabata2019_PRX,SSH_non_herm}:
\begin{equation}
 \sigma_{z}\tilde{H}(k)\sigma_{z}=-\tilde{H}^{\dag}(k).\label{Chiral_symetry}
\end{equation}
It is obvious that this expression reduces to a typical chiral-symmetry condition when the condition of Hermiticity is demanded. This kind of symmetry is at times referred to as pseudo-anti-Hermiticity as well. Solving for $\tilde{H}$ we get the following general expression:
\begin{equation}
 \tilde{H}=\begin{pmatrix}
 iH_{1}|Q| & Q^{*}\\
 Q & iH_{2}|Q|
 \end{pmatrix},
\end{equation}
where $Q$ is an arbitrary complex number, while $H_{1}$ and $H_{2}$ are real numbers. Solving the eigenvalue problem, we get:
\begin{equation}
 E_{\pm}=|Q|\left[i\frac{H_{1}+H_{2}}{2}\pm\sqrt{1-\left(\frac{H_{1}-H_{2}}{2}\right)^{2}}\right].
\end{equation}
Notice that there are two different cases appearing here. In the first case we have $\left|\frac{H_{1}-H_{2}}{2}\right|<1$. The second term is real, and the energy eigenstates are as follows:
\begin{equation}
 |\pm\rangle=\frac{1}{\sqrt{2}}\begin{pmatrix}
 1\\\pm e^{i\left(\arg{\{Q\}}+\theta\right)}
 \end{pmatrix},|\tilde{\pm}\rangle=\frac{1}{\sqrt{2}}\begin{pmatrix}
 1\\\pm e^{i\left(\arg{\{Q\}}-\theta\right)}
 \end{pmatrix}.
\end{equation}
The phase $\theta$ in this expression is given by $\theta=\arcsin\left(\frac{H_{2}-H_{1}}{2}\right)$. Notice that although the Hamiltonian itself does not satisfy $\mathcal{P}\mathcal{T}_{+}$ symmetry (\ref{PT_symetry}), the eigenstates do, in the sense of (\ref{PT_symetric_states_unbroken}). For that reason we call this case a $\mathcal{P}\mathcal{T}_{+}$ unbroken phase. As in the $\mathcal{P}\mathcal{T}_{+}$-symmetric scenario analyzed in the previous section, the resulting real Berry connection is quantized, while the complex connection is not.
The case when the $\mathcal{P}\mathcal{T}_{+}$ symmetry is broken corresponds to the condition $\left|\frac{H_{1}-H_{2}}{2}\right|>1$. In this scenario, the eigenvalues are purely imaginary. The eigenstates are given by:
\begin{align}
 |\pm\rangle&=\frac{1}{\sqrt{1+e^{\pm2\theta}}}\begin{pmatrix}
 1\\i\mathrm{sgn}\{H_{2}-H_{1}\}e^{i\arg\{Q\}\pm\theta}
 \end{pmatrix},\\
 |\tilde{\pm}\rangle&=\frac{1}{\sqrt{1+e^{\pm2\theta}}}\begin{pmatrix}
 1\\-i\mathrm{sgn}\{H_{2}-H_{1}\}e^{i\arg\{Q\}\pm\theta}
 \end{pmatrix},
\end{align}
where $\theta=\mathrm{sgn}\{H_{2}-H_{1}\}\mathrm{arccosh}\left(\frac{H_{2}-H_{1}}{2}\right)$. Notice that this states posses the same symmetry as in the broken case of the $\mathcal{P}\mathcal{T}_{+}$ symmetric system (\ref{PT_symetric_states_broken}). The chiral symmetry acts on the right and left states, in the unbroken and the broken phases, respectively, in the following manner:
\begin{equation}
 |\tilde{\mp}\rangle=\sigma_{z}|\pm\rangle\hspace{2mm}\land\hspace{2mm}|\tilde{\pm}\rangle=\sigma_{z}|\pm\rangle.
\end{equation}
Since the system displays the same behavior as the $\mathcal{P}\mathcal{T}_{+}$-symmetric one in both phases, we therefore infer that the Berry phases are quantized in the same way as in the $\mathcal{P}\mathcal{T}_{+}$-symmetric case. In summary, chiral symmetry leads to an energy spectrum characterized by two separate phases. In the unbroken $\mathcal{P}\mathcal{T}_{+}$-symmetric phase, the real parts of the energy eigenvalues behave as they do in the Hermitian case, appearing in $\pm$ pairs. In contrast, once the symmetry is broken, the full set of eigenvalues becomes purely imaginary.
\subsubsection{Embedding parity and conjugated pseudo-Hermiticity}
\label{sec:CPH_sym}
Following the approach in Ref.~\cite{SSH_non_herm}, we demonstrate that, in a two-dimensional Hilbert space, a sufficient condition for the quantization of the complex Berry phase is that the Hamiltonian exhibits CPH symmetry, defined by:
\begin{equation}
    \sigma_{x}\widetilde{H}(k)^{*}\sigma_{x}=\widetilde{H}^{\dag}(k).\label{conjugated-pseudo-Hermiticity}
\end{equation}
The most general $2\times2$ matrix that satisfies this condition is:
\begin{equation}
    \widetilde{H}(k)=\begin{pmatrix}
        H_{0}(k) & Q_{1}(k)\\
        Q_{2}(k) & H_{0}(k)
    \end{pmatrix},\label{Eq:CPH_Ham}
\end{equation}
where $H_{0}(k)$, $Q_{1}(k)$ and $Q_{2}(k)$ are arbitrary complex functions. Therefore, the only condition imposed by this symmetry is the absence of a term proportional to $\sigma_{z}$ in the Hamiltonian.

In the periodic odd-\(N\) sector, embedding parity is represented by
\(\mathcal P=\sigma_x\) and obeys
\begin{equation}
    \mathcal P\widetilde H(k)\mathcal P^{-1}
    =\widetilde H(-k).
    \label{eq:embedding_parity}
\end{equation}
For the general CPH Hamiltonian \eqref{Eq:CPH_Ham}, Eq.~\eqref{eq:embedding_parity} then gives
\begin{equation}
    H_0(k)=H_0(-k),\, {\rm and}\,
    Q_1(k)=Q_2(-k).
    \label{eq:embedding_Q_relation}
\end{equation}
Thus the opposite-momentum relation between the off-diagonal amplitudes
is a direct consequence of parity.

The same sector satisfies TRS\(^{\dagger}\),
\begin{equation}
    \widetilde H^T(k)=\widetilde H(-k),
    \label{eq:TRS_dagger_CPH}
\end{equation}
which is equivalently
\(\widetilde H^*(k)=\widetilde H^\dagger(-k)\). Combining it with
Eq.~\eqref{eq:embedding_parity} yields CPH,
\begin{equation}
    \sigma_x\widetilde H(k)^*\sigma_x
    =\widetilde H(k)^\dagger.
\end{equation}
Parity and CPH therefore have distinct roles: parity produces
Eq.~\eqref{eq:embedding_Q_relation} and pairs the EPs,
while parity together with TRS\(^{\dagger}\) induces CPH and constrains
the biorthogonal eigenstates.

The eigenvalues are
\begin{equation}
    E_\pm(k)=H_0(k)\pm\sqrt{Q_1(k)Q_2(k)}.
    \label{eq:CPH_eigenvalues}
\end{equation}
Away from degeneracies, define
\begin{equation}
    q(k)=\sqrt{\frac{Q_2(k)}{Q_1(k)}}
    =\rho(k)e^{i\theta(k)},\qquad \rho(k)>0.
    \label{eq:q_ratio}
\end{equation}
A convenient biorthogonal basis is
\begin{equation}
    |\pm\rangle=\frac{1}{\sqrt{1+\rho^2}}
    \begin{pmatrix}1\\ \pm\rho e^{i\theta}\end{pmatrix},
    \quad
    |\widetilde\pm\rangle=\frac{1}{\sqrt{1+\rho^2}}
    \begin{pmatrix}\rho\\ \pm e^{i\theta}\end{pmatrix}.
    \label{eq:CPH_eigenstates}
\end{equation}
With an appropriate phase convention,
\(|\widetilde\pm\rangle=\sigma_x|\pm\rangle^*\). The complex Berry
invariant becomes
\begin{equation}
\begin{split}
    \widetilde\nu_\pm={}&-\frac{1}{2\pi}\oint d\theta\\
    &-\frac{1}{2\pi i}\oint
    d\ln\!\left(\frac{\rho}{1+\rho^2}\right).
\end{split}
    \label{nu_inv}
\end{equation}
The second term is an exact differential on a closed nondegenerate
cycle, so
\begin{equation}
    \widetilde\nu_\pm
    =-\frac{1}{2\pi i}\oint d\ln q.
    \label{eq:Berry_q_winding}
\end{equation}

To compare the CPH Berry invariant with spectral braiding, we introduce
the discriminant
\begin{equation}
    \Delta(k)
    =
    [E_+(k)-E_-(k)]^2
    =
    4Q_1(k)Q_2(k),
    \label{eq:discriminant_CPH}
\end{equation}
and consider any closed contour \(\mathcal C\) on which
\(\Delta(k)\neq0\). In a regular sector, \(\mathcal C\) can be chosen
as the physical Brillouin zone, whereas at an EP it may
be taken as an infinitesimally displaced or local enclosing contour.
The off-diagonal windings are
\begin{equation}
    w_a[\mathcal C]
    =
    \frac{1}{2\pi i}
    \oint_{\mathcal C}d\ln Q_a,
    \qquad
    a=1,2.
    \label{eq:individual_Q_windings}
\end{equation}
The spectral-braid invariant is then
\begin{equation}
    \nu_{\rm br}[\mathcal C]
    =
    \frac{1}{4\pi i}
    \oint_{\mathcal C}d\ln\Delta
    =
    \frac{1}{2}
    \left(
        w_1[\mathcal C]+w_2[\mathcal C]
    \right),
    \label{eq:braid_sum_windings}
\end{equation}
whereas Eq.~\eqref{eq:Berry_q_winding} gives
\begin{equation}
    \widetilde{\nu}[\mathcal C]
    =
    \frac{1}{2}
    \left(
        w_1[\mathcal C]-w_2[\mathcal C]
    \right).
    \label{eq:Berry_difference_windings}
\end{equation}
Thus, the braid and Berry invariants are, respectively, the symmetric
and antisymmetric combinations of the two off-diagonal windings.

Embedding parity relates a contour \(\mathcal C\) to its
momentum-inverted partner \(\overline{\mathcal C}\). With both contours
assigned the same orientation convention, the relation
\(Q_2(k)=Q_1(-k)\) implies
\begin{equation}
    w_2[\mathcal C]
    =
    -w_1[\overline{\mathcal C}].
    \label{eq:general_parity_winding_relation}
\end{equation}
Equations~\eqref{eq:braid_sum_windings} and
\eqref{eq:Berry_difference_windings} therefore become
\begin{align}
    \nu_{\rm br}[\mathcal C]
    &=
    \frac{1}{2}
    \left[
        w_1[\mathcal C]
        -
        w_1[\overline{\mathcal C}]
    \right],
    \label{eq:general_braid_parity}
    \\
    \widetilde{\nu}[\mathcal C]
    &=
    \frac{1}{2}
    \left[
        w_1[\mathcal C]
        +
        w_1[\overline{\mathcal C}]
    \right].
    \label{eq:general_Berry_parity}
\end{align}

In a nondegenerate sector, the physical Brillouin zone is itself an
admissible parity-invariant contour,
\(\overline{\mathcal C}=\mathcal C=\mathrm{BZ}\). Hence,
\begin{equation}
    \nu_{\rm br}[\mathrm{BZ}]=0,
    \qquad
    \widetilde{\nu}[\mathrm{BZ}]
    =
    w_1[\mathrm{BZ}].
    \label{eq:Berry_braid_parity_result}
\end{equation}
The quantized phase labels of the nondegenerate regions are therefore
given by the CPH Berry invariant.

When an EP lies on the physical Brillouin zone, the
latter passes through a zero of \(\Delta\) and is no longer an
admissible contour. The same construction remains applicable upon
choosing the parity-related one-sided contours
\begin{equation}
    \mathcal C_{\pm}:
    \qquad
    k=x\pm i0^{+},
    \qquad
    -\pi\leq x\leq\pi.
    \label{eq:regularized_contours}
\end{equation}
Momentum inversion exchanges \(\mathcal C_+\) and
\(\mathcal C_-\), so that
\begin{equation}
    w_2^{(+)}=-w_1^{(-)},
    \qquad
    w_2^{(-)}=-w_1^{(+)}.
    \label{eq:one_sided_winding_relation}
\end{equation}
The corresponding invariants are
\begin{align}
    \nu_{\rm br}^{(+)}
    &=
    \frac{1}{2}
    \left(
        w_1^{(+)}-w_1^{(-)}
    \right),
    \nonumber\\
    \nu_{\rm br}^{(-)}
    &=
    -\nu_{\rm br}^{(+)},
    \label{eq:braid_winding_jump}
\end{align}
and
\begin{equation}
    \widetilde{\nu}^{(+)}
    =
    \widetilde{\nu}^{(-)}
    =
    \frac{1}{2}
    \left(
        w_1^{(+)}+w_1^{(-)}
    \right).
    \label{eq:Berry_average_windings}
\end{equation}
The one-sided contours therefore separate the same parity-constrained winding data into a spectral-braid contribution and a CPH Berry contribution.

Equivalently, an EP can be characterized locally by a
small closed contour \(\gamma_{\rm EP}\) surrounding the degeneracy:
\begin{equation}
    \nu_{\rm EP}
    =
    \frac{1}{4\pi i}
    \oint_{\gamma_{\rm EP}}d\ln\Delta.
    \label{eq:local_EP_charge}
\end{equation}
For a simple EP,
\(\nu_{\rm EP}=\pm\tfrac{1}{2}\), with the sign determined by the
orientation of the local spectral winding. This half-integer charge is
the local counterpart of the difference between the two one-sided
windings in Eq.~\eqref{eq:braid_winding_jump}. The coalescence of the
eigenvalues therefore does not make the braiding trivial. Rather, the
EP connects the two sheets of the complex spectrum, and
the contours \(\mathcal C_{+}\) and \(\mathcal C_{-}\) regularize the
singular crossing into oppositely oriented half-twists of the spectral
strands, as expressed by
\(\nu_{\rm br}^{(-)}=-\nu_{\rm br}^{(+)}\).

Locally, one circuit around a simple EP exchanges the
two eigenvalue sheets, while a second circuit returns to the original
sheet. For the compact Brillouin zone, the resulting global monodromy is
determined by the combined contribution of all EPs
encountered during one traversal. EPs thus mediate spectral reconnections between neighboring nondegenerate sectors, which are characterized by the CPH Berry invariant in Eq.~\eqref{eq:general_Berry_parity}. The regular and
exceptional sectors are therefore described within the same contour
framework: the physical Brillouin zone is used when the spectrum is
nondegenerate, and an infinitesimally displaced contour provides the
corresponding resolution when an EP lies on it.

Finally, the CPH Berry invariant can be written solely in terms of
right eigenstates as
\begin{equation}
    \nu_{\rm CPH}
    =
    \frac{1}{\pi}
    \int_{\widetilde{\mathcal C}}\!dk\,
    \frac{
        \langle\pm|
        \sigma_x K\,i\partial_k
        |\pm\rangle
    }{
        \langle\pm|
        \sigma_x K
        |\pm\rangle
    },
    \label{nu_CPH}
\end{equation}
where \(\widetilde{\mathcal C}\) denotes the appropriate closed
spectral cycle, including band sewing when a single Brillouin-zone
traversal exchanges the two bands. 
This expression yields the CPH Berry invariant of
Eq.~\eqref{eq:Berry_difference_windings}, whereas
Eq.~\eqref{eq:braid_sum_windings} characterizes the local or
one-sided spectral-braid contribution associated with an exceptional
crossing.

\section{Hamiltonian in coordinate representation}\label{App:H_coord_rep}

Let us first address the case of the open boundary conditions. The diagonal matrix elements of $(n,n')$ block are given by:
\begin{align}
 \langle n,m|H^{OBC}_{PTB}|n',m\rangle&\equiv\varepsilon^{(N_{m+1})}_{n-n'}\label{int_F}\\
 &=\int_{-\pi}^{\pi}\frac{\mathrm{d}k}{2\pi}\left(-i|q|F_{N_{m+1}}\right)e^{i(n-n')k},\nonumber
\end{align}
while the off-diagonal elements are easy to calculate, and are as follows:
\begin{align}
 \langle n,0|H_{PTB}^{OBC}|n',1\rangle&=-t_y\delta_{nn'}-t_x\delta_{n,n'+1},\\
 \langle n,1|H_{PTB}^{OBC}|n',0\rangle&=-t_y\delta_{nn'}-t_x\delta_{n,n'-1}.
\end{align}
The integral (\ref{int_F}) can be calculated using complex integration methods. Notice that by definition (\ref{int_F}) we have $\varepsilon_{n}^{(N)}=\varepsilon^{(N)}_{-n}$, using the change of variables $k\mapsto-k$. Making the substitution $z=e^{ik}$ results into:
\begin{equation}
 \varepsilon_{n}^{(N)}=-\frac{1}{2\pi}\oint_{\Gamma}\mathrm{d}zz^{n-1}|q(z)|F_{N}(\phi(z)),\label{eps_n}
\end{equation}
where $\Gamma$ is a unit circle, which will be evaluated using the Cauchy residue theorem. The pole structure of integral (\ref{eps_n}) is given by the poles of the $F_{N}$ together with the possible pole at $z=0$. Once again, the cases of even and odd $N$ must be investigated separately. For even $N$, the poles of $F_{N}$ satisfy $\cosh((N+1)\phi_{s})=0$, which implies $\phi_{s}=i\frac{2s-1}{N+1}\frac{\pi}{2}$ with $s\in\mathbb{Z}$. Employing the relation among $\phi$, $|q|$, and $z$, we obtain $N/2$ distinct poles inside the unit circle, given by:
\begin{equation}
 z_{s}=-\zeta_{s}+\sqrt{\zeta_{s}^{2}-1},\hspace{2mm}\zeta_{s}=\frac{t_{x}^{2}+t_{y}^{2}+\frac{\omega^{2}}{4\sin^{2}\left(\frac{2s-1}{N+1}\frac{\pi}{2}\right)}}{2t_{x}t_{y}},\label{poles_even_N}
\end{equation}
where $s\in\left\{1,2,...,\frac{N}2{}\right\}$. For odd values of $N$, the function $F_N$ has poles defined by $\sinh((N+1)\phi_s)=0$, that is $\phi_s = i\frac{s\pi}{N+1}$ with $s\in\mathbb{Z}$. Because the pole at $z=0$ must be treated separately, we omit it and obtain:

\begin{equation}
 z_{s}=-\zeta_{s}+\sqrt{\zeta_{s}^{2}-1},\hspace{2mm}\zeta_{s}=\frac{t_{x}^{2}+t_{y}^{2}+\frac{\omega^{2}}{4\sin^{2}\left(\frac{s\pi}{N+1}\right)}}{2t_{x}t_{y}},\label{poles_odd_N}
\end{equation}
where $s\in\left\{1,2,..,\frac{N-1}{2}\right\}$. Using the Cauchy residue theorem the integral (\ref{eps_n}) is given by the following expression:
\begin{widetext}
\begin{align}
 \varepsilon_{n-n'}^{(N)}&=-i\frac{N+2}{2}\omega\delta_{n,n'}+\frac{i\omega^{3}}{8(N+1)t_{x}t_{y}}\sum_{s=1}^{\frac{N}{2}}\frac{\cos^{2}\left(\frac{2s-1}{N+1}\frac{\pi}{2}\right)}{\sin^{4}\left(\frac{2s-1}{N+1}\frac{\pi}{2}\right)}\frac{z_{s}^{|n-n'|}}{\sqrt{\zeta_{s}^{2}-1}},\label{eps_n-n'_even}\\
 \varepsilon_{n-n'}^{(N)}&=-i\left(\frac{(N+2)(N+3)}{6(N+1)}\omega+2\frac{t_{x}^{2}+t_{y}^{2}}{(N+1)\omega}\right)\delta_{n,n'}-\frac{2it_{x}t_{y}}{(N+1)\omega}\delta_{|n-n'|,1}
 +\frac{i\omega^{3}}{8(N+1)t_{x}t_{y}}\sum_{s=1}^{\frac{N-1}{2}}\frac{\cos^{2}\left(\frac{s\pi}{N+1}\right)}{\sin^{4}\left(\frac{s\pi}{N+1}\right)}\frac{z_{s}^{|n-n'|}}{\sqrt{\zeta_{s}^{2}-1}},\label{eps_n-n'_odd}
\end{align}
\end{widetext}
where the first line corresponds to the case of even $N$, and the second line applies when $N$ is odd. The terms appearing outside of the sums are associated with the pole at $z=0$, while the pole $z_{s}$ corresponds to the $s$-th term in the sums. Note that $\varepsilon_{n-n'}^{(N)}$ depends only on the difference $n-n'$, implying that the Hamiltonian exhibits translational symmetry. Now, the physical interpretation of the poles $z_{s}$ becomes apparent; 
\begin{equation}
 \xi_{s}=-\frac{1}{d}\ln{|z_{s}|}=\frac{1}{d}\ln\left(\zeta_{s}+\sqrt{\zeta_{s}^{2}-1}\right)
\end{equation}
represent the exponential decaying rates of the hopping parameter associated to the diagonal part of the $(n,n')$ block of the Hamiltonian. Moreover, it is made clear that, for odd $N$, the only Hamiltonian matrix elements that diverge as $\omega \to 0$ are those of the form $(n,n\pm1)$ and $(n,n)$. There is also a pole for $s=\left\lfloor\frac{N}{2}\right\rfloor+1$, but its residue is equal to zero, which means that it does not contribute to the calculation of the integral. For the sake of simplicity, we rewrite this expression in the following form:
\begin{equation}
 \varepsilon_{n-n'}^{(N)}=\varepsilon\delta_{n,n'}+\tilde{\varepsilon}\delta_{|n-n'|,1}+\sum_{s}\varepsilon_{s}z_{s}^{|n-n'|},\label{eps_sol}
\end{equation}
where $\varepsilon$, $\tilde{\varepsilon}$ and $\varepsilon_{s}$ are determined by direct comparison between (\ref{eps_n-n'_even}-\ref{eps_n-n'_odd}) and (\ref{eps_sol}). Figure \ref{fig:eps_n} illustrates how $\varepsilon_{n}^{(N)}$ varies as a function of $n$.

Next we move on to the computation of the hopping parameters in the case of the periodic boundary conditions Hamiltonian (\ref{H_PTB_periodic_BC}). For the diagonal part of the $(n,n')$ block we have the same result:
\begin{equation}
 \langle n,m|H^{PBC}_{PTB}|n',m\rangle=\varepsilon^{(N)}_{n-n'},\hspace{2mm}\forall m\in\{0,1\},
\end{equation}
while the off-diagonal elements are given by:
\begin{align}
 \langle n,0|H_{PTB}^{PBC}|n',1\rangle&=-t_y\delta_{nn'}-t_x\delta_{n,n'+1}-t^{(N)}_{n'-n},\\
 \langle n,1|H_{PTB}^{PBC}|n',0\rangle&=-t_y\delta_{nn'}-t_x\delta_{n,n'-1}-t^{(N)}_{n-n'},
\end{align}
where the hopping parameters $t_{n}^{(N)}$ are defined by:
\begin{equation}
 t_{n}^{(N)}=\int_{-\pi}^{\pi}\frac{\mathrm{d}k}{2\pi}\frac{\left(-q^{*}\right)^{N+1}}{|q|^{N}}G_{N}(\phi)e^{ink}.\label{t_int}
\end{equation}
The integral (\ref{t_int}) can be evaluated by means of complex integration. It has singularities at $z=0$ and at the poles of the function $G_{N}$. These poles are the same as the poles of function $F_{N}$ defined in (\ref{poles_even_N}) and (\ref{poles_odd_N}) when $N$ is even or odd respectively. Application of the Cauchy residue theorem yields:
\begin{widetext}
 \begin{align}
 t_{n}^{(N)}&=\begin{dcases}
 \frac{\omega^{2-N}}{2^{2-N}(N+1)t_{x}t_{y}}\sum_{s=1}^{\frac{N}{2}}(-1)^{s-1}\frac{\cos^{2}\left(\frac{2s-1}{N+1}\frac{\pi}{2}\right)}{\sin^{4}\left(\frac{2s-1}{N+1}\frac{\pi}{2}\right)}\left[\sin{\left(\frac{2s-1}{N+1}\frac{\pi}{2}\right)}\left(-t_{y}-t_{x}z_{s}^{-1}\right)\right]^{N+1}\frac{z_{s}^{n}}{\sqrt{\zeta_{s}^{2}-1}},&n\geqslant\frac{N}{2}+2,\\
 \frac{\omega^{2-N}}{2^{2-N}(N+1)t_{x}t_{y}}\sum_{s=1}^{\frac{N}{2}}(-1)^{s-1}\frac{\cos^{2}\left(\frac{2s-1}{N+1}\frac{\pi}{2}\right)}{\sin^{4}\left(\frac{2s-1}{N+1}\frac{\pi}{2}\right)}\left[\sin{\left(\frac{2s-1}{N+1}\frac{\pi}{2}\right)}\left(-t_{y}-t_{x}z_{s}\right)\right]^{N+1}\frac{z_{s}^{-n}}{\sqrt{\zeta_{s}^{2}-1}}\\
 \hspace{2cm}+t_{y}\left(-\frac{t_{y}}{t_{x}}\right)^{\frac{N}{2}}\delta_{n,\frac{N}{2}}-\frac{1}{2t_{x}}\left(-\frac{t_{y}}{t_{x}}\right)^{\frac{N}{2}}\left[\frac{N(N+2)}{4}\omega^{2}-(N+2)t_{x}^{2}+Nt_{y}^{2}\right]\delta_{n,\frac{N}{2}+1},&n\leqslant\frac{N}{2}+1,
 \end{dcases}\label{tau_n-n'_even}
\end{align}

\begin{align}
 t_{n}^{(N)}&=\begin{dcases}
 \frac{i\omega^{2-N}}{2^{2-N}(N+1)t_{x}t_{y}}\sum_{s=1}^{\frac{N-1}{2}}(-1)^{s}\frac{\cos^{2}\left(\frac{s\pi}{N+1}\right)}{\sin^{4}\left(\frac{s\pi}{N+1}\right)}\left[\sin{\left(\frac{s\pi}{N+1}\right)}\left(-t_{y}-t_{x}z_{s}^{-1}\right)\right]^{N+1}\frac{z_{s}^{n}}{\sqrt{\zeta_{s}^{2}-1}},\hspace{1.7cm}n\geqslant\frac{N+3}{2}
 \\
 \hspace{0.7cm}+\frac{2it_{x}^{2}}{\omega(N+1)}\left(-\frac{t_{x}}{t_{y}}\right)^{\frac{N-1}{2}}\delta_{n,\frac{N+3}{2}},\\
 \frac{i\omega^{2-N}}{2^{2-N}(N+1)t_{x}t_{y}}\sum_{s=1}^{\frac{N-1}{2}}(-1)^{s}\frac{\cos^{2}\left(\frac{s\pi}{N+1}\right)}{\sin^{4}\left(\frac{s\pi}{N+1}\right)}\left[\sin{\left(\frac{s\pi}{N+1}\right)}\left(-t_{y}-t_{x}z_{s}\right)\right]^{N+1}\frac{z_{s}^{-n}}{\sqrt{\zeta_{s}^{2}-1}},\hspace{1.8cm}n\leqslant\frac{N+1}{2}\\
 \hspace{0.7cm}+\frac{2it_{y}^{2}}{\omega(N+1)}\left(-\frac{t_{y}}{t_{x}}\right)^{\frac{N-1}{2}}\delta_{n,\frac{N-1}{2}}+i\left(\frac{(N-1)(N+3)}{12(N+1)}\omega+\frac{(N-1)t_{y}^{2}-(N+3)t_{x}^{2}}{\omega(N+1)}\right)\left(-\frac{t_{y}}{t_{x}}\right)^{\frac{N+1}{2}}\delta_{n,\frac{N+1}{2}}.
 \end{dcases}\label{tau_n-n'_odd}
 \end{align}
\end{widetext}
The first line applies to the case where $N$ is even, while the second line corresponds to the scenario with odd $N$. Similarly to the case of open boundary conditions, for large $n$, we have an exponentially decreasing hopping parameter $t_{n}^{(N)}$. In contrast, for $0\leqslant n\leqslant\frac{N+1}{2}$ $t_{n}$ is an increasing function of $n$. The $\omega\to0$ limit is well defined when $N$ is even ($\omega^{2-N}$ cancels with the terms inside the sum), while when $N$ is odd, the divergence occurs due to the zeroth, and first order hopping parameters. For the sake of simplicity, we rewrite this expression in the following form:
\begin{equation}
 t_{n}=\begin{cases}
 \sum_{s}\tau_{s}z_{s}^{n}+\tilde{\tau}'\delta_{n,\frac{N+3}{2}},n\geqslant\left\lceil\frac{N+3}{2}\right\rceil,\\
 \sum_{s}\tau_{s}'z_{s}^{-n}+\tilde{\tau}\delta_{n,\left\lceil\frac{N-1}{2}\right\rceil}+\tau\delta_{n,\left\lceil\frac{N+1}{2}\right\rceil},n\leqslant\left\lceil\frac{N+1}{2}\right\rceil,
 \end{cases}\label{tau_sol}
\end{equation}
where $\tau_{s}$, $\tau_{s}'$, $\tilde{\tau}$, $\tilde{\tau}'$ and $\tau$ are determined by direct comparison between (\ref{tau_n-n'_even}-\ref{tau_n-n'_odd}) and (\ref{tau_sol}). The behavior of $t_{n}^{(N)}$ as a function of $n$ is shown in Figure \ref{fig:t_n}.

Applying the inverse Fourier transform $f(k)=\sum_{n}f_{n}e^{-ink}$ of a function $f_{n}$, we obtain the explicit $k$-dependence of the hopping parameters $\varepsilon_{n}^{(N)}$ and $t_{n}^{(N)}$, which is required to demonstrate the validity of expression (\ref{beta_eigen_value_problem}). Direct calculation yields:
\begin{widetext}
\begin{align}
 -i|q|F_{N}&=\varepsilon+2\tilde{\varepsilon}\cos(k)+\sum_{s}\varepsilon_{s}\frac{2\sinh{\theta_{s}}e^{ik}}{1-2\cosh\theta_{s}e^{ik}+e^{2ik}},\label{F(k)}\\
 \frac{\left(-q^{*}\right)^{N+1}}{|q|^{N}}G_{N}&=\sum_{s}\left[\tau_{s}'\frac{\left(z_{s}e^{ik}\right)^{-\left\lceil\frac{N+1}{2}\right\rceil}}{1-z_{s}e^{ik}}-\tau_{s}\frac{\left(z_{s}^{-1}e^{ik}\right)^{-\left\lceil\frac{N+1}{2}\right\rceil}}{1-z_{s}^{-1}e^{ik}}\right]+\tau e^{-i\left\lceil\frac{N+1}{2}\right\rceil k}+\tilde{\tau}e^{-i\left\lceil\frac{N-1}{2}\right\rceil k}+\tilde{\tau}'e^{-i\frac{N+3}{2}k}\delta_{N-odd}.\label{G(k)}
\end{align}
\end{widetext}
where $\theta_{s}$ are defined via $e^{\theta_{s}}\equiv z_{s}$. These expressions are formulated so that they remain valid whether $N$ is even or odd.

\section{\texorpdfstring{{Bulk--boundary response: Technical details}}{Spectral braid structure and bulk-boundary response: Technical details}}
\label{app:BBC_details}

{In this appendix we collect the technical details underlying the bulk--boundary analysis discussed in Sec.~\ref{sec:BBC}. We first show that the projected brane Hamiltonians do not exhibit an NHSE. Then, we describe the coordinate-space formulation used to analyze boundary modes and distinguish SSH-like sectors, where a conventional edge-mode bulk--boundary correspondence is recovered, from the odd-$N$ periodic sector, where the boundary response is refined and termination-sensitive.}

\subsection{Generalized Brillouin zone and absence of NHSE}\label{app:GBZ_NHSE}

The eigenvalues (\ref{Energy_open_BC}) and (\ref{Energy_periodic_BC}) of the projected Hamiltonians (\ref{H_PTB_open_BC}) and (\ref{H_PTB_periodic_BC}) can be expressed in terms of the variable $\phi$, since $|q|=\omega/(2\sinh\phi)$ and $\cos\alpha=f(|q|)$. Because $\phi=\phi(E)$ may be multi-valued, the eigenvalue problem for the non-Bloch variable $\beta=e^{ik}$ reduces to
\begin{equation}
 \prod_{s=1}^{M}
 \left(\beta^{2}-
 \frac{\dfrac{\omega^{2}}{4\sinh^{2}\phi_{s}(E)}-t_{x}^{2}-t_{y}^{2}}
 {t_{x}t_{y}}\beta+1\right)=0,
 \label{beta_eigen_value_problem}
\end{equation}
where the product runs over all distinct branches $\phi_s(E)$. The reciprocal pairing of roots follows from the reciprocal   coordinate-space hopping amplitudes of the projected Hamiltonian: the characteristic polynomial is palindromic in $\beta$, so roots occur in pairs $\beta$ and $\beta^{-1}$. The solutions $\beta_s^{\pm}(E)$ therefore satisfy $\beta_s^+\beta_s^-=1$. Ordering the roots as
\[
|\beta_1^-|<|\beta_2^-|<\cdots<|\beta_M^-|<|\beta_M^+|<\cdots<|\beta_2^+|<|\beta_1^+|,
\]
the condition for a continuous generalized Brillouin zone is $|\beta_M^-|=|\beta_M^+|$~\cite{non-Bloch_Band_theory}. Together with $\beta_M^+\beta_M^-=1$, this gives $|\beta|=1$. Hence the generalized Brillouin zone coincides with the conventional Brillouin zone, and the NHSE is absent in the projected brane model.

\begin{figure}
 \begin{center}
 \captionsetup{justification=justified,singlelinecheck=false}
 \includegraphics[width=0.45\textwidth]{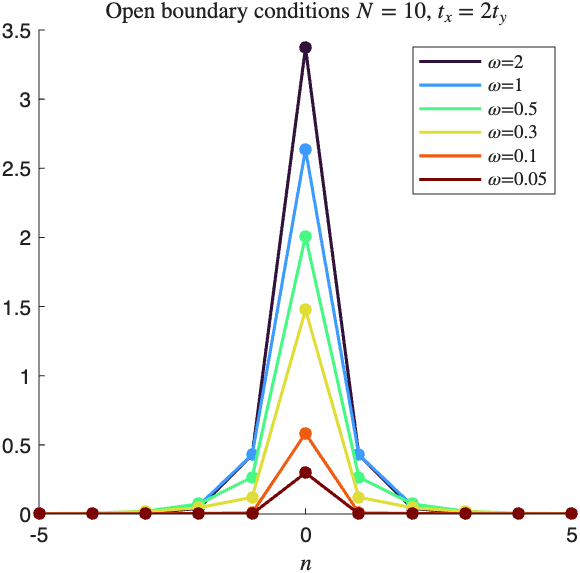}
 \caption{Dependence of the hopping parameter $\varepsilon_{n}^{(N)}$ in Eq.~ (\ref{eps_sol_1}) on the inter-site distance $n$ in the scenario with even $N$. A clear exponential decay is observed, with peak appearing at $n=0$. For even $N$, this maximum diminishes as the Matsubara frequency is decreased. In contrast, for odd $N$, the peak becomes larger as $\omega$ is decreased. Additional cases, including the impact of increasing $t_x$, are presented in Sec. S.3. of the SM.}\label{fig:eps_n}
 \end{center}
\end{figure}

To analyze the edge states and to confirm the validity of the claim in (\ref{beta_eigen_value_problem}), it is useful to rewrite the Hamiltonian in the coordinate representation. Computation details of the direct-space hopping parameters can be found in Appendix \ref{App:H_coord_rep}. The open boundary conditions Hamiltonian is given by:
\begin{align}
 H_{PTB}&=\sum_{n,m}\bigg[\varepsilon_{n-m}^{(N_{1})}|n,0\rangle\langle n,0|+\varepsilon_{n-m}^{(N_{2})}|n,1\rangle\langle n,1|\bigg]\label{H_OBC_coord_rep}\\
 &-\sum_{n}\bigg[t_{y}|n,0\rangle\langle n,1|+t_{x}|n+1,0\rangle\langle n,1|+h.c.\bigg],\nonumber
\end{align}
where hopping parameter $\varepsilon_{n-m}^{(N)}$ depends on the distance between sites $n-m$ as follows:
\begin{equation}
 \varepsilon_{n-m}^{(N)}=\varepsilon\delta_{n,m}+\tilde{\varepsilon}\delta_{|n-m|,1}+\sum_{s=1}^{N/2}\varepsilon_{s}z_{s}^{|n-m|},\label{eps_sol_1}
\end{equation}
where $z_{s}$, $\varepsilon$, $\tilde{\varepsilon}$ and $\varepsilon_{s}$ are defined in App.~\ref{App:H_coord_rep}. As expected, equation (\ref{eps_sol_1}) explicitly shows that the projection generates a long-range hopping terms that decay exponentially, as illustrated in Figure \ref{fig:eps_n}. 

\begin{figure}
 \begin{center}
 \captionsetup{justification=justified,singlelinecheck=false}
 \includegraphics[width=0.45\textwidth]{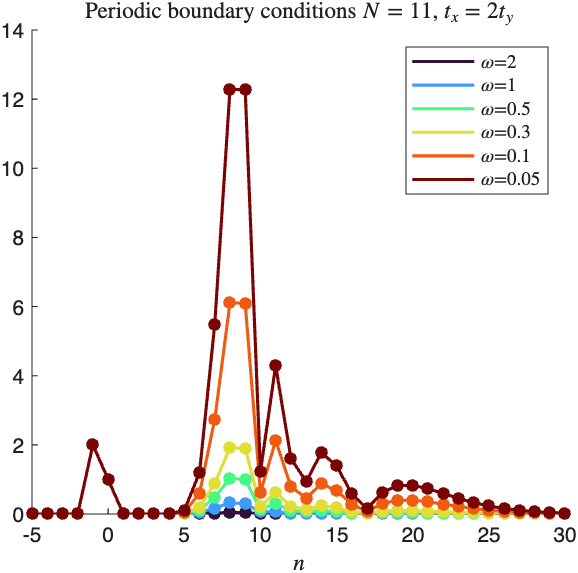}
 \caption{Dependence of the hopping parameter $t_{n}^{(N)}$ in Eq.~(\ref{tau_sol_1}) on the inter-site distance $n$ for odd $N$. There are two distinct peaks: one located at the usual SSH point ($n = -1$), and another appearing at a finite inter-site separation. As the Matsubara frequency decreases, the SSH peak remains unchanged, whereas the new peak increases in magnitude and, beyond a certain critical value, surpasses the SSH peak. Additional examples are presented in Sec. S.3. of the SM.}\label{fig:t_n}
 \end{center}
\end{figure}

The periodic boundary conditions Hamiltonian has a following coordinate representation:
\begin{align}
 H_{PTB}&=\sum_{n,m}\bigg[\varepsilon_{n-m}^{(N)}|n,0\rangle\langle n,0|+\varepsilon_{n-m}^{(N)}|n,1\rangle\langle n,1|\bigg]\nonumber\\
 &-\sum_{n,m}t^{(N)}_{n-m}\bigg[|m,0\rangle\langle n,1|+|n,1\rangle\langle m,0|\bigg]\label{H_PBC_coord_rep}\\
 &-\sum_{n}\bigg[t_{y}|n,0\rangle\langle n,1|+t_{x}|n+1,0\rangle\langle n,1|+h.c.\bigg],\nonumber
\end{align}
where hopping parameters $t_{n}^{(N)}$ are given by:
\begin{equation}
 t_{n}^{(N)}=\begin{dcases}
 \sum_{s}\tau_{s}z_{s}^{n}+\tilde{\tau}'\delta_{n,\frac{N+3}{2}},\hspace{5mm}n\geqslant\left\lceil\frac{N+3}{2}\right\rceil,\\
 \sum_{s}\tau_{s}'z_{s}^{-n}+\tilde{\tau}\delta_{n,\left\lceil\frac{N-1}{2}\right\rceil}+\tau\delta_{n,\left\lceil\frac{N+1}{2}\right\rceil},
 \end{dcases}\label{tau_sol_1}
\end{equation}
where $\tau$, $\tilde{\tau}$, $\tilde{\tau}'$, $\tau_{s}$ and $\tau_{s}'$ are defined in Sec.S.3. of the Supplemental Material, while $z_{s}$ are the same as in the case of the open boundary conditions. Once again, these hopping parameters exhibit the expected exponential decay. This behavior is shown in Figure \ref{fig:t_n}.

Now we proceed to prove the claim (\ref{beta_eigen_value_problem}). The periodic eigenvalue problem for $\beta=e^{ik}$ is defined through the eigenvalues (\ref{Energy_open_BC}), or:
\begin{equation}
 \left(E+i|q|F_{N_{1}}\right)\left(E+i|q|F_{N_{2}}\right)=|q|^{2}.\label{a1}
\end{equation}
By substituting (\ref{F(k)}) into equation (\ref{a1}), we obtain a polynomial equation in $\mu=\frac{1}{2}\left(\beta+\beta^{-1}\right)$, because:
\begin{align}
 -i|q|F_{N_{1,2}}&=\varepsilon_{1,2}+2\tilde{\varepsilon}_{1,2}\mu+\sum_{s}\frac{\varepsilon^{1,2}_{s}\sinh\theta^{1,2}_{s}}{\mu-\cosh\theta^{1,2}_{s}}\\
 |q|^{2}&=t_{x}^{2}+t_{y}^{2}+2t_{x}t_{y}\mu.
\end{align}
Upon multiplying by $\prod_{s,r}(\mu-\cosh\theta^{1}_{s})(\mu-\cosh\theta^{1}_{r})$, one can clearly see that equation (\ref{a1}) is indeed a polynomial equation in $\mu$, or in other words:
\begin{equation}
 \prod_{s=1}^{\left\lfloor\frac{N_{1}+N_{2}+2}{2}\right\rfloor}\left(\mu-\mu_{s}\right)=0.\label{a2}
\end{equation}
Returning $\beta$, equation (\ref{a2}) assumes precisely the same form as expression (\ref{beta_eigen_value_problem}). Thus we have successfully proven the initial claim. For the case of periodic boundary conditions, the eigenvalue equation is:
\begin{equation}
 \begin{aligned}
 \left(q-\frac{\left(-q^{*}\right)^{N+1}}{|q|^{N}}G_{N}\right)&\left(q^{*}-\frac{\left(-q\right)^{N+1}}{|q|^{N}}G_{N}\right)\\
 &\hspace{1.5cm}=\left(E+i|q|F_{N}\right)^{2}.
 \end{aligned}\label{b1}
\end{equation}
\smallskip
The right-hand side of equation (\ref{b1}) depends on $\mu$ in exactly the same manner as in the case with open boundary conditions. The expression in the first bracket depends on $\beta$, whereas the second bracket is the corresponding expression with $\beta$ replaced by $\beta^{-1}$. Using some algebra and relation (\ref{G(k)}), we deduce that the left-hand side of (\ref{b1}) can likewise be written solely as a function of $\mu$. Multiplying both sides by $\prod_{s}(\mu-\cosh\theta_{s})^{2}$, we arrive at:
\begin{equation}
 \prod_{s=1}^{{N+2}}\left(\mu-\mu_{s}\right)=0.\label{b2}
\end{equation}
Thus, we have also proved the initial claim (\ref{beta_eigen_value_problem}) in the case of the periodic boundary conditions. In equations (\ref{a2}) and (\ref{b2}) we see the exact values of $M$ appearing in (\ref{beta_eigen_value_problem}).
\subsection{Open-boundary edge problem}\label{Sec:OBC_Bulk_boundary_corr}
In this section, we obtain the eigenstates of the Hamiltonian in the coordinate representation, subject to open boundary conditions at the brane ends, expressed in terms of $\varepsilon_{n}$ as defined in (\ref{eps_n-n'_even}-\ref{eps_n-n'_odd}). The brane boundaries are located at $n+1=-M_{1}$ and $n+1=M_{2}$. The state can be written as $|\psi\rangle=(a_{-M_{1}}\;b_{-M_{1}}\;a_{-M_{1}+1}\;b_{-M_{1}+1}\cdots a_{M_{2}}\; b_{M_{2}})$. In the case of open boundary conditions Hamiltonian (\ref{H_PTB_open_BC}), the time-independent Schrödinger equation reduces to:
\begin{widetext}
 \begin{equation}
 \begin{aligned}
 &\sum_{k=-M_{1}}^{n}a_{k}\varepsilon_{n+1-k}^{a}+\sum_{k=n+1}^{^{M_{2}}}a_{k}\varepsilon_{k-n-1}^{a}-t_{x}b_{n}-t_{y}b_{n+1}=Ea_{n+1},\\
 &\sum_{k=-M_{1}}^{n}b_{k}\varepsilon_{n+1-k}^{b}+\sum_{k=n+1}^{^{M_{2}}}b_{k}\varepsilon_{k-n-1}^{b}-t_{y}a_{n+1}-t_{x}a_{n+2}=Eb_{n+1},
 \end{aligned}\label{difference_eq_OBC}
 \end{equation}
\end{widetext}
where indices $a$ refer to the function $F_{N_{1}}$, whereas indices $b$ correspond to $F_{N_{2}}$. Using the ansatz $a_{n}=\sum_{i}A_{i}\lambda_{i}^{n}$ and $b_{n}=\sum_{i}B_{i}\lambda_{i}^{n}$ and setting the coefficients of the $\lambda_{i}^{n+1}$ terms in the equation to zero, we obtain the following system of equations:
\begin{widetext}
 \begin{equation}
 \begin{aligned}
 &\left(\sum_{s}\frac{2\varepsilon_{s}^{a}\lambda_{i}\sinh{\theta_{s}^{a}}}{1-2\lambda_{i}\cosh\theta_{s}^{a}+\lambda_{i}^{2}}+\tilde{\varepsilon}_{a}\left(\lambda_{i}+\lambda_{i}^{-1}\right)-(E-\varepsilon_{a})\right)A_{i}=(t_{y}+t_{x}\lambda_{i}^{-1})B_{i}\\
 &\left(\sum_{s}\frac{2\varepsilon_{s}^{b}\lambda_{i}\sinh{\theta_{s}^{b}}}{1-2\lambda_{i}\cosh\theta_{s}^{b}+\lambda_{i}^{2}}+\tilde{\varepsilon}_{b}\left(\lambda_{i}+\lambda_{i}^{-1}\right)-(E-\varepsilon_{b})\right)B_{i}=(t_{y}+t_{x}\lambda_{i})A_{i}.
 \end{aligned}\label{dispersion_relation}
 \end{equation}
\end{widetext}
The quantities $\tilde{\varepsilon}_{a,b}$ are defined only when $N_{1}$ and $N_{2}$ are odd. These two equations are interpreted as the dispersion relation, along with the expression of $B_{i}$ in terms of $A_{i}$ (or conversely, $A_{i}$ in terms of $B_{i}$). In addition, the equation (\ref{difference_eq_OBC}) contains terms proportional to $\left(z_{s}^{a,b}\right)^{n+1}$ and $\left(z_{s}^{a,b}\right)^{-n-1}$. Requiring these terms to vanish yields a set of constraints that must be placed upon the coefficients $A_{i}$ and $B_{i}$:
\begin{equation}
 \begin{aligned}
 &\sum_{i}A_{i}\frac{\lambda_{i}^{-M_{1}}\left(z_{s}^{a}\right)^{M_{1}}}{1-\lambda_{i}\left(z_{s}^{a}\right)^{-1}}=\sum_{i}A_{i}\frac{\left(\lambda_{i}z_{s}^{a}\right)^{M_{2}+1}}{1-\lambda_{i}z_{s}^{a}}=0,\\
 &\sum_{i}B_{i}\frac{\lambda_{i}^{-M_{1}}\left(z_{s}^{b}\right)^{M_{1}}}{1-\lambda_{i}\left(z_{s}^{b}\right)^{-1}}=\sum_{i}B_{i}\frac{\left(\lambda_{i}z_{s}^{b}\right)^{M_{2}+1}}{1-\lambda_{i}z_{s}^{b}}=0,
 \end{aligned}\label{conditions_z_s}
\end{equation}
where, in the first line, $s\in\left\{1,2,\ldots,\left\lfloor\frac{N_{1}}{2}\right\rfloor\right\}$, and in the second line, $s\in\left\{1,2,\ldots,\left\lfloor\frac{N_{2}}{2}\right\rfloor\right\}$. To this set of constraints, we need to add four more boundary conditions at $n+1=-M_{1}$ and $n+1=M_{2}$:
\begin{equation}
 \begin{aligned}
 &\sum_{i}\lambda_{i}^{-M_{1}-1}\left(t_{x}B_{i}-\tilde{\varepsilon}_{a}A_{i}\right)=\tilde{\varepsilon}_{b}\sum_{i}\lambda_{i}^{-M_{1}-1}B_{i}=0,\\
 &\sum_{i}\lambda_{i}^{M_{2}+1}\left(t_{x}A_{i}-\tilde{\varepsilon}_{b}B_{i}\right)=\tilde{\varepsilon}_{a}\sum_{i}\lambda_{i}^{M_{2}+1}A_{i}=0,
 \end{aligned}\label{boundary_conditions}
\end{equation}
The total number of conditions depends on the parity of $N_{1}$ and $N_{2}$ because the boundary conditions (\ref{boundary_conditions}) explicitly involve $\tilde{\varepsilon}_{a,b}$, which is non-zero only for odd numbers. This leads to three separate cases:
\begin{itemize}
 \item If both $N_{1}$ and $N_{2}$ are even, then $\tilde{\varepsilon}_{a,b}=0$, so the four boundary conditions reduce to two, giving a total of $N_{1}+N_{2}+2$ conditions. When equation (\ref{dispersion_relation}) is recast as a polynomial in $\lambda$, its maximal degree is precisely $N_{1}+N_{2}+2$, so the number of $\lambda_{i}$ coincides with the number of conditions required for consistency.
 \item If $N_{1}$ and $N_{2}$ have opposite parity, then exactly one of $\tilde{\varepsilon}_{a,b}$ is zero. In this case, the number of boundary conditions again drops by two: one condition is automatically satisfied and two others become identical. The resulting number of independent conditions is $N_{1}+N_{2}+1$, which matches the degree of the polynomial defining $\lambda_{i}$.
 \item If both $N_{1}$ and $N_{2}$ are odd, all boundary conditions must be enforced, yielding $N_{1}+N_{2}+2$ independent conditions. This is equal to the degree of the polynomial (\ref{disp_rel_dd1}) in $\lambda$.
\end{itemize}
Across all three cases, the total number of solutions $\lambda_{i}$ to the dispersion relation (\ref{dispersion_relation}) can be expressed as $2\left(\left\lfloor\frac{N_{1}+N_{2}}{2}\right\rfloor+1\right)$. Therefore, the algorithm for solving the eigenvalue problem can be summarized as follows:
\begin{itemize}
 \item Determine all $2\left(\left\lfloor\frac{N_{1}+N_{2}}{2}\right\rfloor+1\right)$ solutions for $\lambda_{i}$ as functions of the energy $E$ by using the dispersion relation (\ref{dispersion_relation}).
 \item Insert these solutions into the constraints (\ref{conditions_z_s}) and the boundary conditions (\ref{boundary_conditions}).
 \item Using (\ref{dispersion_relation}), express $B_{i}$ in terms of $A_{i}$.
 \item The resulting system of homogeneous linear equations for the coefficients $A_{i}$ admits nontrivial solutions only if its determinant vanishes. Since this determinant depends solely on the energy $E$, setting it equal to zero yields the final equation from which the allowed energy values are obtained.
\end{itemize}

The first consistency check of this approach is whether the continuum bands are obtained in the infinite-system limit, i.e., for $M_{1}\to\infty$ and $M_{2}\to\infty$. In this limit, all constraints (\ref{conditions_z_s}), together with the boundary conditions (\ref{boundary_conditions}), are automatically satisfied, so that a single $\lambda=e^{ik}$ suffices. Consequently, the original energy bands are recovered by matching (\ref{Energy_open_BC}) with the dispersion relation (\ref{dispersion_relation}) by using (\ref{F(k)}).

To test whether the bulk-boundary correspondence holds, we analyze how the dispersion relation (\ref{dispersion_relation}) behaves near the critical point $t_{x}=t_{y}$, as indicated by the bulk topological invariants (\ref{top_inv_even}) and (\ref{top_inv_odd_1}). The dispersion relation is written in terms of $\mu_{i}=\frac{1}{2}\left(\lambda_{i}+\lambda_{i}^{-1}\right)$. Because the topological phase transition takes place at $t_{x}=t_{y}$, we anticipate a solution close to $\lambda_{0}\approx-\frac{t_{y}}{t_{x}}$, which can be equivalently expressed as $\mu_{0}=-\frac{t_{x}^{2}+t_{y}^{2}}{2t_{x}t_{y}}+\delta_{0}$, with $\delta_{0}$ a small correction proportional to $\delta_{t}$ defined via $\frac{t_{x}}{t_{y}}=1+\delta_{t}$. We therefore adopt a perturbative approach in which $\delta_{t}$ serves as the expansion parameter. In this framework, the mode energy is written as $E=E_{0}+\delta_{E}$, where $E_{0}$ is the energy associated with $\mu_{0}=-1$. Under these assumptions, the dispersion relation simplifies to:
\begin{equation}
 \begin{aligned}
 \delta_{0}=\frac{1}{2t_{x}t_{y}}&\left[\delta_{E}+E_{0}-\varepsilon_{a}+2\tilde{\varepsilon}_{a}-\frac{i\omega}{N_{1}+1}\sum_{s}\cot^{2}\phi_{s}^{a}\right]\\
 \times&\left[\delta_{E}+E_{0}-\varepsilon_{b}+2\tilde{\varepsilon}_{b}-\frac{i\omega}{N_{2}+1}\sum_{s}\cot^{2}\phi_{s}^{b}\right],
 \end{aligned}\label{disp_rel_dd1}
\end{equation}
where $\phi_{s}=\frac{2s-1}{N+1}\frac{\pi}{2}$ for even $N$ and $\phi_{s}=\frac{s\pi}{N+1}$ for odd $N$. These sums have closed forms given by:
\begin{align}
 \sum_{s=1}^{\frac{N}{2}}\cot^{2}\left(\frac{2s-1}{N+1}\frac{\pi}{2}\right)&=\frac{N(N+1)}{2},\\
 \sum_{s=1}^{\frac{N-1}2{}}\cot^{2}\left(\frac{s\pi}{N+1}\right)&=\frac{N(N-1)}{6}.
\end{align}
Placing these sums in the dispersion relation (\ref{disp_rel_dd1}) yields:
\begin{equation}
 \delta_{0}=\frac{1}{2t_{x}t_{y}}\left(E_{0}+i\omega+\delta_{E}\right)^{2},
\end{equation}
for both parities of $N_{1}$ and $N_{2}$. Since this quantity must be small, the zeroth-order mode energy is given by $E_{0} = -i\omega$. To find the zeroth-order of the remaining $\mu_{i}$, we rewrite the dispersion relation for $E=E_{0}=-i\omega$ using expression (\ref{F(k)}):
\begin{equation}
 \frac{\sinh(N_{1}\phi)}{\cosh\left(\left(N_{1}+1\right)\phi\right)}\frac{\sinh(N_{2}\phi)}{\cosh\left(\left(N_{2}+1\right)\phi\right)}=-1,
\end{equation}
where the specific case in which both $N_{1}$ and $N_{2}$ are even is considered. In general, this type of equation admits $\left\lfloor\frac{N_{1}+N_{2}}{2}\right\rfloor$ solutions, which can be written in the following form:
\begin{equation}
 \phi_{r}^{a\land b}\equiv i\alpha_{r}=\begin{dcases}i\frac{2r-1}{N_{1}+N_{2}+1}\frac{\pi}{2}, &N_{1}+N_{2}\text{ - even}\\
 i\frac{r}{N_{1}+N_{2}+1}\pi,&N_{1}+N_{2}\text{ - odd}
 \end{dcases}
\end{equation}
where $r\in\left\{1,2,...,\left\lfloor\frac{N_{1}+N_{2}}{2}\right\rfloor\right\}$. Therefore, the complete set of solutions for $\mu_{i}$ at energy $E=-i\omega$, valid for arbitrary $t_{x}$ and $t_{y}$, is given by:
\begin{equation}
 \begin{aligned}
 \mu_{r}&=\cosh\theta^{a\land b}_{r}\equiv-\frac{t_{x}^{2}+t_{y}^{2}+\frac{\omega^{2}}{4\sin^{2}\alpha_{r}}}{2t_{x}t_{y}},\\
 \mu_{0}&=\cosh\theta^{a\land b}_{0}\equiv-\frac{t_{x}^{2}+t_{y}^{2}}{2t_{x}t_{y}},
 \end{aligned}\label{mu_r_sol}
\end{equation}
where we have written $\mu_{r}$ in terms of $z_{r}^{a\land b}=\exp\left(\theta_{r}^{a\land b}\right)$ introduced in Sec. S.3. of the SM. The first-order perturbative corrections, $\mu_{i}=\cosh\theta_{i}^{a\land b}+\delta_{i}$ are as follows:
\begin{equation}
 \begin{aligned}
 \delta_{r}&=i\frac{\delta_{E}\omega}{t_{y}^{2}}\frac{1+(-1)^{\tilde{r}}\sin\alpha_{r}\cos\left[\left(N_{2}-N_{1}\right)\left(\alpha_{r}+\frac{\pi}{2}\right)\right]}{2\left(N_{1}+N_{2}+1\right)\sin^{2}\alpha_{r}},\\
 \delta_{0}&=\frac{\delta_{E}^{2}}{2t_{x}t_{y}},
 \end{aligned}\label{delta_r_sol}
\end{equation}
where $\tilde{r}=r+\left\lceil\frac{N_{1}+N_{2}}{2}\right\rceil$. Finally, the values of $\lambda_{i}$ and $\lambda_{i}^{-1}$ are obtained by solving the equation $\lambda_{i}^{2}-2\mu_{i}\lambda_{i}+1=0$, which yields:
\begin{equation}
 \begin{aligned}
 \lambda_{r}&=z_{r}^{a\land b}\left(1+\frac{\delta_{r}}{\sinh\theta^{a\land b}_{r}}\right),\\
 \lambda_{0}&=-\frac{t_{y}}{t_{x}}\left[1+\delta_{t}\left(1-\sqrt{1-2\frac{\delta_{0}}{\delta_{t}^{2}}}\right)\right],
 \end{aligned}\label{lambda_r_sol}
\end{equation}
The final expressions for $\mu_{i}$ and $\lambda_{i}$ are written in terms of $\delta_{E}$ and $\delta_{t}$. The remaining relation connecting $\delta_{E}$ and $\delta_{t}$ is obtained from the constraints (\ref{conditions_z_s}) and boundary conditions (\ref{boundary_conditions}) by requiring that the determinant of this system of equations vanishes. Because these constraints depend on the type of the system, namely, whether the system is infinite, half-infinite or finite, as well as on the values of $N_{1}$ and $N_{2}$, these cases must be treated separately.

\subsubsection{Half-infinite system}

Next, we consider the half-infinite setup with a boundary at $n=1$, which implies $M_{1}=-1$; while the other boundary is sent to infinity, i.e. $M_{2}\to\infty$. The bulk states are obtained by requiring that one pair $\lambda_{i}$, $\lambda_{i}^{-1}$ equals $e^{ik}$, $e^{-ik}$. Because the states must be $\delta$-function normalizable, all remaining $\lambda_{i}$ must satisfy $|\lambda_{i}|\leqslant1$, so for each pair we select one of the two possibilities $\lambda_{i}$ or $\lambda_{i}^{-1}$. Consequently, there are $\left\lfloor\frac{N_{1}+N_{2}}{2}\right\rfloor+2$ admissible values of $\lambda_{i}$. On the other hand, when we take the limit $M_{2}\to\infty$, half of the constraints in (\ref{conditions_z_s}), together with the boundary condition at infinity, disappear. As a result, we end up with $\left\lfloor\frac{N_{1}+N_{2}}{2}\right\rfloor+1$ independent conditions. Solving this system allows us to express all coefficients $A_{i}$ and $B_{i}$ in terms of a single remaining free coefficient, whose value is then fixed by normalizing the states.

To determine whether any edge modes exist, we adopt a different method. We require that no states penetrate into the bulk, which means that all $K+1=\left\lfloor\frac{N_{1}+N_{2}}{2}\right\rfloor+1$ solutions for $\lambda_{i}$ with $|\lambda_{i}|<1$ are allowed. An equal number of constraints must therefore be applied, which leads to the condition that the determinant of the corresponding system of equations must vanish. This yields the equation governing the possible edge modes:
\begin{equation}
 \begin{vmatrix}
 1+\frac{t_{x}^{2}\lambda_{0}+\tilde{\varepsilon}_{a}\tilde{E}_{b}^{0}}{t_{x}t_{y}} & 1+\frac{t_{x}^{2}\lambda_{1}+\tilde{\varepsilon}_{a}\tilde{E}_{b}^{1}}{t_{x}t_{y}} & \cdots & 1+\frac{t_{x}^{2}\lambda_{K}+\tilde{\varepsilon}_{a}\tilde{E}_{b}^{K}}{t_{x}t_{y}}\\
 \tilde{\varepsilon_{b}}\left(t_{y}+t_{x}\lambda_{0}\right) & \tilde{\varepsilon_{b}}\left(t_{y}+t_{x}\lambda_{1}\right) & \cdots & \tilde{\varepsilon_{b}}\left(t_{y}+t_{x}\lambda_{K}\right)\\
 \frac{\lambda_{0}(t_{y}+t_{x}\lambda_{0})}{z_{1}^{b}-\lambda_{0}} & \frac{\lambda_{1}(t_{y}+t_{x}\lambda_{1})}{z_{1}^{b}-\lambda_{1}} & \cdots & \frac{\lambda_{K}(t_{y}+t_{x}\lambda_{K})}{z_{1}^{b}-\lambda_{K}}\\
 \frac{\lambda_{0}(t_{y}+t_{x}\lambda_{0})}{z_{2}^{b}-\lambda_{0}} & \frac{\lambda_{1}(t_{y}+t_{x}\lambda_{1})}{z_{2}^{b}-\lambda_{1}} & \cdots & \frac{\lambda_{K}(t_{y}+t_{x}\lambda_{K})}{z_{2}^{b}-\lambda_{K}}\}\\
 \vdots & \vdots & \ddots &\vdots\\
 \frac{\lambda_{0}(t_{y}+t_{x}\lambda_{0})}{z_{\lfloor N_{2}/2\rfloor}^{b}-\lambda_{0}} & \frac{\lambda_{1}(t_{y}+t_{x}\lambda_{1})}{z_{\lfloor N_{2}/2\rfloor}^{b}-\lambda_{1}} & \cdots & \frac{\lambda_{K}(t_{y}+t_{x}\lambda_{K})}{z_{\lfloor N_{2}/2\rfloor}^{b}-\lambda_{K}}\\
 \frac{\lambda_{0}(E+i|q|F_{b}^{0})}{z_{1}^{a}-\lambda_{0}} & \frac{\lambda_{1}(E+i|q|F_{b}^{1})}{z_{1}^{a}-\lambda_{1}} & \cdots & \frac{\lambda_{K}(E+i|q|F_{b}^{K})}{z_{1}^{a}-\lambda_{K}}\\
 \frac{\lambda_{0}(E+i|q|F_{b}^{0})}{z_{2}^{a}-\lambda_{0}} & \frac{\lambda_{1}(E+i|q|F_{b}^{1})}{z_{2}^{a}-\lambda_{1}} & \cdots & \frac{\lambda_{K}(E+i|q|F_{b}^{K})}{z_{2}^{a}-\lambda_{K}}\\
 \vdots & \vdots & \ddots &\vdots\\
 \frac{\lambda_{0}(E+i|q|F_{b}^{0})}{z_{\lfloor N_{1}/2\rfloor}^{a}-\lambda_{0}} & \frac{\lambda_{1}(E+i|q|F_{b}^{1})}{z_{\lfloor N_{1}/2\rfloor}^{a}-\lambda_{1}} & \cdots & \frac{\lambda_{K}(E+i|q|F_{b}^{K})}{z_{\lfloor N_{1}/2\rfloor}^{a}-\lambda_{K}}
 \end{vmatrix}=0,\label{mode_eq_OBC}
\end{equation}
where $\tilde{E}_{b}^{i}=E+i|q|F_{b}(\lambda_{i})$. The first two rows correspond to the two boundary conditions (\ref{boundary_conditions}), while the remaining rows are associated with constraints (\ref{conditions_z_s}). Observe that if either $\tilde{\varepsilon}_{a}$ or $\tilde{\varepsilon}_{b}$ vanishes, one of the boundary conditions is automatically fulfilled. In that situation, we remove the row associated with that condition. This is consistent with the simultaneous reduction in the number of $\lambda_i$ by one. 

In general, computing this determinant exactly is a difficult task. We therefore limit our analysis to the case in which the model is in near-critical behavior. In this regime, $\lambda_i$ are expressed by (\ref{lambda_r_sol}) in terms of the small parameter $\delta_t$. The only unknown quantity that remains is the small energy shift $\delta_E$. Using $t_y + t_x \lambda_0 = -t_y \delta_t \left(1 - \sqrt{1 - 2\left(\frac{\delta_0}{\delta_t}\right)^2}\right)$ and $E + i|q| F_b(\lambda_0) = \delta_E$, we see that the first column is first-order in the perturbative parameter. Expanding the determinant along the first column results in a small quantity, and the remaining determinants are calculated in the zeroth order in perturbative expansion. Thus, we arrive at:
\begin{equation}
 -i\eta\frac{\delta_{E}}{t_{y}\delta_{t}}=1-\sqrt{1-\left(\frac{\delta_{E}}{t_{y}\delta_{t}}\right)^{2}},\label{delta_E_equation}
\end{equation}
where $\eta=D_{1}/D_{2}\in\mathbb{R}$ is a fraction of the following two zeroth-order determinants:
\begin{equation}
 D_{1}=\begin{vmatrix}
 -\frac{i\tilde{\varepsilon}_{a}}{t_{y}} & \frac{t_{y}(1+z_{1}^{a\land b})+i\tilde{\varepsilon}_{a}\tilde{F}_{b}^{1}}{t_{y}}& \cdots & \frac{t_{y}(1+z_{K}^{a\land b})+i\tilde{\varepsilon}_{a}\tilde{F}_{b}^{K}}{t_{y}}\\
 0 & \frac{i\tilde{\varepsilon}_{b}}{t_{y}}(1+z_{1}^{a\land b}) & \cdots & \frac{i\tilde{\varepsilon}_{b}}{t_{y}}(1+z^{a\land b}_{K})\\
 0 & \frac{z_{1}^{a\land b}(1+z_{1}^{a\land b})}{z_{1}^{b}-z_{1}^{a\land b}} & \cdots & \frac{z_{K}^{a\land b}(1+z_{K}^{a\land b})}{z_{1}^{b}-z_{K}^{a\land b}}\\
 \vdots & \vdots & \ddots &\vdots\\
 0 & \frac{z_{1}^{a\land b}(1+z_{1}^{a\land b})}{z_{\lfloor N_{2}/2\rfloor}^{b}-z_{1}^{a\land b}} & \cdots & \frac{z_{K}^{a\land b}(1+z_{K}^{a\land b})}{z_{\lfloor N_{2}/2\rfloor}^{b}-z_{K}^{a\land b}}\\
 \frac{1}{z_{1}^{a}+1} & \frac{\tilde{F}^{1}_{b}z_{1}^{a\land b}}{z_{1}^{a}-z_{1}^{a\land b}} & \cdots & \frac{\tilde{F}^{K}_{b}z_{K}^{a\land b}}{z_{1}^{a}-z_{K}^{a\land b}}\\
 \vdots & \vdots & \ddots &\vdots\\
 \frac{1}{z_{\lfloor N_{1}/2\rfloor}^{a}+1} & \frac{\tilde{F}^{1}_{b}z_{1}^{a\land b}}{z_{\lfloor N_{1}/2\rfloor}^{a}-z_{1}^{a\land b}} & \cdots & \frac{\tilde{F}^{K}_{b}z_{K}^{a\land b}}{z_{\lfloor N_{1}/2\rfloor}^{a}-z_{K}^{a\land b}}
 \end{vmatrix},\label{DET_D1}
\end{equation}

\begin{equation}
 D_{2}=\begin{vmatrix}
 -1 & \frac{t_{y}(1+z_{1}^{a\land b})+i\tilde{\varepsilon}_{a}\tilde{F}_{b}^{1}}{t_{y}}& \cdots & \frac{t_{y}(1+z_{K}^{a\land b})+i\tilde{\varepsilon}_{a}\tilde{F}_{b}^{K}}{t_{y}}\\
 -\frac{i\tilde{\varepsilon}_{b}}{t_{y}} & \frac{i\tilde{\varepsilon}_{b}}{t_{y}}(1+z_{1}^{a\land b})&\cdots&\frac{i\tilde{\varepsilon}_{b}}{t_{y}}(1+z_{K}^{a\land b})\\
 \frac{1}{z_{1}^{b}+1} & \frac{z_{1}^{a\land b}(1+z_{1}^{a\land b})}{z_{1}^{b}-z_{1}^{a\land b}} & \cdots & \frac{z_{K}^{a\land b}(1+z_{K}^{a\land b})}{z_{1}^{b}-z_{K}^{a\land b}}\\
 \vdots & \vdots & \ddots &\vdots\\
 \frac{1}{z_{\lfloor N_{2}/2\rfloor}^{b}+1} & \frac{z_{1}^{a\land b}(1+z_{1}^{a\land b})}{z_{\lfloor N_{2}/2\rfloor}^{b}-z_{1}^{a\land b}} & \cdots & \frac{z_{K}^{a\land b}(1+z_{K}^{a\land b})}{z_{\lfloor N_{2}/2\rfloor}^{b}-z_{K}^{a\land b}}\\
 0 & \frac{\tilde{F}^{1}_{b}z_{1}^{a\land b}}{z_{1}^{a}-z_{1}^{a\land b}} & \cdots & \frac{\tilde{F}^{K}_{b}z_{K}^{a\land b}}{z_{1}^{a}-z_{K}^{a\land b}}\\
 \vdots & \vdots & \ddots &\vdots\\
 0 & \frac{\tilde{F}^{1}_{b}z_{1}^{a\land b}}{z_{\lfloor N_{1}/2\rfloor}^{a}-z_{1}^{a\land b}} & \cdots & \frac{\tilde{F}^{K}_{b}z_{K}^{a\land b}}{z_{\lfloor N_{1}/2\rfloor}^{a}-z_{K}^{a\land b}}
 \end{vmatrix},\label{DET_D2}
\end{equation}
where, for even $N_{2}$, one has $\tilde{F}_{b}^{r}=\frac{\omega}{2t_{y}\sin\alpha_{r}}\frac{\sin\left(N_{2}\alpha_{r}\right)}{\cos\left(\left(N_{2}+1\right)\alpha_{r}\right)}$, and for odd $N_{2}$, $\tilde{F}_{b}^{r}=-\frac{\omega}{2t_{y}\sin\alpha_{r}}\frac{\cos\left(N_{2}\alpha_{r}\right)}{\sin\left(\left(N_{2}+1\right)\alpha_{r}\right)}$. By definition $\eta$ is a real number. Once again, the second row is excluded from both determinants if either $\tilde{\varepsilon}_{b}$ or $\tilde{\varepsilon}_{a}$ are equal to zero. The only viable solution to the quadratic equation (\ref{delta_E_equation}) is $\delta_{E}=-\frac{2i\eta t_{y}}{1-\eta^{2}}\delta_{t}$. Thus, the energy of the mode and $\lambda_{0}$ are given by:
\begin{align}
 E&=-i\omega-\frac{2i\eta t_{y}}{1-\eta^{2}}\delta_{t},\\
 \lambda_{0}&=-1+\frac{1+\eta^{2}}{1-\eta^{2}}\delta_{t}.
\end{align}
The criterion for edge localization of the modes is $|\lambda_{i}|<1$. By construction, this is automatically satisfied for $\lambda_{r}$, since there are always two solutions, one greater than 1 and the other less than 1, and we simply select the one with magnitude below 1. The remaining condition is therefore $|\lambda_{0}|<1$. This holds when $\delta_{t}>0$ and $|\eta|<1$ or $\delta_{t}<0$ and $|\eta|>1$. Because the topological invariant predicts edge states only for $\delta_{t}>0$, verifying the bulk-boundary correspondence requires us to evaluate $\eta$ and check if $|\eta|<1$. We therefore proceed to calculate $\eta$ in several special cases.
\subsubsection{Half-infinite system with edge brane}
The edge-brane is a configuration in which either $N_{1}$ or $N_{2}$ is equal to zero. In this case, determinant (\ref{mode_eq_OBC}) can be computed exactly. We choose the case when $N_{2}=0$ and arrive at the following closed from expression for the determinant:
\begin{equation}
 D\equiv\frac{\prod\limits_{s<r}\left(z_{s}-z_{r}\right)}{\prod\limits_{s,i}\left(z_{s}-\lambda_{i}\right)}\prod_{i<j}\left(\lambda_{i}-\lambda_{j}\right)\left[\tilde{t}_{y}\prod_{s}z_{s}+t_{x}\prod_{i}\lambda_{i}\right],
\end{equation}
where $\tilde{t}_{y}=t_{y}+\frac{\tilde{\varepsilon}_{a}}{t_{x}}(E+i\omega)$ and the index $a$ has been dropped from $z_{s}^{a}$. Since $\lambda_{i}\neq\lambda_{j}$ when $i\neq j$, the condition for the edge modes is given by:
\begin{equation}
 \prod_{i}\lambda_{i}=-\frac{t_{y}}{t_{x}}\left(1+\frac{\tilde{\varepsilon}_{a}(E+i\omega)}{t_{x}t_{y}}\right)\prod_{s}z_{s}.
\end{equation}
Substituting ansatz for $\lambda_{i}$ from (\ref{lambda_r_sol}) we get an equation equivalent to (\ref{delta_E_equation}) where $\eta$ is given by the following expression:
\begin{equation}
 \eta=\frac{1}{N+1}\left[\sum_{r}\frac{\cos^{2}\alpha_{r}}{\sqrt{\sin^{2}\alpha_{r}+\left(\frac{\omega}{4t_{y}}\right)^{2}}}+\frac{2t_{y}}{\omega}\delta_{N-odd}\right].\label{eta_half_inf}
\end{equation}
This sum is an increasing function of $N$ and a decreasing function of $\omega$. In the large system limit $N \to \infty$, it can be rewritten in terms of the complete elliptic integrals of the first ($K$) and second ($E$) kind:
\begin{equation}
 \eta=\frac{\sqrt{\tilde{\omega}^{2}+4}}{2\pi}\left[K\left(\frac{2}{\sqrt{\tilde{\omega}^{2}+4}}\right)-E\left(\frac{2}{\sqrt{\tilde{\omega}^{2}+4}}\right)\right],
\end{equation}
where $\tilde{\omega}=\frac{\omega}{t_{y}}$. It is important to note that for $\tilde{\omega}<0.256$ we find $\eta>1$, which would suggest, at first glance, the absence of edge modes at small $\omega$ and large $N$. For finite even $N$, however, the lowest admissible value of $\tilde{\omega}$ decreases as $N$ is reduced, and for $N\leqslant12$ the sum drops below 1 for every $\omega$. This behavior contrasts with our numerical simulations, in which the edge mode consistently emerges whenever $\delta_{t}>0$, which is in agreement with the prediction of the bulk topological invariant. This discrepancy may be due to a possible breakdown of the perturbative approach at small $\omega$. In particular, as $\omega\to0$ one finds $\sinh\theta^{a\land b}_{r}\to0$, suggesting that the series expansions in (\ref{lambda_r_sol}) are no longer reliable. Conversely, for odd $N$ the term outside of the sum scales as $1/\omega$. Hence, for finite $N$ and sufficiently small $\omega$, we obtain $|\eta|>1$ and the mode vanishes. For the same $\omega$ and $t_x<t_y$, however, an edge mode appears in a parameter regime that is not anticipated by the bulk topological invariant. This observation is consistent with our numerical simulations. In the simplest case for even $N=2$ we have:
\begin{equation}
 \begin{aligned}
 E&=-i\omega-\frac{2it_{y}\delta_{t}}{\tilde{\omega}^{2}+3}\sqrt{\tilde{\omega}^{2}+4},\\
 \lambda_{0}&=-\frac{t_{y}}{t_{x}}\left(1-\frac{2\delta_{t}}{\tilde{\omega}^{2}+3}\right),\\
 \lambda_{1}&=z\left(1+\frac{2\delta_{t}}{\tilde{\omega}^{2}+3}\right).
 \end{aligned}
\end{equation}
We can see that for $N=2$, $|\lambda_{0}|<1$ for all values of Matsubara frequency when $\delta_{t}>0$. This is the exact behavior predicted by the topological invariant. For smallest possible odd $N=1$ the relevant edge mode parameters are given by:
\begin{equation}
 \begin{aligned}
 E&=-i\omega\left(1+\frac{t_{x}^{2}/{t_{y}^{2}}-1}{\tilde{\omega}^{2}-1}\right),\\
 \lambda_{0}&=-\frac{t_{y}}{t_{x}}\left(1-\frac{t_{x}^{2}/t_{y}^{2}-1}{\tilde{\omega}^{2}-1}\right).
 \end{aligned}
\end{equation}
These two equations are not a perturbative result; instead, they describe the exact edge mode for the case $N=1$. Upon expansion, we recover the expected expression $\eta=1/{\tilde{\omega}}$ given in (\ref{eta_half_inf}). We observe that for $\tilde{\omega}>1$, the standard bulk-boundary correspondence is valid, and the edge mode emerges in the topologically nontrivial phase. In contrast, when $\tilde{\omega}<1$, the mode appears in a topologically trivial phase.

In the alternative setup, where the boundary is fixed at $M_{2}=-1$ and $M_{1}\to-\infty$, we obtain a different boundary condition, and the determinant takes the form
\begin{equation}
 D\equiv\frac{\prod\limits_{s<r}\left(z_{s}-z_{r}\right)}{\prod\limits_{s,i}\left(1-\lambda_{i}z_{s}\right)}\prod_{i<j}\left(\lambda_{i}-\lambda_{j}\right).
\end{equation}
Because $\lambda_{i}\neq\lambda_{j}$ for $i\neq j$, the equation $D=0$ has no solutions, which forces $A_{i}=0$. Using the dispersion relation (\ref{dispersion_relation}), we obtain the corresponding mode parameters:
\begin{equation}
 E=-i\omega,\hspace{2mm}\lambda=-\frac{t_{x}}{t_{y}}.
\end{equation}
This mode is realized for $|\lambda|>1$ since the site index $n<0$. Hence, we infer that this mode present in the topological phase $t_{x}>t_{y}$, consistent with the prediction of the topological invariant. If, instead of choosing $N_{2}=0$, we set $N_{1}=0$, the situation would be reversed. In that case, the mode fixed at $E=-i\omega$ would emerge when the boundary is set at $M_{1}=-1$ and $M_{2}\to\infty$, while the other mode would occur in the second scenario.
\subsubsection{Half-infinite symmetric system}
The symmetric setup arises when the brane is positioned at the center of the system, so that $N_{1}=N_{2}=N$. In this scenario, both cases when boundary is at $n=-M_{1}$ and $n=M_{2}$ are equivalent. We present the two simplest examples. First, let us look into the even $N=2$ case. The determinants $D_{1}$ and $D_{2}$ are given by:
\begin{equation}
 \begin{aligned}
 D_{1}&=\frac{z(1+z_{2})(1+z_{1})\left(z_{2}-z_{1}\right)}{(1+z)(z-z_{1})(z-z_{2})},\\
 D_{2}&=\frac{z\left(z_{1}(1+z_{2})^{2}\tilde{F}_{1}-z_{2}(1+z_{1})^{2}\tilde{F}_{2}\right)}{(1+z)(z-z_{1})(z-z_{2})},
 \end{aligned}
\end{equation}
where $z$ denotes the unique pole of the system, whereas $z_{1}$ and $z_{2}$ represent the zeroth-order solutions for $\lambda_{i}$ in the perturbative expansion. Then, calculation of the $\eta=D_{1}/D_{2}$ parameter yields:
\begin{equation}
 \eta=\frac{(1+z_{2})(1+z_{1})\left(z_{2}-z_{1}\right)}{z_{1}(1+z_{2})^{2}\tilde{F}_{1}-z_{2}(1+z_{1})^{2}\tilde{F}_{2}}.
\end{equation}
For the even case $N=2$, we obtain $\alpha_{1}=\frac{\pi}{10}$ and $\alpha_{2}=\frac{3\pi}{10}$. By substituting these values of $\alpha_{r}$ into $z_{r}$ and evaluating $\tilde{F}_{r}$ explicitly, we find that $\eta<1$ for all $\omega$. Hence, the bulk-boundary correspondence holds in this case.

\begin{figure}[t]
 \begin{center}
 \captionsetup{justification=justified,singlelinecheck=false}
\includegraphics[width=0.47\textwidth]{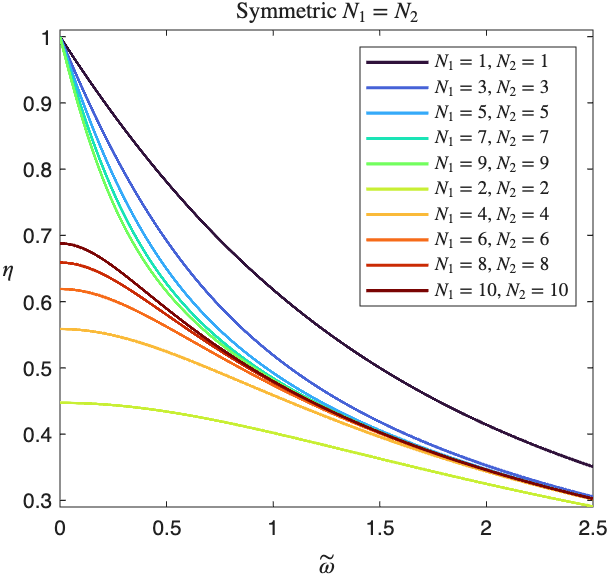}
 \caption{This figure shows how the parameter $\eta$, defined in (\ref{delta_E_equation}), behaves as $\omega$ is varied for both even and odd values of $N_{1}=N_{2}=N$. We observe that, as $\omega$ decreases, $\eta$ approaches a value smaller than 1 for even $N$, indicating that the mode is always present. In contrast, for odd $N$, $\eta \to 1$ as $\omega \to 0$. This also means that the mode remains present, since $\eta<1$, but the perturbative approach breaks as $\omega$ is lowered. Both of these conclusions are consistent with the numerical simulations on a finite lattice.}
 \label{fig:eta_symmetruc}
 \end{center}
\end{figure}

Next, we investigate the simplest symmetric odd case, namely $N=1$. Because both $N_{1}$ and $N_{2}$ are odd, we must impose two boundary conditions, which in turn leads to two parameters $\lambda_{i}$ and a single pole $z$ in this configuration. Direct calculation of the determinants yields the following expression for $\eta$:
\begin{equation}
 \eta=\frac{1}{2}\left(\sqrt{\tilde{\omega}^{2}+4}-\tilde{\omega}\right).
\end{equation}
We clearly have $0<\eta<1$ for all values of $\omega$, indicating that the mode is always present. The corresponding mode parameters are:
\begin{equation}
 \begin{aligned}
 E&=-i\omega\left(1+\frac{2\delta_{t}}{\tilde{\omega}^{2}}\right),\\
 \lambda_{0}&=-\frac{t_{y}}{t_{x}}\left(1-\frac{\sqrt{\tilde{\omega}^{2}+4}-\tilde{\omega}}{2\tilde{\omega}}\delta_{t}\right),\\
 \lambda_{1}&=z+2\left(1-\frac{\tilde{\omega}}{\sqrt{\tilde{\omega}^{2}+4}}\right)\delta_{t}.
 \end{aligned}
\end{equation}
Note that the energy and $\lambda_{0}$ diverge as $\tilde{\omega}\to0$, while $\eta\to 1$ within the first-order perturbative correction. This suggests that the perturbative approach breaks at small values of Matsubara frequency as we discussed before. A more detailed illustration of how $\eta$ behaves as $\omega$ is varied for different values of $N$ is provided in Figure \ref{fig:eta_symmetruc}.
\subsubsection{Half-infinite asymmetric system}
\begin{figure*}[htbp]
 \centering
 \includegraphics[width=\textwidth]{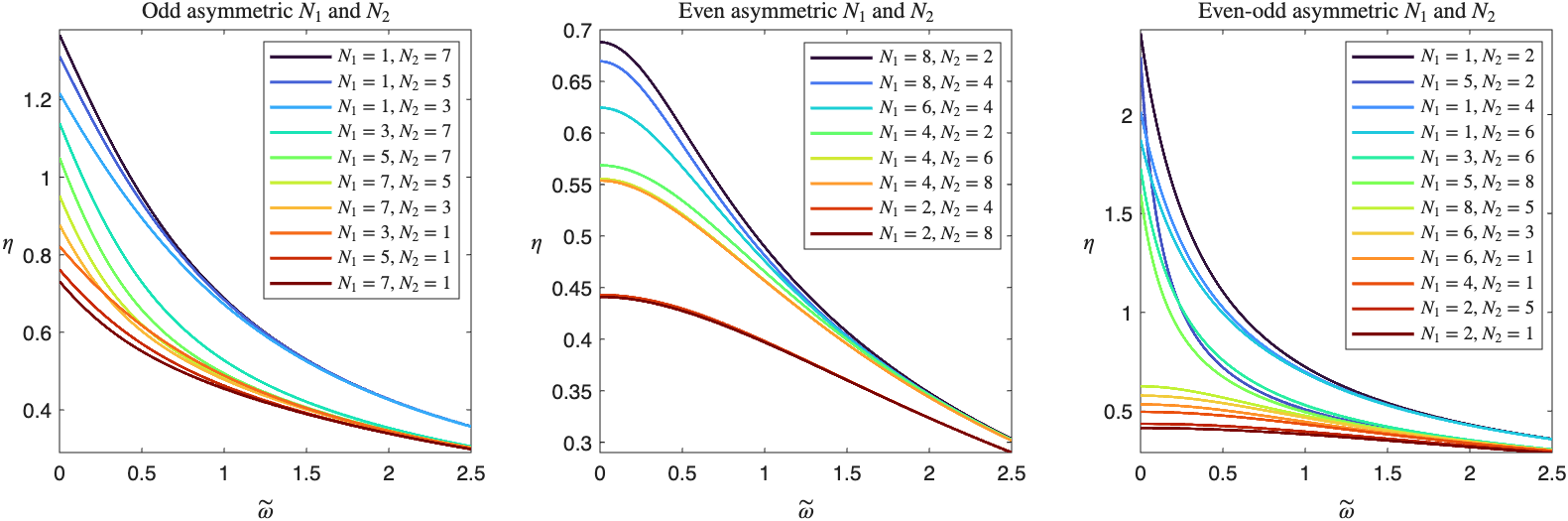}
 \captionsetup{justification=justified,singlelinecheck=false}
 \caption{The behavior of $\eta$, defined in (\ref{delta_E_equation}), as a function of Matsubara frequency, is depicted in this figure. Three distinct scenarios are examined. From the perspective of the bulk-boundary correspondence, the expected behavior $\eta<1$ is obtained when both $N_{1}$ and $N_{2}$ are even, which is shown in the second panel. First panel corresponds to a scenario when both $N_{1}$ and $N_{2}$ are odd. We see that when $N_{1}>N_{2}$, $\eta<1$ for all values of $\omega$, while, when $N_{1}<N_{2}$, $\eta>1$ for sufficiently small $\omega$. Hence, in this regime there is no edge mode when $t_{x}>t_{y}$, it appears when $t_{x}<t_{y}$, breaking the bulk-boundary correspondence. Similar behavior is observed in the third panel, where the scenario with $N_{1}$ and $N_{2}$ of different parity is shown. When $N_{1}$ is even, $\eta<1$ always; while when $N_{1}$ is odd $\eta>1$ when $\omega\to0$. This behavior is observed in numerical simulations on the finite lattice as well}
 \label{fig:eta_asymetric}
\end{figure*}

We now examine the more general case in which both $N_{1}$ and $N_{2}$ are arbitrary but nonzero. For clarity, we restrict our analytical discussion of edge modes to small values of $N_{1}$ and $N_{2}$, while larger values are handled numerically. Because the equations governing $\eta$ become increasingly complicated, we instead characterize the dependence of $\eta$ on $\omega$ by numerically evaluating the determinants $D_{1}$ and $D_{2}$. The corresponding results are presented in Figure \ref{fig:eta_asymetric}.

We conclude that for a half-infinite system, the bulk–boundary correspondence is satisfied for all Matsubara frequencies only when the $\mathcal{PT}_{+}$ symmetry of the states remains intact. This corresponds to the situation in which $N_{1}$ and $N_{2}$ are both even, or both odd and equal. If the boundary is placed on the left side of the brane, i.e., at site $n+1=-M_{1}=1$, and either (i) $N_{1}$ and $N_{2}$ are both odd with $N_{1}>N_{2}$, or (ii) $N_{1}$ is odd and $N_{2}$ is even, then there exists a sufficiently small $\omega_{0}$ such that $\eta>1$ for $\omega<\omega_{0}$. In this regime, the mode disappears for $t_{x}>t_{y}$ and emerges for $t_{x}<t_{y}$, signaling a breakdown of bulk–boundary correspondence. Conversely, when $N_{1}$ and $N_{2}$ are both odd with $N_{1}<N_{2}$, or when $N_{2}$ is odd and $N_{1}$ is even, the analogous behavior occurs in the complementary configuration where the boundary is on the right side of the brane. Therefore, in a finite system there is always at least one mode that violates the bulk–boundary correspondence whenever the states lose their $\mathcal{PT}_{+}$ symmetry, which is equivalent to the existence of a region in the Brillouin zone where the bands become flat. This is clearly illustrated in Figures \ref{fig:eta_symmetruc} and \ref{fig:eta_asymetric} for a brane which occupies the positive half of the real axis.

Finally, we point out that for sufficiently large even values of $N$, the parameter $\eta$ can exceed 1. As discussed earlier, this is interpreted as an artifact of the perturbative approach, since such behavior is not observed in numerical simulations on finite lattices.

\subsubsection{Finite lattice with open boundary conditions}
In this section, we briefly outline how edge modes emerge in a finite system. When both $M_{1}$ and $M_{2}$ are finite, all constraints in (\ref{conditions_z_s}) as well as all boundary conditions in (\ref{boundary_conditions}) must be taken into account. The algorithm for obtaining all eigenstates is the one introduced at the beginning of this section. One can readily verify that if all $\lambda_{i}$ satisfy $|\lambda_{i}|\neq1$ the determinant that characterizes the modes factorizes into two diagonal blocks which yield precisely the two edge modes localized at the left and at the right ends of the system. This modes coincide with the ones obtained for the half-infinite system discussed in the previous section.

Our numerical simulations reveal up to $2(N_{1}+N_{2})$ additional edge modes. For large values of $\omega$, no such extra modes are found. As $\omega$ is decreased, these modes initially appear gradually as gapped edge modes. With a further decrease of the Matsubara frequency, more modes emerge and progressively move toward the imaginary axis, thereby becoming gapless edge modes. Nevertheless, since they do not arise in the half-infinite system limit and are absent at large $\omega$, they must be attributed to the specific truncation of the Hamiltonian’s long-range hopping. 

A thorough examination of this behavior is provided in Sec. S.4. of the SM, where numerically computed complex energy spectra are shown for various choices of the Matsubara frequency and different values of $N_{1}$ and $N_{2}$.

\subsection{Periodic-boundary edge response}\label{app:PBC_edge_problem}
Now we turn to the case where the periodic boundary conditions are applied. Using the same setup as for the open boundary conditions, the eigenvalue problem for the Hamiltonian (\ref{H_PTB_periodic_BC}) in coordinate representation is given by:
\begin{widetext}
 \begin{equation}
 \begin{aligned}
 &\sum_{k=-M_{1}}^{n}a_{k}\varepsilon_{n+1-k}+\sum_{k=n+1}^{^{M_{2}}}a_{k}\varepsilon_{k-n-1}-\sum_{k=-M_{1}}^{M_{2}}b_{k}t_{k-n-1}-t_{x}b_{n}-t_{y}b_{n+1}=Ea_{n+1},\\
 &\sum_{k=-M_{1}}^{n}b_{k}\varepsilon_{n+1-k}+\sum_{k=n+1}^{^{M_{2}}}b_{k}\varepsilon_{k-n-1}-\sum_{k=-M_{1}}^{M_{2}}a_{k}t_{n+1-k}-t_{y}a_{n+1}-t_{x}a_{n+2}=Eb_{n+1}.
 \end{aligned}\label{difference_eq_PBC}
 \end{equation}
\end{widetext}
Here we do not need the superscripts $a$ and $b$ to distinguish between the two sub-lattices, since they share the same $z_{s}$ values. By inserting expressions (\ref{eps_sol}) and (\ref{tau_sol}) into (\ref{difference_eq_PBC}) and employing the standard ansatz $a_{n}=\sum_{i}A_{i}\lambda_{i}^{n}$ and $b_{n}=\sum_{i}B_{i}\lambda_{i}^{n}$, we get the system of equations for the coefficients $A_{i}$ and $B_{i}$, by demanding that the coefficient in front of $\lambda_{i}^{n+1}$ vanishes:
\begin{widetext}
 \begin{equation}
 \begin{aligned}
 &\left(\sum_{s}\frac{2\varepsilon_{s}\lambda_{i}\sinh{\theta_{s}}}{1-2\lambda_{i}\cosh\theta_{s}+\lambda_{i}^{2}}+\tilde{\varepsilon}\left(\lambda_{i}+\lambda_{i}^{-1}\right)-(E-\varepsilon)\right)A_{i}\\
 &\hspace{1cm}=\left[\sum_{s}\left(\tau_{s}'\frac{\left(z_{s}^{-1}\lambda_{i}\right)^{\left\lceil\frac{N+1}{2}\right\rceil}}{1-z_{s}\lambda_{i}^{-1}}-\tau_{s}\frac{\left(z_{s}\lambda_{i}\right)^{\left\lceil\frac{N+1}{2}\right\rceil}}{1-z_{s}^{-1}\lambda_{i}^{-1}}\right)+\tau \lambda_{i}^{\left\lceil\frac{N+1}{2}\right\rceil}+\tilde{\tau}\lambda_{i}^{\left\lceil\frac{N-1}{2}\right\rceil}+\tilde{\tau}'\lambda_{i}^{\frac{N+3}{2}}\delta_{N-odd}+t_{y}+t_{x}\lambda_{i}^{-1}\right]B_{i},\\
 &\left(\sum_{s}\frac{2\varepsilon_{s}\lambda_{i}\sinh{\theta_{s}}}{1-2\lambda_{i}\cosh\theta_{s}+\lambda_{i}^{2}}+\tilde{\varepsilon}\left(\lambda_{i}+\lambda_{i}^{-1}\right)-(E-\varepsilon)\right)B_{i}\\
 &\hspace{1cm}=\left[\sum_{s}\left(\tau_{s}'\frac{\left(z_{s}\lambda_{i}\right)^{-\left\lceil\frac{N+1}{2}\right\rceil}}{1-z_{s}\lambda_{i}}-\tau_{s}\frac{\left(z_{s}^{-1}\lambda_{i}\right)^{-\left\lceil\frac{N+1}{2}\right\rceil}}{1-z_{s}^{-1}\lambda_{i}}\right)+\tau \lambda_{i}^{-\left\lceil\frac{N+1}{2}\right\rceil}+\tilde{\tau}\lambda_{i}^{-\left\lceil\frac{N-1}{2}\right\rceil}+\tilde{\tau}'\lambda_{i}^{-\frac{N+3}{2}}\delta_{N-odd}+t_{y}+t_{x}\lambda_{i}\right]A_{i}.
 \end{aligned}\label{disp_rel_PBC}
 \end{equation}
\end{widetext}
Analogously to the situation with open boundary conditions (\ref{dispersion_relation}), these two equations express the relationship between the coefficients $A_{i}$ and $B_{i}$, along with the dispersion relation $E = E(\lambda_{i})$, which is obtained by requiring that the determinant of this system of equations vanishes. Imposing that the coefficients next to $z_{s}^{n+1}$ and $z_{s}^{-n-1}$ vanish, we derive the following set of constraints for $A_{i}$ and $B_{i}$:
\begin{equation}
 \begin{aligned}
 &\sum_{i}\frac{\lambda_{i}^{-M_{1}}z_{s}^{M_{1}}}{1-\lambda_{i}z_{s}^{-1}}\left(\varepsilon_{s}A_{i}-\tau_{s}'B_{i}\right)=0,\\
 &\sum_{i}\frac{\left(\lambda_{i}z_{s}\right)^{M_{2}+1}}{1-\lambda_{i}z_{s}}\left(\varepsilon_{s}A_{i}-\tau_{s}B_{i}\right)=0,\\
 &\sum_{i}\frac{\lambda_{i}^{-M_{1}}z_{s}^{M_{1}}}{1-\lambda_{i}z_{s}^{-1}}\left(\varepsilon_{s}B_{i}-\tau_{s}A_{i}\right)=0,\\
 &\sum_{i}\frac{\left(\lambda_{i}z_{s}\right)^{M_{2}+1}}{1-\lambda_{i}z_{s}}\left(\varepsilon_{s}B_{i}-\tau_{s}'A_{i}\right)=0.
 \end{aligned}
\end{equation}
It turns out that the first and third conditions (and likewise the second and fourth) are equivalent, since $\varepsilon_{s}^{2}=\tau_{s}\tau_{s}'$. Consequently, there are $2\left\lfloor\frac{N}{2}\right\rfloor$ independent constraints. Besides these constraints, appropriate boundary conditions must also be imposed. Unlike the situation with open boundary conditions, the present case admits a much larger set of boundary conditions; in fact, their total number is $2\left(1+\left\lfloor\frac{N+3}{2}\right\rfloor\right)$. Thus, the total number of conditions is $2\left(1+\left\lfloor\frac{N+3}{2}\right\rfloor+\left\lfloor\frac{N}{2}\right\rfloor\right)=2(N+2)$. 

Because Eq.~\eqref{disp_rel_PBC} contains terms of order
\(\lambda_i^{(N+3)/2}\), the characteristic polynomial has maximal
degree \(2(N+\lfloor N/2\rfloor+2)\), with roots arranged in reciprocal
pairs \((\lambda_i,\lambda_i^{-1})\). The apparent solutions
\(\mu=\mu_s\), generated only after multiplying by
\(\prod_s(\mu-\mu_s)^2\), must be discarded because they do not solve
the original equation. This leaves \(2(N+2)\) admissible roots, matching
the number of boundary conditions. The eigenvalues are then obtained by
the same root-matching procedure used for open boundary conditions.

For periodic boundary conditions we focus on finite-lattice spectra.
Gapless boundary modes on the imaginary axis are guaranteed only for
even \(N\), in agreement with the SSH-type invariant of
Sec.~\ref{sec:Main_top_inv}; these modes preserve
\(\mathcal{PT}_{+}\). For odd \(N\), boundary-localized modes may occur
at large \(\omega\), but are absorbed into the bulk as \(\omega\) is
lowered. Thus Fig.~\ref{fig:Phase_diagram_odd_N} is a bulk phase
diagram of parity-quantized crystalline phases, not an edge-mode
counting diagram.

Sec. S.4. of the SM shows representative finite-lattice
spectra. These results show that the absence of the NHSE and the
quantization of the complex Berry phase ensure a well-defined Bloch
bulk invariant, while conventional edge-mode correspondence is recovered
only in the SSH-type sectors. In the odd-\(N\) periodic sector, the
boundary response is symmetry- and termination-sensitive rather than
SSH-like.
\section{Circuit admittance and exceptional-point
transition frequencies: an odd-$N$ periodic case}
\label{app:circuit}

This appendix specifies a topolectrical realization of the odd-\(N\)
projected response. It shows the reduced admittance matrix obtained by
Kron reduction and the conditions required for a physical circuit to
reproduce the projected finite-frequency kernel.

\subsection*{H.1\quad Circuit Laplacian from Kron Reduction}

Kron reduction of a circuit-node Laplacian is a Schur-complement
elimination of the internal-node block~\cite{Chua1987,Dorfler2013_Kron,
Caliskan2014_Kron}. For a five-node unit cell with two external brane
nodes and three internal complement nodes, we write
\begin{equation}
 J_{\rm full}(k,\Omega)=
 \begin{pmatrix}
 J_{BB}(k,\Omega) & J_{BC}(k,\Omega)\\
 J_{CB}(k,\Omega) & J_{CC}(k,\Omega)
 \end{pmatrix}.
 \label{eq:J_full_block}
\end{equation}
Eliminating the internal nodes gives the reduced admittance
\begin{equation}
 J_{\rm eff}(k,\Omega)
 =
 J_{BB}
 -
 J_{BC}J_{CC}^{-1}J_{CB}.
 \label{eq:J_eff_kron}
\end{equation}
The circuit realizes the projected theory when
\begin{equation}
 J_{\rm eff}(k,\Omega)
 =
 \mathcal N(\Omega)
 H_{\rm PTB}[k,z(\Omega)] ,
 \label{eq:circuit_mapping}
\end{equation}
where \(\mathcal N(\Omega)\neq0\) is an admittance scale and
\(z(\Omega)\) is fixed by the unreduced network. The scalar factor
leaves the eigenvectors and exceptional-point condition unchanged,
whereas \(z(\Omega)\) determines the physical transition frequencies.
Thus the circuit elements entering \(J_{BB}\), \(J_{BC}\), and
\(J_{CC}\) must be specified before the imaginary-frequency parameter is
mapped to a drive frequency.

For \(N=3\), the projected kernel entering Eq.~\eqref{eq:circuit_mapping}
is
\begin{equation}\label{eq:Htarget_N3}
  H_{\rm target}(k,\omega)
  \equiv
  H_{\rm PTB}(k,i\omega)
  =
  \begin{pmatrix}
    d(k,\omega) & Q_1(k,\omega) \\[4pt]
    Q_2(k,\omega) & d(k,\omega)
  \end{pmatrix}.
\end{equation}
Equivalently,
\begin{equation}
  J_{\rm eff}(k,\Omega)
  =
  \mathcal N(\Omega)
  H_{\rm target}\bigl(k,\omega(\Omega)\bigr),
  \end{equation}
  where 
  \begin{equation}
  z(\Omega)=i\omega(\Omega).
\end{equation}
Substituting \(N=3\) into Eqs.~\eqref{H_PTB_periodic_BC} gives
\begin{equation}\label{eq:d_N3}
  d(k,\omega)
  =
  -i|q|\frac{\cosh 5\phi}{\sinh 4\phi},
\end{equation}
and
\begin{align}
  Q_1(k,\omega)
  &=
  q^*
  +
  \frac{i|q|\cosh\phi}{\sinh 4\phi}\,e^{4i\arg q},
  \label{eq:Q1_N3}\\
  Q_2(k,\omega)
  &=
  q
  +
  \frac{i|q|\cosh\phi}{\sinh 4\phi}\,e^{-4i\arg q}.
  \label{eq:Q2_N3}
\end{align}
Here
\begin{equation}\label{eq:phi_def}
  q=-(t_y+t_xe^{ikd}),
  \qquad
  \phi=\operatorname{arcsinh}\!\left(\frac{\omega}{2|q|}\right).
\end{equation}
The diagonal term is purely imaginary on the 
imaginary-frequency slice and encodes the back-action of the eliminated
complement. The off-diagonal terms combine the bare SSH hopping with the
resonant correction proportional to \(\cosh\phi/\sinh4\phi\), and their
product gives the spectral discriminant \(\Delta=4Q_1Q_2\).

\subsection*{H.2\quad Two-Terminal Admittance at \texorpdfstring{$k=0$}{k=0}}

\begin{figure*}[t!]
  \centering
\includegraphics[width=\linewidth]{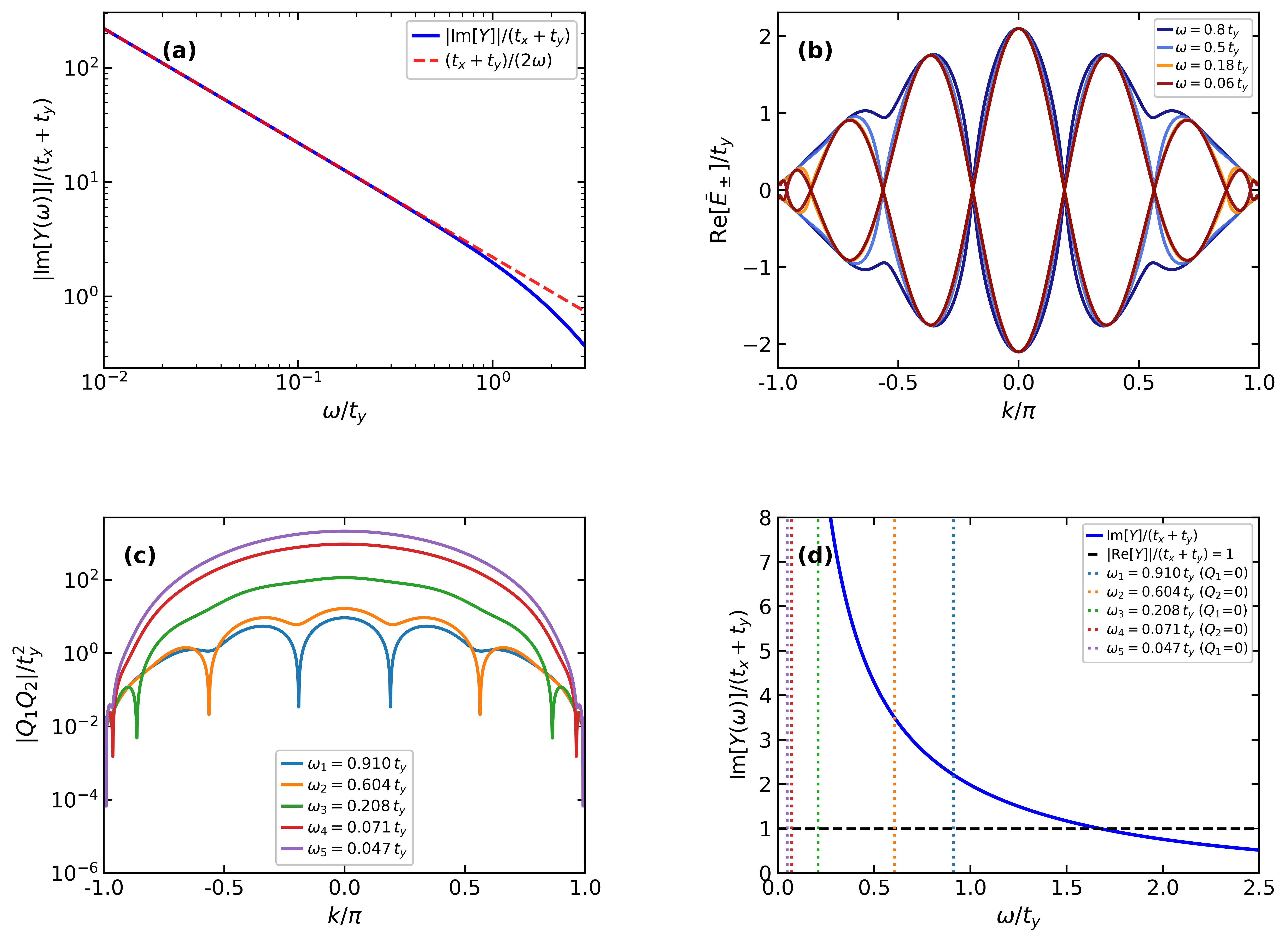}
  \caption{$N=3$ circuit admittance and exceptional-point transition structure
  for $t_x=1.1\,t_y$.
  (a)~$|\operatorname{Im}[Y(\omega)]|/(t_x+t_y)$ versus $\omega/t_y$
  on a log--log scale, confirming the $1/\omega$ divergence
  (Eq.~\eqref{eq:ImY_lowfreq}) and its saturation to the bare-hopping
  value at high frequency.
  (b)~Real part of the bulk spectrum
  $\operatorname{Re}[\bar E_\pm(k,\omega)]/t_y$ at four representative
  frequencies; vertical dashed lines mark the five transition momenta
  $\pm k_\ell$.
  (c)~Discriminant $|Q_1 Q_2|/t_y^2$ as a function of $k/\pi$
  evaluated at each transition frequency $\omega_\ell$ of
  Table~\ref{tab:transitions}, confirming that the zero occurs
  precisely at $k=\pm k_\ell$.
  (d)~$\operatorname{Im}[Y(\omega)]/(t_x+t_y)$ versus $\omega/t_y$
  at $k=0$ with the five predicted transition frequencies
  superimposed; the $1/\omega$ resonant background against which
  $k$-resolved measurements detect the individual transitions is
  clearly visible.
  All frequencies in units of $t_y$.}
  \label{fig:circuit_admittance}
\end{figure*}

At \(k=0\), \(q_0=-(t_x+t_y)\), so \(\arg q_0=\pi\) and
\(e^{4i\arg q_0}=1\). The target kernel reduces to
\(H_{\rm target}(0,\omega)=d_0(\omega)\mathbb I+
Q(0,\omega)\sigma_x\), with
\begin{equation}\label{eq:Y_explicit}
  Y_{\rm target}(\omega)
  \equiv
  [H_{\rm target}]_{12}(0,\omega)
  =
  (t_x+t_y)
  \left[
  -1+\frac{i\cosh\phi_0}{\sinh4\phi_0}
  \right],
\end{equation}
where
\[
  \phi_0=
  \operatorname{arcsinh}
  \!\left[\frac{\omega}{2(t_x+t_y)}\right].
\]
Thus
\begin{equation}\label{eq:ReY_ImY}
  \operatorname{Re}Y_{\rm target}=-(t_x+t_y),
  \, {\rm and}\,
  \operatorname{Im}Y_{\rm target}
  =
  (t_x+t_y)\frac{\cosh\phi_0}{\sinh4\phi_0}.
\end{equation}
For \(\omega\gg |q_0|\), the resonant term vanishes and
\(Y_{\rm target}\to-(t_x+t_y)\), recovering the bare brane hopping. For
\(\omega\ll |q_0|\),
\begin{equation}
  \operatorname{Im}Y_{\rm target}(\omega)
  \xrightarrow{\;\omega\to0^+\;}
  \frac{(t_x+t_y)^2}{2\omega}.
  \label{eq:ImY_lowfreq}
\end{equation}
This \(1/\omega\) divergence is the reduced-response signature of the
complement zero mode. A physical circuit reproduces it when the
internal complement admittance contains the corresponding inverse-
frequency dependence, for example through inductive or engineered
active elements. For \(t_x=1.1t_y\), Eq.~\eqref{eq:ImY_lowfreq} gives
\(\operatorname{Im}Y_{\rm target}\simeq2.205\,t_y/\omega\), consistent
with Fig.~\ref{fig:circuit_admittance}(a).

In the even-\(N\) regular SSH sector, where the complement has no zero
mode, the low-frequency response scales instead as
\(\operatorname{Im}Y\sim\omega\). The \(1/\omega\) and \(\omega\)
scalings therefore distinguish the resonant and regular projection
regimes at the single-unit-cell level.

\begin{table}[t!]
  \centering
  \renewcommand{\arraystretch}{1.35}
  \begin{tabular}{|c|c|c|c|c|}
    \hline
    \(\ell\) & \(|k_\ell|\) & \(\omega_{eff,\ell}^+\)
    & \(\omega_\ell/t_y\) & Vanishing element \\
    \hline\hline
    1 & \(0.599\) & \(2.007\) & \(0.910\) & \(Q_1(k_\ell)=0\) \\
    \hline
    2 & \(1.768\) & \(1.333\) & \(0.604\) & \(Q_2(k_\ell)=0\) \\
    \hline
    3 & \(2.712\) & \(0.458\) & \(0.208\) & \(Q_1(k_\ell)=0\) \\
    \hline
    4 & \(3.025\) & \(0.158\) & \(0.0714\) & \(Q_2(k_\ell)=0\) \\
    \hline
    5 & \(3.112\) & \(0.105\) & \(0.0474\) & \(Q_1(k_\ell)=0\) \\
    \hline
  \end{tabular}
  \caption{Exceptional-point transition values for \(N=3\) and
  \(t_x=1.1t_y\). The momenta are listed in \([0,\pi]\), with partner
  transitions at \(\pm k_\ell\). Here
  \(s_{cr}=\sinh\phi_{cr}\approx0.2267\)
  [Eq.~\eqref{eq:cubic}], independent of \(t_x/t_y\). The listed
  \(\omega_\ell\) are target spectral parameters; the physical drive
  frequencies \(\Omega_\ell\) follow from
  \(z(\Omega_\ell)=i\omega_\ell\).}
  \label{tab:transitions}
\end{table}
\subsection*{H.3\quad Exceptional-point transition frequencies}
            
For $N=3$, $t_x=1.1\,t_y$, the five exceptional-point transitions arise from the condition
$\Delta(k_\ell,\omega_\ell)=4Q_1(k_\ell,\omega_\ell)
Q_2(k_\ell,\omega_\ell)=0$, which requires simultaneously
$|\tilde G_3(\phi_{cr})|=1$ (fixing $\phi$ to a universal
critical value) and $\alpha(k_\ell)=(2\ell-1)\pi/2$ (fixing the
momentum).

The condition $|\tilde G_3(\phi_{cr})|=1$, i.e.\
$\cosh\phi_{cr}=\sinh(4\phi_{cr})$, reduces  to a cubic equation of the form, 
 \begin{equation}\label{eq:cubic}
  8s^3 + 4s - 1 = 0,
  \end{equation}
  where
  \begin{equation}
  s\equiv\sinh\phi_{cr},
\end{equation}
with unique positive real root $s_{cr}=\sinh\phi_{cr}\approx 0.2267$.
This value is independent of $k$, $t_x$, and $t_y$.

For general odd \(N\) the critical condition takes the same form,
\(|\tilde G_N(\phi_{cr})|=1\), which from Eq.~\eqref{G_N_functions} reads
\(\cosh\phi_{cr}=|\sinh[(N+1)\phi_{cr}]|\). Writing \(s\equiv\sinh\phi_{cr}\)
and expanding \(\sinh[(N+1)\phi]\) as a polynomial in \(s\) yields a
single algebraic equation with a unique positive root
\(s_{cr}(N)\), which decreases monotonically with \(N\) (the \(N=3\)
case reduces to the cubic above). Each odd \(N\) therefore fixes its own
universal \(\phi_{cr}\), independent of \(t_x,t_y\); the number of
physical transitions then follows from combining this \(\phi_{cr}\) with
the angular condition \(\alpha=(2\ell-1)\pi/2\), reproducing the count
\(N_+^{(\mathrm{max})}\) of Eq.~\eqref{N_plus_max}.

At each transition, one off-diagonal element vanishes. For \(N=3\),
Eqs.~\eqref{eq:Q1_N3}--\eqref{eq:Q2_N3} show that the resonant phase
\(e^{\pm4i\arg q}\) combines with the phase of \(q\) or \(q^*\), so the
relevant phase is \(5\arg q\). The two conditions are
\begin{align}
    Q_1(k_\ell)&=0:\;5\arg q(k_\ell)\equiv \frac{\pi}{2}\quad (\mathrm{mod}\;2\pi)\\
    Q_2(k_\ell)&=0:\;5\arg q(k_\ell)\equiv \frac{3\pi}{2}
    \quad (\mathrm{mod}\;2\pi).
\end{align}
Using \(\arg q(k_\ell)=\pi-\vartheta_\ell\), these conditions alternate
with \(\ell\), giving the sequence
\(Q_1,Q_2,Q_1,Q_2,Q_1\) in Table~\ref{tab:transitions}.

Table~\ref{tab:transitions} lists the five predicted transition
frequencies, transition momenta (in radians), and vanishing elements.
The transitions span nearly two decades in frequency
($0.047\,t_y\lesssim\omega_\ell\lesssim 0.910\,t_y$) and sample
the full Brillouin zone.
The spacing is set by the SSH winding: transitions near $|k|\approx 0$
(large $|q|$, large $\omega$) correspond to the topologically
nontrivial regime $t_x>t_y$, while transitions near $|k|\approx\pi$
(small $|q|$, small $\omega$) probe the near-critical region.

\subsection*{H.4\quad Spectral Signatures and Measurement Protocol}

A single brane unit cell with three complement nodes reduces to the
two-terminal network described by Eq.~\eqref{eq:Y_explicit}. The imaginary part of the admittance is the susceptance and can
be reconstructed from the measured complex two-terminal response. The
corresponding imaginary part of the impedance is the reactance.  It
diverges as $1/\omega$ at low frequency (Eq.~\eqref{eq:ImY_lowfreq})
in the odd-$N$ sector and vanishes linearly in the even-$N$ sector,
providing an unambiguous low-frequency diagnostic of the zero-mode
resonance.  The exceptional-point transitions do not produce features in
$Y(k=0,\omega)$ directly, since each transition occurs at
$k_\ell\neq 0$.
 
For a chain of \(M\) brane unit cells, the momentum-resolved
transfer response can be obtained either by coherently driving the unit
cells with programmable relative phases \(e^{ikn}\) or by Fourier
transforming the measured real-space response matrix.  At each transition value \(\omega_\ell\) and momentum \(k_\ell\), one
of the target off-diagonal elements \(Q_1\) or \(Q_2\) vanishes. For a
physical two-port network with equal reference impedances \(Z_0\), the
scattering matrix is obtained from the reduced admittance as
\begin{equation}
 S(\Omega)=
 [I-Z_0J_{\rm eff}(\Omega)]
 [I+Z_0J_{\rm eff}(\Omega)]^{-1}.
 \label{eq:admittance_to_scattering}
\end{equation}
The vanishing projected coupling therefore produces a pronounced
minimum in the momentum-resolved transfer response. It is an exact
transmission zero only when Eq.~\eqref{eq:admittance_to_scattering},
together with the port impedances and the circuit mapping, gives
\(S_{21}(k_\ell,\Omega_\ell)=0\). Component tolerances and parasitic
series resistance broaden the minimum in a physical realization.
 
At each transition $(\pm k_\ell, \omega_\ell)$, the discriminant
$|Q_1(k,\omega_\ell)Q_2(k,\omega_\ell)|$ reaches zero at
$k=\pm k_\ell$ while remaining nonzero at all other momenta
(panel~(c) of Fig.~\ref{fig:circuit_admittance}).  The two complex
eigenvalues $\bar E_\pm = d \pm\sqrt{Q_1Q_2}$ coalesce at these
points.  In a circuit satisfying
Eq.~\eqref{eq:circuit_mapping}, the corresponding response eigenvalues
approach, exchange, and reconnect. This produces a reorganization of
the transmission or impedance at the frequency
\(\Omega_\ell\) determined by \(z(\Omega_\ell)=i\omega_\ell\).
The five predicted critical frequencies in
Table~\ref{tab:transitions} are the prediction
of the exceptional-point  structure and its associated Berry-phase reconstruction for this parameter choice.

\subsection*{H.5\quad Circuit-implementation requirements}

The minimal network contains two external brane nodes and three internal
complement nodes per unit cell. Its unreduced Laplacian must be chosen
so that Eqs.~\eqref{eq:J_full_block}--\eqref{eq:circuit_mapping}
hold over the frequency interval of interest. Capacitive links provide
admittances proportional to \(i\Omega\), inductive links provide
admittances proportional to \(1/(i\Omega)\), and resistive or active
elements may be included to control losses and signs. A purely
capacitive network has an overall \(i\Omega\) scaling that is preserved
by Kron reduction and therefore cannot by itself generate the required
\(1/\Omega\) pole. The internal complement must contain inductive or
engineered frequency-dependent admittances so that its reduced
back-action reproduces the zero-mode pole.

\clearpage
\bibliographystyle{apsrev4-2}
\nocite{apsrev42Control}
\bibliography{reference}

\clearpage
\onecolumngrid
\begin{center}
{\large\bfseries Supplemental Material for\\
``Non-Hermitian Crystalline Braid Topology from Hermitian Projection:\\
A Zero-Mode Resonance Mechanism''}\\[1em]
Stefan {\DJ}or{\dj}evi{\'c}\\
{\it Faculty of Physics, University of Belgrade, Studentski Trg 12-16, 11000 Belgrade, Serbia}\\[0.5em]
Vladimir Juri\v{c}i\'c\\
{\it Departamento de F\'{i}sica, Universidad T\'{e}cnica Federico Santa Mar\'{i}a, Casilla 110, Valparaíso, Chile}\\[0.5em]
\end{center}
\twocolumngrid

\appendix
\setcounter{section}{0}
\setcounter{subsection}{0}
\setcounter{subsubsection}{0}
\renewcommand{\thesection}{S.\arabic{section}}
\renewcommand{\thesubsection}{S.\arabic{section}.\arabic{subsection}}
\renewcommand{\thesubsubsection}{S.\arabic{section}.\arabic{subsection}.\arabic{subsubsection}}
\counterwithout{equation}{section}
\setcounter{equation}{0}
\renewcommand{\theequation}{S.\arabic{equation}}

\setcounter{figure}{0}
\setcounter{table}{0}
\renewcommand{\thetable}{S\arabic{table}}
\renewcommand{\thefigure}{S\arabic{figure}}

This Supplemental Material contains: (S.1.) notation from the main text; (S.2.) additional spectral results; (S.3.)  density-of-states calculations, and (S.4.) further edge-mode spectra.

\section{Core formulas and notation from the main text}
\label{SM:sec:Main_top_inv}
\label{SM:sec:SM_core_formulas}

For completeness, we collect here the main formulas cited throughout the Supplemental Material.  This makes the Supplemental Material self-contained and fixes the equation labels inherited from the main manuscript.

The brane is translation invariant along the vector \(\vec e=\vec e_1+\vec e_2\).  In the symmetry-adapted basis, the corresponding translation operators are
\begin{equation}
 T_m=\sum_n |n,n+m\rangle\langle n+1,n+m+1|.
 \label{SM:trans_op_m}
\end{equation}
We use
\begin{align}
 q&=-(t_y+t_x e^{ik}),\label{SM:q_deff}\\
 \phi&=\operatorname{arcsinh}\left(\frac{\omega}{2|q|}\right),\label{SM:phi_deff}
\end{align}
and define
\begin{equation}
 F_{N}(\phi)=
 \begin{cases}
 \dfrac{\sinh[(N+2)\phi]}{\cosh[(N+1)\phi]}, & N\;\text{even},\\[0.6em]
 \dfrac{\cosh[(N+2)\phi]}{\sinh[(N+1)\phi]}, & N\;\text{odd}.
 \end{cases}
 \label{SM:F_N_functions}
\end{equation}
For open boundary conditions in the eliminated direction, the projected brane Hamiltonian is
\begin{equation}
 \begin{aligned}
 \tilde{H}_{\rm PTB}^{\rm (OBC)}=&-i|q|\big[F_{N_{1}}(\phi)|0\rangle\langle0|+F_{N_{2}}(\phi)|1\rangle\langle1|\big]\\
 &+q|1\rangle\langle0|+q^{*}|0\rangle\langle1|,
 \end{aligned}
 \label{SM:H_PTB_open_BC}
\end{equation}
with spectrum
\begin{equation}
 \bar{E}_{\pm}=|q|\left[-i\frac{F_{N_{1}}+F_{N_{2}}}{2}\pm\sqrt{1-\left(\frac{F_{N_{1}}-F_{N_{2}}}{2}\right)^{2}}\right].
 \label{SM:Energy_open_BC}
\end{equation}
The SSH part appearing in the zero-frequency diagnostic is
\begin{equation}
 H_{\rm SSH}=q|1\rangle\langle0|+q^{*}|0\rangle\langle1|.
 \label{SM:SSH_hamiltonian}
\end{equation}

For periodic boundary conditions in the eliminated direction, the projected brane Hamiltonian is
\begin{equation}
 \begin{aligned}
 \widetilde H_{\rm PTB}
 =&-i|q|F_N(\phi)\left(|0\rangle\langle 0|+|1\rangle\langle 1|\right)\\
 &+\left[q^{*}-\frac{(-q)^{N+1}}{|q|^N}G_N(\phi)\right]|0\rangle\langle 1|\\
 &+\left[q-\frac{(-q^{*})^{N+1}}{|q|^N}G_N(\phi)\right]|1\rangle\langle 0|.
 \end{aligned}
 \label{SM:H_PTB_periodic_BC}
\end{equation}
Here
\begin{equation}
 G_N(\phi)
 =
 \begin{cases}
 (-1)^{N/2}\dfrac{\cosh\phi}{\cosh[(N+1)\phi]},
 & N\ \mathrm{even},\\[1.1em]
 (-1)^{(N-1)/2}\dfrac{i\cosh\phi}{\sinh[(N+1)\phi]},
 & N\ \mathrm{odd}.
 \end{cases}
 \label{SM:G_N_functions}
\end{equation}

The corresponding complex spectrum is
\begin{equation}
 \bar E_{\pm}
 =
 |q|
 \left[
 -iF_N
 \pm
 \sqrt{
 1+G_N^2+2(-1)^N G_N\cos\alpha
 }
 \right],
 \label{SM:Energy_periodic_BC}
\end{equation}
with $\alpha=(N+2)\arg q$. 
In the large-complement limit,
\begin{equation}
 \tau(E)=\frac{2\hbar}{\omega+\sqrt{\omega^2+4E^2}}.
 \label{SM:tau_large_system}
\end{equation}

The biorthogonal complex Berry phase used to characterize the projected non-Hermitian bands is
\begin{align}
 &\tilde{U}^{(n)}_{\mathrm{B}}[\Gamma]=\exp{\left(i \oint_{\Gamma}\tilde{\mathcal{A}}^{(n)}_{i}\mathrm{d}k^{i}\right)},\nonumber\\
&\tilde{\mathcal{A}}^{(n)}_{i}(\vec{k})=i\frac{\langle\tilde{\psi}_{n}(\vec{k})|\partial_{i}|\psi_{n}(\vec{k})\rangle}{\langle\tilde{\psi}_{n}(\vec{k})|\psi_{n}(\vec{k})\rangle}.
 \label{SM:complex_berry_phase}
\end{align}
For the odd-\(N\) periodic sector, the crystalline invariant is
\begin{equation}
 \tilde{\nu}_{\pm}
 =
 -\frac{1}{2\pi}
 \bigl(\arg\{q\}+\theta\bigr)\Big|_{-\pi}^{\pi},
 \label{SM:top_inv_odd}
\end{equation}
where
\begin{equation}
 e^{2i\theta}
 =
 \frac{
 1+\tilde{G}^{2}_{N}e^{-2i\alpha}
 }{
 \sqrt{
 1+\tilde{G}^{4}_{N}
 +2\tilde{G}^{2}_{N}\cos(2\alpha)
 }
 } .
 \label{SM:Theta_N_ODD}
\end{equation}
The SSH-type comparison sectors are described by
\begin{equation}
 \nu_{\pm}=-\frac{1}{2\pi}\bigl(\arg\{q\}\pm\theta\bigr)\Big|_{-\pi}^{\pi},
 \label{SM:top_inv_even}
\end{equation}
which at finite \(\omega\) reduces to
\begin{equation}
 \nu_{\pm}=-\eta(t_x-t_y).
 \label{SM:nu_sol_even_finate_omega}
\end{equation}
For even \(N\) with periodic boundary conditions, the zero-frequency descendant gives
\begin{equation}
 \nu_{\pm}=\frac{N}{2}\eta(t_x-t_y).
 \label{SM:nu_sol_even_zero_omega_periodic_BC}
\end{equation}
For open boundary conditions with odd \(N\), the partially \(\mathcal{PT}_{+}\)-broken regime is instead described by
\begin{align}
 \nu_{\pm}&=-\frac{1}{2\pi}\bigl(\arg\{q\}+\theta\bigr)\Big|_{-\pi}^{-k_{\rm cr}}
 -\frac{1}{2\pi}\arg\{q\}\Big|_{-k_{\rm cr}}^{k_{\rm cr}}\nonumber\\
 &\quad -\frac{1}{2\pi}\bigl(\arg\{q\}+\theta\bigr)\Big|^{\pi}_{k_{\rm cr}}.
 \label{SM:top_inv_odd_1}
\end{align}

The non-Bloch eigenvalue problem used to diagnose the absence of a non-Hermitian skin effect can be written as
\begin{equation}
 \prod_{s=1}^{M}
 \left(\beta^{2}-
 \frac{\dfrac{\omega^{2}}{4\sinh^{2}\phi_{s}(E)}-t_{x}^{2}-t_{y}^{2}}
 {t_{x}t_{y}}\beta+1\right)=0,
 \label{SM:beta_eigen_value_problem}
\end{equation}
where the product runs over all distinct branches \(\phi_s(E)\).

The coordinate-space Hamiltonians used in the bulk-boundary analysis are, for open boundary conditions,
\begin{align}
 H_{PTB}&=\sum_{n,m}\bigg[\varepsilon_{n-m}^{(N_{1})}|n,0\rangle\langle n,0|+\varepsilon_{n-m}^{(N_{2})}|n,1\rangle\langle n,1|\bigg]\label{SM:H_OBC_coord_rep}\\
 &-\sum_{n}\bigg[t_{y}|n,0\rangle\langle n,1|+t_{x}|n+1,0\rangle\langle n,1|+h.c.\bigg],\nonumber
\end{align}
and, for periodic boundary conditions,
\begin{align}
 H_{PTB}&=\sum_{n,m}\bigg[\varepsilon_{n-m}^{(N)}|n,0\rangle\langle n,0|+\varepsilon_{n-m}^{(N)}|n,1\rangle\langle n,1|\bigg]\nonumber\\
 &-\sum_{n,m}t^{(N)}_{n-m}\bigg[|m,0\rangle\langle n,1|+|n,1\rangle\langle m,0|\bigg]\label{SM:H_PBC_coord_rep}\\
 &-\sum_{n}\bigg[t_{y}|n,0\rangle\langle n,1|+t_{x}|n+1,0\rangle\langle n,1|+h.c.\bigg].\nonumber
\end{align}
The hoppings appearing in (\ref{SM:H_OBC_coord_rep}) and (\ref{SM:H_PBC_coord_rep}) are given by:
\begin{widetext}
\begin{align}
 \varepsilon_{n-n'}^{(N)}&=-i\frac{N+2}{2}\omega\delta_{n,n'}+\frac{i\omega^{3}}{8(N+1)t_{x}t_{y}}\sum_{s=1}^{\frac{N}{2}}\frac{\cos^{2}\left(\frac{2s-1}{N+1}\frac{\pi}{2}\right)}{\sin^{4}\left(\frac{2s-1}{N+1}\frac{\pi}{2}\right)}\frac{z_{s}^{|n-n'|}}{\sqrt{\zeta_{s}^{2}-1}},\label{SM:eps_n-n'_even}\\
 \varepsilon_{n-n'}^{(N)}&=-i\left(\frac{(N+2)(N+3)}{6(N+1)}\omega+2\frac{t_{x}^{2}+t_{y}^{2}}{(N+1)\omega}\right)\delta_{n,n'}-\frac{2it_{x}t_{y}}{(N+1)\omega}\delta_{|n-n'|,1}
 +\frac{i\omega^{3}}{8(N+1)t_{x}t_{y}}\sum_{s=1}^{\frac{N-1}{2}}\frac{\cos^{2}\left(\frac{s\pi}{N+1}\right)}{\sin^{4}\left(\frac{s\pi}{N+1}\right)}\frac{z_{s}^{|n-n'|}}{\sqrt{\zeta_{s}^{2}-1}},\label{SM:eps_n-n'_odd}
\end{align}

 \begin{align}
 t_{n}^{(N)}&=\begin{dcases}
 \frac{\omega^{2-N}}{2^{2-N}(N+1)t_{x}t_{y}}\sum_{s=1}^{\frac{N}{2}}(-1)^{s-1}\frac{\cos^{2}\left(\frac{2s-1}{N+1}\frac{\pi}{2}\right)}{\sin^{4}\left(\frac{2s-1}{N+1}\frac{\pi}{2}\right)}\left[\sin{\left(\frac{2s-1}{N+1}\frac{\pi}{2}\right)}\left(-t_{y}-t_{x}z_{s}^{-1}\right)\right]^{N+1}\frac{z_{s}^{n}}{\sqrt{\zeta_{s}^{2}-1}},&n\geqslant\frac{N}{2}+2,\\
 \frac{\omega^{2-N}}{2^{2-N}(N+1)t_{x}t_{y}}\sum_{s=1}^{\frac{N}{2}}(-1)^{s-1}\frac{\cos^{2}\left(\frac{2s-1}{N+1}\frac{\pi}{2}\right)}{\sin^{4}\left(\frac{2s-1}{N+1}\frac{\pi}{2}\right)}\left[\sin{\left(\frac{2s-1}{N+1}\frac{\pi}{2}\right)}\left(-t_{y}-t_{x}z_{s}\right)\right]^{N+1}\frac{z_{s}^{-n}}{\sqrt{\zeta_{s}^{2}-1}}\\
 \hspace{2cm}+t_{y}\left(-\frac{t_{y}}{t_{x}}\right)^{\frac{N}{2}}\delta_{n,\frac{N}{2}}-\frac{1}{2t_{x}}\left(-\frac{t_{y}}{t_{x}}\right)^{\frac{N}{2}}\left[\frac{N(N+2)}{4}\omega^{2}-(N+2)t_{x}^{2}+Nt_{y}^{2}\right]\delta_{n,\frac{N}{2}+1},&n\leqslant\frac{N}{2}+1,
 \end{dcases}\label{SM:tau_n-n'_even}
\end{align}
\begin{align}
 t_{n}^{(N)}&=\begin{dcases}
 \frac{i\omega^{2-N}}{2^{2-N}(N+1)t_{x}t_{y}}\sum_{s=1}^{\frac{N-1}{2}}(-1)^{s}\frac{\cos^{2}\left(\frac{s\pi}{N+1}\right)}{\sin^{4}\left(\frac{s\pi}{N+1}\right)}\left[\sin{\left(\frac{s\pi}{N+1}\right)}\left(-t_{y}-t_{x}z_{s}^{-1}\right)\right]^{N+1}\frac{z_{s}^{n}}{\sqrt{\zeta_{s}^{2}-1}},\hspace{1.7cm}n\geqslant\frac{N+3}{2}
 \\
 \hspace{0.7cm}+\frac{2it_{x}^{2}}{\omega(N+1)}\left(-\frac{t_{x}}{t_{y}}\right)^{\frac{N-1}{2}}\delta_{n,\frac{N+3}{2}},\\
 \frac{i\omega^{2-N}}{2^{2-N}(N+1)t_{x}t_{y}}\sum_{s=1}^{\frac{N-1}{2}}(-1)^{s}\frac{\cos^{2}\left(\frac{s\pi}{N+1}\right)}{\sin^{4}\left(\frac{s\pi}{N+1}\right)}\left[\sin{\left(\frac{s\pi}{N+1}\right)}\left(-t_{y}-t_{x}z_{s}\right)\right]^{N+1}\frac{z_{s}^{-n}}{\sqrt{\zeta_{s}^{2}-1}},\hspace{1.8cm}n\leqslant\frac{N+1}{2}\\
 \hspace{0.7cm}+\frac{2it_{y}^{2}}{\omega(N+1)}\left(-\frac{t_{y}}{t_{x}}\right)^{\frac{N-1}{2}}\delta_{n,\frac{N-1}{2}}+i\left(\frac{(N-1)(N+3)}{12(N+1)}\omega+\frac{(N-1)t_{y}^{2}-(N+3)t_{x}^{2}}{\omega(N+1)}\right)\left(-\frac{t_{y}}{t_{x}}\right)^{\frac{N+1}{2}}\delta_{n,\frac{N+1}{2}}.
 \end{dcases}\label{SM:tau_n-n'_odd}
 \end{align}
\end{widetext}

\section{Further energy spectrum results}\label{SM:app:Energy_bands}
Here we present all possible different shapes of the energy bands and state lifetimes which may occur depending on the parameters of the system $N$, $N_{1}$, $N_{2}$ and $\omega$ for both cases of open and periodic boundary conditions. 
\par
\begin{figure*}[htbp]
 \centering
 \begin{subfigure}[b]{0.45\textwidth}
 \centering
 \includegraphics[width=\textwidth]{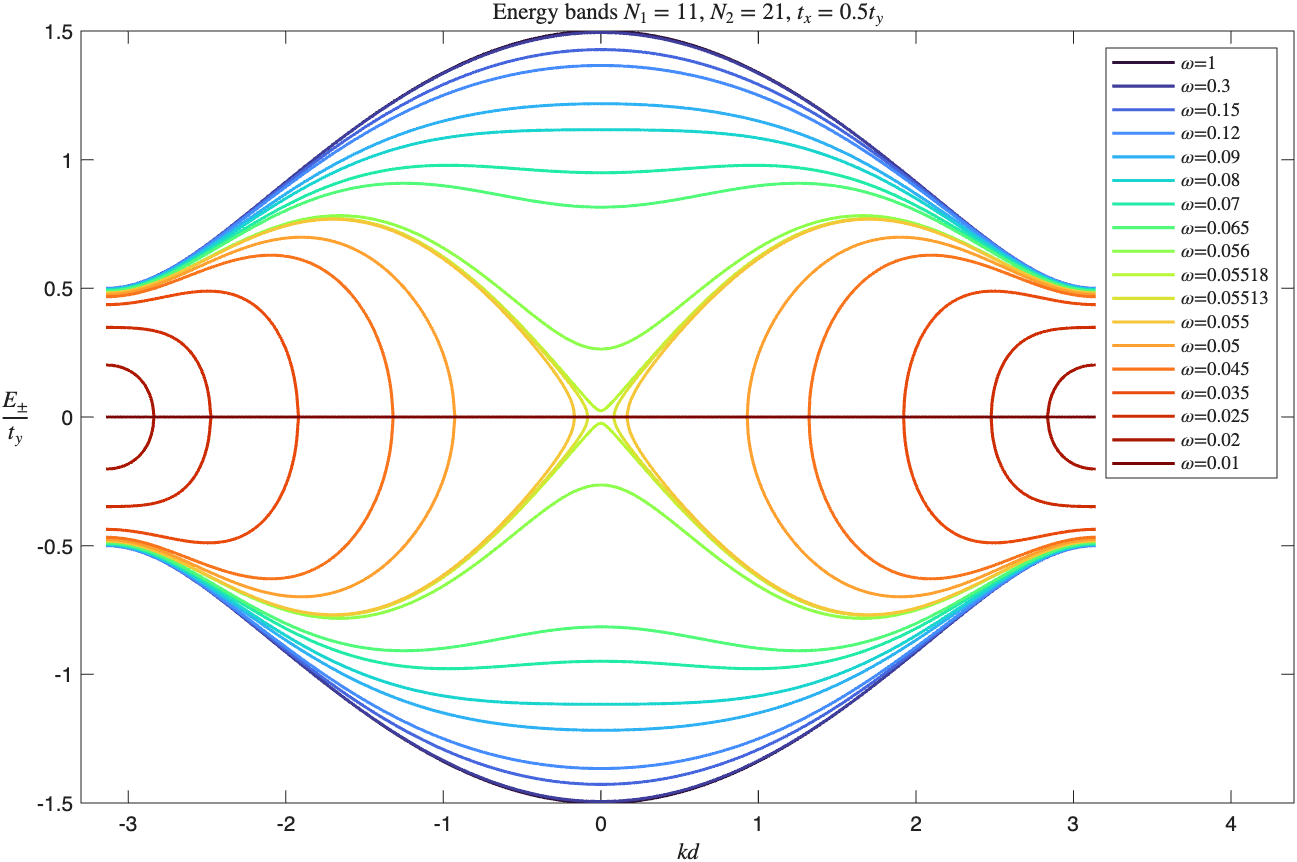}
 \caption{}
 \label{SM:fig:E_tau_odd-a}
 \end{subfigure}
 \hfill
 \begin{subfigure}[b]{0.45\textwidth}
 \centering
 \includegraphics[width=\textwidth]{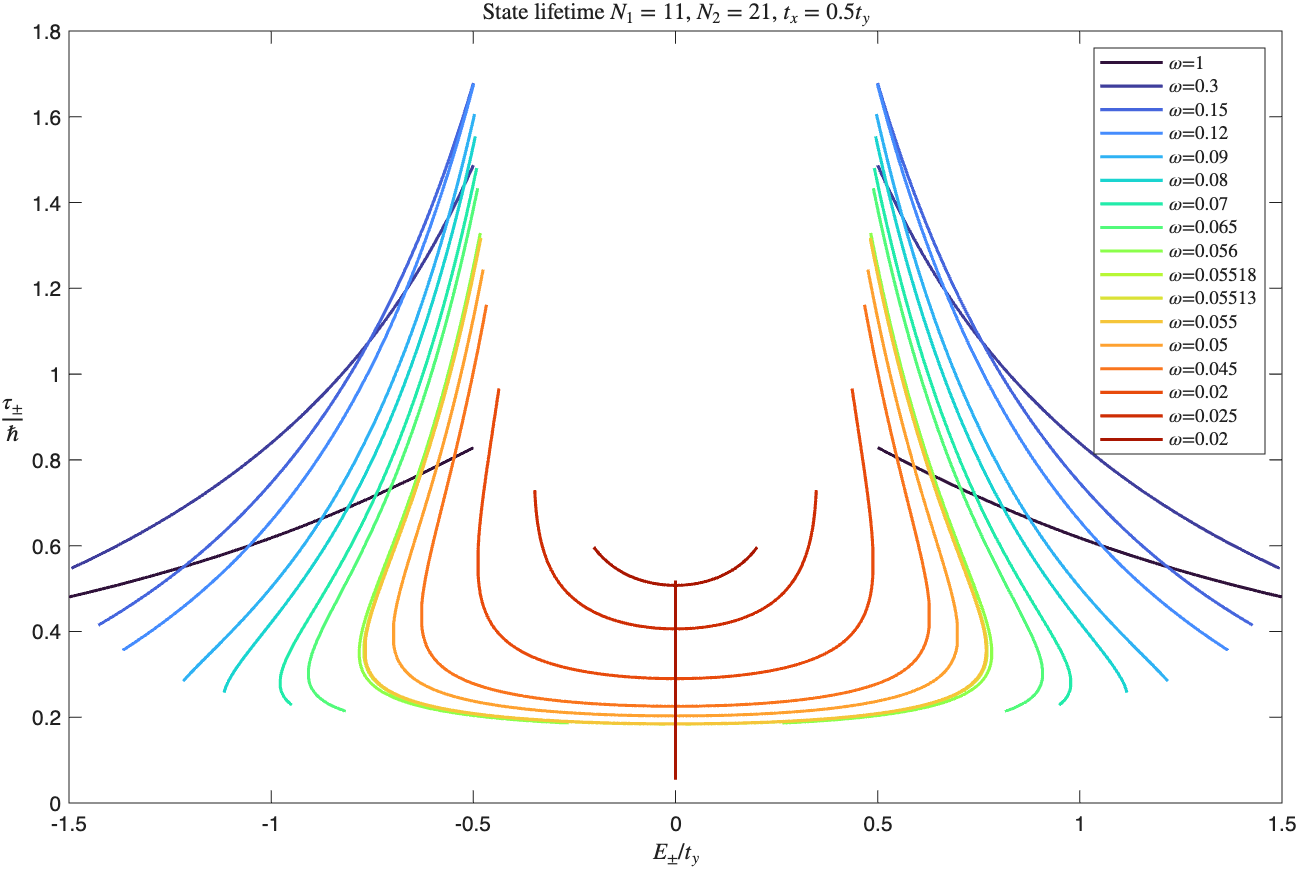}
 \caption{}
 \label{SM:fig:E_tau_odd-b}
 \end{subfigure}
 \vspace{1cm}
 \begin{subfigure}[b]{0.45\textwidth}
 \centering
 \includegraphics[width=\textwidth]{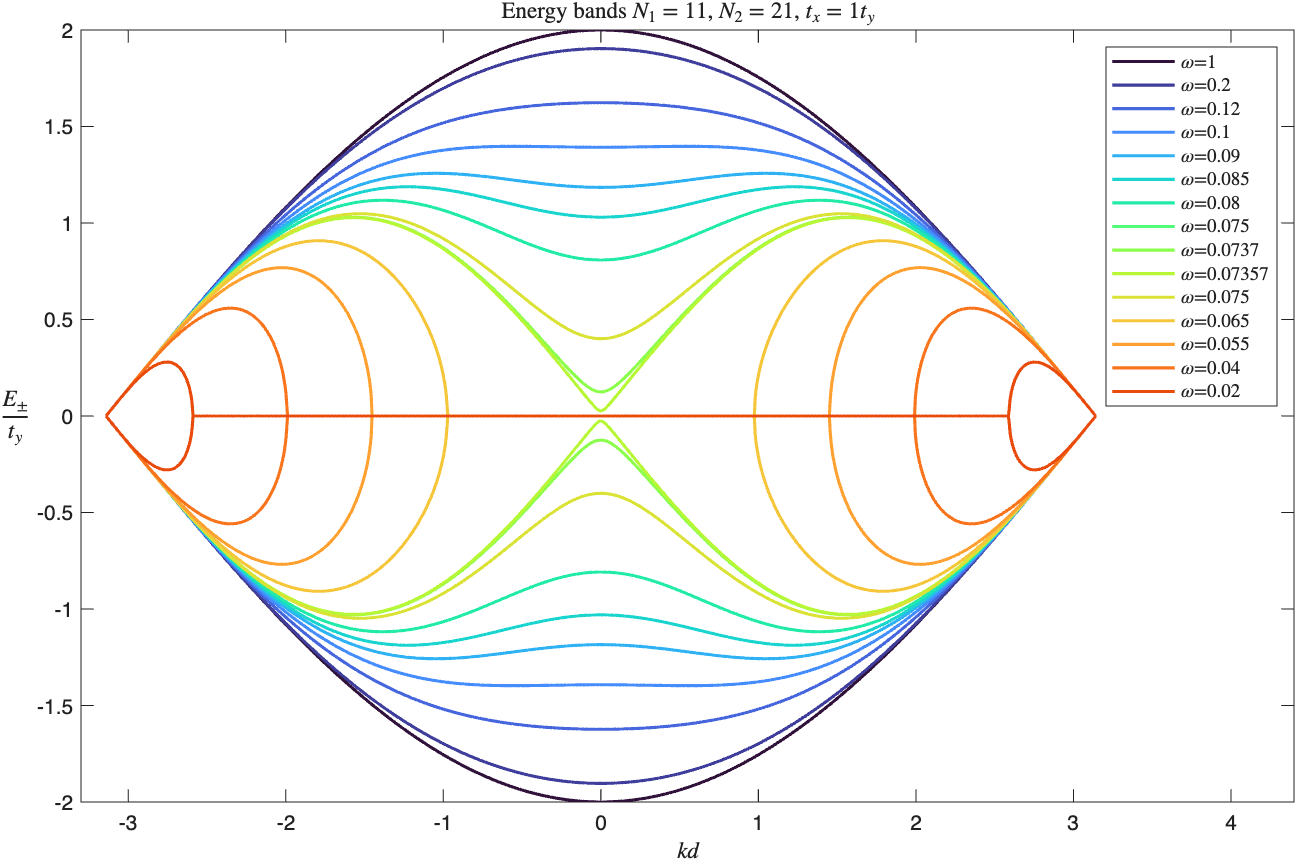}
 \caption{}
 \label{SM:fig:E_tau_odd-c}
 \end{subfigure}
 \hfill
 \begin{subfigure}[b]{0.45\textwidth}
 \centering
 \includegraphics[width=\textwidth]{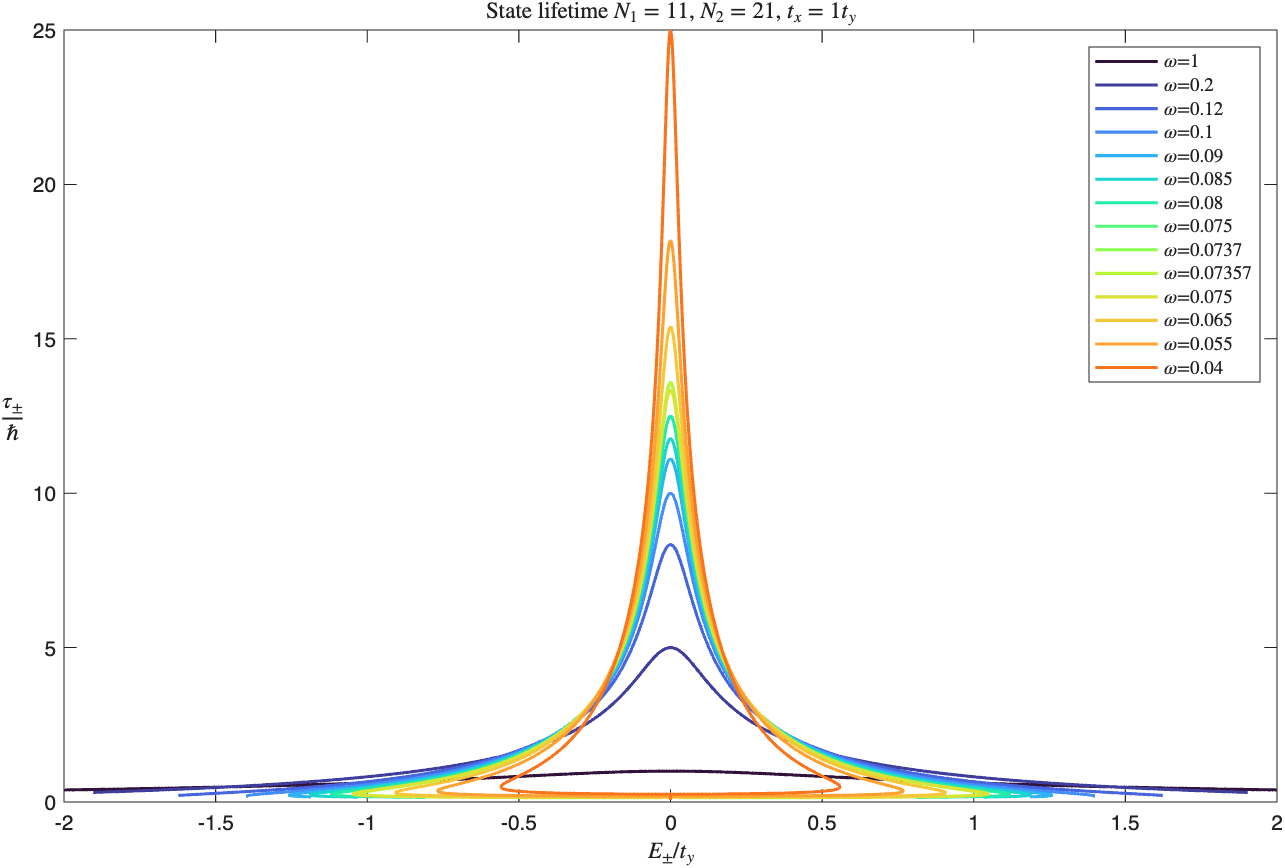}
 \caption{}
 \label{SM:fig:E_tau_odd-d}
 \end{subfigure}
 \vspace{1cm}
 \begin{subfigure}[b]{0.45\textwidth}
 \centering
 \includegraphics[width=\textwidth]{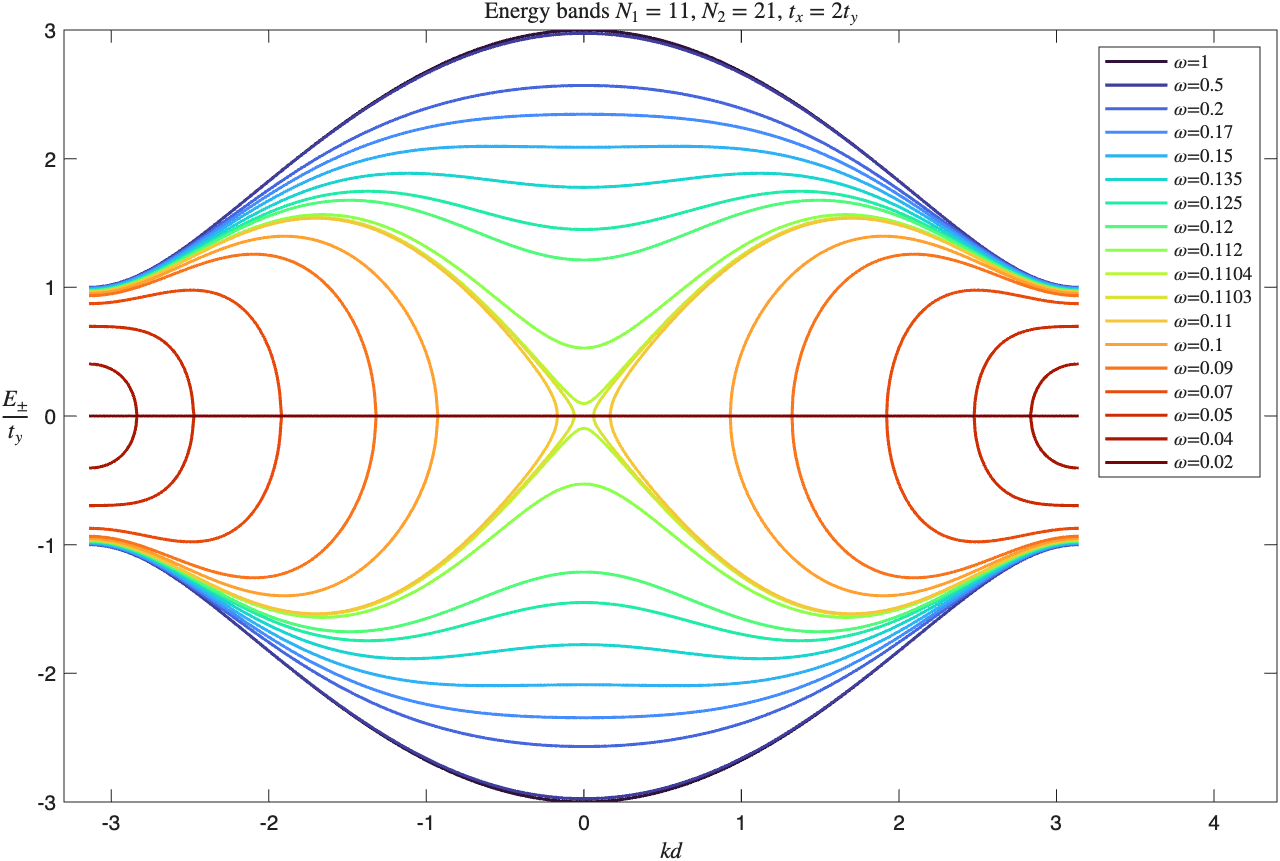}
 \caption{}
 \label{SM:fig:E_tau_odd-e}
 \end{subfigure}
 \hfill
 \begin{subfigure}[b]{0.45\textwidth}
 \centering
 \includegraphics[width=\textwidth]{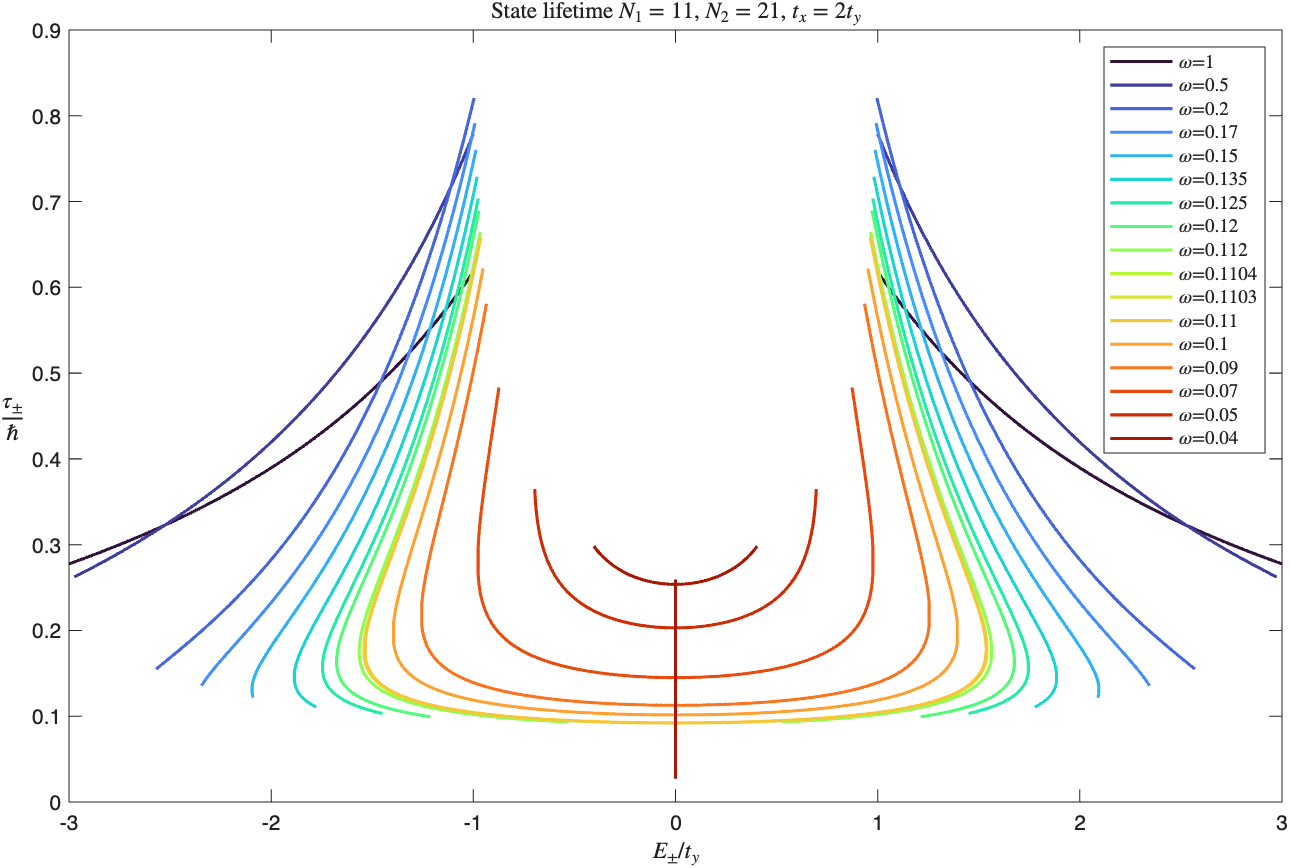}
 \caption{}
 \label{SM:fig:E_tau_odd-f}
 \end{subfigure} 
 \captionsetup{justification=justified,singlelinecheck=false}
 \caption{Figures (a), (c), and (e) illustrate how the energy bands evolve with decreasing Matsubara frequency $\omega$, whereas Figs.~(b), (d), and (f) depict how the state lifetime varies with decreasing $\omega$. The first row corresponds to the regime $t_{x}<t_{y}$, the second to the critical point $t_{x}=t_{y}$, and the third to the regime $t_{x}>t_{y}$. Two distinct critical values of $\omega$ can be identified: one at which zero-energy modes first emerge, and another where the entire band becomes flat. For sufficiently large Matsubara frequencies, $\omega>1$, the bands reduce to those of the SSH model.}
 \label{SM:fig:E_tau_odd}
\end{figure*}

\begin{figure*}
 \centering
 \includegraphics[width=0.82\textwidth]{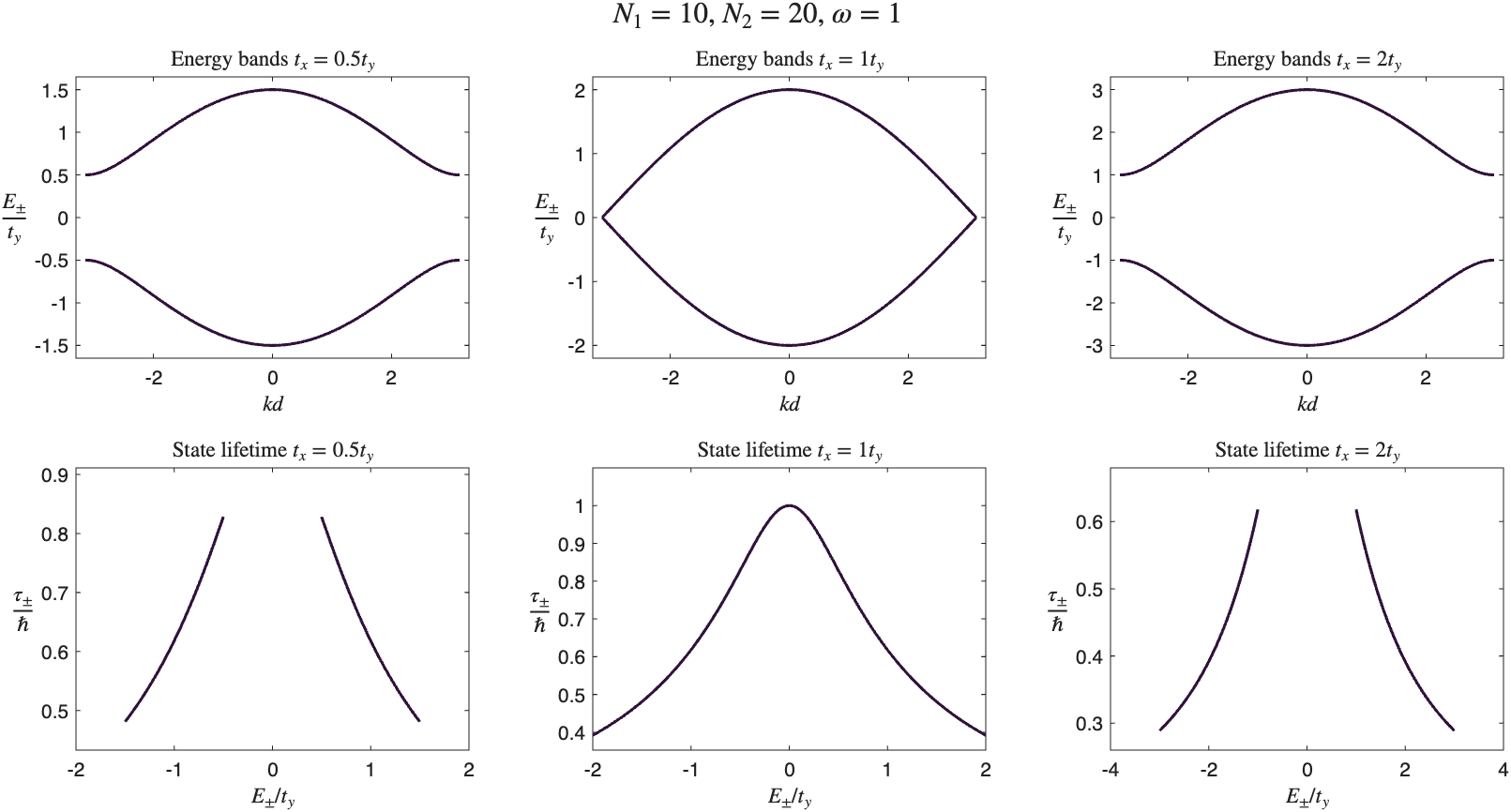}
 \caption{Energy bands and state lifetimes for the case where $N_{1}$ and $N_{2}$ are even integers}\label{SM:fig:E_tau_even}
\end{figure*}

 First, we examine how the complex energy spectrum behaves with decreasing Matsubara frequency in the case of the open boundary conditions. The behavior for even values of $N_{1}$ and $N_{2}$ is shown in Figures \ref{SM:fig:E_tau_even}, while the case where either $N_{1}$ or $N_{2}$ is odd is presented in Figure \ref{SM:fig:E_tau_odd}. For even values of the numbers, there is no significant change as $\omega$ decreases. In contrast, for odd values, the transitions from the $\mathcal{P}\mathcal{T}_{+}$-unbroken phase to the $\mathcal{P}\mathcal{T}_{+}$-broken phase in part of the Brillouin zone and finally to the completely $\mathcal{P}\mathcal{T}_{+}$-broken phase are clearly visible in all three SSH regimes. The corresponding critical frequencies for a fixed $t_{x}$ are indicated in Figure \ref{SM:fig:E_tau_odd} for each regime. When $t_{x}\neq t_{y}$ two critical values exist. The first critical point corresponds to the value of the Matsubara frequency when the gap closes at $k=0$. The further decrease in the Matsubara frequency creates a flat band in the part of the Brillouin zone around $kd=0$. The second critical point occurs when the entire energy band becomes flat. When $t_{x}=t_{y}$ the second transition never occurs, i.e., the bands never become completely flat. An additional illustration of how the energy bands evolve as the Matsubara frequency decreases is presented in Figure \ref{SM:fig:Bands_t_x>t_y} for $t_{x}>t_{y}$ and in Figure \ref{SM:fig:Bands_t_x=t_y} for $t_{x}=t_{y}$.
\par
\begin{figure*}[htbp]
 \centering
 \includegraphics[width=0.82\textwidth]{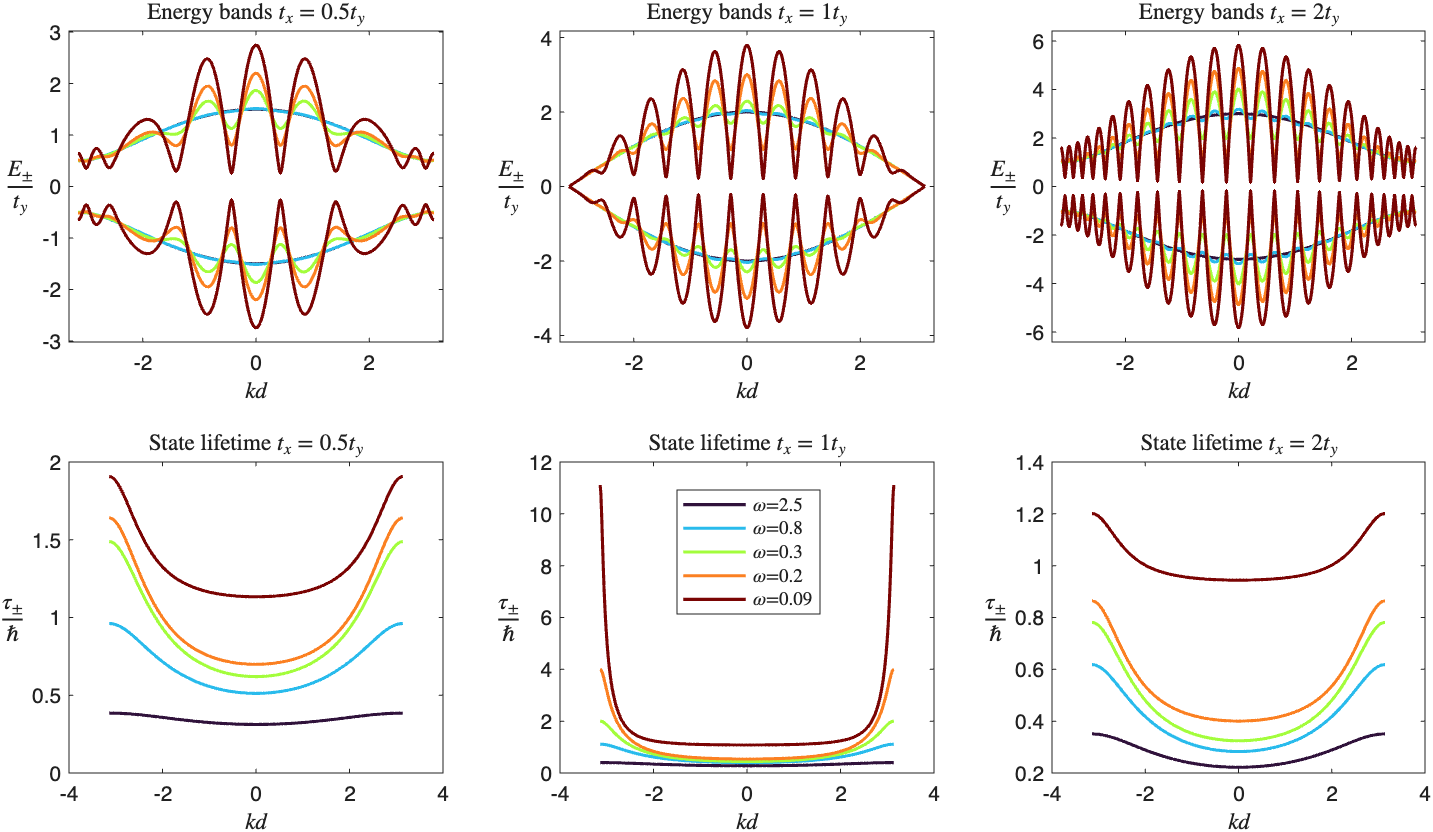}
 \captionsetup{justification=justified,singlelinecheck=false}
 \caption{The energy bands for $N=20$ are presented in the top row, and the corresponding state lifetimes are shown in the bottom row. In each row, three regimes $t_{x}<t_{y}$, $t_{x}=t_{y}$, and $t_{x}>t_{y}$ are displayed. The different colors represent decreasing values of the Matsubara frequency $\omega$. For large $\omega$, the dispersion relation coincides with that of the SSH model. As $\omega$ is reduced, the bands become more oscillatory creating new gaps. For any positive Matsubara frequency, this gap remains open and never closes.}
 \label{SM:fig:E_tau_even_periodic}
\end{figure*}

\begin{figure*}[htbp]
 \centering
 \includegraphics[width=0.82\textwidth]{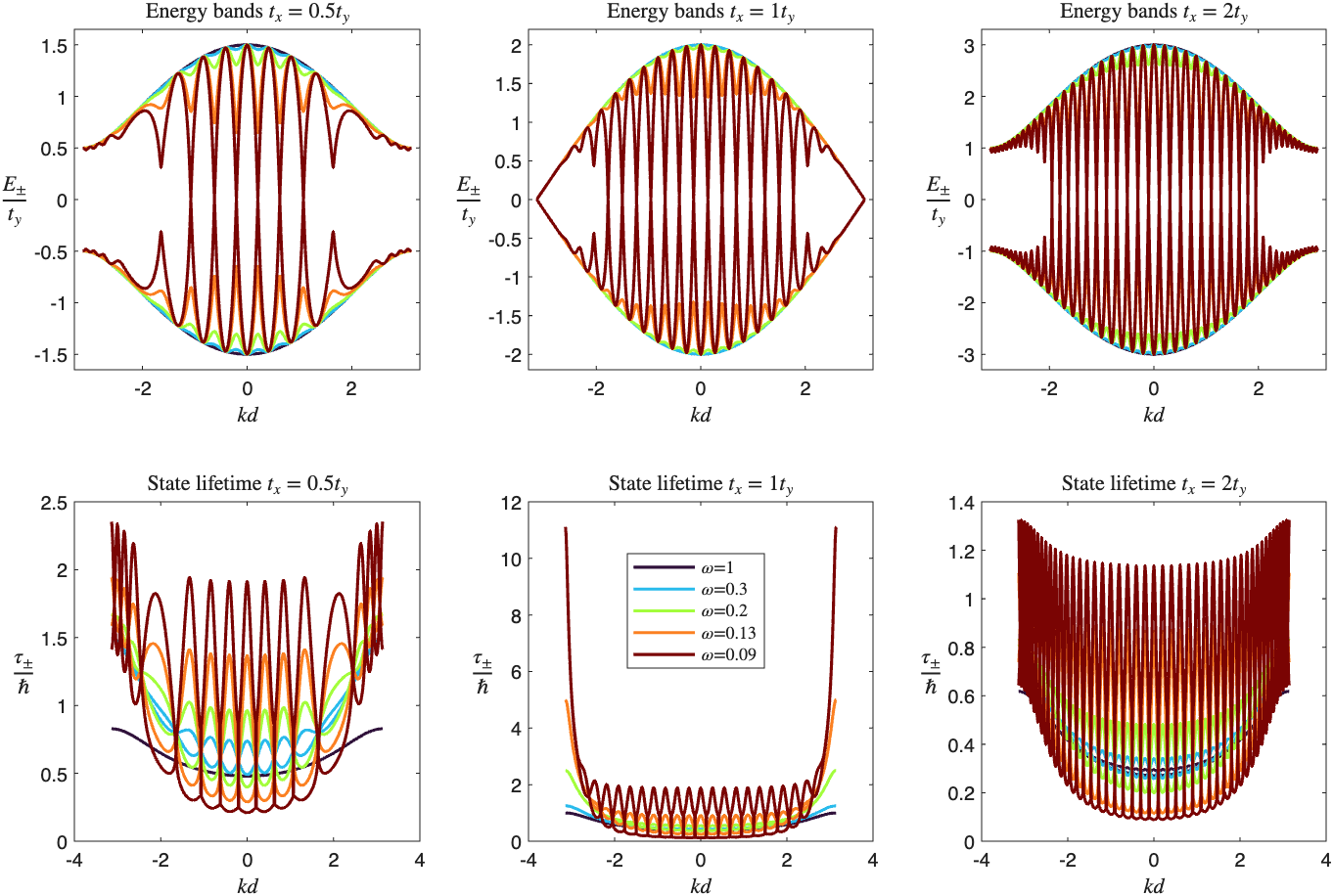}
 \captionsetup{justification=justified,singlelinecheck=false}
 \caption{The energy bands for $N=21$ are presented in the top row, and the corresponding state lifetimes are shown in the bottom row. In each row, three regimes $t_{x}<t_{y}$, $t_{x}=t_{y}$, and $t_{x}>t_{y}$ are displayed. The different colors represent decreasing values of the Matsubara frequency $\omega$. For large $\omega$, the dispersion relation coincides with that of the SSH model. As $\omega$ is reduced, the bands become more oscillatory creating new gaps. When $\omega$ falls below some critical value, the gaps start closing. As $\omega$ is further decreased the region in which the gaps close, widens.}
 \label{SM:fig:E_tau_odd_periodic}
\end{figure*}

 Next, we move on to the periodic boundary conditions. The energy bands as well as the state lifetimes in all three SSH regimes are shown in Figure \ref{SM:fig:E_tau_even_periodic} for the even value of $N$, and Figure \ref{SM:fig:E_tau_odd_periodic} for the odd value of $N$. In both cases, as $\omega$ decreases, the energy bands exhibit more oscillatory behavior creating periodic gaps. Although for even $N$, those gaps remain open, for odd $N$ they close as the Matsubara frequency is further decreased. The number of closed gaps increases with the decreasing of $\omega$ for fixed $t_{x}$. 
\par The state lifetime is depicted next to the corresponding energy bands as a function of energy. It clearly displays a Lorentzian-like profile centered at $E=0$, as explicitly obtained for the large system in (\ref{SM:tau_large_system}).


\section{Density of states}
\label{SM:app:DOS}
The density of states is defined as a simple sum of Dirac delta functions evaluated at the eigen-energies. In a case of a 1D system it can be much simplified to a following form:
\begin{equation}
 g_{\pm}(E)=\frac{1}{\pi}\sum_{E_{\pm}(k)=E}\frac{1}{\left|\frac{\mathrm{d}E_{\pm}}{\mathrm{d}k}\right|},\label{SM:DOS_deff}
\end{equation}
where the sum has been multiplied by 2 for the two possible intrinsic spin states. Both limits, $\omega \to0$ and $N \to \infty$, reduce the system to the simple SSH model (\ref{SM:SSH_hamiltonian}). In these cases, the density of states is given by:
\begin{equation}
 g(E)=\frac{2|E|}{\pi t_{x}t_{y}}\left(1-\left(\frac{E^{2}-t_{x}^{2}-t_{y}^{2}}{2t_{x}t_{y}}^{2}\right)\right)^{-\frac{1}{2}}.\label{SM:DOS_SSH}
\end{equation}
From this expression, we identify four energies at which the SSH density of states diverges. These are located at $E = \pm (t_{x} \pm t_{y})$ and are referred to as Van Hove singularities. The more general expression for finite $\omega$ and $N_{1}$ and $N_{2}$ in Eq.~(\ref{SM:Energy_open_BC}) is not shown explicitly here. Instead, the complete result is illustrated in the figures. 
\par
\begin{figure*}[htbp]
 \centering
 \includegraphics[width=0.81\textwidth]{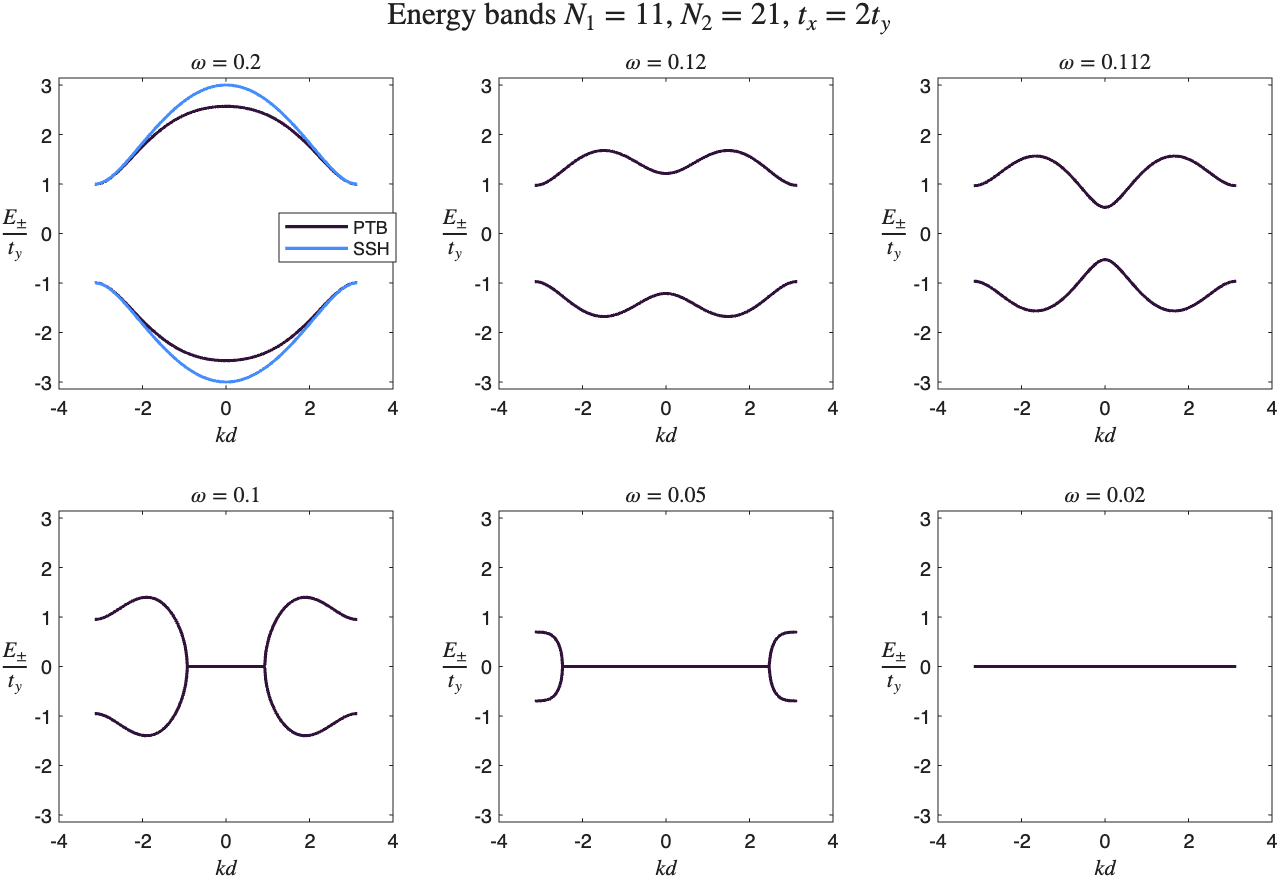}
 \captionsetup{justification=justified,singlelinecheck=false}
 \caption{This figure displays the energy bands with open boundary conditions for $N_{1}=11$ and $N_{2}=21$ when $t_{x}>t_{y}$. The trivial SSH phase, $t_{x}<t_{y}$ exhibits the same characteristics in regards to the shape of the energy bands. Moving from left to right, one can trace how the bands evolve as the Matsubara frequency is gradually decreased. The specific Matsubara frequency values match those used in the corresponding panels of Figure \ref{SM:fig:DOS_odd_x>y}, where the density of states is shown. For large $\omega$, the bands coincide with those of the SSH model. As $\omega$ is lowered, the local maximum at $k=0$ turns into a local minimum, generating five Van Hove singularities, which are clearly visible in the corresponding density of states. In the next panel, a further reduction of $\omega$ makes the minimum at $k=0$ the global minimum. These three panels represent the $\mathcal{PT}_{+}$-unbroken phase. The first two panels in the second row illustrate the partially $\mathcal{PT}_{+}$-broken phase, where the energy bands become flat over a portion of the Brillouin zone centered around $k=0$. In the first of these panels, the bands display three extrema, directly associated with three Van Hove singularities in the corresponding density of states, whereas in the second panel there is only a single extremum at $k=\pi$. The final panel shows completely flat bands, indicating that the system has entered the fully $\mathcal{PT}_{+}$-broken phase.}
 \label{SM:fig:Bands_t_x>t_y}
\end{figure*}

\begin{figure*}[htbp]
 \centering
 \includegraphics[width=0.81\textwidth]{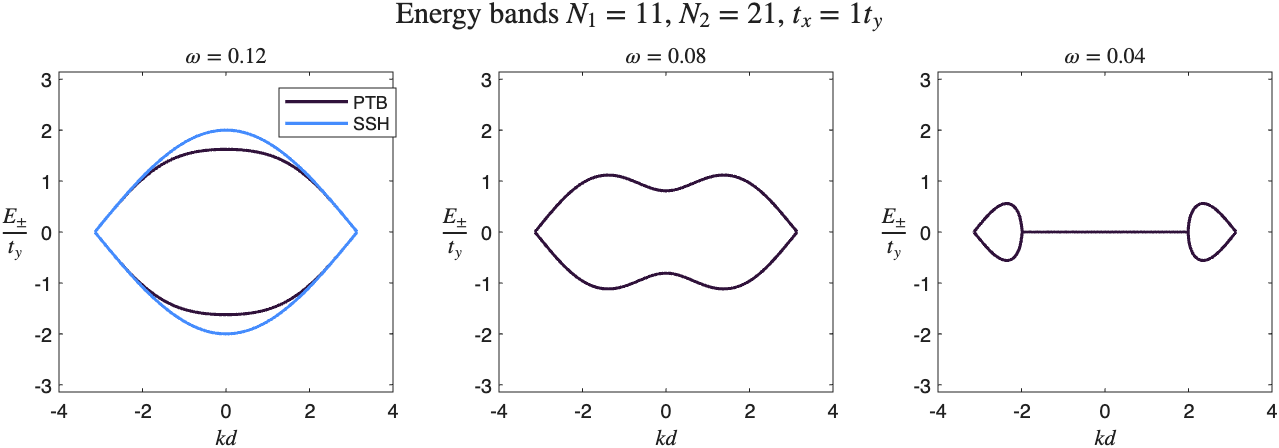}
 \captionsetup{justification=justified,singlelinecheck=false}
 \caption{This figure displays the energy bands with open boundary conditions for $N_{1}=11$ and $N_{2}=21$ at SSH critical point $t_{x}=t_{y}$. In comparison with Figure \ref{SM:fig:Bands_t_x>t_y}, there are only three distinct types of energy-band behavior, which match exactly those found in the corresponding density of states in Figure \ref{SM:fig:DOS_odd_x=y}. As the Matsubara frequency decreases, additional Van Hove singularities emerge. The key difference from the $t_{x}>t_{y}$ case is that the system never reaches a fully $\mathcal{PT}_{+}$-broken phase.}
 \label{SM:fig:Bands_t_x=t_y}
\end{figure*}

\begin{figure*}[htbp]
 \centering
 \includegraphics[width=0.88\textwidth]{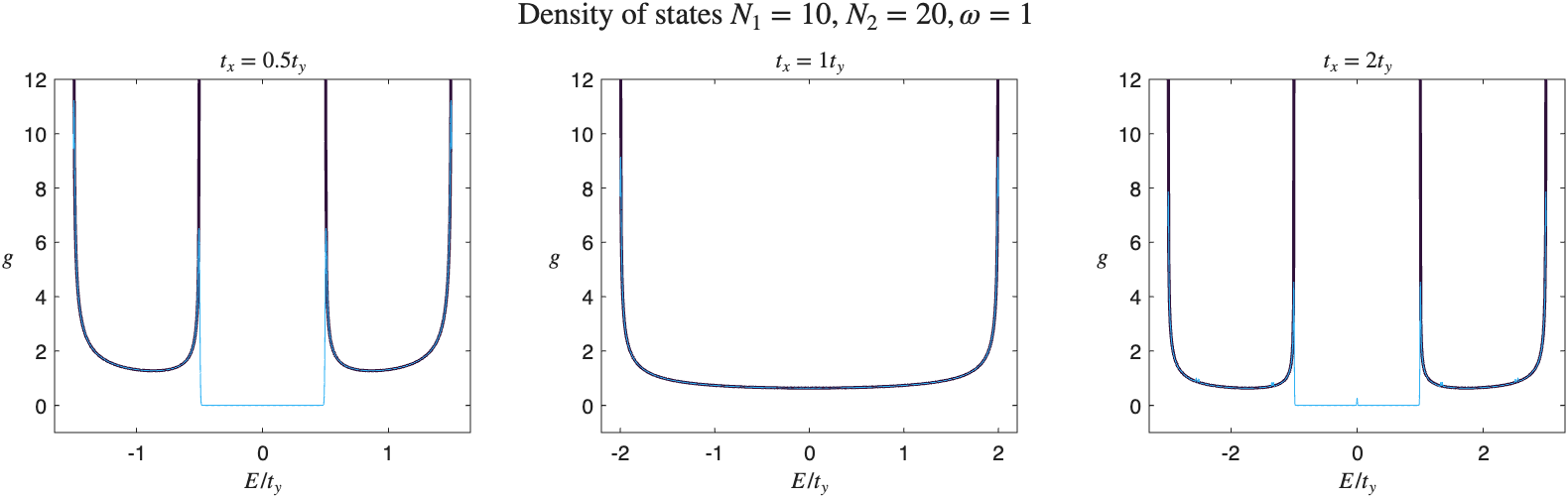}
 \captionsetup{justification=justified,singlelinecheck=false}
 \caption{Density of states for the case of even $N_{1}$ and $N_{2}$. The first panel shows the situation with $t_{x}<t_{y}$, while the next two correspond to $t_{x}=t_{y}$ and $t_{x}>t_{y}$, respectively. The plots clearly indicate that the gap closes when $t_{x}=t_{y}$, whereas it remains open in the other two cases. The purple curve shows the analytical prediction, whereas the blue curve corresponds to the numerical simulation. In the simulation, we consider $K=1200$ sites along the direction parallel to the $\vec{e}$ vector defined via (\ref{SM:trans_op_m}).}
 \label{SM:fig:DOS_even}
\end{figure*}

\begin{figure*}[htbp]
 \centering
 \includegraphics[width=0.88\textwidth]{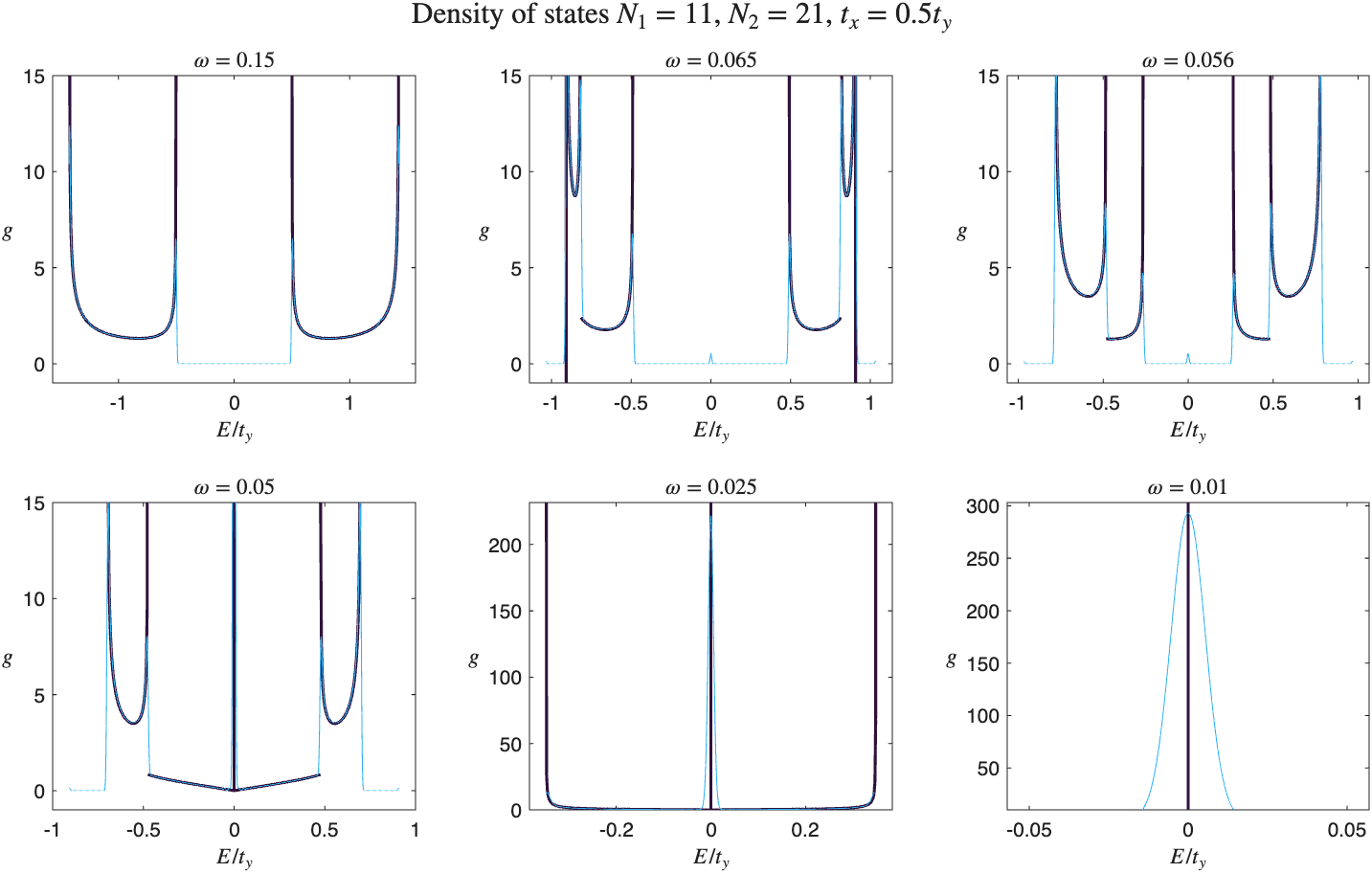}
 \captionsetup{justification=justified,singlelinecheck=false}
 \caption{Density of states for odd $N_{1}$ and $N_{2}$ with $t_{x}<t_{y}$. We examine the different regimes that arise as the Matsubara frequency $\omega$ is decreased.. By directly comparing with the energy bands shown in figure \ref{SM:fig:E_tau_odd-a}, we identify six cases determined by how many extrema the bands possess, since the density of states is inversely proportional to the derivative of the dispersion. Initially, there is an energy gap with four extrema at $E_{\mathrm{ext}}=\pm(t_{x}\pm t_{y})$, as illustrated in the first panel. As $\omega$ is lowered, an additional gap opens at $E=0$, generating extra extrema in both bands, which is depicted in the second and third panels. The first plot in the second row corresponds to the situation after the Matsubara frequency falls below the first critical value: the gap at $E=0$ closes and then broadens, so that the bands become flat over a finite energy interval. In this regime there are still two additional extrema, giving rise to two divergent points in the density of states. The following panel corresponds to the case where no further extrema remain apart from those at the band edges. Finally, the last figure shows the density of states for completely flat bands at $E=0$, once $\omega$ has decreased below the second critical point. The purple curve shows the analytical prediction, whereas the blue curve corresponds to the numerical simulation. In the simulation, we consider $K=1200$ sites along the direction parallel to the $\vec{e}$ vector defined via (\ref{SM:trans_op_m}).}
 \label{SM:fig:DOS_odd_x<y}
\end{figure*}

\begin{figure*}[htbp]
 \centering
 \includegraphics[width=0.9\textwidth]{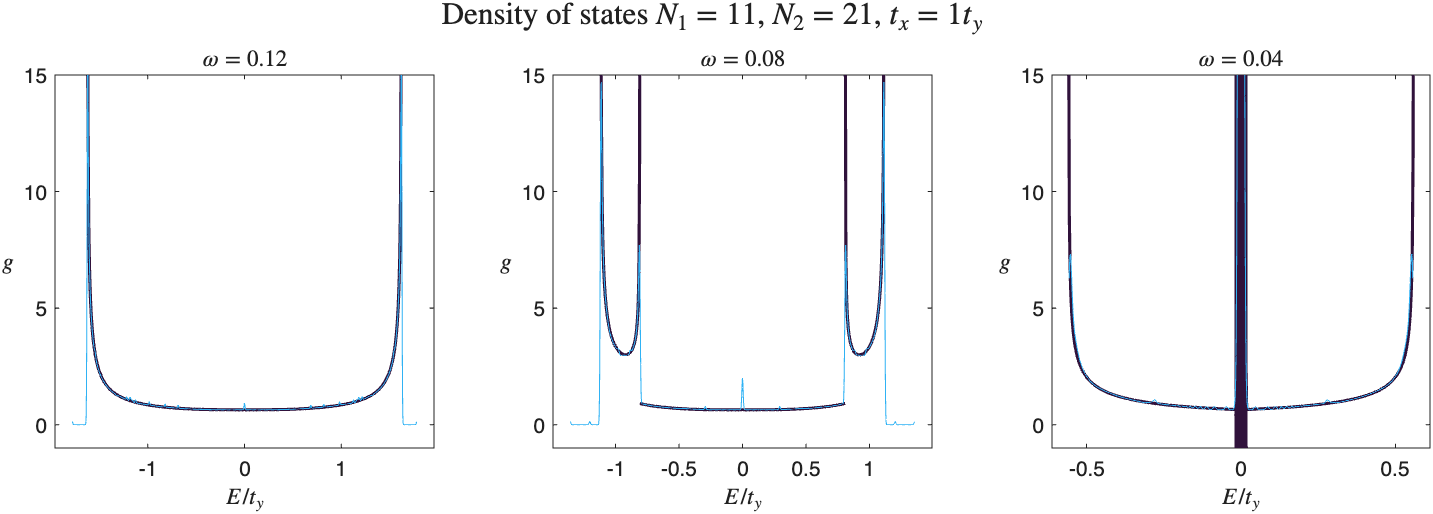}
 \captionsetup{justification=justified,singlelinecheck=false}
 \caption{Density of states for odd $N_{1}$ and $N_{2}$ at the critical point $t_{x}=t_{y}$. Using the same line of argument as in the case $t_{x}<t_{y}$ (Figure \ref{SM:fig:DOS_odd_x<y}), we conclude that there are three distinct regimes as the Matsubara frequency decreases. As expected, the key difference here is that the gap remains closed at the SSH critical point, which is clearly visible in the figure. The purple curve shows the analytical prediction, whereas the blue curve corresponds to the numerical simulation. In the simulation, we consider $K=1200$ sites along the direction parallel to the $\vec{e}$ vector defined via (\ref{SM:trans_op_m}). The energy bands corresponding to the same values of $\omega$ are shown in Figure \ref{SM:fig:Bands_t_x=t_y}.}
 \label{SM:fig:DOS_odd_x=y}
\end{figure*}

\begin{figure*}[htbp]
 \centering
 \includegraphics[width=0.9\textwidth]{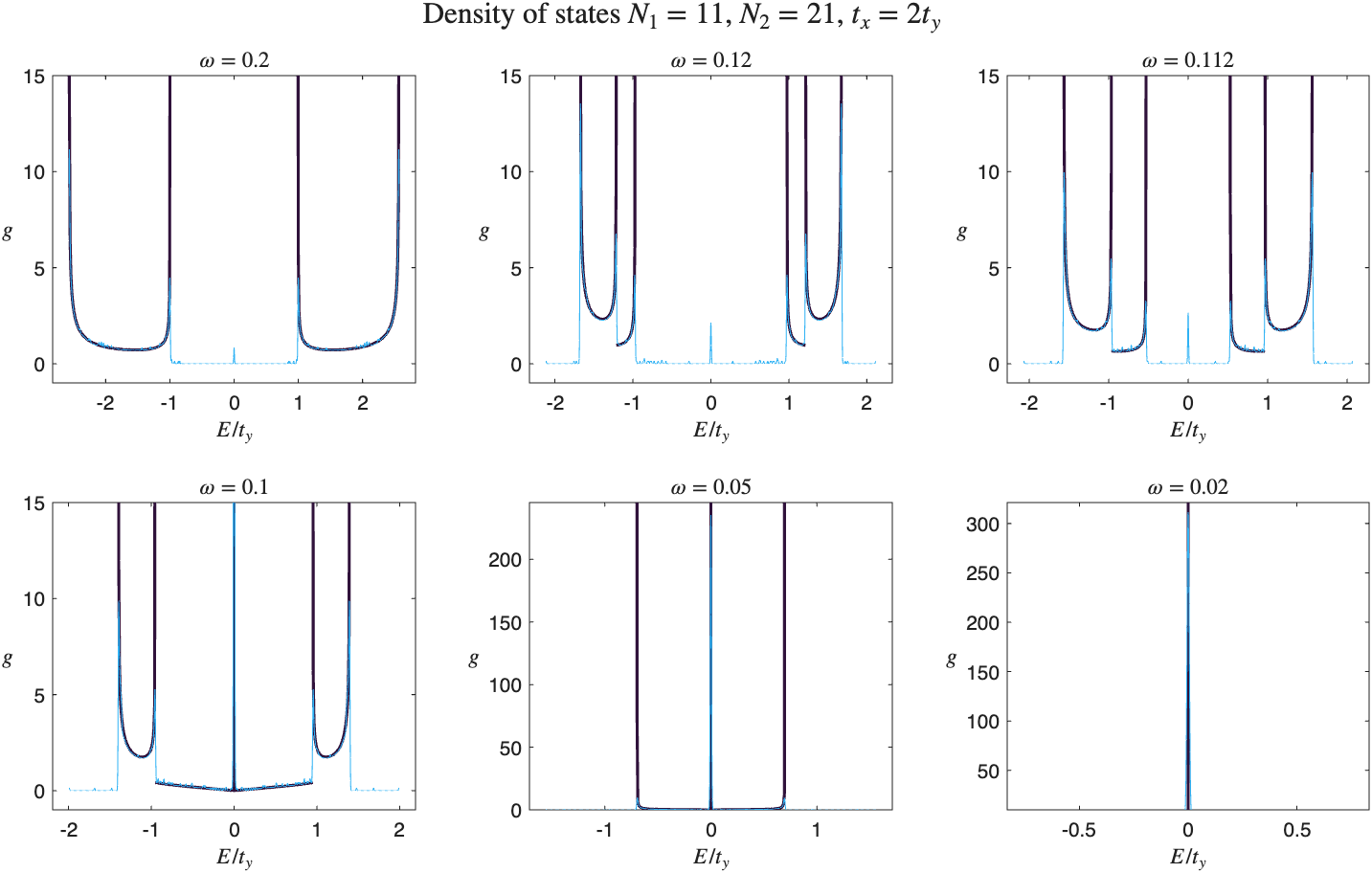}
 \captionsetup{justification=justified,singlelinecheck=false}
 \caption{Density of states for odd $N_{1}$ and $N_{2}$ with $t_{x}>t_{y}$. This figure is similar to the situation with $t_{x}<t_{y}$ shown in Figure \ref{SM:fig:DOS_odd_x<y}, and we apply the same line of reasoning as in that case. The purple curve shows the analytical prediction, whereas the blue curve corresponds to the numerical simulation. In the simulation, we consider $K=1200$ sites along the direction parallel to the $\vec{e}$ vector defined via (\ref{SM:trans_op_m}). The energy bands corresponding to the same values of $\omega$ are shown in Figure \ref{SM:fig:Bands_t_x>t_y}.}
 \label{SM:fig:DOS_odd_x>y}
\end{figure*}

 First we address the case of open boundary conditions. The scenarios with even $N_{1}$ and $N_{2}$ are depicted in Figure \ref{SM:fig:DOS_even}, whereas the cases with odd $N_{1}$ and $N_{2}$ are presented in Figures \ref{SM:fig:DOS_odd_x<y}, \ref{SM:fig:DOS_odd_x=y}, and \ref{SM:fig:DOS_odd_x>y}, corresponding to $t_{x}<t_{y}$, the SSH critical point $t_{x}=t_{y}$ and $t_{x}>t_{y}$ respectively. For reference, the energy bands at the same Matsubara frequencies as in Figures \ref{SM:fig:DOS_odd_x=y} and \ref{SM:fig:DOS_odd_x>y} are displayed in Figures \ref{SM:fig:Bands_t_x=t_y} and \ref{SM:fig:Bands_t_x>t_y}, respectively. The density of states is found both using the analytical expression (\ref{SM:DOS_deff}) as well as numerical methods.
\par
\begin{figure*}[htbp]
 \centering
 \includegraphics[width=0.9\textwidth]{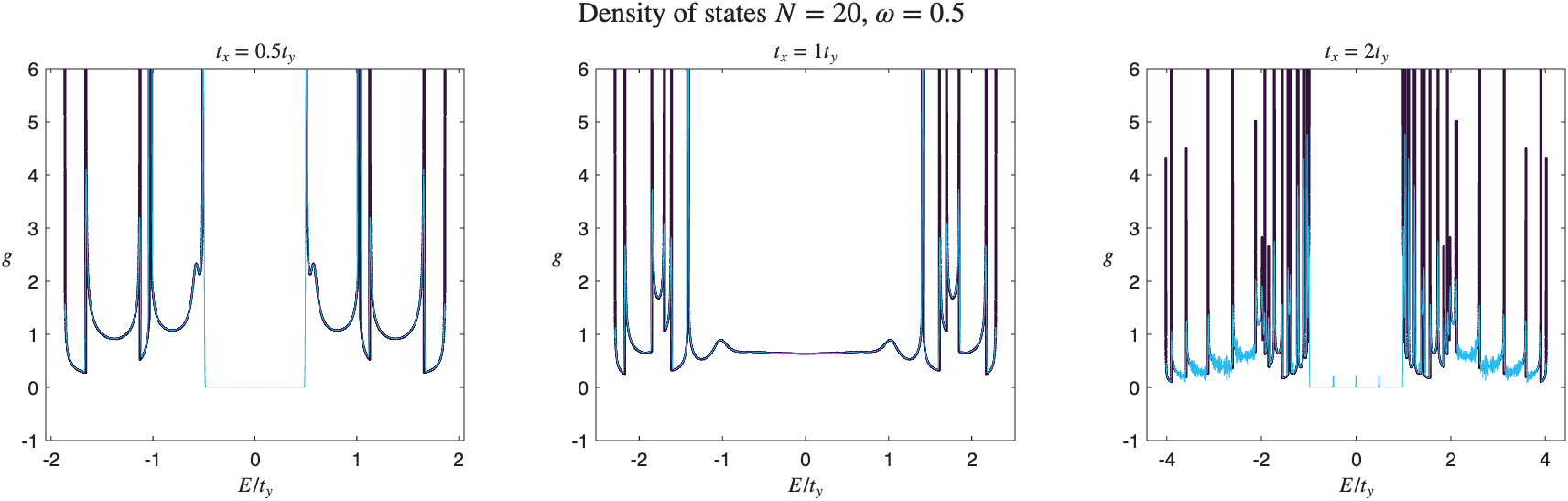}
 \captionsetup{justification=justified,singlelinecheck=false}
 \caption{Density of states for the case of even $N$. The first panel shows the situation with $t_{x}<t_{y}$, while the next two correspond to $t_{x}=t_{y}$ and $t_{x}>t_{y}$, respectively. The plots clearly indicate that the gap closes when $t_{x}=t_{y}$, whereas it remains open in the other two cases. The purple curve shows the analytical prediction, whereas the blue curve corresponds to the numerical simulation. In the simulation, we consider $K=1500$ sites along the direction parallel to the $\vec{e}$ vector defined via (\ref{SM:trans_op_m}).}
 \label{SM:fig:DOS_even_periodic}
\end{figure*}

\begin{figure*}[htbp]
 \centering
 \includegraphics[width=0.9\textwidth]{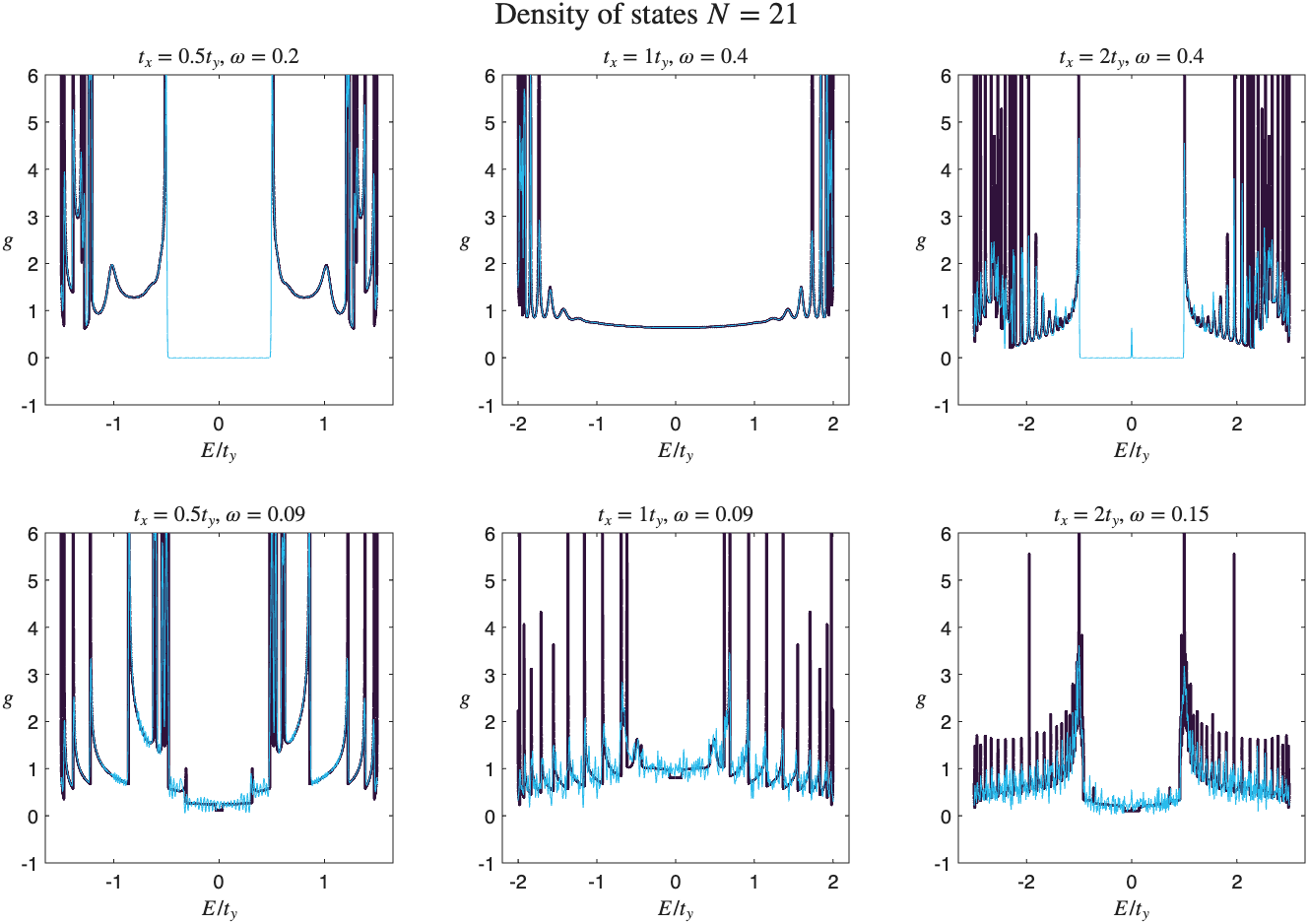}
 \captionsetup{justification=justified,singlelinecheck=false}
 \caption{Density of states for the case of odd $N$. In both rows, the three regimes are displayed in sequence: $t_{x}<t_{y}$, the critical point $t_{x}=t_{y}$, and $t_{x}>t_{y}$. The first row represents the case where the Matsubara frequency satisfies $\omega>\omega_{cr}$. In this situation, the gap closes only at the SSH critical point. The second row, in contrast, exhibits a non-zero density of states at zero energy even away from criticality. These panels correspond to $\omega$ values below the critical threshold. The purple curve shows the analytical prediction, whereas the blue curve corresponds to the numerical simulation. In the simulation, we consider $K=1500$ sites along the direction parallel to the $\vec{e}$ vector defined via (\ref{SM:trans_op_m}).}
 \label{SM:fig:DOS_odd_periodic}
\end{figure*}

 For the case of the periodic boundary conditions we find the density of states using formula (\ref{SM:DOS_deff}) and numerical approaches. The corresponding results for even $N$ are presented in Figure \ref{SM:fig:DOS_even_periodic}. For odd $N$, the gap closure becomes apparent once the Matsubara frequency is sufficiently small, as illustrated in Figure \ref{SM:fig:DOS_odd_periodic}. 
\par Our results show that number of Van Hove singularities increases for finite values of $\omega$ and $N$. This follows from formula (\ref{SM:DOS_deff}), which implies that singularities occur at the extrema of the energy bands. When $\omega$ and $N$ are non-zero, the energy bands becomes more oscillatory, leading to a significantly larger number of extrema comparable to $N$.
\par We observe that the numerically computed density of states deviates more from the analytical result when periodic boundary conditions are used. The blue curves in Figures \ref{SM:fig:DOS_even_periodic} and \ref{SM:fig:DOS_odd_periodic} display stronger oscillations. This behavior stems from the more complex oscillatory energy spectrum (\ref{SM:Energy_periodic_BC}) that appears for periodic boundary conditions. To obtain a smoother numerical density of states that more closely matches the analytical prediction, one must consider a larger lattice with $K>1500$ sites.

\section{Further results regarding the edge modes}\label{SM:app:edge_modes}

We here examine the various types of edge modes that emerge in the complex energy spectra of Hamiltonians with both open and periodic boundary conditions. The section is divided into three subsections. In the first, we study Hermitian-like edge modes, which are obtained by expanding the Hamiltonian around the point in the Brillouin zone where the energy gap closes and then introducing spatial dependence via the substitution $k \mapsto -i\partial_{x}$. We investigate these modes for both open and periodic boundary conditions. The remaining two subsections are devoted to a detailed analysis of the numerically computed complex energy spectra, with an emphasis on characterizing the different kinds of edge modes that appear.

\subsection{Hermitian-like edge modes}

When a real line gap exists in the complex eigenvalue spectrum, Hermitian-like edge modes emerge. These modes are directly linked to the topological invariant associated with the phase of the SSH function $q$, as shown in (\ref{SM:top_inv_even}) and (\ref{SM:top_inv_odd_1}). In what follows, we compute these modes explicitly, beginning with open boundary conditions and subsequently addressing the case of periodic boundary conditions.\\
\subsubsection{Open boundary conditions}
\par From a topological perspective, the modes manifest as gapless edge modes. Under open boundary conditions, the energy gap closes at the point $k = \pi$ in the Brillouin zone, just as in the SSH model. To obtain the edge modes, the translational symmetry must be broken. A common approach is to expand the Hamiltonian around the point in the Brillouin zone where the gap closes and then make the substitution $k \mapsto -i\partial_x$. The continuum Hamiltonian reads:
\begin{widetext}
 \begin{equation}
 H_{PTB}^{k\approx\pi}=\begin{pmatrix}
 -i|t_{x}-t_{y}|F_{01}\left(1-\alpha_{1}\partial_{x}^{2}\right) & (t_{x}-t_{y})\left[1+\frac{t_{x}}{t_{x}-t_{y}}\left(-\partial_{x}+\frac{1}{2}\partial_{x}^{2}\right)\right]\\
 (t_{x}-t_{y})\left[1+\frac{t_{x}}{t_{x}-t_{y}}\left(\partial_{x}+\frac{1}{2}\partial_{x}^{2}\right)\right] & -i|t_{x}-t_{y}|F_{02}\left(1-\alpha_{2}\partial_{x}^{2}\right)
 \end{pmatrix},
 \end{equation}
\end{widetext}
where $F_{01}=F_{N_{1}}(k=\pi)$ and $F_{02}=F_{N_{2}}(k=\pi)$, while $\alpha_{i}$, $i\in\{1,2\}$, are defined by:
\begin{equation}
 \alpha_{i}=\frac{t_{x}t_{y}}{2(t_{x}-t_{y})^{2}}\left[1-\frac{\omega}{\sqrt{\omega^{2}+4(t_{x}-t_{y})^{2}}}\frac{\mathrm{d}\ln{F_{N_{i}}(\phi_{0})}}{\mathrm{d}\phi}\right],\label{SM:alpha_i}
\end{equation}
where $\phi_{0}=\phi(k=\pi)=\mathrm{arcsinh}\left(\frac{\omega}{2|t_{x}-t_{y}|}\right)$. The system occupies the positive half of the $x$-axis and extends to infinity. Because we seek edge states that are localized near the boundary at $x=0$, we adopt an exponentially decaying ansatz for the wavefunction, subject to the boundary condition $\psi(0)=0$:
\begin{equation}
 \psi(x)=C_{1}\begin{pmatrix}
 a_{+}\\b_{+}
 \end{pmatrix}e^{-\lambda_{+}x}-C_{2}\begin{pmatrix}
 a_{-}\\b_{-}
 \end{pmatrix}e^{-\lambda_{-}x},\label{SM:ansatz_edge_mode}
\end{equation}
where $\lambda_{\pm}>0$, implying the normalizable solution. The boundary condition demands $\frac{b_{+}}{a_{+}}=\frac{b_{-}}{a_{-}}$ and $C_{2}=\frac{a_{+}}{a_{-}}C_{1}$. In the non-Hermitian projected problem, boundary modes need not appear as strictly zero-energy eigenstates. They can instead be pinned to $\mathrm{Re}\,E=0$ while acquiring a finite imaginary part, which encodes their finite lifetime in the reduced brane description. We therefore search for boundary solutions with eigenvalue $E=i\varepsilon$. The corresponding Schrödinger equation, together with the boundary condition, reduces to the following set of equations:
\begin{align}
 \frac{t_{x}}{t_{x}-t_{y}}\frac{\frac{\varepsilon}{|t_{x}-t_{y}|}+F_{20}}{2\alpha_{2}F_{20}}&=\frac{\lambda_{+}+\lambda_{-}-\frac{t_{x}}{t_{x}-t_{y}}\lambda_{+}\lambda_{-}}{2-\lambda_{+}-\lambda_{-}},\\
 \frac{t_{x}}{t_{x}-t_{y}}\frac{\frac{\varepsilon}{|t_{x}-t_{y}|}+F_{10}}{2\alpha_{1}F_{10}}&=-\frac{\lambda_{+}+\lambda_{-}+\frac{t_{x}}{t_{x}-t_{y}}\lambda_{+}\lambda_{-}}{2+\lambda_{+}+\lambda_{-}}.
\end{align}
This system of equations can be solved exactly, but it yields rather complicated expressions for $\varepsilon$ and $\lambda_{\pm}$, which offer little insight. In contrast, here we present a symmetric case where $N_{1}=N_{2}$. The solution for the decaying constants $\lambda_{\pm}$ is given by:
\begin{equation}
 \lambda_{\pm}=\frac{1}{\sqrt{1+\eta^{2}}}\left[1\pm\sqrt{1-2(1+\eta^{2})\frac{t_{x}-t_{y}}{t_{x}}}\right],
\end{equation}
where $\eta=-2\alpha F_{0}\frac{|t_{x}-t_{y}|}{t_{x}}$, while the lifetime of this state is given by:
\begin{equation}
 \frac{\hbar}{\tau}\equiv-\varepsilon=F_{0}|t_{x}-t_{y}|\left[1+2\alpha\frac{t_{x}-t_{y}}{t_{x}}\right]
\end{equation}
Finally, the wave function is given by the following expression:
\begin{equation}
 \psi(x)=\frac{1}{\sqrt{2}}\begin{pmatrix}
 \sqrt{\sqrt{1+\eta^{2}}+1}\\\mathrm{sgn}\{\eta\}\sqrt{\sqrt{1+\eta^{2}}-1}
 \end{pmatrix}\bigg{(}e^{-\lambda_{+}x}-e^{-\lambda_{-}x}\bigg{)}.
\end{equation}
The condition of positive $\lambda_{\pm}$ is equivalent to $\lambda_{+}\lambda_{-}>0$, implying:
\begin{equation}
 \lambda_{+}\lambda_{-}=2\frac{t_{x}-t_{y}}{t_{x}}>0.
\end{equation}
We therefore find that edge states are present when $t_{x}>t_{y}$. In this regime, there are in fact two modes sharing the same eigenvalue $i\varepsilon$. This precisely matches the prediction of the topological invariant (\ref{SM:nu_sol_even_finate_omega}) for any value of $N$. Furthermore, when $t_{x}=t_{y}$ we obtain $\lambda_{-}=0$, indicating that the modes penetrate into the bulk, analogous to the situation in Hermitian systems.
\par The most general asymmetric configuration with $N_{1}\neq N_{2}$ shows the same qualitative behavior as the symmetric one, in the sense that the condition $\lambda_{+}\lambda_{-}>0$ again leads to the requirement $t_{x}>t_{y}$ for the presence of modes, i.e., the system must lie in the SSH topological phase. In addition, the two modes no longer have degenerate eigenvalue, in contrast equations admit two distinct solutions $\varepsilon_{\pm}$. Comparing these analytical predictions with numerical results shows good agreement within a certain range of parameters, as the model exhibits a highly nontrivial dependence on $\omega$ when expanded around $k=\pi$ point in the Brillouin zone.\\
\subsubsection{Periodic boundary conditions}
The continuum Hamiltonian for the case of periodic boundary conditions is given by:
\begin{widetext}
 \begin{equation}
 H_{PTB}^{k\approx\pi}=\begin{pmatrix}
 -i|t_{x}-t_{y}|F_{0}\left(1-\alpha\partial_{x}^{2}\right) & (t_{x}-t_{y})\left(\gamma_{0}+\gamma_{1}\partial_{x}-\gamma_{2}\partial_{x}^{2}\right)\\
 (t_{x}-t_{y})\left(\gamma_{0}-\gamma_{1}\partial_{x}-\gamma_{2}\partial_{x}^{2}\right)] & -i|t_{x}-t_{y}|F_{0}\left(1-\alpha\partial_{x}^{2}\right)
 \end{pmatrix},
 \end{equation}
where $\alpha$ and $F_{0}$ are the same as for the open boundary conditions (\ref{SM:alpha_i}), while $\gamma_{i}$ are given by:
\begin{align}
 \gamma_{0}&=1+G_{0},\\
 \gamma_{1}&=\frac{t_{x}}{t_{x}-t_{y}}\bigg{(}(N+1)G_{0}-1\bigg{)},\\
 \gamma_{2}&=-\frac{t_{x}}{2(t_{x}-t_{y})}(1+G_{0})-G_{0}\left(\gamma_{G}+\frac{N(N+2)t_{x}^{2}}{2(t_{x}-t_{y})^{2}}\right),\\
 \gamma_{G}&=\frac{t_{x}t_{y}}{2(t_{x}-t_{y})^{2}}\frac{\omega}{\sqrt{\omega^{2}+4(t_{x}-t_{y})^{2}}}\frac{\mathrm{d}\ln{G_{N}(\phi_{0})}}{\mathrm{d}\phi},
\end{align}
where $G_{0}=(-\mathrm{sgn}\{t_{x}-t_{y}\})^{N}G_{N}(\phi_{0})$. Applying the same ansatz (\ref{SM:ansatz_edge_mode}) as in the case of open boundary conditions, and inserting it into the Schrödinger equation with energy $i\varepsilon$, while imposing the boundary condition $\psi(0)=0$, we obtain:
\begin{align}
 \frac{\frac{\varepsilon}{|t_{x}-t_{y}|}+F_{0}}{\alpha F_{0}}&=\frac{\gamma_{0}(\lambda_{+}+\lambda_{-})-\gamma_{1}\lambda_{+}\lambda_{-}}{\gamma_{2}(\lambda_{+}+\lambda_{-})+\gamma_{1}},\\
 \frac{\frac{\varepsilon}{|t_{x}-t_{y}|}+F_{0}}{\alpha F_{0}}&=\frac{\gamma_{0}(\lambda_{+}+\lambda_{-})+\gamma_{1}\lambda_{+}\lambda_{-}}{\gamma_{2}(\lambda_{+}+\lambda_{-})-\gamma_{1}}
\end{align}
\end{widetext}
The solution for the decaying rates $\lambda_{\pm}$ is as follows:
\begin{equation}
\lambda_{\pm}^{(\pm')}=\pm'\left[\frac{\gamma_{1}}{2\sqrt{\alpha^{2}F_{0}^{2}+\gamma_{2}^{2}}}\pm\sqrt{\frac{\gamma_{1}^{2}}{4\left(\alpha^{2}F_{0}^{2}+\gamma_{2}^{2}\right)}+\frac{\gamma_{0}}{\gamma_{2}}}\right].
\end{equation}
Notice that there are four solutions. We should choose the sign $\pm'$ such that both $\lambda_{\pm}$ have a positive real part. This leaves only one set of $\lambda_{\pm}$ as expected. The eigenvalues are now given by:
\begin{equation}
 \frac{\hbar}{\tau}\equiv-\varepsilon=F_{0}|t_{x}-t_{y}|\left(1-\alpha\frac{\gamma_{0}}{\gamma_{2}}\right).
\end{equation}
Since $G_{0}$ is imaginary when $N$ is odd, this case must be treated separately. We begin by considering the situation where $N$ is even. In the topological phase we have $\lambda_{+}\lambda_{-}=-\frac{\gamma_{0}}{\gamma_{2}}>0$, which leads to the following condition for the existence of the edge modes:
\begin{equation}
 \frac{t_{x}-t_{y}}{1+\frac{G_{0}}{1+G_{0}}\frac{1}{t_{x}-t_{y}}\left[t_{x}N(N+2)+\frac{\omega t_{y}\frac{\mathrm{d}\ln{G_{0}}}{\mathrm{d}\phi}}{\sqrt{\omega^{2}+4(t_{x}-t_{y})^{2}}}\right]}>0.
\end{equation}
This requirement is met exclusively when $t_{x}>t_{y}$, meaning that the system resides in the topological phase. Thus, we again find that SSH-like edge modes emerge, in agreement with the prediction of the topological invariant (\ref{SM:nu_sol_even_finate_omega}). Finally, we mention for completeness that for an $N$ odd, $G_{0}$ becomes imaginary, resulting in the gapped edge modes.


\subsection{Complex energy spectrum of the OBC Hamiltonian}

\par
\begin{figure*}[htbp]
 \centering
 \includegraphics[width=0.97\textwidth]{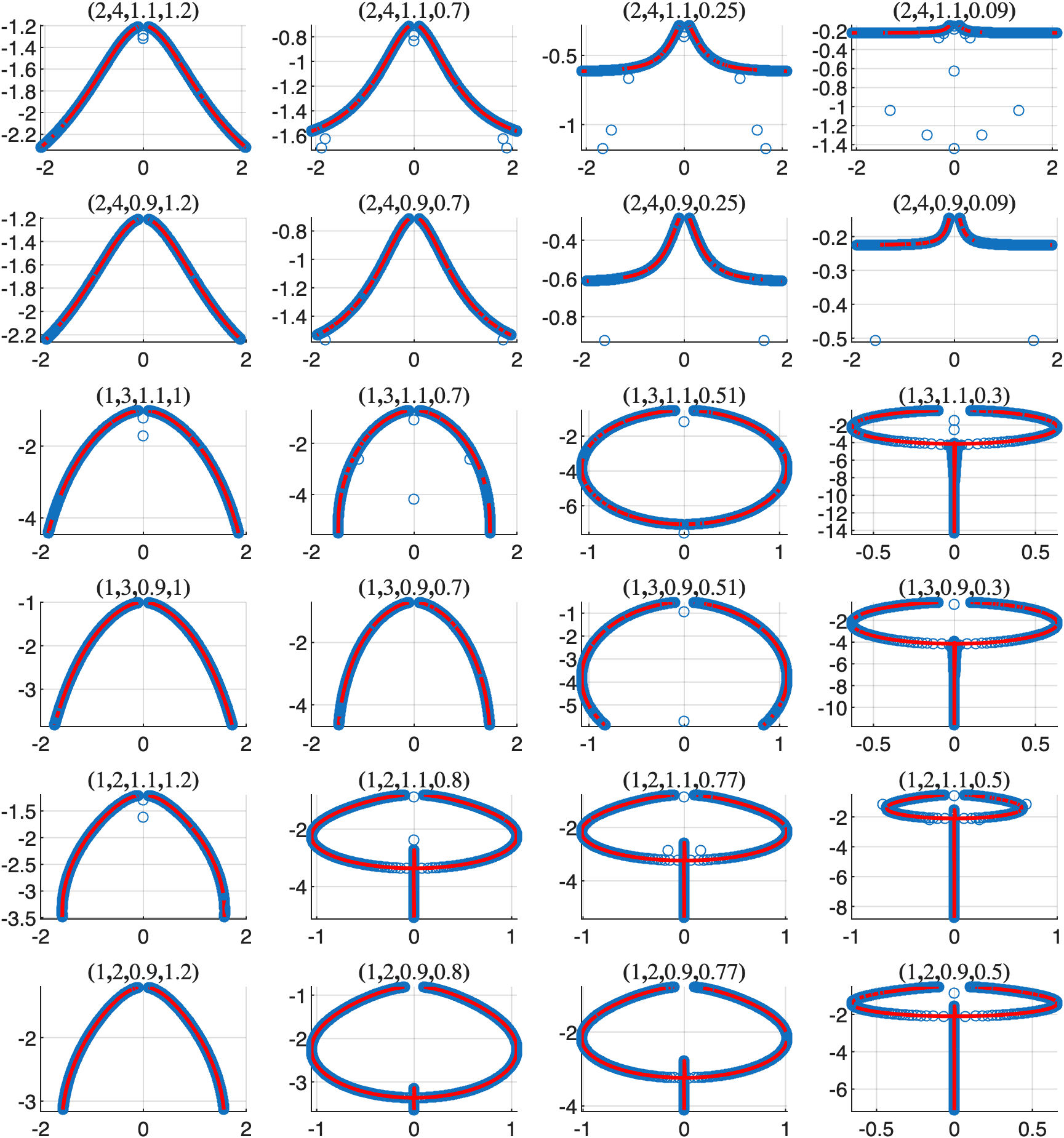}
 \captionsetup{justification=justified,singlelinecheck=false}
 \caption{Complex spectra of the projected Hamiltonian with open
 boundary conditions. Each panel is labeled by
 $(N_{1},N_{2},t_{x}/t_{y},\omega/t_{y})$. The rows compare even,
 odd, and mixed parities of the two eliminated complements, while
 the columns show the evolution as $\omega$ decreases. Red curves
 denote the analytical bulk spectrum from Eq.~(\ref{SM:Energy_open_BC}),
 and blue circles denote numerical eigenvalues of the finite
 projected chain. Detached blue points identify boundary-localized
 modes. Some are the SSH-like modes associated with the {SSH-type
 edge correspondence}, while others originate from the
 finite-frequency truncation of the projection-induced long-range
 hopping.}
\label{SM:fig:Complex_energy_spectrum_OBC}
\end{figure*}

 The complex energy spectrum of the OBC Hamiltonian is depicted in Fig.~\ref{SM:fig:Complex_energy_spectrum_OBC}. We begin by examining the case where both $N_{1}$ and $N_{2}$ are even, for the specific choice $N_{1} = 2$ and $N_{2} = 4$. This particular choice does not entail any loss of generality. The first row depicts how the system behaves in the topologically nontrivial phase as $\omega$ is progressively decreased (from left to right), whereas the second row displays the evolution of the complex energy spectrum in the topologically trivial phase. In the rightmost panel, for the largest value of $\omega$, the system is in the nontrivial phase and exhibits two gapless edge modes, each localized at one end of the system. In the second panel, for both the trivial and nontrivial phases, additional gapped edge modes appear. These modes arise due to the specific truncation of the Hamiltonian’s long-range hopping and therefore do not play any role in the topological characterization of the system. Further decreasing $\omega$ leads to an even larger number of gapped modes, as seen in the third panel. Some of these gapped modes turn gapless in the last panel, corresponding to the smallest value of $\omega$ shown. Importantly, the topological gapless edge modes persist for all values of $\omega$ only in the topological phase. We therefore conclude that the bulk-boundary correspondence is satisfied when both $N_{1}$ and $N_{2}$ are even. This agrees with the perturbative analysis of the \(\eta\) parameter for asymmetric complements.
\par Next, we analyze the situation where both $N_{1}$ and $N_{2}$ are odd, choosing $N_{1}=1$ and $N_{2}=3$ as a representative example. We observe the same qualitative behavior as in the case where $N_{1}$ and $N_{2}$ are even. At first, there are only two gapless edge modes, each localized at a different boundary of the system. As $\omega$ is reduced, additional gapped modes associated with truncation emerge. Once the $\mathcal{PT}_{+}$ symmetry of the states is broken, and $\omega$ is lowered further, the bulk-boundary correspondence fails: gapless edge modes now also appear in the topologically trivial phase. This behavior is consistent with our perturbative analysis of the model. However, the two modes appearing in the topologically nontrivial phase are localized on different ends of the system which is not predicted by the $\eta$ parameter analysis. This may be due to the breakdown of the perturbative approach when $\eta\approx1$. On the other hand, by explicitly plotting the trivial-phase modes, we find that for $N_{1}<N_{2}$ they localize on the left boundary of the system, whereas for $N_{1}>N_{2}$ they localize on the right boundary. This precisely matches the behavior predicted in our discussion of the $\eta$ parameter.
\par We finally consider the mixed odd--even configuration, with $N_{1}=1$ and $N_{2}=2$. Initially, the system hosts two standard, gapless edge modes. Because $N_{1}$ is odd, lowering $\omega$ eventually causes the $\mathcal{PT}_{+}$ symmetry of the states to break at one point. Once this occurs, the mode localized at the left end of the system hybridizes with another state from the flat band, and together they form a gapped edge mode. As the Matsubara frequency is reduced further, this composite mode gradually merges into the bulk states. In the final panel, only one mode remains, localized at the right boundary of the system. If we continue to decrease $\omega$ while the system is in the topologically trivial phase, we observe the reverse sequence of gap closings. Specifically, when $\omega$ drops below a certain threshold, a gapped edge mode appears, and as $\omega$ is lowered further, this mode de-hybridizes, giving rise to a single gapless edge mode localized at the left boundary. If we instead exchange $N_{1}$ and $N_{2}$ so that $N_{2}$ is odd, the same sequence unfolds, except that the mode persisting at small $\omega$ is now localized at the left boundary of the system. This result aligns with the analysis of the $\eta$ parameter. In this situation, $\eta$ is sufficiently far from 1, so the perturbative method remains applicable.
\par The general trend of the complex energy spectrum when $N_{1}$ and $N_{2}$ are even is that the states cluster along the real axis as $\omega$ is decreased. This is consistent with the $\omega\to0$ limit discussed in Sec.~V.A of the main text. In this limit the system tends to the Hermitian SSH model. The only artifact of the intricate NH structure of the system is left with the scattered edge modes with large energy imaginary values. In contrast, when $N_{1}$ and $N_{2}$ are odd, is is clear that the whole spectrum becomes purely imaginary as $\omega\to0$, as predicted by the exact solution featuring flat energy bands for small $\omega$.
\subsection{Complex energy spectrum of the PBC Hamiltonian}

\par
\begin{figure*}[htbp]
 \centering
 \includegraphics[width=0.95\textwidth]{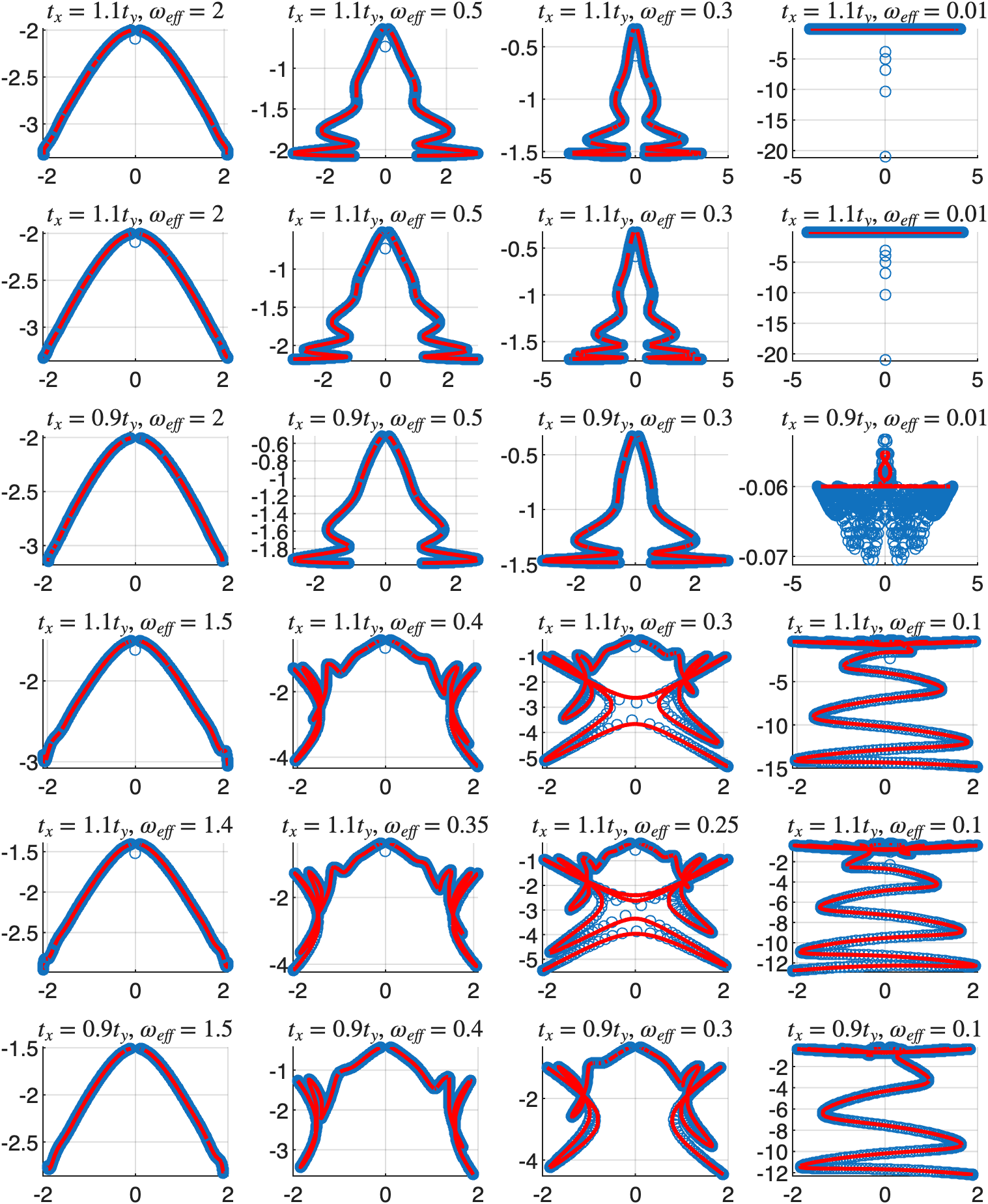}
 \captionsetup{justification=justified,singlelinecheck=false}
 \caption{Complex spectra of the projected Hamiltonian with
 periodic boundary conditions in the eliminated direction and open
 boundaries along the brane. Each panel is labeled by
 $(N,t_{x}/t_{y},\omega/t_{y})$. The first three rows correspond to
 even $N$, and the last three rows to odd $N$. Red curves denote
 the analytical bulk spectrum from Eq.~(\ref{SM:Energy_periodic_BC}),
 while blue circles denote finite-chain eigenvalues. The figure
 illustrates the contrast between SSH-like boundary modes in the
 even-$N$ sectors and the refined, termination-sensitive boundary
 response of the odd-$N$ crystalline sector.}
 \label{SM:fig:Complex_energy_spectrum_PBC}
\end{figure*}

 The key parameter that determines the edge modes for the Hamiltonian with periodic boundary conditions is the integer $N$. We first consider the situation for even \(N\). The complex energy spectrum of the PBC Hamiltonian is shown in (\ref{SM:fig:Complex_energy_spectrum_PBC}). In the first and third rows we take $N=10$, whereas in the second row we set $N=12$. These two choices exemplify two distinct cases associated with the parity of $N/2$. This distinction matters because the functions $G$ (\ref{SM:G_N_functions}) change sign depending on whether $N/2$ is even or odd. This behavior manifests itself in the orientation of the energy bands: when we follow the eigenvalues in the direction of increasing imaginary part, we observe that for even $N/2$ the bands initially bend inward toward the imaginary axis, whereas for odd $N/2$ they first bend outward.

\par In the first two rows, the system is in a topologically nontrivial phase and supports two gapless edge modes with degenerate eigenvalues, localized on the two boundaries of the system, as clearly seen in the figure. As $\omega$ is reduced, gaps form, yet these modes persist. When $\omega$ becomes sufficiently small (rightmost panel in each row), additional modes appear. The total number of modes in this limit is exactly $N$. In contrast, the third row illustrates the topologically trivial phase, in which no such modes are present. This behavior agrees with the predictions of the bulk topological invariants in (\ref{SM:nu_sol_even_finate_omega}) for finite $\omega$ and (\ref{SM:nu_sol_even_zero_omega_periodic_BC}) for vanishing for $\omega$. Notice that in the last panel of the third row, it appears that many states do not follow the analytical bands. On closer inspection, we can see that they are all concentrated around the same value of the imaginary part of the complex energy. Therefore, this behavior is merely an artifact of the numerical calculation performed on a finite lattice.

\par The last three rows illustrate the intricate structure of the complex energy spectrum of the Hamiltonian under periodic boundary conditions with odd $N$. A detailed analysis of this behavior for $N=3$ is presented in Appendix D.3. of the main text. Here, we focus on the emergence of edge modes in two different cases, distinguished by the parity of $(N-1)/2$. This distinction is crucial because the functions $G$ for odd $N$ in (\ref{SM:G_N_functions}) change sign when this parity is flipped. In the spectrum, this fact is reflect by the fact that the bands move in the opposite directions as imaginary part is increased. 
\par In the topologically nontrivial phase, the system supports two degenerate gapped modes that disappear when the system enters the trivial phase. These modes are initially localized near the edge, but as $\omega$ is reduced they become localized in the bulk. They should therefore not be interpreted as protected edge modes in the conventional bulk--boundary sense. Instead, they provide a finite-size manifestation of the parity-quantized bulk invariant, consistent with a refined, symmetry-sensitive boundary response in the odd-$N$ periodic sector.

\begin{figure*}[htbp]
 \centering
 \includegraphics[width=0.82\textwidth]{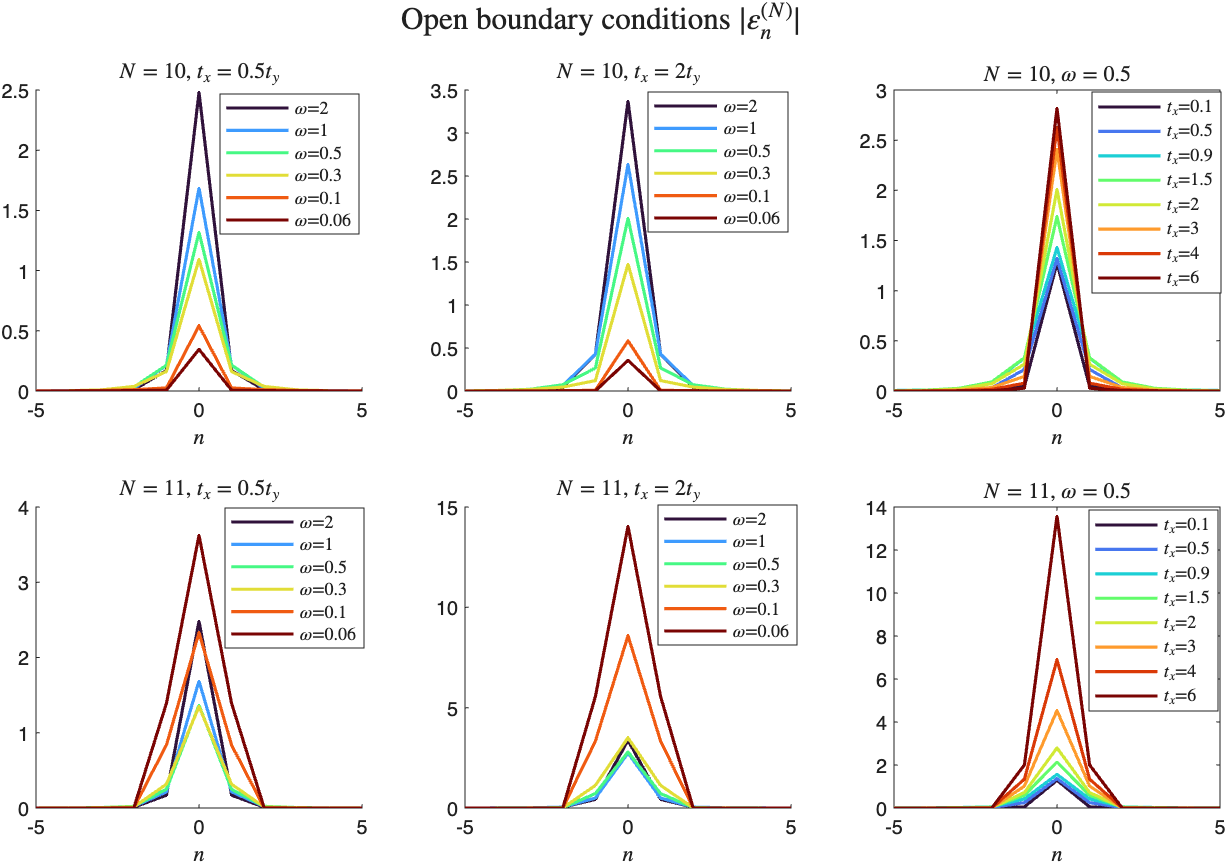}
 \captionsetup{justification=justified,singlelinecheck=false}
 \caption{Dependence of the hopping parameter $\varepsilon_{n}^{(N)}$ (\ref{SM:eps_n-n'_even}-\ref{SM:eps_n-n'_odd}) on the inter-site distance $n$. This parameter appears in both the PBC Hamiltonian (\ref{SM:H_PBC_coord_rep}) and the OBC Hamiltonian (\ref{SM:H_OBC_coord_rep}). It reaches its maximum at $n=0$ and then decays exponentially as $n$ tends to infinity. The first row shows the case of even $N$, while the second row shows the case of odd $N$. In each row, the first two panels correspond, respectively, to the two distinct SSH phases $t_{x}<t_{y}$ and $t_{x}>t_{y}$, with different colors indicating decreasing values of $\omega$. The third panel in each row illustrates how the hopping parameter evolves as $t_{x}$ is increased. A pronounced difference emerges between the even and odd $N$ cases as $\omega$ is varied: for even $N$, the peak at $n=0$ grows as $\omega$ is increased, whereas for odd $N$ it diminishes. In contrast, when $t_{x}$ is increased, both even and odd $N$ exhibit the same trend of an increasing peak. These behaviors can be verified from the exact analytical expressions for the hopping parameters.}
 \label{SM:fig:Hoping_eps_n}
\end{figure*}
\begin{figure*}[htbp]
 \centering
 \includegraphics[width=0.82\textwidth]{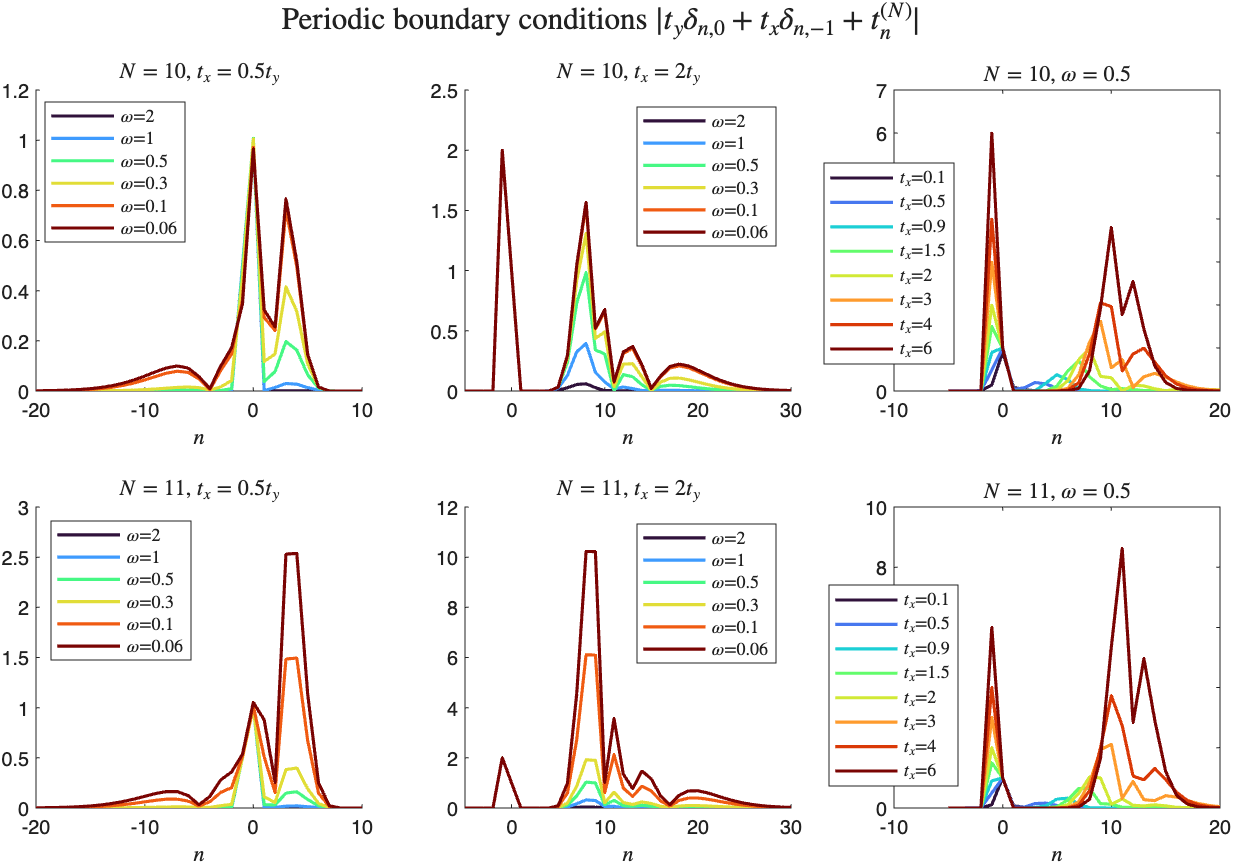}
 \captionsetup{justification=justified,singlelinecheck=false}
 \caption{Dependence of the PBC Hamiltonian (\ref{SM:H_PBC_coord_rep}) hopping parameter $t_{n}^{(N)}$ (\ref{SM:tau_n-n'_even}-\ref{SM:tau_n-n'_odd}) on the inter-site distance $n$. The arrangement of the panels is identical to that used for $\varepsilon_{n}^{(N)}$ in Figure \ref{SM:fig:Hoping_eps_n}. We can clearly see that there are two prominent peaks in these plots. One peak corresponds to the SSH nearest-neighbor hopping amplitude, while the other appears at a certain finite inter-site distance. Examining the rightmost panels in each row, we observe that increasing $t_{x}$ enhances the SSH peak, as it scales directly with $t_{x}$. The second peak also grows and shifts toward larger $n$, ultimately saturating at a finite $n$ as $t_{x}\to\infty$. In contrast, when $\omega$ is decreased, only the second peak increases, regardless of whether $N$ is even or odd, whereas the SSH peak remains unchanged. The key difference between the even- and odd-$N$ cases is that, for even $N$, the new peak never exceeds the SSH peak, while for odd $N$ there exists a critical value of $\omega$ at which the two peaks coincide, below which the new peak becomes dominant.\vspace{2cm}}
 \label{SM:fig:Hoping_t_n}
\end{figure*}

\end{document}